\documentclass[11pt,a4paper]{article}

\pdfoutput=1 

\usepackage{jcappub} 

\usepackage[T1]{fontenc} 

\usepackage[varg]{txfonts}
\usepackage{graphicx}
\usepackage{breakurl}
\usepackage[disable,colorinlistoftodos]{todonotes}
\usepackage{siunitx}
 
\newcommand*\degr{\ensuremath{^\circ}}
\newcommand*\arcmin{\ensuremath{^\prime}}
\newcommand*\tens[1]{\ensuremath{\mathsf{#1}}}
\newcommand*\arcsec{\ensuremath{^{\prime\prime}}}

\newcommand{\newc}[1]{{\textcolor[rgb]{0, 0, 0}{ #1}}}

\newcommand{\newb}[1]{{\textcolor[rgb]{0, 0, 0}{ #1}}}
\newcommand{\new}[1]{{\textcolor[rgb]{0, 0, 0}{ #1}}}

\newcommand{\Strip}{Strip}

\def\NHUNIT{\ifmmode {\rm \,cm^{-2}} \else $\rm \,cm^{-2}$ \fi} 

\def\muKcmb{\ifmmode \,\mu$K$_{\rm CMB}$\else \,$\mu$K$_{\rm CMB}$\fi}
\newcommand{\OmegaM}{\ifmmode\Omega_{\rm M}\else $\Omega_{\rm M}$\fi}

\providecommand{\Planck}{\textit{Planck}}

\providecommand{\text}[1]{\rm{#1}}

\providecommand{\muK}{\mu\rm{K}}

\newcommand{\begm}{\begin{pmatrix}}
\newcommand{\enm}{\end{pmatrix}}

\def\pmb#1{\setbox0=\hbox{#1}%
    \kern-.025em\copy0\kern-\wd0
    \kern.05em\copy0\kern-\wd0
    \kern-.025em\raise.0433em\box0}

\def\p2Y{\;_2Y}
\def\m2Y{\;_{-2}Y}
\def\beglet{
  \addtocounter{equation}{1}%
  \setcounter{parentequation}{\value{equation}}%
  \setcounter{equation}{0}%
  \def\theequation{\arabic{parentequation}\alph{equation}}%
  \ignorespaces
}
\def\endlet{
  \setcounter{equation}{\value{parentequation}}%
  \def\theequation{\arabic{equation}}%
}
\providecommand{\beglet}{\begin{subequations}}
\providecommand{\endlet}{\end{subequations}}

\newcommand{\mksym}[1]{\ifmmode {\rm #1}\else #1\fi}

\providecommand{\text}[1]{\rm{#1}}

\providecommand{\muK}{\mu\rm{K}}

\providecommand{\healpix}{\texttt{HEALPix}}

\newcommand\ba{\begin{eqnarray}}
\newcommand\ea{\end{eqnarray}}
\newcommand\bea{\begin{eqnarray}}
\newcommand\eea{\end{eqnarray}}

\newcommand\be{\begin{equation}}
\newcommand\ee{\end{equation}}

\newcommand{\degree}{\ensuremath{^\circ}}

\newcommand{\Nside}{\ensuremath{N_{\mathrm{side}}}}

\def\setsymbol#1#2{\expandafter\def\csname #1\endcsname{#2}}
\def\getsymbol#1{\csname #1\endcsname}

\def\Planck{\textit{Planck}}

\newbox\tablebox    \newdimen\tablewidth
\def\leaderfil{\leaders\hbox to 5pt{\hss.\hss}\hfil}
\def\tablenote#1 #2\par{\begingroup \parindent=0.8em
    \abovedisplayshortskip=0pt\belowdisplayshortskip=0pt
    \noindent
    $$\hss\vbox{\hsize\tablewidth \hangindent=\parindent \hangafter=1 \noindent
    \hbox to \parindent{$^#1$\hss}\strut#2\strut\par}\hss$$
    \endgroup}

\def\L2{\ifmmode L_2\else $L_2$\fi}

\def\DeltaT{\ifmmode \Delta T\else $\Delta T$\fi}
\def\deltat{\ifmmode \Delta t\else $\Delta t$\fi}
\def\fknee{\ifmmode f_{\rm knee}\else $f_{\rm knee}$\fi}
\def\Fmax{\ifmmode F_{\rm max}\else $F_{\rm max}$\fi}
\def\solar{\ifmmode{\rm M}_{\mathord\odot}\else${\rm M}_{\mathord\odot}$\fi}
\def\Msolar{\ifmmode{\rm M}_{\mathord\odot}\else${\rm M}_{\mathord\odot}$\fi}
\def\Lsolar{\ifmmode{\rm L}_{\mathord\odot}\else${\rm L}_{\mathord\odot}$\fi}
\def\inv{\ifmmode^{-1}\else$^{-1}$\fi}
\def\mo{\ifmmode^{-1}\else$^{-1}$\fi}
\def\sup#1{\ifmmode ^{\rm #1}\else $^{\rm #1}$\fi}
\def\expo#1{\ifmmode \times 10^{#1}\else $\times 10^{#1}$\fi}
\def\,{\thinspace}
\def\lsim{\mathrel{\raise .4ex\hbox{\rlap{$<$}\lower 1.2ex\hbox{$\sim$}}}}
\def\gsim{\mathrel{\raise .4ex\hbox{\rlap{$>$}\lower 1.2ex\hbox{$\sim$}}}}

\def\simprop{\mathrel{\raise .4ex\hbox{\rlap{$\propto$}\lower 1.2ex\hbox{$\sim$}}}}
\def\deg{\ifmmode^\circ\else$^\circ$\fi}
\def\pdeg{\ifmmode $\setbox0=\hbox{$^{\circ}$}\rlap{\hskip.11\wd0 .}$^{\circ}
          \else \setbox0=\hbox{$^{\circ}$}\rlap{\hskip.11\wd0 .}$^{\circ}$\fi}
\def\arcs{\ifmmode {^{\scriptstyle\prime\prime}}
          \else $^{\scriptstyle\prime\prime}$\fi}
\def\arcm{\ifmmode {^{\scriptstyle\prime}}
          \else $^{\scriptstyle\prime}$\fi}
\newdimen\sa  \newdimen\sb
\def\parcs{\sa=.07em \sb=.03em
     \ifmmode \hbox{\rlap{.}}^{\scriptstyle\prime\kern -\sb\prime}\hbox{\kern -\sa}
     \else \rlap{.}$^{\scriptstyle\prime\kern -\sb\prime}$\kern -\sa\fi}
\def\parcm{\sa=.08em \sb=.03em
     \ifmmode \hbox{\rlap{.}\kern\sa}^{\scriptstyle\prime}\hbox{\kern-\sb}
     \else \rlap{.}\kern\sa$^{\scriptstyle\prime}$\kern-\sb\fi}
\def\ra[#1 #2 #3.#4]{#1\sup{h}#2\sup{m}#3\sup{s}\llap.#4}
\def\dec[#1 #2 #3.#4]{#1\deg#2\arcm#3\arcs\llap.#4}
\def\deco[#1 #2 #3]{#1\deg#2\arcm#3\arcs}
\def\rra[#1 #2]{#1\sup{h}#2\sup{m}}

\def\dots{\relax\ifmmode \ldots\else $\ldots$\fi}
\def\WHzsr{\ifmmode $W\,Hz\mo\,sr\mo$\else W\,Hz\mo\,sr\mo\fi}
\def\mHz{\ifmmode $\,mHz$\else \,mHz\fi}
\def\GHz{\ifmmode $\,GHz$\else \,GHz\fi}
\def\mKs{\ifmmode $\,mK\,s$^{1/2}\else \,mK\,s$^{1/2}$\fi}
\def\muKs{\ifmmode \,\mu$K\,s$^{1/2}\else \,$\mu$K\,s$^{1/2}$\fi}
\def\muKRJs{\ifmmode \,\mu$K$_{\rm RJ}$\,s$^{1/2}\else \,$\mu$K$_{\rm RJ}$\,s$^{1/2}$\fi}
\def\muKHz{\ifmmode \,\mu$K\,Hz$^{-1/2}\else \,$\mu$K\,Hz$^{-1/2}$\fi}
\def\MJysr{\ifmmode \,$MJy\,sr\mo$\else \,MJy\,sr\mo\fi}
\def\MJysrmK{\ifmmode \,$MJy\,sr\mo$\,mK$_{\rm CMB}\mo\else \,MJy\,sr\mo\,mK$_{\rm CMB}\mo$\fi}
\def\microns{\ifmmode \,\mu$m$\else \,$\mu$m\fi}

\def\muK{\ifmmode \,\mu$K$\else \,$\mu$\hbox{K}\fi}
\def\microK{\ifmmode \,\mu$K$\else \,$\mu$\hbox{K}\fi}
\def\muW{\ifmmode \,\mu$W$\else \,$\mu$\hbox{W}\fi}
\def\kms{\ifmmode $\,km\,s$^{-1}\else \,km\,s$^{-1}$\fi}
\def\kmsMpc{\ifmmode $\,\kms\,Mpc\mo$\else \,\kms\,Mpc\mo\fi}
\providecommand{\sorthelp}[1]{}

\begin{document}

\title{The large scale polarization explorer (LSPE) for CMB measurements: performance forecast}
\author[a]{G.~Addamo,}
\author[b]{ P.~A.~R.~Ade,}
\author[c]{C.~Baccigalupi,}
\author[d]{A.~M.~Baldini,}
\author[e]{P.~M.~Battaglia,} 
\author[f,g]{E.~S.~Battistelli,}
\author[h]{A.~Ba\`{u},}
\author[f,g]{ P.~de~Bernardis,}
\author[i,j]{ M.~Bersanelli,}
\author[k,l]{ M.~Biasotti,}
% ...

\affiliation[a]{CNR--IEIIT, Corso Duca degli Abruzzi, 24, 10129 Torino TO, Italy} 
\affiliation[b]{School of Physics and Astronomy, Cardiff University, Queens Buildings, The Parade, Cardiff, CF24 3AA, U.K.} 
\affiliation[c]{SISSA, Astrophysics Sector, via Bonomea 265, 34136, Trieste, Italy}
\affiliation[d]{INFN--Sezione di Pisa, Largo B.~Pontecorvo~3, 56127 Pisa (Italy)}
\affiliation[e]{INAF/IASF Milano, Via E. Bassini 15, Milano, Italy}
\affiliation[f]{Dipartimento di Fisica, Sapienza Universit\`a di Roma, P.le A. Moro 5, 00185, Roma, Italy}
\affiliation[g]{INFN--Sezione di Roma1, P.le A. Moro 5, 00185, Roma, Italy} 
\affiliation[h]{Universit\`{a} degli studi di Milano-Bicocca, Dipartimento di Fisica, Piazza delle Scienze 3, 20126 Milano, Italy}
\affiliation[i]{Dipartimento di Fisica, Universit\`{a} degli Studi di Milano, Via Celoria, 16, Milano, Italy}
\affiliation[j]{INFN--Sezione di Milano, Via Celoria 16, Milano, Italy}
\affiliation[k]{Dipartimento di Fisica - Universit\`{a} di Genova, Via Dodecaneso, 33, 16146 Genova GE, Italy}
\affiliation[l]{INFN--Sezione di Genova, Via Dodecaneso, 33, 16146 Genova GE, Italy}
% ...

\emailAdd{francesco.piacentini@roma1.infn.it}
\date{}
\collaboration{The LSPE collaboration}

\arxivnumber{2008.11049}

\abstract{
The measurement of the polarization of the 
Cosmic Microwave Background (CMB) radiation is one 
of the current frontiers in cosmology. In particular, 
the detection of the primordial divergence-free component of the 
polarization field, the B-mode, could reveal the presence of
gravitational waves in the early Universe. 
The detection of such \newc{a} component is at the moment the 
most promising technique to probe the inflationary theory 
describing the very early evolution of the Universe.
We present the updated performance forecast of the Large Scale Polarization Explorer (LSPE),
a program dedicated to the measurement of the CMB polarization. 
LSPE is composed of two instruments: LSPE-\Strip{}, a radiometer-based telescope on the ground in Tenerife-Teide observatory, 
and LSPE-SWIPE (Short-Wavelength Instrument for 
the Polarization Explorer) a bolometer-based instrument designed to fly on a winter arctic stratospheric 
long-duration balloon.
The program is among the few dedicated to observation 
of the Northern Hemisphere, while most of 
the international effort is focused into ground-based 
observation in the Southern Hemisphere.  
Measurements are currently scheduled in Winter \newb{2022/23} for 
LSPE-SWIPE, with a flight duration up to 15\,days, and in Summer \newb{2022} with two years observations for LSPE-\Strip{}. 
We describe the main features of the two instruments, identifying 
the most critical aspects of the design, in terms of impact \newc{on the} 
performance forecast. We estimate the expected sensitivity of
each instrument and propagate their combined observing power to the sensitivity to cosmological 
parameters, including the effect of scanning strategy, component separation, 
residual foregrounds and partial sky coverage. We also set requirements on the control of
the most critical systematic effects
and describe techniques to mitigate 
their impact. 
LSPE will reach a sensitivity in tensor-to-scalar ratio of $\sigma_r<0.01$, \new{set an upper limit $r<0.015$ at 95\% confidence level,}
and improve constraints on other cosmological parameters. 
}


\maketitle
\flushbottom


\section{Introduction}

The Large Scale Polarization Explorer (LSPE) is designed to measure the polarization of the 
Cosmic Microwave Background (CMB) at large
angular scales, and in particular to constrain the curl component of CMB polarization (B-mode).
This is produced by tensor perturbations generated during cosmic inflation in the very early Universe~\cite{PhysRevLett.78.2054, doi:10.1146/annurev-astro-081915-023433}.
The level of this signal is unknown: current inflation models are unable to provide a firm reference
value. However, the detection of this signal would be of utmost importance, providing a way to
measure the energy-scale of inflation and a window on the physics at extremely high energies.
While the level of CMB temperature anisotropy is of the order of 
\SI{100}{\micro\kelvin} r.m.s. and the level of the gradient
component of CMB polarization (E-mode generated by scalar - density perturbations) is of the order
of \SI{3}{\micro\kelvin}, the current upper limits for the level of B-mode polarization are a fraction of $\SI{}{\micro\kelvin}$,
corresponding to a ratio between the amplitude of tensor perturbations and the amplitude of scalar
perturbations \newb{(tensor-to-scalar ratio)
$r<0.044$ at 95\% confidence level, combining data from the Planck satellite and
the BICEP/Keck ground telescopes~\citep{planck2016-l01,2021tristram,bicep2018}}.
The B-mode of inflationary origin is observable at large angular 
scales, greater than 1.5\deg.

The main scientific target of LSPE is to improve this limit.
This and the additional scientific targets of the mission are
reported in the following list:
\begin{itemize}
\item a detection of B-mode of CMB polarization 
at a level corresponding to a tensor-to-scalar 
ratio $r=0.03$ with 99.7\% confidence level (CL);
\newb{or} an upper limit to tensor-to-scalar ratio \new{$r=0.015$ at 95\% CL}; 
\item an improved measurement of the optical depth to the 
cosmic microwave background $\tau$, measured from the large scale
E-mode CMB polarization; \newb{a measurement of 
$\tau$ is also critical to constraints on the sum of neutrino masses, from 
large-scale-structure probes~\cite{2010_arcidiacono,2015_allison}}; 
\item investigation of the so called
{\em low-$\ell$ \newb{anomalies}}, a series of anomalies observed in the
large angular scales of the CMB polarization, including
lack of power \newb{on the largest angular scales}, asymmetries and alignment of multipole moments; 
\item wide maps of foreground polarization produced in our galaxy
by synchrotron emission and interstellar dust emission, which will be important to \newb{mapping} the magnetic
field in our Galaxy and to \newc{studying} the properties of the ionized gas and of the diffuse interstellar dust
in the Milky Way; 
\item improved limits or detection of cosmic birefringence;
\item \new{an improved measurement of} the quality of the atmosphere at Teide Observatory (Tenerife) for CMB polarization measurements. \end{itemize}

The observational cosmology community is carrying on 
a global effort to improve the measurement of the CMB polarization, aiming 
\newc{at} a detection, 
or an improved upper limit, on the 
tensor-to-scalar ratio $r$.  
A list of \newb{the main} experiments 
observing CMB polarization at large scales includes\footnote{\newb{For a complete list and 
data, see \url{https://lambda.gsfc.nasa.gov/product/expt/}.}}:
the BICEP/Keck \newc{array program~\cite{2016SPIE.9914E..0SG,2020_moncelsi} deployed} at South Pole, 
aiming at improving the current upper limit 
\newb{(multipole range $21<\ell<335$)};
CLASS~\cite{2016SPIE.9914E..1KH}, in operation in \newc{the} Atacama, aiming at detecting $r=0.01$
\newb{($2<\ell<200$)};
Polarbear-2/Simons Array~\cite{2016JLTP..184..805S}, \newc{beginning operations in the} Atacama, aiming at 
$\sigma(r) = 0.006$ if $r=0.1$
\newb{($30<\ell<3000$)};
\newb{SPT-Pol~\cite{2020PhRvD.101l2003S}, operated at \newc{the} South Pole, measured $r<0.44$ at 95\% c.l. ($52<\ell<2301$)}
\newc{and the third SPT generation SPT-3G~\cite{2014_spt-3g}, aiming at 
$\sigma(r)=0.01$ ($50<\ell<11000$)};
\newc{ACT~\cite{2020_aiola_act,2020_choi_actC}, operated in \newc{the} Atacama, providing relevant constraints at smaller angular scales ($225<\ell<8725$);} 
Simons Observatory~\cite{2019JCAP_SO}, in preparation in \newc{the} Atacama for early 2020s, aiming at $\sigma(r) = 0.003$
\newb{($30<\ell<8000$)};
\newb{GroundBIRD~\cite{groundbird}, in preparation in \newc{the} Tenerife-Teide observatory, aiming at $\sigma(r) \simeq 0.01$
($6<\ell<300$);}
QUBIC~\cite{2020arXiv201102213H}, in preparation for installation in Alto Chorrillos (Argentina, altitude 4869\,m a.s.l), aiming at $\sigma(r)=0.021$
\newb{($30<\ell<200$)}; 
CMB-S4~\cite{2020arXiv200812619T}, in preparation for ground-based observations in 2027, aiming at detecting $r>0.003$ at greater than 5$\sigma$, or $r<0.001$ at 95\% c.l.;
SPIDER~\cite{2018JLTP..193.1112G}, balloon-based, waiting for the second flight, aiming at detecting $r>0.03$ at 99.7\% c.l. 
\newb{($2<\ell<200$)};
PIPER~\cite{2016SPIE.9914E..1JG}, balloon-based, aiming at constraining $r<0.007$ after 8 flights; 
\newb{PICO~\cite{2019PICO}, a satellite-based instrument currently in study phase 
aiming at detection of $r=5\times 10^{-4}$ at 5$\sigma$ c.l. (full sky);}
and LiteBIRD~\cite{2019JLTP..194..443H,2020LiteBIRD-JLTP}, which is currently the only approved satellite-based mission, planned for a launch in early 2028, aiming at $\delta r <0.001$, where $\delta r$ is the total error on $r$, including statistical, systematic error, and margin
\newb{($2<\ell<200$)}.

The overall design of the LSPE program \new{has} largely evolved
since \newc{its} first proposal \citep{2012_pdb_lspe,2012_aiola, 2012SPIE_bersanelli,ltd18lamagna}, and  
this paper presents its final design and expected
performance. 
Section~\ref{sec:instruments} describes the two instruments in detail; 
section~\ref{sec:sensitivity} reports the expected instrumental sensitivities; section~\ref{sec:systematics} describes the major systematic effects, mitigation techniques and calibration; 
section~\ref{sec:results} presents the methods used in the foreground cleaning and likelihood evaluation and reports the expected performances on cosmological parameters. 
Finally, section~\ref{sec:conclusion} draws conclusions.

\section{The instruments}\label{sec:instruments}

Since the expected B-mode signal is smaller than the polarized foreground from our
Galaxy, a wide frequency coverage is needed to monitor precisely the foregrounds at frequencies
where they are most important, and to subtract them, in order to estimate the cosmological part of the detected
B-mode signal. 
For the synchrotron foreground, prominent at frequencies below \SI{\sim 100}{\giga\hertz}, where atmospheric transmission and noise are favorable, a ground based instrument is the most effective strategy, 
while for the CMB and the interstellar dust foreground, prominent at higher frequencies, a stratospheric balloon mission is preferred.
For this reason, the LSPE program
is based on the combination of two independent instruments: the~\Strip{} ground-based telescope, 
observing at \SI{44}{\giga\hertz}, plus a \SI{95}{\giga\hertz} channel for  atmospheric measurements, to be implemented at the Teide Observatory (Tenerife); and the SWIPE balloon-borne mission, observing at 145, 210 and \SI{240}{\giga\hertz} in a winter arctic stratospheric flight.

Table~\ref{tab:instruments} reports basic parameters for the 
two instruments, in the baseline configuration. 
Map sensitivity is an approximated value, computed as the square root of
$
\sigma^2_{Q,U} = p\,\text{NET}^2 \, {4 \pi f_\text{sky}}/(T_\text{obs} N_\text{det})
$, 
where $p=1$ for \Strip\ and $p=2$ for SWIPE, to take into account that each \newb{SWIPE} detector
is instantaneously sensitive to one polarization only, 
$T_\text{obs}$ is the effective integration time, NET is the noise
equivalent temperature of each detector, $f_{sky}$ is the observed sky fraction, 
and $N_\text{det}$ is the number of detectors. 
The power spectrum of the noise in polarization can be approximated by 
$\mathcal{N}^{E,B}_\ell=\sigma^2_{Q,U}/f_\text{sky,cmb}$, \new{where  $f_\text{sky,cmb}$
is the sky fraction used for CMB analysis, after masking the 
\newb{G}alactic plane.}
More accurate performance is estimated using the instrument simulators, 
component separation, and cosmological parameters extraction 
algorithms, as described in sections~\ref{sec:sensitivity} and \ref{sec:results}.

\begin{table*}[!tb]
    \begin{center}
        \begin{tabular}{l |c c |c c c}
            \hline
            \hline
            Instrument & \multicolumn{2}{c|}{Strip} & \multicolumn{3}{c}{SWIPE} \\
            \hline
            Site \dotfill & \multicolumn{2}{c|}{Tenerife} & \multicolumn{3}{c}{balloon}\\
            Freq (GHz) \dotfill & 43 & 95 & 145 & 210 & 240 \\
            Bandwidth \dotfill& 17\%  & 8\%   &  30\%  & 20\%  & 10\% \\
            Angular resolution FWHM \dotfill& 20\arcmin & 10\arcmin  & \multicolumn{3}{c}{85\arcmin}\\
            \new{Field of view \dotfill}  & \multicolumn{2}{c|}{$\pm 5\degr$}& \multicolumn{3}{c}{$\pm 11\degr$ }\\
            Detector technology\dotfill & \multicolumn{2}{c|}{HEMT}& \multicolumn{3}{c}{Multi-moded TES }\\
            Number of polarimeters (\Strip) / detectors (SWIPE) \dotfill & 49 & 6 & 162 &82 & 82\\
            NET  
            (\SI{}{\micro K_{CMB}.s^{1/2}})
            \dotfill& 515 & 1139 & 12.6 & 15.6 & 31.4\\
            Observation time \dotfill&  \multicolumn{2}{c|}{\SI{2}{years}} &  \multicolumn{3}{c}{8 -- \SI{15}{days}}\\
            \newb{Observing efficiency}  \dotfill&  \multicolumn{2}{c|}{50\%$^1$}&  \multicolumn{3}{c}{90\%}\\
            \new{Sky coverage$^2$ (nominal) $f_\text{sky,0}$} \dotfill& \multicolumn{2}{c|}{28\%}&  \multicolumn{3}{c}{38\%}\\
            Sky coverage$^2$ (this paper) $f_\text{sky}$ \dotfill& \multicolumn{2}{c|}{50\%}&  \multicolumn{3}{c}{38\%}\\
          \new{Masked sky coverage for CMB analysis $f_\text{sky,cmb}$} \dotfill& \multicolumn{2}{c|}{25\%}&  \multicolumn{3}{c}{25\%}\\
            Map sensitivity (nominal) $\sigma_{Q,U,0}$ 
            (\SI{}{\micro K_{CMB}.arcmin})
            \dotfill& 102 & 777 & 10&  17 & 34  \\
            \new{Map sensitivity (this paper)} $\sigma_{Q,U}$ 
            (\SI{}{\micro K_{CMB}.arcmin})
            \dotfill& 130 & 990 & 10&  17 & 34  \\
            Noise power spectrum $(\mathcal{N}^{E,B}_\ell)^{1/2}$ 
            (\SI{}{\micro K_{CMB}.arcmin})
            \dotfill&260 &1980 & 20 & 34 & 68\\
            \hline
            \multicolumn{6}{p{14 cm}}
            {\footnotesize $^1$We estimate as 50\% the time dedicated to sky observations, including calibration sources. We split the remaining 50\% as follows: (i) 15\% of lost time due to bad weather, (ii) 15\% of unusable data when the Sun will have an angular distance from the nearest feed less than 10$^\circ$ \citep{Incardona2020}, 20\% of time dedicated to relative calibration (see section~\ref{sec_strip_calibration}).}\\
            \multicolumn{6}{p{14 cm}}
            {\footnotesize
            \new{
            $^2$ We consider two cases for \Strip{} coverage, the nominal case
            with zenith angle $\beta_\text{nominal} = 20$\deg,
            and the case specific to this 
            paper with zenith angle $\beta = 35$\deg, which maximise\newb{s} the overlap as discussed in section~\ref{sec:observation_strategy} and illustrated in figure~\ref{fig:trade-off}}.
            }
        \end{tabular}
    \end{center}
    \caption{LSPE baseline instrumental parameters.
    Details are reported in tables~\ref{tab_strip_white_noise_properties} and~\ref{tab:swipe-noise}.
    \label{tab:instruments}}
\end{table*}

\subsection{\new{Observation strategy and sky coverage}}
\label{sec:observation_strategy}

\begin{figure}[!t]
   \centering
   \begin{tabular}{p{0.5\textwidth} p{0.5\textwidth}}
  \vspace{0pt} \includegraphics[width=0.54\textwidth]{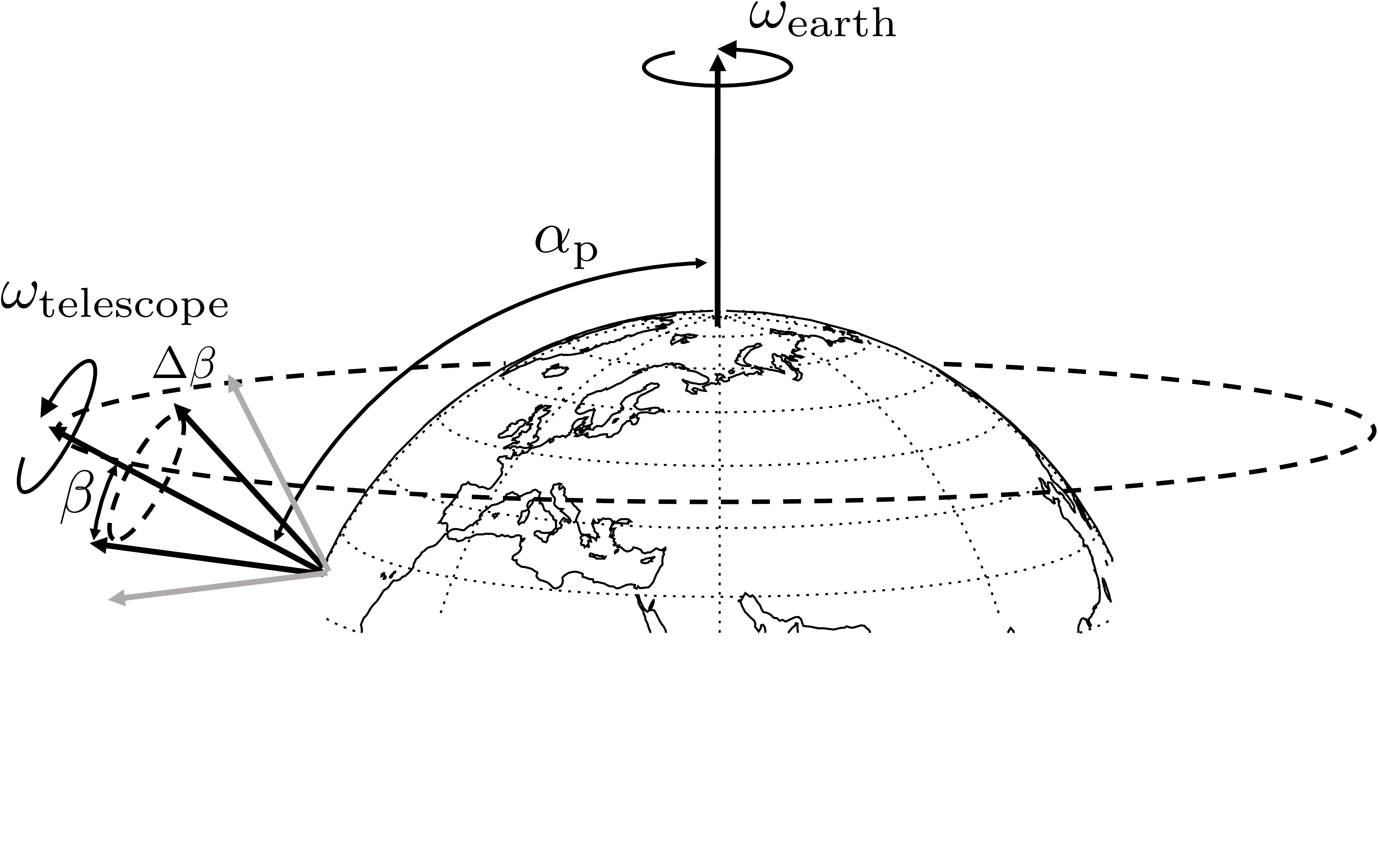} &
  \vspace{0pt} \includegraphics[width=0.44\textwidth]{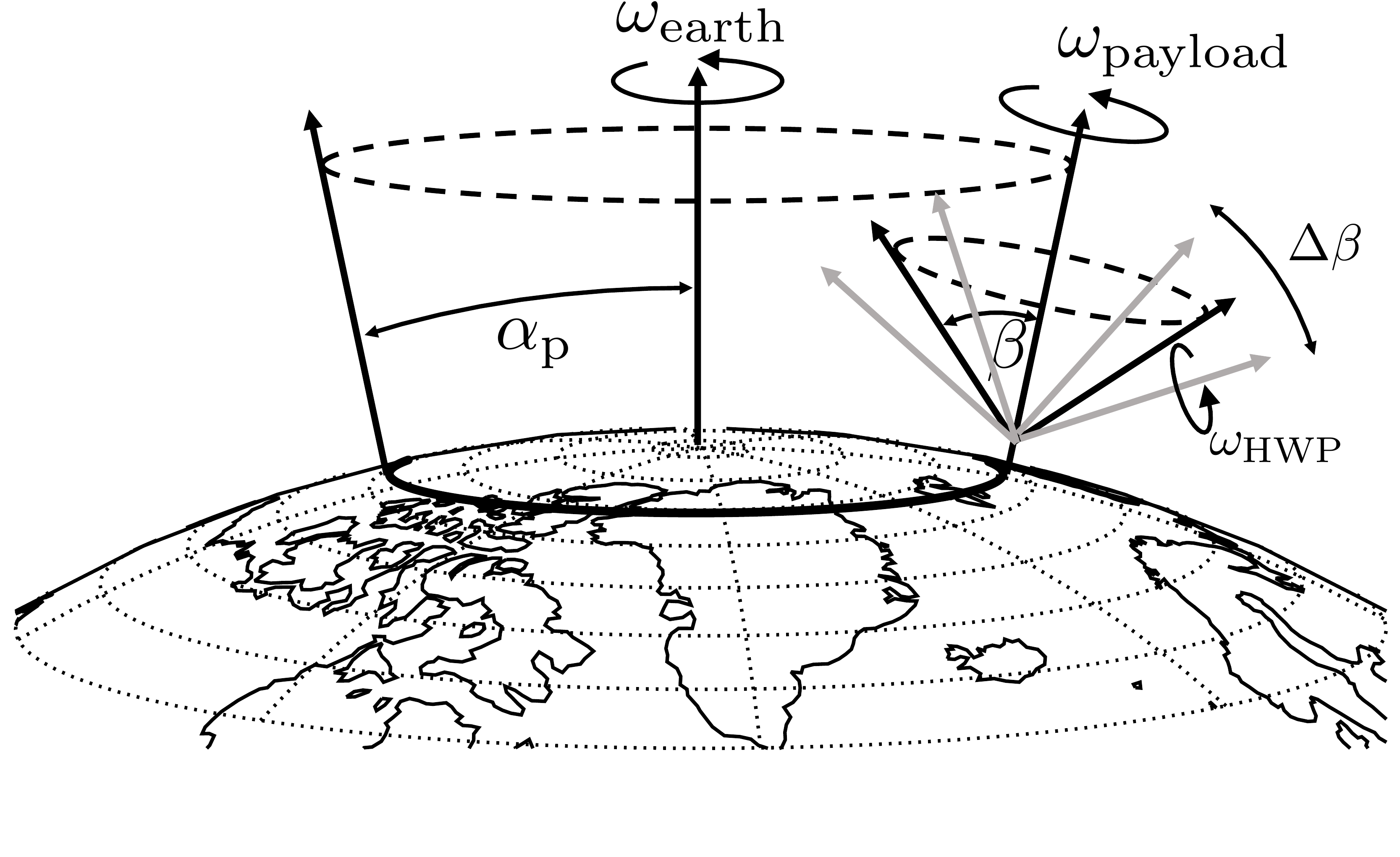}
\end{tabular}
\vspace{-1.6cm}
   \caption{\new{Observing strategy of the two instruments. 
   {\emph Left}: 
   the LSPE-\Strip{} telescope spins around the local zenith
   at an adjustable angle $\beta$. The 
   spin velocity is set at 1\,r.p.m. }
   {\emph Right}: the LSPE-SWIPE 
   payload spin axis precesses around the North Pole, 
   with precession angle equal to the co-latitude $\alpha_\text{p}$, with a 
   velocity that is a combination of the daily Earth rotation with the natural wind drift. 
   The telescope spins around the local 
   zenith at an angle $\beta$ which can 
   vary as $\Delta \beta$. 
   The polarized signal is modulated 
   by an Half-Wave Plate spinning at $\omega_\text{HWP}$.
   Numerical values are reported in table~\ref{tab:swipe-obspar}.
  }
              \label{fig:strategies}%
    \end{figure}

\new{
Figure~\ref{fig:strategies} illustrates the observation
strategies for the two instruments.
The \Strip{} telescope will scan the sky at a constant zenith angle $\beta$ with $\omega_\text{telescope}=\SI{1}{r.p.m.}$ spin rate. 
With this strategy, the observations cover a strip in equatorial declination
$\delta$ ranging $\text{lat}_\text{telescope} - \beta <\delta < \text{lat}_\text{telescope}+\beta$,
\newb{where $\text{lat}_\text{telescope}=\ang{28;18;00}$\,N}.
This strategy minimizes atmospheric effects and,
in combination with Earth rotation, to cover a large sky fraction.
}

The SWIPE observation strategy consists in 
continuous spinning of the payload, around the local zenith axis (spin axis),
at fixed angular velocity $\omega_\text{payload}$.
This is 
combined with steps in telescope 
zenith angle $\beta$   
(a few steps per day), to cover 
an altitude range from 35\degr to 55\degr. 
The Earth rotation, combined with the drift of the payload around the Arctic, 
ensures a slow precession of the vertical spin axis around the Equatorial North Pole 
(precession axis). Precession angle $\alpha_\text{p}$ (co-latitude) and precession angular
velocity $\omega_\text{Earth}$ are not 
exactly defined, due the partially random motion of the balloon, drifted by 
stratospheric winds. 
This strategy is combined with a Half-Wave Plate (HWP) based polarization 
modulator continuously spinning at rate $f_\text{HWP}$.
\new{The optimal payload spinning velocity 
and HWP rate are derived in 
Appendix~\ref{app:swipe_scanning_params} from 
detectors time constant and telescope angular response, and are found to be 
$\omega_\text{payload}\simeq \SI{0.7}{\degree.s^{-1}}$ and $f_\text{HWP}=0.5$\,Hz.
}
If the latitude remains
constant, the observation covers a strip in equatorial declination $\delta$
in the range $\ang{90}-(\alpha_\text{p}+\beta_\text{max})<\delta<\ang{90}-(\beta_\text{min}-\alpha_\text{p})$ (see figure~\ref{fig:strategies}).
The values of $\beta_\text{min}$ and 
$\beta_\text{max}$ also 
take into account the wide field of view $\pm10^\circ$.

SWIPE is expected to have a fixed sky coverage of about $38\%$ of the Northern Sky, 
with the precise value
depending on the choice of the launching station \new{and effective
trajectory}. 
The sky fraction observed by LSPE-\Strip{} can be adjusted by changing the telescope zenith angle \citep[][]{incardona_spie2018}, resulting in different sensitivity per sky pixel at the end of the survey. 
The final \Strip{} strategy will be defined to trade-off the sky coverage with the noise per pixel distribution and to maximize the overlap between the sky regions observed by the two LSPE instruments.


  \begin{figure}[t]
  \centering
  \includegraphics[width=12.5cm]{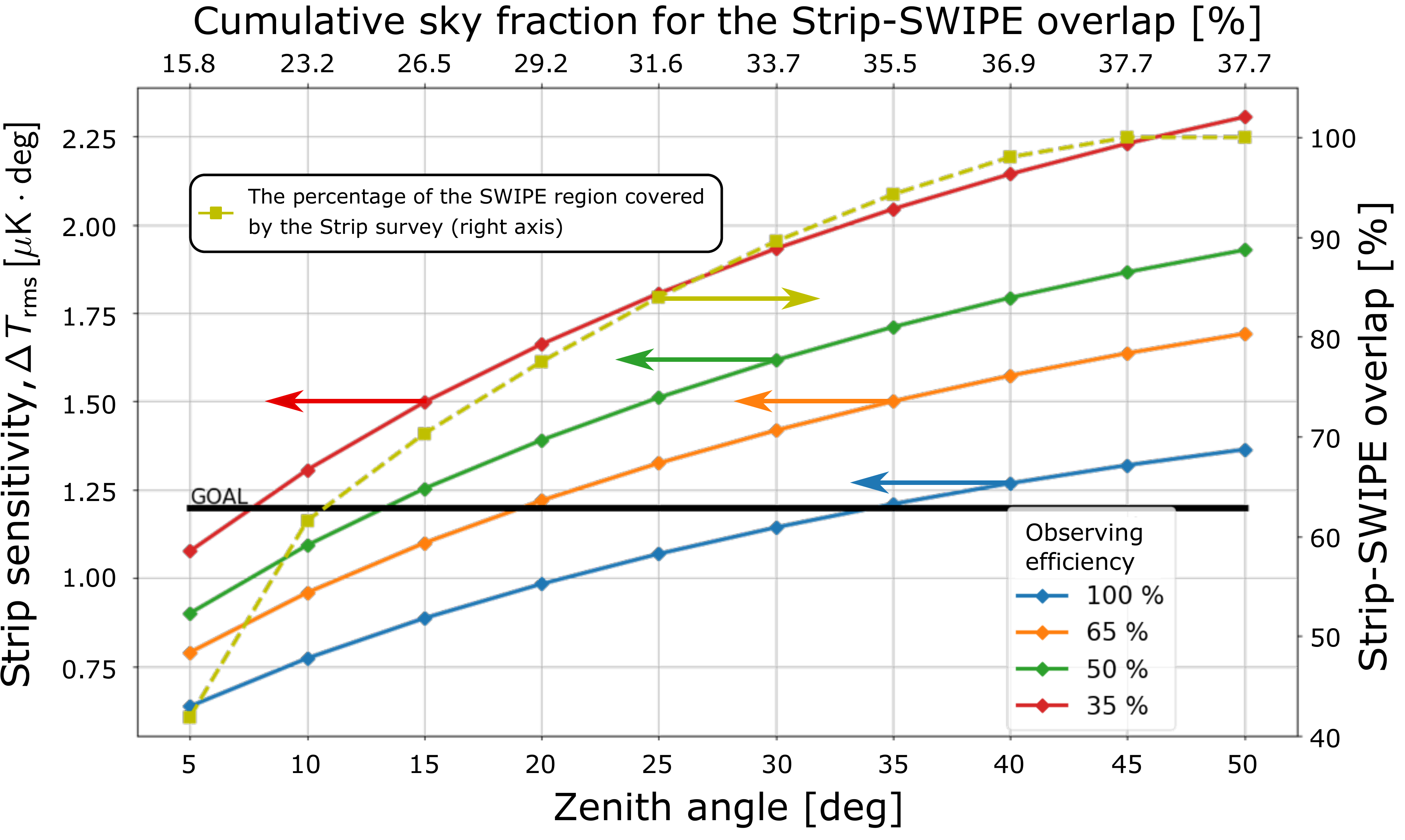}
  \caption{LSPE-\Strip{} map sensitivity  \new{(left axis)} as a function of the zenith angle (bottom axis) and 
  \newb{observing efficiency (colors)}. The black line (left axis) represents the value of the map sensitivity as reported in table~\ref{tab:instruments}. The dashed gold line shows the \Strip{}-SWIPE overlap percentage (right axis). The top axis reports the corresponding cumulative sky fraction. \new{The arrows point, for each curve, towards the corresponding ordinate axis.}
  }
              \label{fig:trade-off}%
    \end{figure}


The baseline configuration of the \Strip\ observation strategy assumes a constant zenith 
angle $\beta =  \SI{20}{\degree}$. 
Such configuration yields a map average noise $\sigma_{Q,U}=\SI{102}{\micro K.arcmin}$ 
\newb{at 43\,GHz}.
Assuming two years of observation time, we can calculate the sensitivity with respect to this baseline value as a function of the zenith angle and of the usable fraction of time (\newb{observing efficiency}). This is shown in figure~\ref{fig:trade-off}, together with the percentage of overlap,
and the total sky fraction as a function of the zenith angle.

\begin{figure}[!t]
   \centering
   \includegraphics[width=12cm]{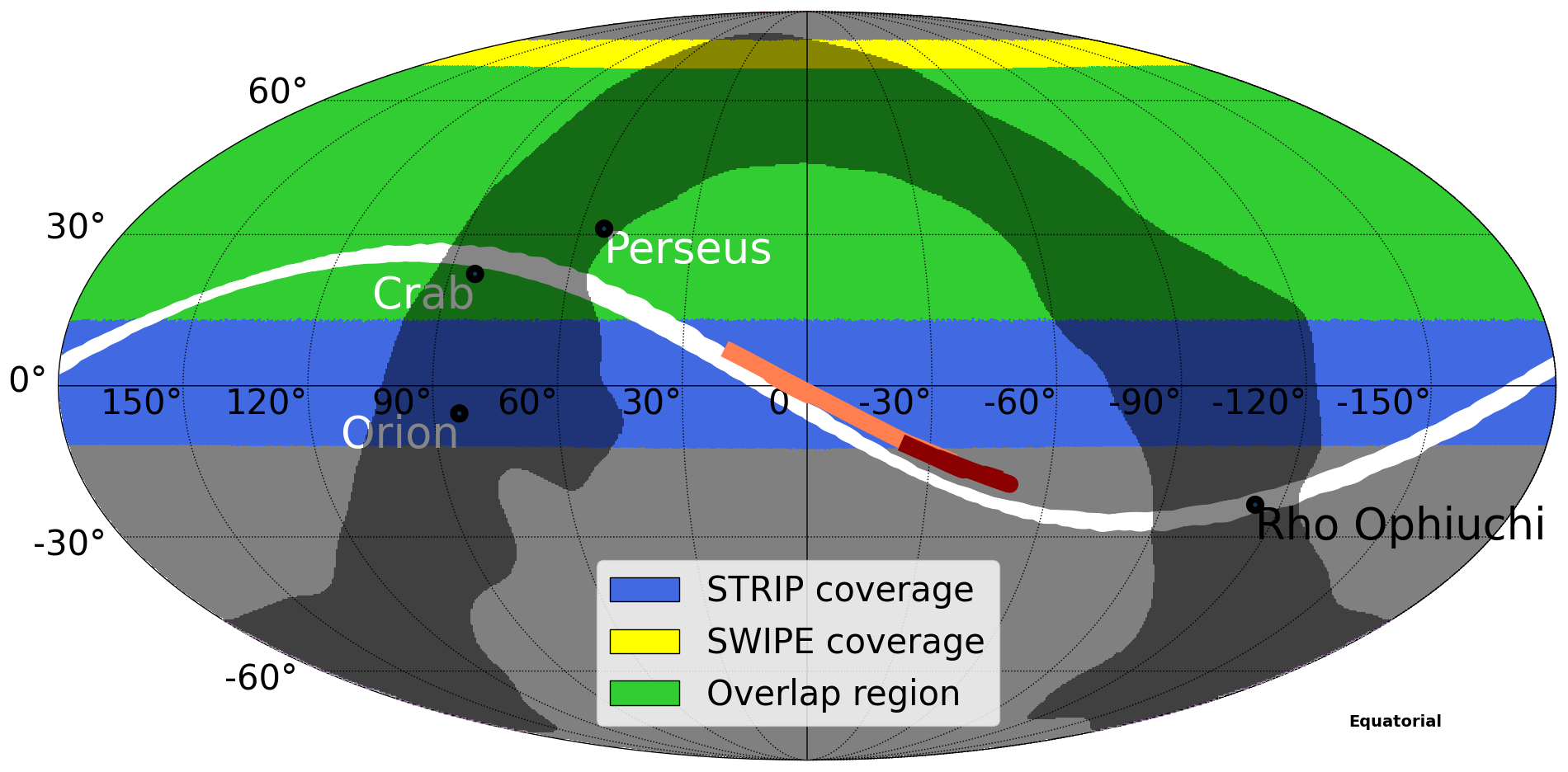}
   \caption{Map in Equatorial coordinates of the \Strip{}-SWIPE coverage. 
   The yellow area represents the SWIPE sky coverage; the blue area represents the \Strip{} sky coverage, in the case of 35\degr zenith angle; the green area shows the overlap \new{and the grey area represent a Galactic mask that covers 30\% of the whole sky}. 
   \new{The \Strip coverage ranges from -7 to 63\degree in latitude, and the SWIPE coverage from 13 to 77\degree.}
   The map also shows the position of the Crab and Orion nebula, of the Perseus molecular cloud and the trajectories of Jupiter (orange), Saturn (dark red) and the Moon (white) from April 2021 to April 2023. 
  }
              \label{fig:overlap20}%
    \end{figure}

In the analysis reported in this paper, we assume a standard coverage for SWIPE, with a launch
from Longyearbyen. In this case, the optimal overlap is obtained with a \Strip\ zenith 
angle of $\beta = \SI{35}{\degree}$, resulting in a
full-frequency coverage over 37\% of the sky, as shown in figure~\ref{fig:overlap20}. 
The map noise is in this case
$\sigma_{Q,U}=\SI{130}{\micro K.arcmin}$ \newb{at 43\,GHz}, with a wider coverage, providing 
the best trade-off for final results reported in sections~\ref{sec:results}.
\new{The two cases for \Strip\ zenith angle $\beta=20$\degr\ and $\beta=35$\degr\
are listed in table\ref{tab:instruments} as \textit{nominal} and \textit{this-paper}, \newc{respectively}.
}


\subsection{LSPE-\Strip{}}
LSPE-\Strip{} is a coherent polarimeter array that will observe the microwave sky from the Teide Observatory in Tenerife in two frequency bands centred at \SI{43}{\giga\hertz} (Q-band, 49 receivers) and \SI{95}{\giga\hertz} (W-band, 6 receivers) through a dual-reflector crossed-Dragone telescope of $\sim\SI{1.5}{\metre}$ projected aperture.

The \Strip{} array uses coherent technology exploiting low noise high electron mobility transistor (HEMT) amplifiers, together with high-performance wave-guide components. The instrument is cooled to \SI{20}{\kelvin} by a two-stage Gifford-McMahon (GM) cooling system and integrated at the focal plane of the telescope \new{that is able to rotate continuously in azimuth}. 
The \new{polarimeter's} design allows \newc{\Strip\ to} directly measure the Stokes $Q$ and $U$ parameters through a double-\new{de}modulation scheme \new{that is explained in section~\ref{sec_instrument_cryogenics}}. This design ensures  excellent rejection of $1/f$ noise from amplifier gain fluctuations as well as of temperature-to-polarization leakage, without the need to introduce extra optical elements to modulate the polarized signal. 

The main objective of \Strip{} is to accurately measure Galactic synchrotron emission in the LSPE sky region in Q-band. Recent studies 
\citep{Krachmalnicoff2018} show that the 
polarized synchrotron emission is significantly structured and characterized by non-trivial variations in its spectral index. Deep measurements at \SI{43}{\giga\hertz}, complemented by lower frequency data, are crucial to constrain synchrotron contamination in the foreground minimum accounting for spectral index variations. Furthermore, achieving a resolution of $\sim 20$\,arcmin will provide key information on the 
spatial properties of synchrotron foreground.

The W-band array, composed of 6 modules, will complement the Q-band data in monitoring the atmospheric load and fluctuations (mostly due to water vapor) during the \Strip{} observations. Atmospheric effects in Q-band can be effectively monitored by measurements in W-band, where the water vapor component is significantly higher.
Note anyway that at the Teide Observatory the atmospheric contamination of Q-band data is clearly dominated by O$_2$, which is stable spatially and with time. 
Yet the W-band channels will help to mitigate Q-band atmospheric fluctuations, 
expected to be of the order of $\sim 2$\,K.

\subsubsection{Observation site}
\label{sec:observation_site}



\Strip{} will be deployed at the Teide Observatory in Tenerife, at an altitude of \SI{2400}{\metre} above sea level, 
\newb{coordinates: \ang{28;18;00}\,N,  \ang{16;30;35}\,W}. The site provides excellent observing conditions and has been well-tested for astronomical observations for more than 30 years. 
The median precipitable water vapour is \SI{3.5}{\milli \metre}, reaching values below \SI{2}{\milli\metre} during 30\% of the time \citep{SPIE-OT}. The inversion layer lies below the observatory for approximately 80\% of the time.

The observatory has a long tradition in CMB research, including past experiments like the Tenerife radiometers \citep{tenerife}, the IAC-Bartol \citep{bartol}, the JBO-IAC two-element interferometer \citep{jboiac}, the COSMOSOMAS experiment \citep{cosmosomas} and the Very Small Array interferometer (VSA~\cite[]{vsa}). 
The \Strip{} telescope will be installed inside an aluminium ground screen to limit interference and ground-spill. The telescope will be protected by a sliding roof that will cover the whole enclosure. 

In addition to serving as low frequency monitor for LSPE, \Strip{} also will complement two existing CMB experiments in Tenerife, QUIJOTE \citep{quijote} and GroundBIRD~\citep{2020groundbirdJLTD}, by observing in different frequency bands: 10--\SI{40}{\giga\hertz} for QUIJOTE, 40--\SI{95}{\giga\hertz} for \Strip{}, and 145--\SI{225}{\giga\hertz} for GroundBIRD.
All three Tenerife projects (QUIJOTE, LSPE-\Strip{} and GroundBIRD) aim at measuring approximately the same area in the Northern sky and at degree scales, opening the possibility of future combined analyses, including useful redundancy for cross-checks of systematic effects.
\Strip{} measurements are currently scheduled to start during Summer \newb{2022} and last two years.



    The \Strip{} telescope will scan the sky at a constant zenith angle, nominally 20$^\circ$, with 1\,r.p.m. \new{azimuthal spin rate.} 
This strategy will allow us to minimize atmospheric effects and to cover about 38\% of the Northern sky,
thus ensuring a large overlap with the SWIPE observations.
After two years \new{of} operations with 50\% \newb{observing efficiency} we will reach a sensitivity of \SI{\sim 102}{\micro K_{CMB}.arcmin}
at \SI{43}{\giga\hertz} and 
\SI{\sim 777}{\micro K_{CMB}.arcmin}
at \SI{95}{\giga\hertz} 
\newb{(see section~\ref{sec:observation_strategy}, table~\ref{tab:instruments}
and figure~\ref{fig:trade-off}
for more details)}. 
\newb{The observing efficiency} does not account for down time due to the Moon, glitches, 
Radio-Frequency Interference (RFI), or other unpredictable instrument-specific anomalies, thus moving our estimate somewhat on the optimistic side. A breakdown of our estimated data loss is given in the footnote of table~\ref{tab:instruments}.




\subsubsection{Telescope and mount structure}
\label{sec_telescope}

    The \Strip{} telescope consists of two reflectors, a parabolic primary mirror and hyperbolic secondary mirror, arranged in a Dragonian cross-fed design, originally developed for the CLOVER experiment \citep{TAYLOR2006993}. This configuration preserves polarization purity on the optical axis and gives low aberrations across a wide, flat focal plane. The projected diameter of the main reflector is \SI{1.5}{\metre} and the entire system has an equivalent focal length of \SI{2700}{\milli\metre}, resulting in $\sim$ f/1.8.

\begin{figure}[!t]
   \centering
   \includegraphics[width=15cm]{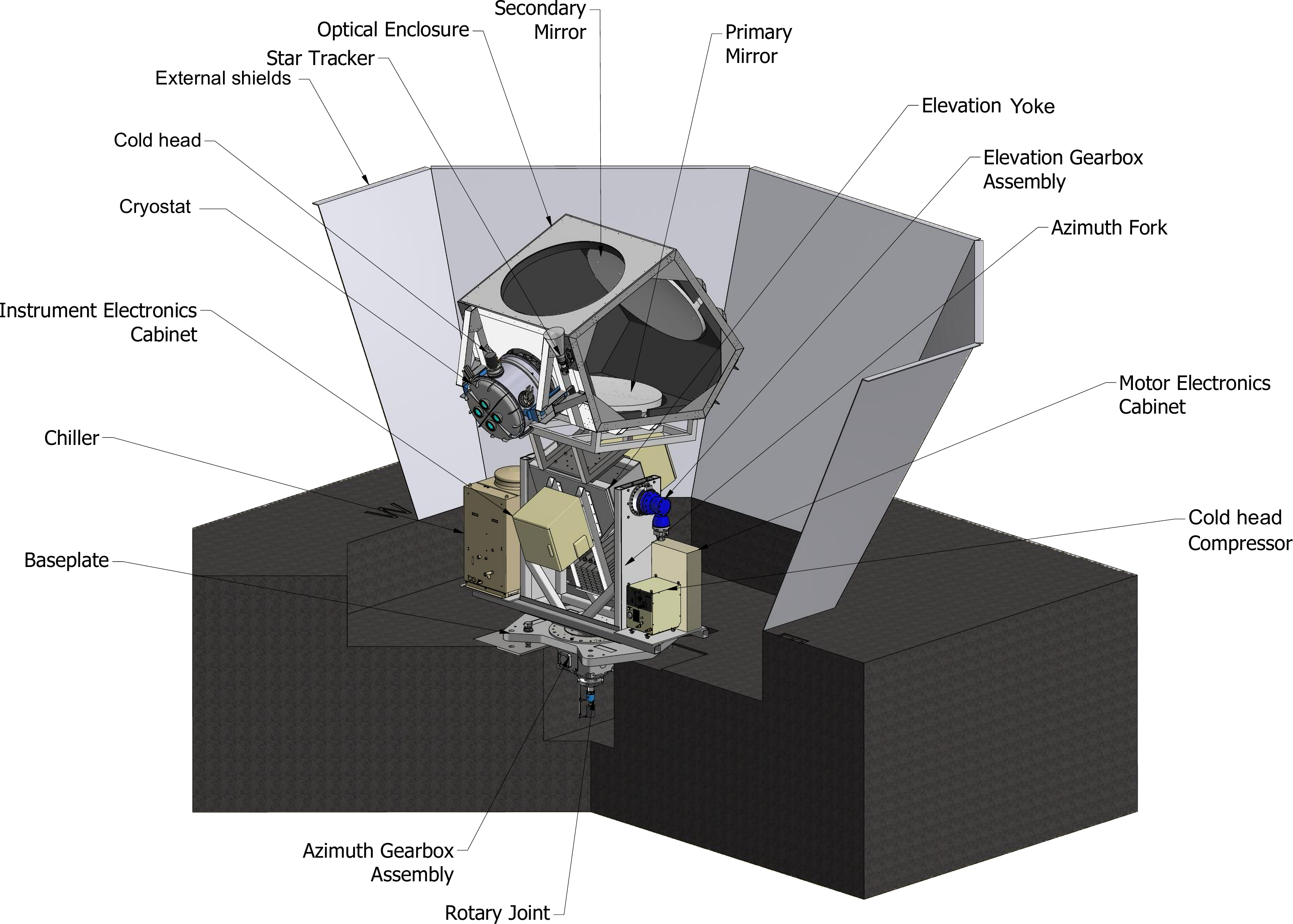}
   \caption{\new{LSPE-\Strip{} optical system overview. The mirrors are held inside a co-moving \newb{optical enclosure}.}}
   \label{fig:strip-overview}%
\end{figure}

The telescope is surrounded by a co-moving baffle made of aluminum plates coated by a millimetre-wave absorber, which reduces the contamination due to stray light. The optical assembly is installed on top of an alt-azimuth mount, which allows the rotation of the telescope around two perpendicular axes to change the azimuth and elevation angle. An integrated
rotary joint will transmit power and data to the telescope and the instrument, and will allow a continuous spin as required by the scanning strategy. A general view of the \Strip{} system is shown in figure~\ref{fig:strip-overview}.

The telescope provides an angular resolution of $\sim$20\arcmin\ in the Q-band and $\sim$10\arcmin\ in the W-band. 
The feedhorn array is placed in the focal region, ensuring no obstruction of the field of view. All the modules are optimally oriented according to the shape of the focal surface, with illumination centred on the primary mirror. 
The two mirrors determine the main beam shapes of the \Strip{} detectors, while the shielding structures  affect the near and far sidelobes \citep{2018_franceschet}.

\paragraph{Optical performance.} 
\new{We have modeled the optical assembly with the}
GRASP\footnote{\url{https://www.ticra.com/software/grasp/}}
software and the model includes the nominal reflectors, the focal plane unit, the IR filters, and the shielding structures. The model is also able to reproduce the dual circular polarization antenna-feed system \citep{2019_realini}.

\new{We have simulated the main beam radiation patterns} using the Physical Optics (PO) method, which is needed to correctly model the detector patterns in the far field. Given the off-axis configuration, the main beams are characterized by several parameters, \newc{such} as the angular resolution, the ellipticity, the main beam directivity, and the cross-polar discrimination factor (XPD). 

\begin{figure}[!t]
   \centering
   \includegraphics[width=15.6cm]{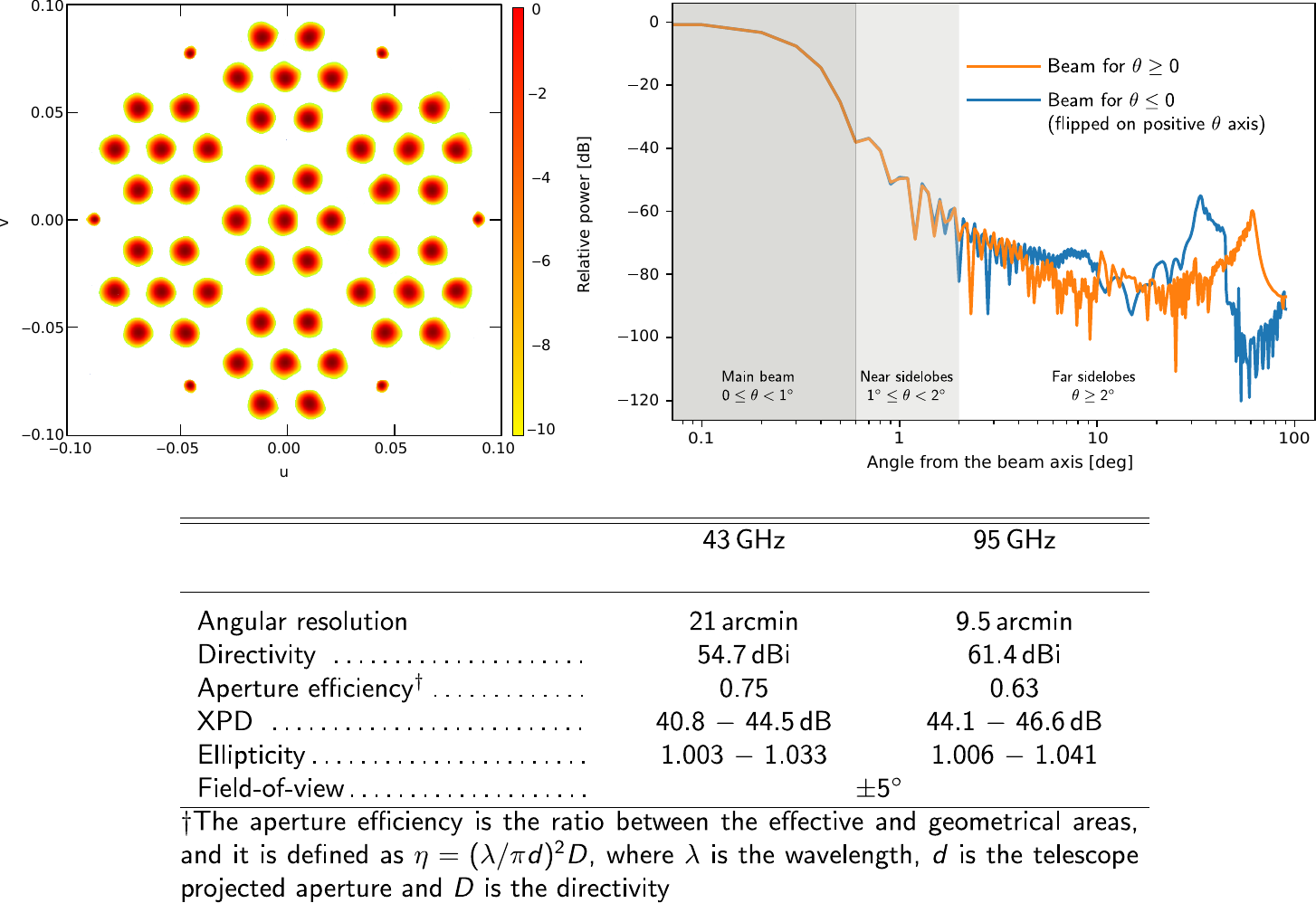}
   \caption{\label{fig_strip_beams}\newb{\textit{Left:} footprint of the LSPE-\Strip{} main beams in the $(u,v)$ plane (large beams, 43\,GHz, small beams, 95\,GHz). \textit{Right:} beam cut for $\phi=0$ of the central horn.  We have flipped the beam section for $\theta< 0$ on the positive axis to better highlight the asymmetries.
   The $(u,v)$ variables are defined as $u=\sin(\theta)\cos(\phi)$,
   $v=\sin(\theta)\sin(\phi)$, where $(\theta,\phi)$ are standard spherical coordinates, with the center
   of the telescope pointing towards $\theta=0$.}
   The inset table reports the averaged main optical parameters.}
\end{figure}

%

Sidelobes have been computed using the Multi-Reflector Geometrical Theory of Diffraction (MrGTD). While less accurate than PO, this ray-tracing technique is much more efficient and it is able to predict the full-sky radiation pattern of complex optical systems.  The $4\pi$ radiation patterns show unevenly distributed features that are due to multiple reflections inside the shielding structure and rays entering the feedhorns without any interaction with the reflectors. Each contribution has been analyzed separately and then combined in an integrated model beam. We find that \newb{the level of sidelobes at angles larger than \SI{1}{\degree} is less than \SI{-55}{dB} at \SI{43}{\giga\hertz} and less than \SI{-65}{dB} at \SI{95}{\giga\hertz}}.

\newb{In the top-left panel of figure~\ref{fig_strip_beams} we show the footprint of the Strip main beams in the $(u,v)$ plane. We can see the 49 Q-band beams grouped in seven hexagonal structures of seven beams each and the six outer W-band beams. In the top-right panel of the same figure we show a cut corresponding to $\phi=0$ of the central beam. We have flipped the beam section for $\theta< 0$ on the positive axis to better highlight the asymmetries. The bottom inset table displays the average main optical parameters}.

\subsubsection{Instrument and cryogenics}
\label{sec_instrument_cryogenics}

\begin{figure}[!t]
    \begin{center}
        \begin{tabular}{c|c}
        \includegraphics[width=7cm]{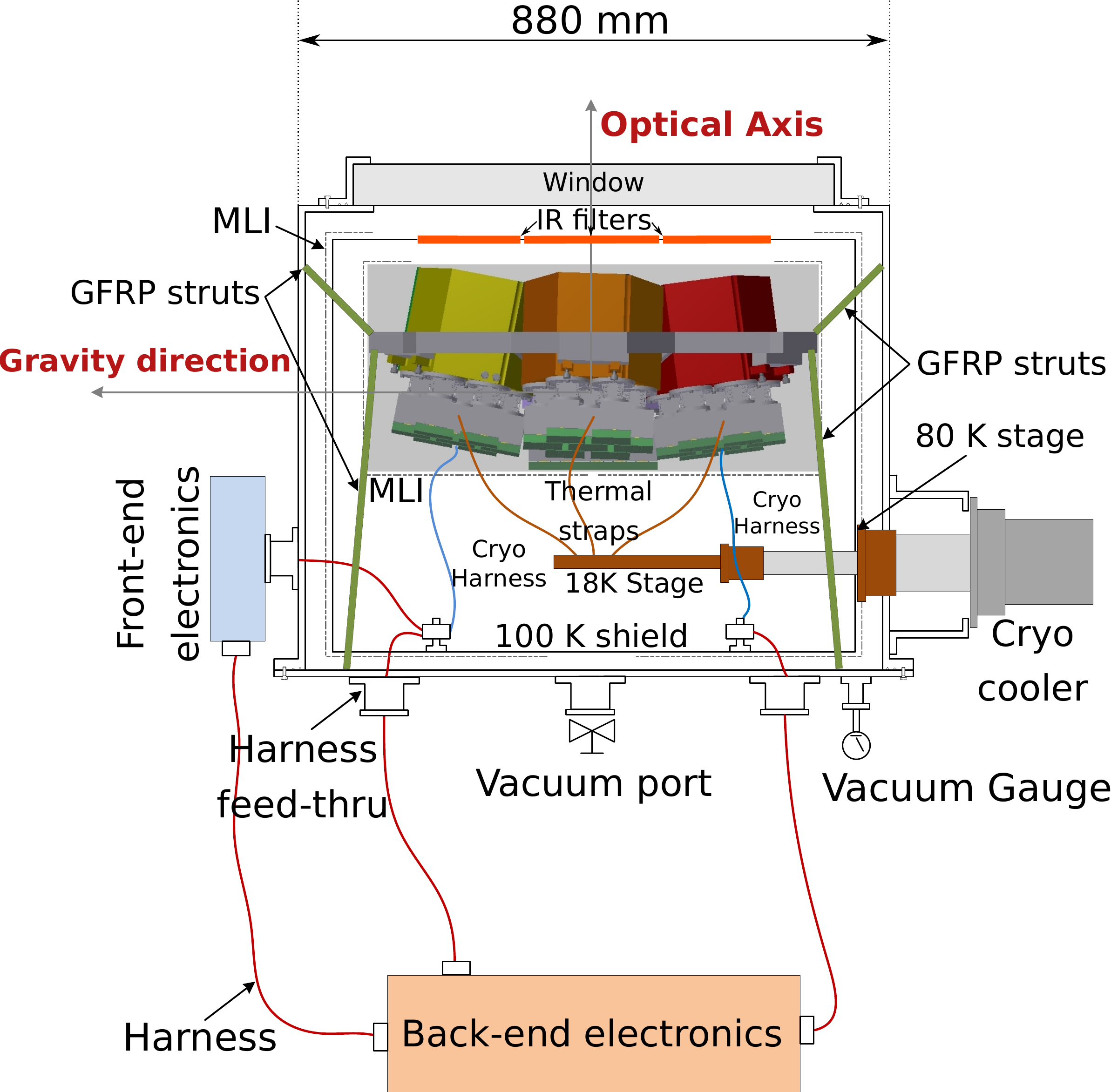}&
        \includegraphics[width = 8cm]{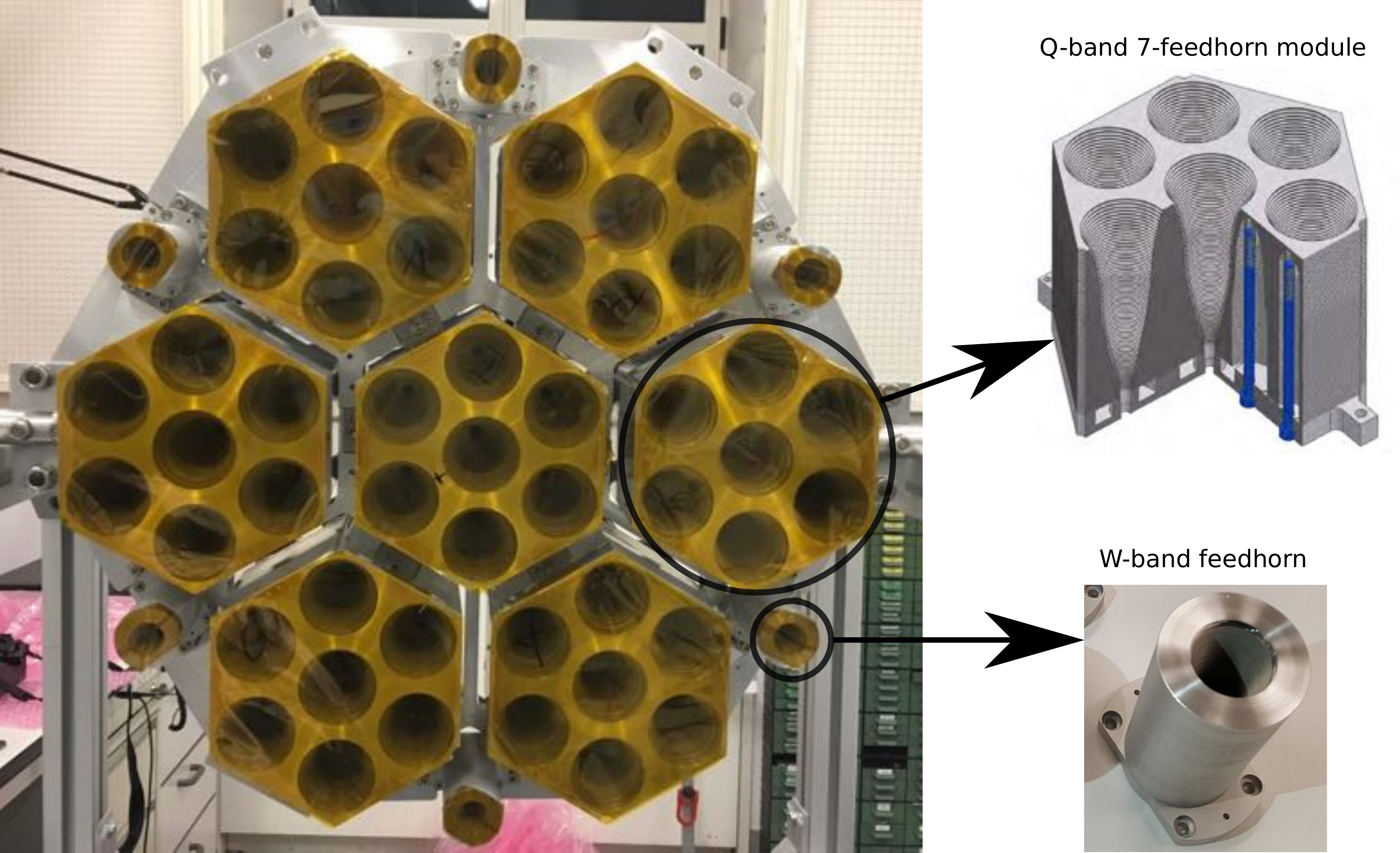}
        \end{tabular}
    \end{center}
    \caption{\label{fig_strip_schematics}\textit{Left}: \new{schematic drawing} of the \Strip{} instrument. The focal plane array is inside the cryostat surrounded by the \SI{80}{\kelvin} shield and thermally connected to the cooler cold head. \new{Note that this drawing does not include the W-band horns that are visible in the real picture on the right.} \textit{Right}: the complete \Strip{} focal plane array with 49 feedhorns in Q band and 6 feedhorns in W band. We also show a cutaway of one Q-band module (\textit{top}) and the detailed view of one of the six W-band feedhorns (\textit{bottom}).}
\end{figure}

     The \Strip{} focal plane array of corrugated feedhorns is placed inside the dewar surrounded by a radiative shield cooled to \SI{80}{\kelvin} by the cooler first stage (see the left panel of figure~\ref{fig_strip_schematics}).

 Copper thermal straps connect the focal plane and the cooler cold head allowing the polarimeter chain to be cooled down to \SI{20}{\kelvin}.
\new{The cryostat window is an ultra-high molecular weight polyethylene (UHMWPE) window with a diameter of 586\,mm and a thickness of 56.34\,mm. We stop the IR radiation from the \SI{300}{\kelvin} environment with 13 polytetrafluoroethylene (PTFE) filters with anti-reflection coating at \SI{150}{\kelvin}. We have one filter for each horn at \SI{95}{\giga\hertz} (diameter 52\,mm and thickness 23\,mm) and one filter for each 7-horns module at \SI{43}{\giga\hertz} (diameter 170\,mm and thickness 23\,mm). The filters are attached to the 100\,K thermal shield in front of the 20\,K feedhorn array.}

The detector assembly is based on coherent polarimeters connected to an optical chain constituted of corrugated feedhorns, each coupled to a polarizer-orthomode transducer (OMT) system at 43\,GHz and to a septum polarizer at 95\,GHz \citep{peverini2015}.

\begin{figure}[t!]
    \begin{center}
        \includegraphics[width=\textwidth]{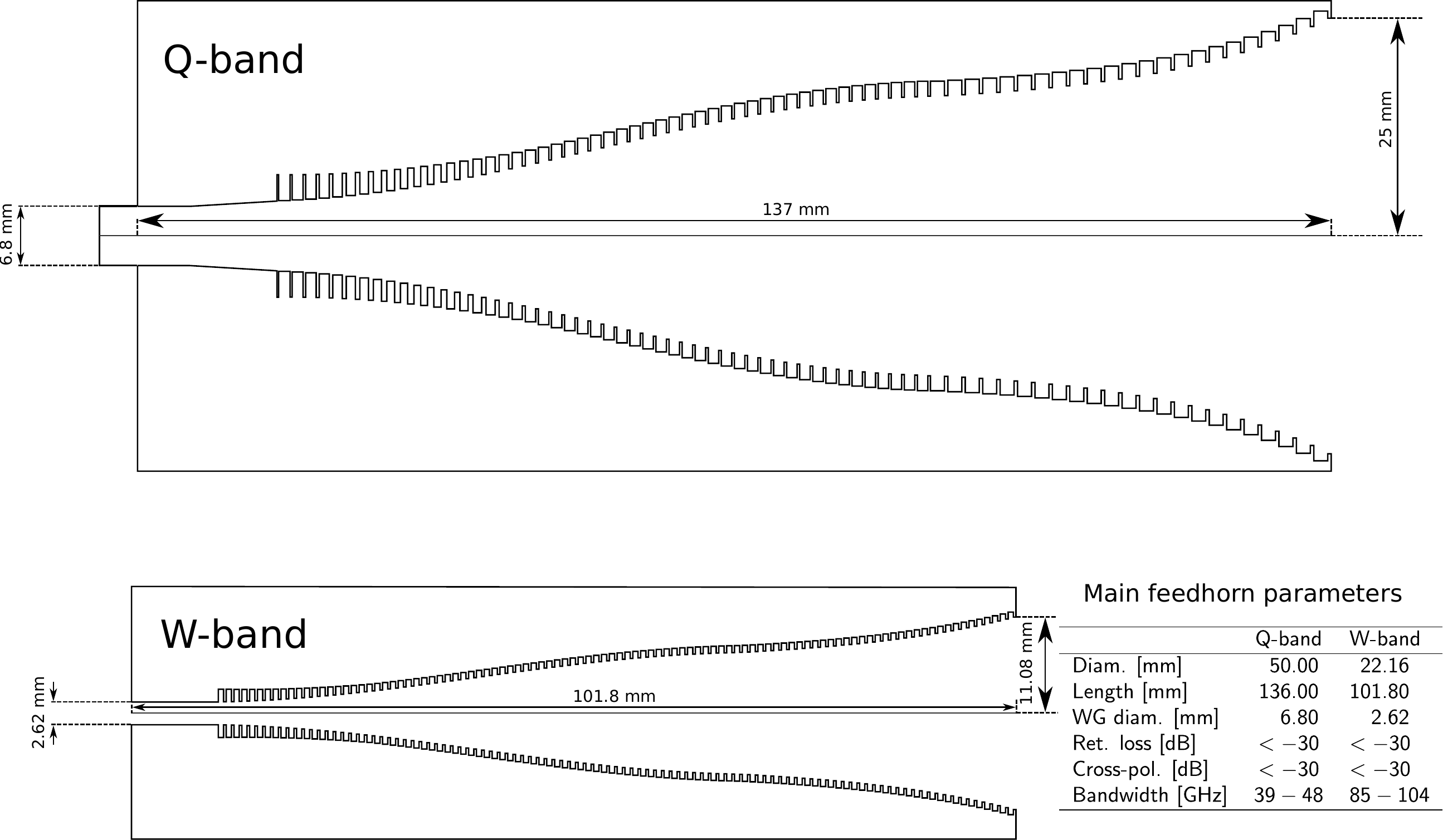}
    \end{center}
    \caption{\label{fig_strip_feedhorns}\new{Corrugation profile of Q- and W-band \Strip{} feedhorns. The inset table summarizes the main feed parameters.}}
\end{figure}

\paragraph{Feedhorns.} The feedhorns are designed implementing a dual profile to obtain an optimal illumination of the secondary with a limited feed size, and are manufactured in aluminum using the platelet technique \citep{deltorto_2011}. The right panel of figure~\ref{fig_strip_schematics} shows a picture of the entire \Strip{} focal plane, with the 49 Q-band feedhorns arranged in 7-unit modules surrounded by the six W-band feedhorns. 
A cutaway of one of the Q-band modules and a detailed view of one of the W-band feedhorns are also presented. In the cutaway it is 
possible to appreciate the platelet structure of the module and the tightening screws that allowed to assemble the horns without the need of bonding material or thermal brazing. \new{In figure~\ref{fig_strip_feedhorns} we show the corrugation profile of the \Strip{} feedhorns in both frequency bands and a summary table of the main parameters.}
 
\paragraph{Polarizers and OMTs.} Each feedhorn is connected to a polarizer system that converts the two orthogonal components of the electric field, $(E_x, E_y)$ into right- and left-circular polarization components, $\left[(E_x+i\,E_y)/\sqrt{2}, (E_x-i\,E_y)/\sqrt{2}\right]$, which propagate through the polarimeter module. This conversion is obtained differently in Q- and W-band.

\begin{figure}[!tb]
    \begin{center}
        \includegraphics[width = 13cm]{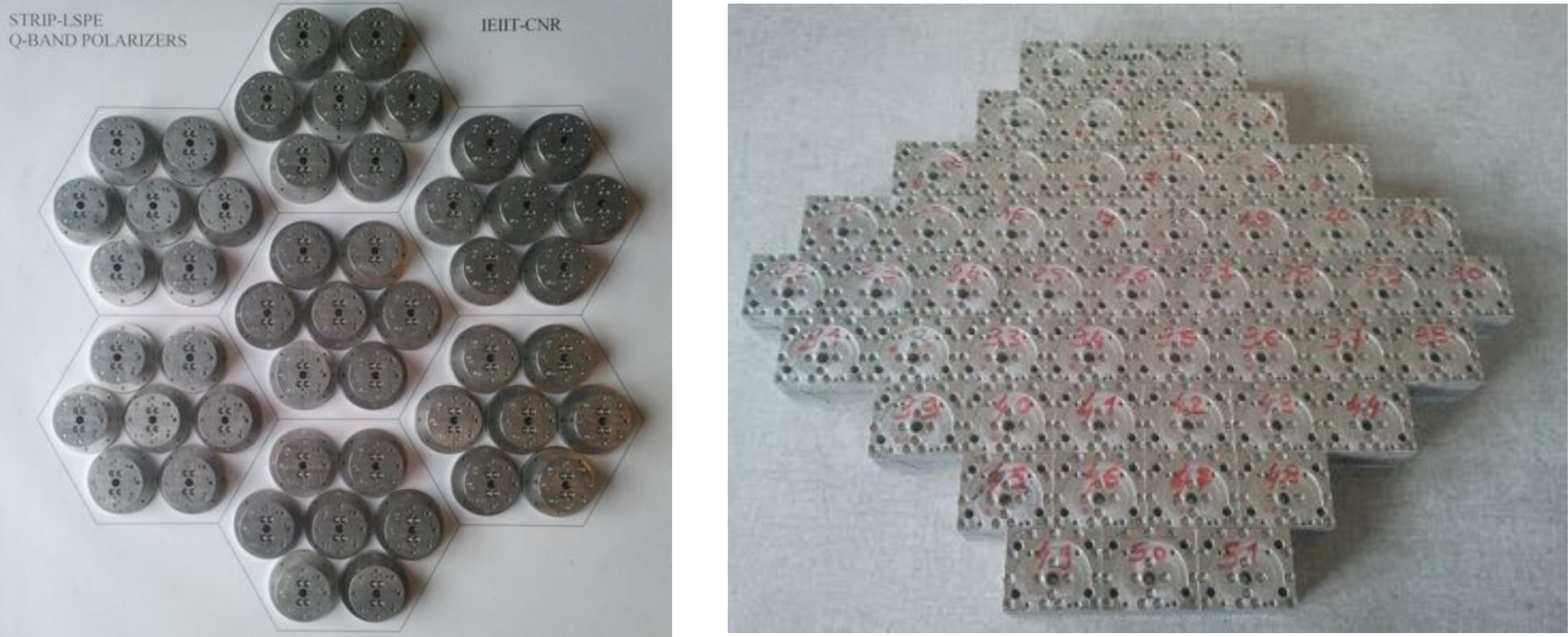}
    \end{center}
    \caption{\label{fig_polarizer_omt}\textit{Left:} the 49 \Strip{} Q-band polarizers arranged according the 7-module feedhorn footprint. \textit{Right:} the 49 \Strip{} Q-band platelet OMTs}
\end{figure}

In Q-band we convert linear to circular polarization using a groove polarizer \citep{eom2006} connected to a platelet OMT \citep{Virone2014}. In figure~\ref{fig_polarizer_omt} we show the complete set of Q-band polarizers (left panel) and OMTs (right panel) implemented in the \Strip{} focal plane.
This solution allowed us to obtain a very good measured performance in terms of transmission ($\gtrsim \SI{-0.5}{dB}$), reflection ($<\SI{-25}{dB}$) and cross-talk ($\sim\SI{-40}{dB}$).

Each W-band channel incorporates a septum \new{polarizer} characterized by a reflection of $<\SI{-20}{dB}$ and a leakage of intensity to polarization of the order of $\sim -\SI{13}{dB}$. The detailed design and performance of these components, originally used by the QUIET \SI{95}{\giga\hertz} instrument \citep{quiet2012b}, can be found in~\citep{chen2014}.

\begin{figure}[!tb]
    \begin{center}
        \includegraphics[width=13cm]{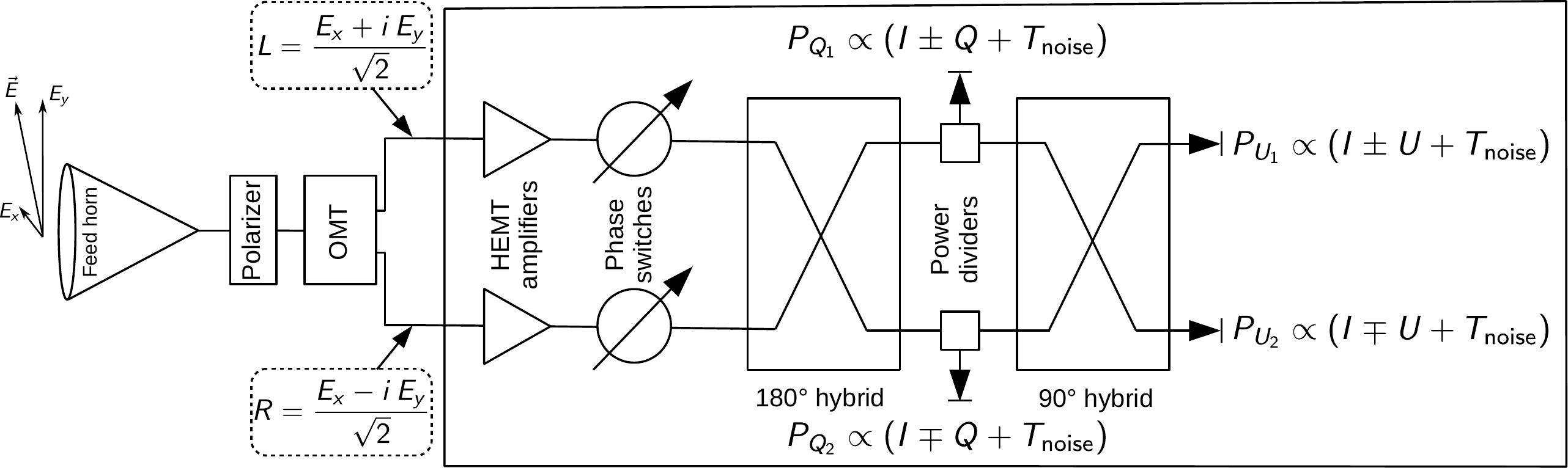}
    \end{center}
    \caption{\label{fig_strip_polarimeter_schematic}The \Strip{} polarimeter's operation principle. The figure also shows the main mathematical relationships among the detected power at the four diodes, $P_{Q_{1,2}}, P_{U_{1,2}}$, the Stokes parameters defined in the polarizer reference frame, $I, Q, U$ and the polarimeter noise temperature, $T_\mathrm{noise}$.}
\end{figure}

\paragraph{Polarimeters.} The \Strip{} Q-band channel uses a combination of the original 19 QUIET Q-band modules \citep{quiet2011} and additional 30 units that were developed according to the same design. The W-band channel uses 6 QUIET polarimeters 
selected among those with the best performance. 
The diagram in figure~\ref{fig_strip_polarimeter_schematic} shows the operation principle. If two circularly polarized signals propagate through a symmetric $180^\circ$ hybrid, the power detected at its output is a combination of $I$ and $Q$ Stokes parameters, with $Q$ having opposite signs at the two detectors. The detected power at the output of a second, 90$^\circ$ hybrid coupler yields a combination of $I$ and $U$, with $U$ appearing with opposite signs. The design takes full advantage of the coherent nature of the signal, implementing a double \new{de}modulation scheme to minimize residual systematic effects. 
This strategy allows \new{\Strip{}} to recover both $Q$ and $U$ from a single measurement, after combining the two linearly polarized components of the input field, $E_x$ and $E_y$, into left and right circular polarization components. 

Ahead of the first hybrid, two multi-stage Indium Phosphide (InP) HEMT amplifiers provide about \SI{50}{dB} amplification while two phase switches shift the signal phase between 0$^\circ$ and 180$^\circ$, and allow 
demodulation.

There are two different kinds of demodulation. A fast ($\sim\SI{4}{\kilo\hertz}$) demodulation, provided by one of the two phase switches that flips the signs of $Q$ and $U$ at each of the four diodes (see figure~\ref{fig_strip_polarimeter_schematic}), removes effectively the effect of amplifier gain fluctuations. 
A slow (\SI{50}{Hz}) demodulation, provided by the second switch that flips the sign between detector pairs, removes any $I\rightarrow Q,U$ leakage arising from asymmetries in the phase switches attenuation. \new{Note that it is irrelevant which of the two phase switches is ``fast'' and which one is ``slow''.}

The correlation units are packaged into square brass modules about \SI{1}{cm} thick and with a footprint of $\sim5\times\SI{5}{cm^2}$ in Q-band and $\sim2.5\times\SI{2.5}{cm^2}$ in W-band. Each complete polarimetric chain from the feed to the detectors will be cooled down to \SI{20}{K} by the \Strip{} cryogenic system.

\begin{figure}[!tb]
    \begin{center}
        \begin{tabular}{c|c}
            \includegraphics[width=6.0cm]{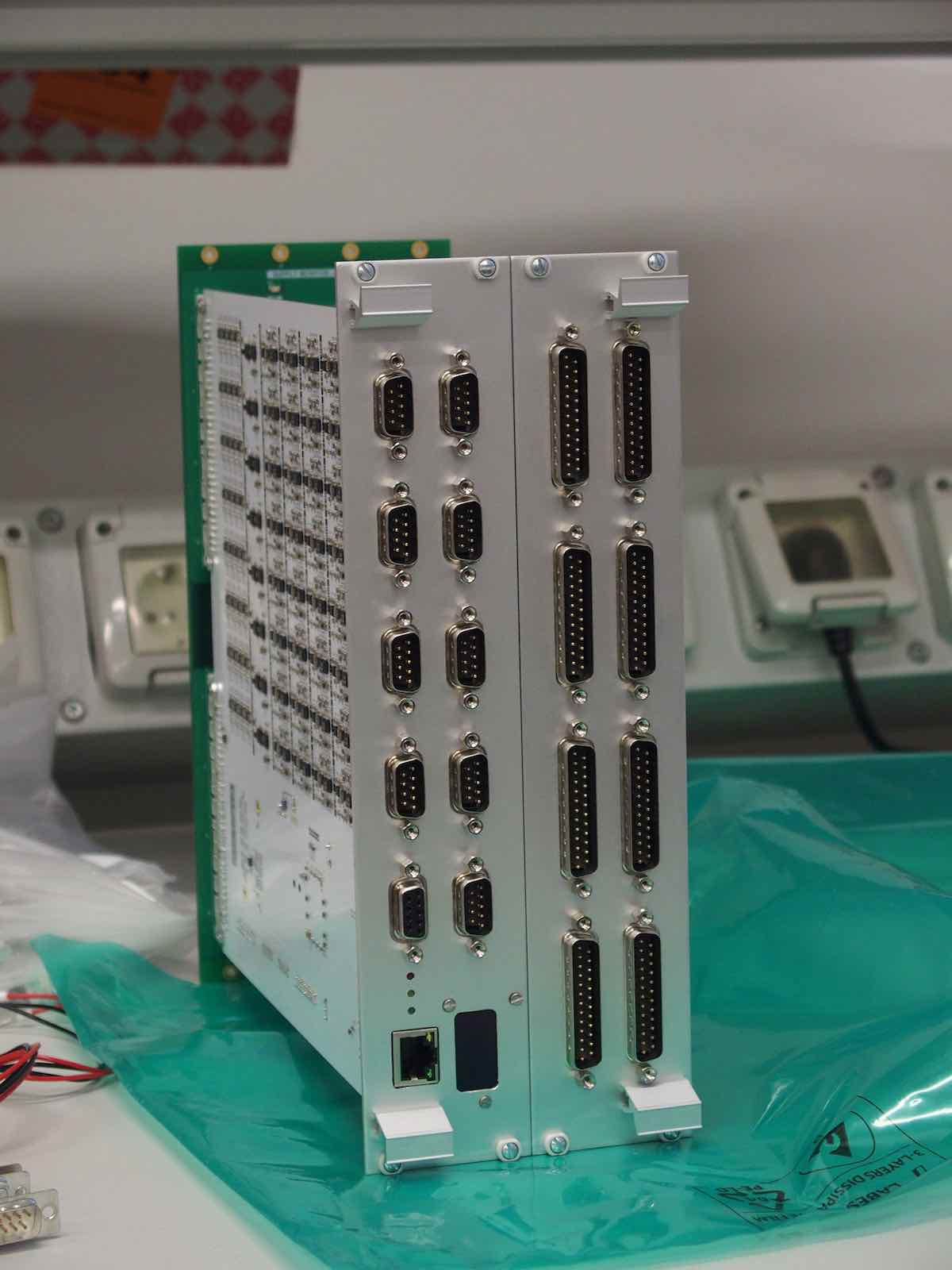} &
            \includegraphics[width=6.0cm]{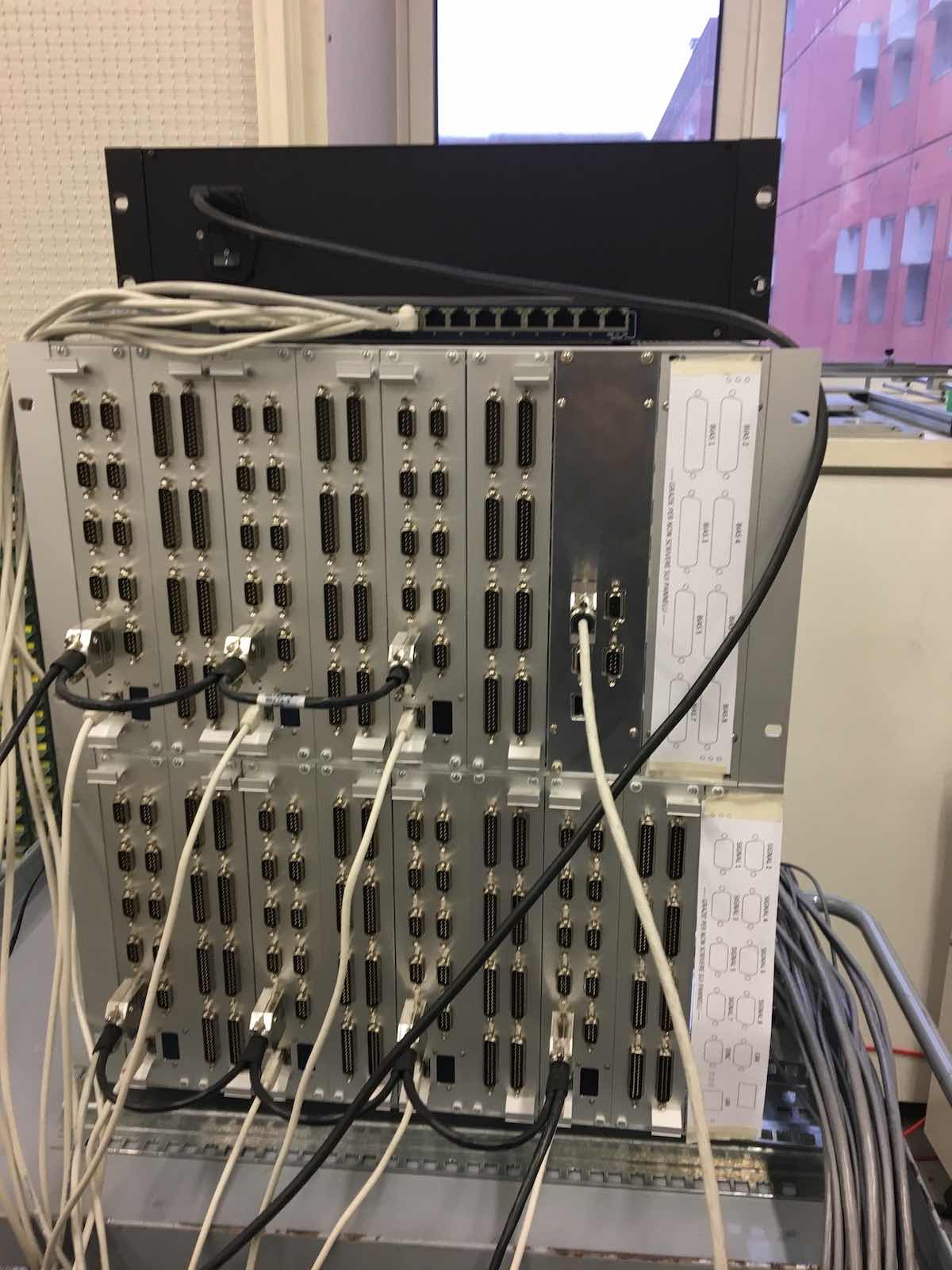}
        \end{tabular}
    \end{center}
    \caption{\label{fig_STRIP_electronics_module}\textit{Left}: one of the 7 modules of \Strip{} electronics composed of one bias board on the right and one data acquisition and logic board (DAQ). Both are connected to a back-plane (in green) for stabilized power supply and data exchange between the two boards. \textit{Right}: the two 6U \SI{19}{inches} racks containing the \Strip{} electronics during the final tests.}
\end{figure}

\paragraph{Electronics.} 
The \Strip{} electronics provides the full biasing and acquisition of the 55 polarimeters on the focal plane. It consists \new{of} 7 pairs of boards that drive and acquire data from 8 polarimeters each. Each pair contains one bias board
and one Data AcQuisition and logic board (DAQ), shown in the left panel of figure~\ref{fig_STRIP_electronics_module}.

The bias voltages are set and monitored by the bias board that controls the HEMT low noise amplifiers (LNAs) and phase switches. 
All the phase switches of all the 7 board pairs are synchronized by a master-clock signal generated and distributed by the GPS and Master-Clock board through a dedicated daisy-chain cable. The bias board can operate the LNAs in open- or closed-loop. In open loop the drain and gate voltages of every transistor are set according to an optimum configuration found during the unit- and system-level tests, and the drain current is simply monitored through the bias house-keeping. In this case bias voltages are susceptible to variations of the focal plane temperature. In closed loop we set the drain voltages and currents, and a completely analogue loop adjusts the gate voltages to keep the desired currents. The closed loop mode is useful in case of excessive temperature instability and its use will be particularly important during the commissioning phase.

The DAQ boards have two functions: they interact with the main computer via telemetry-telecommands and acquire the data generated by the four detectors of each polarimeter. Each board controls 8 polarimeters and receives and stores their bias settings from the main computer via Ethernet network. In this way, 
the operations can autonomously restart in case of communication loss after a black-out. The bias settings are then passed to the bias board. Each DAQ board acquires data from 32 detectors at a rate of \SI{1}{MHz}, demodulates the scientific data at the fast phase switch rate (\SI{4}{kHz}), prepares the data packets with scientific signals, housekeeping data and time tags obtained from the GPS/master clock and sends the data via Ethernet to the main computer for storage. 

A field programmable gate array (FPGA) \new{on the DAQ board} carries out the mathematical operations as well as the digital-to-analog (DAC) and analog-to-digital (ADC) conversions, while a microcontroller handles the communication with the main computer, decodes and routes the commands towards the FPGA and assembles the data packets. The data stream produced by the seven DAQ boards is $\sim\SI{2}{Mb/s}$, well below the maximum Ethernet capability.

The full electronics occupies two 6U \SI{19}{inches} racks (right panel of figure~\ref{fig_STRIP_electronics_module}) that will be positioned close to the dewar and protected by two IP55 grade cabinets.

\subsection{LSPE-SWIPE}
LSPE-SWIPE (Short-Wavelength Instrument for the Polarization Explorer) 
is a mm-wave polarimeter operated onboard a stratospheric balloon. 
The general idea of SWIPE is to use a cryogenic rotating Half-Wave Plate to modulate the incoming polarized radiation and to maximize the sensitivity to CMB  polarization at large scales using a very wide focal plane populated  with multi-moded bolometers. 

The spectral coverage of SWIPE has been optimized to be very sensitive to CMB polarization with one broad-band channel matching the peak of CMB brightness (\SI{145}{GHz}, 
30\% bandwidth), and to be able to monitor and separate the signals from 
interstellar dust (the main polarized foreground at this frequency) 
by means of two ancillary, narrower channels at 210 and \SI{240}{GHz}. These are dedicated to \newb{measuring} the slope of the specific brightness
of interstellar dust.  

The focal planes of SWIPE are large enough that a total of 8800 modes of the incoming radiation are collected by the
multi-moded 326 detectors, 
thus boosting the sensitivity of the polarimeter to unprecedented levels
for such a comparatively low number of detectors. The detectors arrays are cooled to \SI{0.3}{K} by a large wet cryostat, which also cools the polarization modulator and the entire telescope. 

\begin{figure}[!t]
   \centering
   \includegraphics[width=11cm, angle=90]{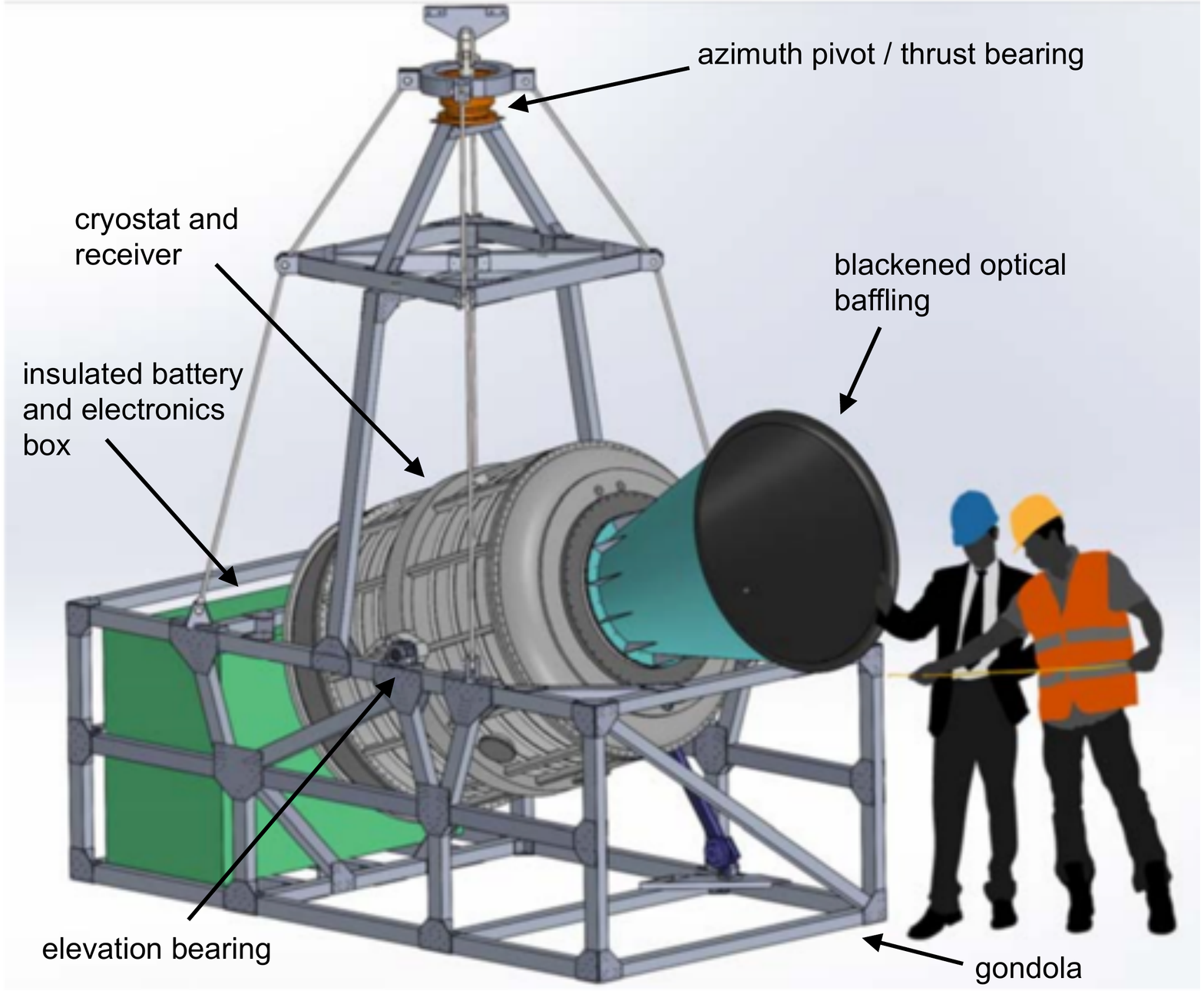}
   \caption{LSPE-SWIPE overview. The instrument is contained in a large 
   liquid Helium cryostat, which also contains the optical elements, including
   the HWP based Stokes polarimeter. The on-board electronics and the 
   Lithium batteries based power system are contained in an Aerogel 
   insulated box, to optimize thermal balance.  
  }
              \label{fig:swipe_overview}%
    \end{figure}

The cryostat is mounted in a frame, the gondola, providing accommodation for an attitude control system, the power system and electronics. The gondola interfaces to the flight train of the stratospheric balloon through an azimuth pivot allowing for azimuth spin and/or scan. 
A general view of the SWIPE instrument is shown in figure~\ref{fig:swipe_overview}.
LSPE-SWIPE measurements are currently scheduled for Winter \newb{2022/23}.

\subsubsection{Winter polar balloon flight}
LSPE-SWIPE is designed to fly on a stratospheric long-duration balloon in the arctic winter. 
Stratospheric balloon  altitudes (about \SI{35}{km} above sea level) are needed to avoid most of the atmospheric emission, which is relevant
at \SI{145}{GHz} and very important at higher frequencies. A winter launch guarantees the possibility to 
exploit the absence of the Sun and cover a large fraction of the sky by spinning the full payload, 
allowing 
\newc{efficient exploration of} 
the CMB polarization anisotropy at large angular scales. It also ensures higher stability of the observing conditions, due both to the thermal stability of the instrument and to the lowest residual turbulence in the atmosphere.

The instrument is designed for a \new{15-day} long flight. 
This long duration is needed to reach the
sensitivity \new{that} matches the scientific goal of the LSPE experiment. 
Options for launching in the polar night are at the moment only possible from the Northern Hemisphere, due to the logistics difficulties related to the access to Antarctic regions during austral winter.
In particular, two possible launching stations are Longyearbyen, in the Svalbard islands (Norway), with a latitude above 78.2$\degr$N, and Kiruna (Sweden) at a latitude of 67.8$\degr$N. 
Several launches have been performed from Longyearbyen, with different balloon and payload sizes, both in Summer and in Winter over the last few years. 
Kiruna offers an established alternative, although at lower latitudes. Stratospheric balloon flights are organized by the Swedish 
Space Corporation in the Esrange Space Center.

\new{
In order to assess the feasibility of winter polar northern hemisphere 
flights, we have developed a trajectory simulator, 
based on the publicly available data from The Research Data Archive (RDA)\footnote{\url{https://rda.ucar.edu/}}, 
managed by the Data Support Section (DSS) of the 
Computational and Information Systems Laboratory (CISL) at National Center for Atmospheric Research (NCAR).
With these data is possible:
(1) to simulate balloon 
trajectories in the past years, for a statistical 
analysis of flight opportunities; 
(2) to predict trajectory in the near future, based on 
a stratospheric wind model, with a prediction 
of 225\,hours in the future; and (3) 
to compare historical predictions and 
historical data, to assess prediction 
reliability. 
   \begin{figure}[!t]
   \centering
   ~~~~\includegraphics[width=.4\textwidth]{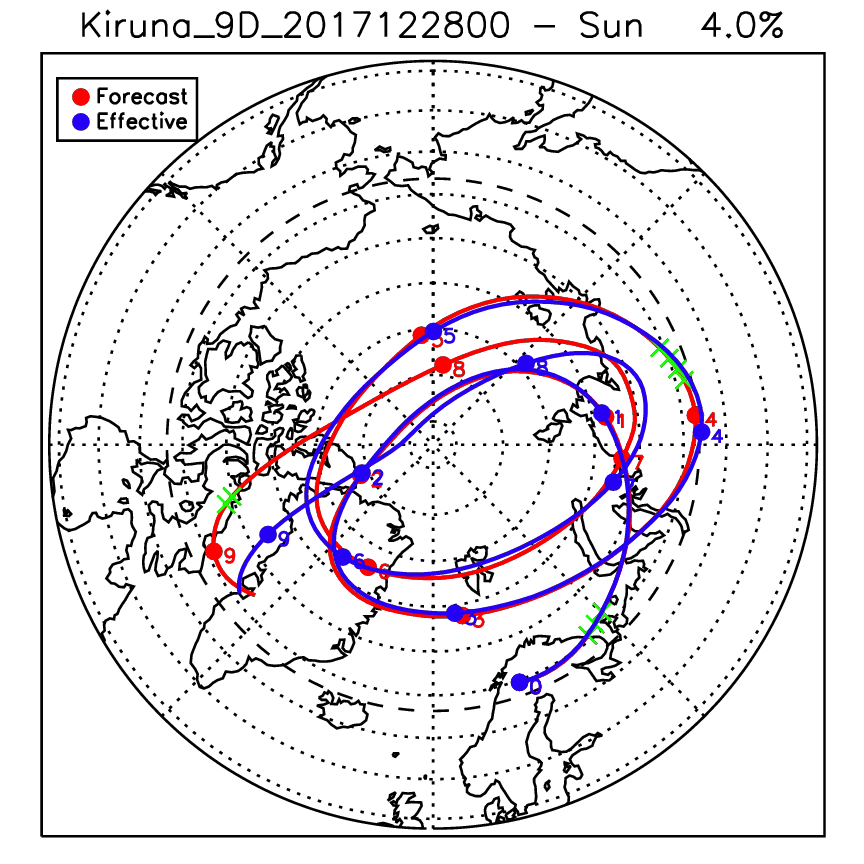}~~~~~~~~~~
   \includegraphics[width=.4\textwidth]{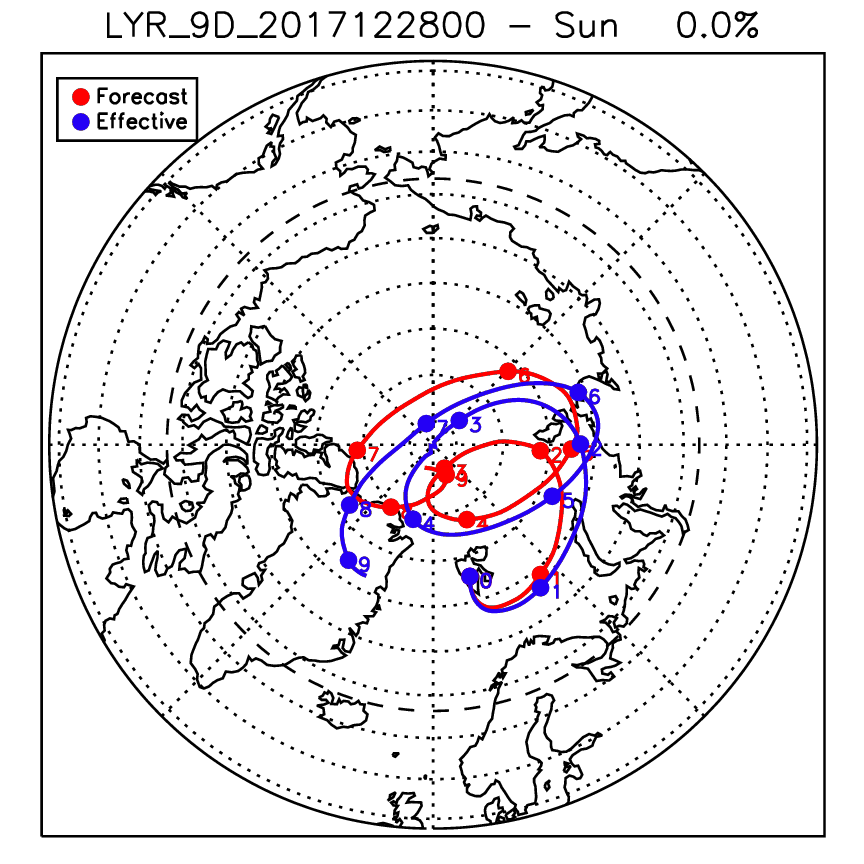}
 \includegraphics[width=.48\textwidth]{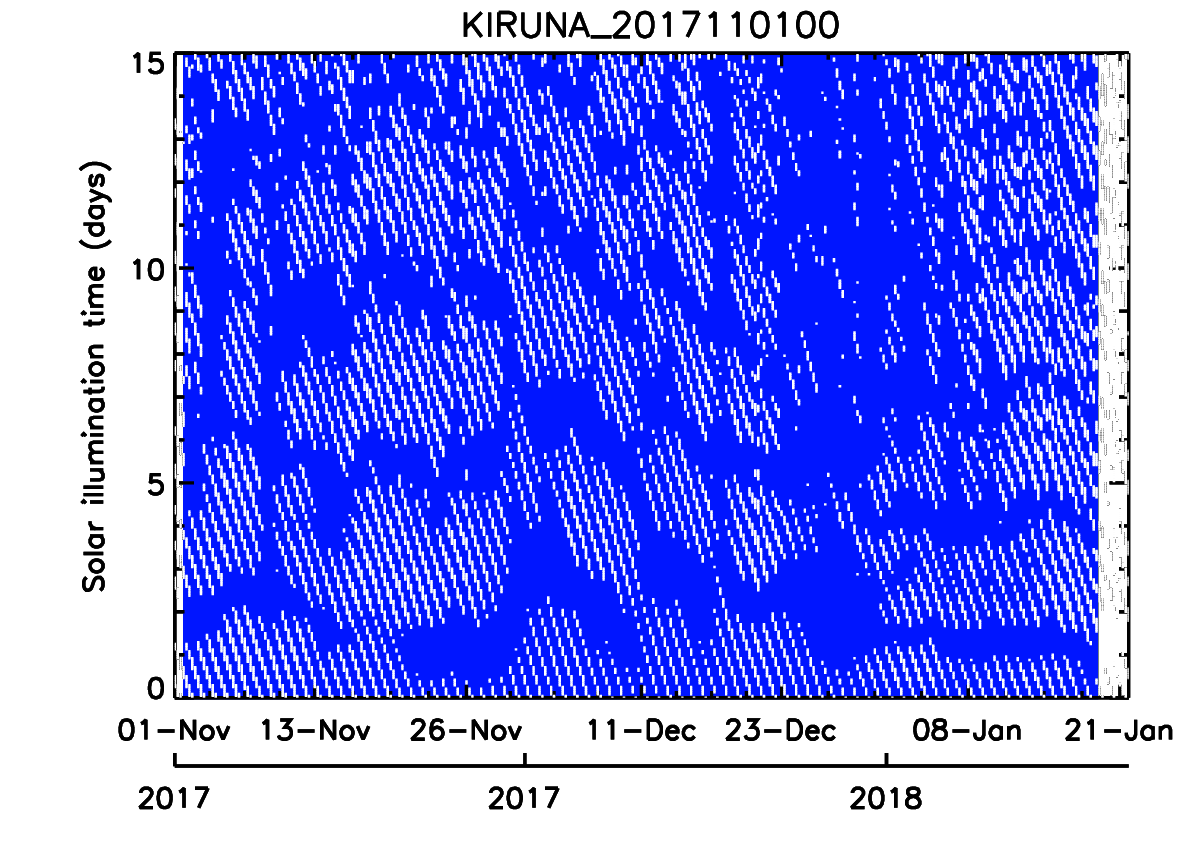}
 \includegraphics[width=.48\textwidth]{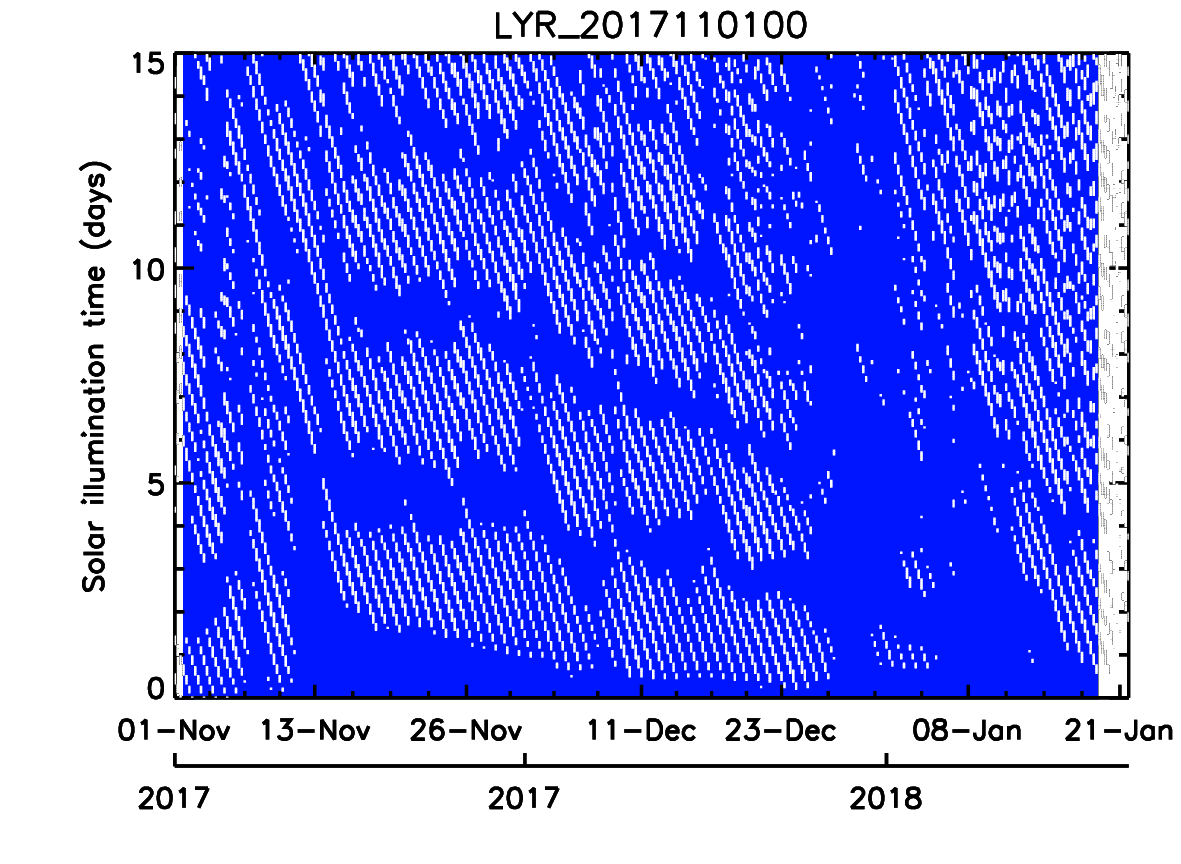}
 \includegraphics[width=.48\textwidth]{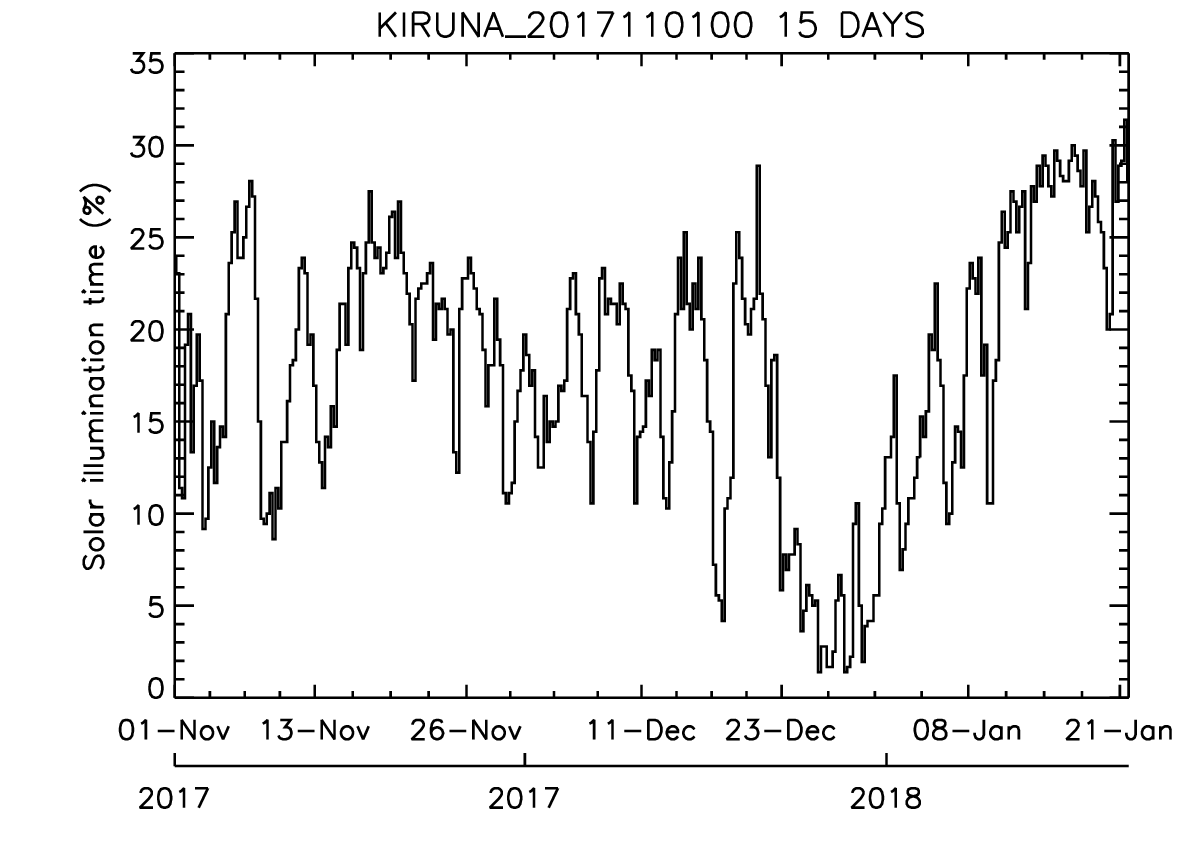}
 \includegraphics[width=.48\textwidth]{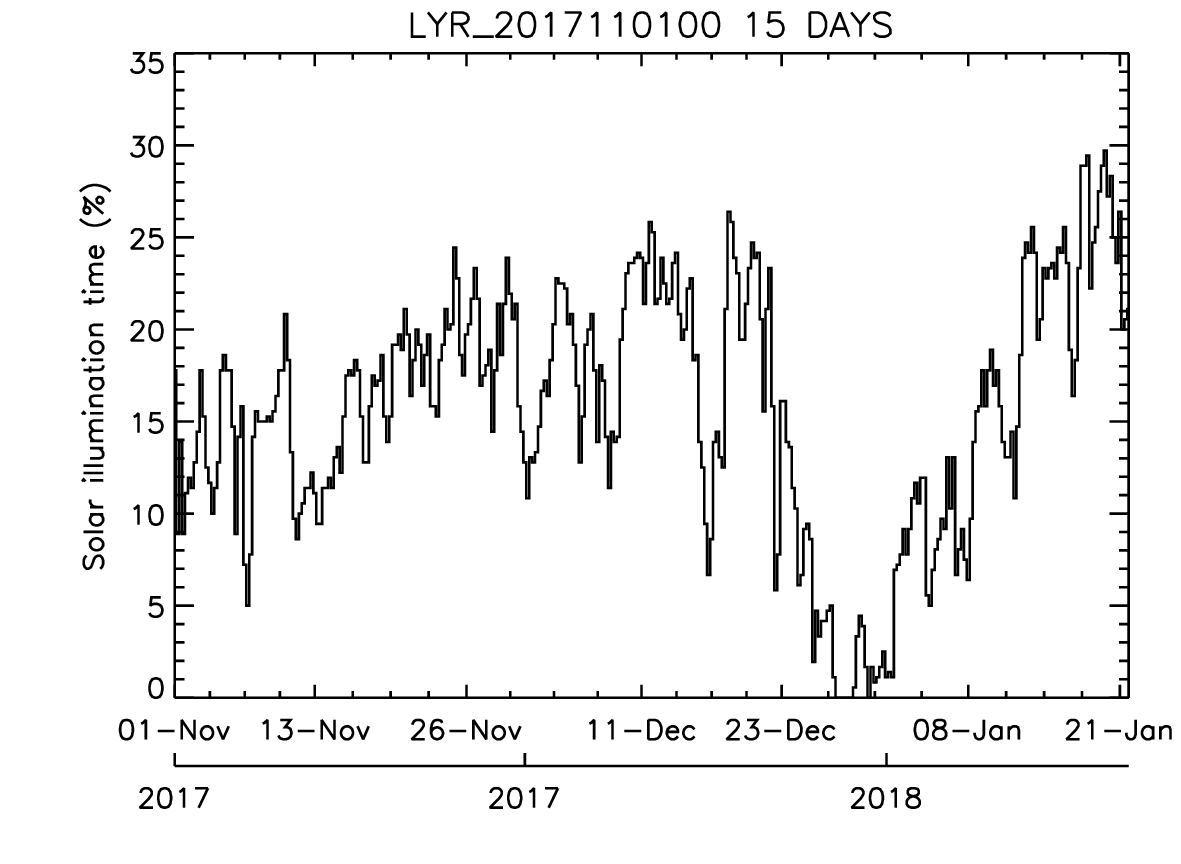}  
   \caption{
\new{   
    Winter polar northern hemisphere trajectories and statistical analysis. Left column plots are for a launch from Kiruna and right plots from Longyearbyen.  
   {\em Top}: in red a simulated trajectory based on 
   historical forecast, for 9 days after a
   launch on December 28, 2017; in blue the simulated trajectory 
   based on historical data, for the same launch day. 
   Dots represent 24\,hours steps.  
   The green dot indicates hours 
   with solar illumination on the payload (Sun higher than $-\SI{4}{^{\circ}}$ above the horizon). 
   The trajectory based on wind prediction is very well reproduced by 
   the trajectory based on real wind data. 
   {\em Center}: the panels show an
   analysis of full winter 2017/18. 
   The abscissa axes indicate the launch date. The 
   ordinate axes indicate the days of flight. For every 
   hour of flight, a blue dot represents an hour in the dark, while a white dot represents an hour with solar illumination.
   A fully blue vertical area indicates a  
   launch day with full flight in the dark. 
   {\em Bottom}: the plots report the fraction of time with 
   solar illumination given a launch day. 
   }
  }
              \label{fig:trajectories}%
    \end{figure}
Figure~\ref{fig:trajectories} illustrates a snapshot 
of trajectories' statistical analysis, that will be 
included in a separate paper.
The simulation tool has also been validated by comparison of predictions with real trajectories, for Summer and  
Winter flights. 
}
\newb{
The payload recovery is essential 
in the case of LSPE-SWIPE, due to the detectors' data-rate 
higher than the possible telemetry rate.
From the top panels of figure~\ref{fig:trajectories}
it is clear that the typical winter trajectory is followed
with a much higher speed with respect to summer 
trajectories. For this reason, the probability 
to have the payload stalled over the ocean is low,
increasing the recovery chances, with unpredictable
difference between the two considered launch sites. 
}

Such a long duration flight in the winter, while being appealing from the scientific point of view, 
is very demanding in terms of power system and thermal balance. 
A series of technological test flights has been carried out over the years, as reported in 
\citep{iarocci2008,peterzen_memsait2008,2019EPJWC_MASI,debernardisIAUS2013,wipica2018,}.  
All the LSPE-SWIPE parts are designed to cope with temperatures as low as \SI{-90}{\celsius}, except the battery pack and part of the electronics, which are contained in a thermally insulated box. 

\subsubsection{Power supply}

For a long-duration night-time flight, a relatively cheap, consolidated, high energy-density power-supply solution is based on lithium batteries. The total power budget of the SWIPE instrument is $\sim \SI{370}{W}$, and the energy necessary for the entire mission is $\sim\SI{0.48}{GJ}$. 
This is stored in a stack of $\sim$ 3500 cells (each \SI{14}{Ah} @ \SI{3.3}{V}). 
Due to the low internal resistance of these cells, and the fact that their capacity decreases at low temperatures, it is necessary to keep the cells warm (at a temperature $>\SI{0}{\celsius}$) during the flight. 
This is obtained \new{by} hosting the batteries in the same box hosting the electronics of the experiment, and in good thermal contact. 
The box is insulated from the cold external environment by a blanket made of three layers of metal reflective foil separated by two thick ($\sim\SI{2.5}{cm}$) layers of aerogel. According to the thermal model, with \SI{200}{W} of power dissipated in the electronics inside the box, and an external temperature of \SI{200}{K}, the internal temperature is maintained at \SI{280}{K}. A prototype of this power and thermal insulation system was flown in a winter arctic balloon in December 2017
\cite{wipica2018}, 
and further tests are planned for the future. 

\subsubsection{Gondola and pointing system}

The gondola is a simple riveted frame of aluminum beams, hosting all the components of the payload and of the flight system, and structurally optimized to withstand an acceleration of 10\,g (${\rm g}={\rm gravitational}$ acceleration) at the opening of the parachute after the flight termination. 
The telescope attitude is controlled 
by the attitude control system (ACS). Its main purpose is to spin in azimuth the entire gondola. The azimuth pivot separates the payload from the flight chain, and is based on thrust bearings and a torque motor. The motor torques directly against the flight chain, to obtain an azimuth spin rate up to $\SI{10}{^{\circ}/s}$,
\new{
much faster than the nominal rate 
of $\SI{0.7}{^{\circ}/s}$
}.

Mechanically and electrically, the system is very similar to the ones used in ARGO \cite{1993AA...271..683D}, BOOMERanG \cite{2003ApJS..148..527C, 2006AA...458..687M}, Archeops \cite{2002APh....17..101B}, OLIMPO \cite{2005ESASP.590..581M, Paiella_2019, 2019JCAP...07..003M}, and described in detail in  \cite{1990SPIE.1341...58B, 1990SPIE.1304..127B, 1994MeScT...5..190B}. 
Given the measured friction of the thrust ball bearing, we expect to use up to $\sim\SI{70}{W}$ to rotate the $\sim\SI{2000}{kg}$ payload at the $\SI{10}{^{\circ}/s}$ scan speed. 
The azimuth speed is sensed by a laser-gyroscope, the signal of which is compared to the desired spin rate, in a feedback loop controlling the current in the torque motor. The elevation of the boresight can be changed by tilting the entire cryostat, using a geared DC motor driving a linear actuator with linear recirculating ball bearing. 
The pointing reconstruction is based on a high altitude GPS receiver to obtain geographical coordinates and on two orthogonal fast star sensors \cite{2003RScI...74.4169N}, the same successfully used for the Archeops flight \cite{2007AA...467.1313M}, for the celestial coordinates of the boresight. The system allows for pointing reconstruction with $\sim$ arcmin accuracy.

\subsubsection{Cryostat}
SWIPE makes use of a custom-designed main cryostat with a bath of \SI{250}{liters} of superfluid helium, connected to the external low-pressure environment to operate at \SI{1.6}{K}. The cryostat shell, the internal shields and the LHe tank are all made of aluminum alloys, to reduce their mass, as developed for the cryostats used in the ARGO \cite{1994Cryo...34.1001P}, BOOMERanG \cite{1999Cryo...39..217M}, PILOT \cite{2016ExA....42..199B} and OLIMPO \cite{Coppo2020}  balloon-borne instruments. 
Two vapor-cooled intermediate shields, separated by super-insulation blankets, are used to minimize the radiative heat load on the LHe bath. The main cryostat provides the base temperature to cool down the polarization modulator and the optical system, and to operate a $^3$He evaporator \cite{2016SPIE.9912E..65C}. The latter cools down to \SI{0.3}{K} the two focal plane arrays, 
as required to operate the SWIPE bolometric detectors.
The hold time forecast for the LHe in the main cryostat is $\sim\SI{20}{days}$, while the $^3$He refrigerator has a hold time of $\sim\SI{7}{days}$, and can be recycled in flight. 
In order to minimize the radiative load on the detectors, the 600\,mm diameter  window has been designed in a similar way as the one used by the EBEX group \citep{2017zilic}, and, less recently, in \citep{1983IAUS..104..135D} and in \citep{macias2007}. In practice, a thick UHMWPE \citep{DAlessandro:UHMW} window used for laboratory tests is removed at float, leaving only a very thin ($\ll \lambda$) Mylar window to withstand the small pressure difference between the cryostat vacuum and the stratospheric pressure. 
The thick window also implements a highly reflective filter to operate the receiver on the ground under radiative loadings representative of the stratospheric environment. Just before the termination of the flight, the motor unit is remotely operated again to put the thick window back in place for a relatively safe receiver landing. 

\subsubsection{Optical system}\label{sec:swipe_optics}
\begin{figure}[!t]
   \centering
   \includegraphics[width=10cm]{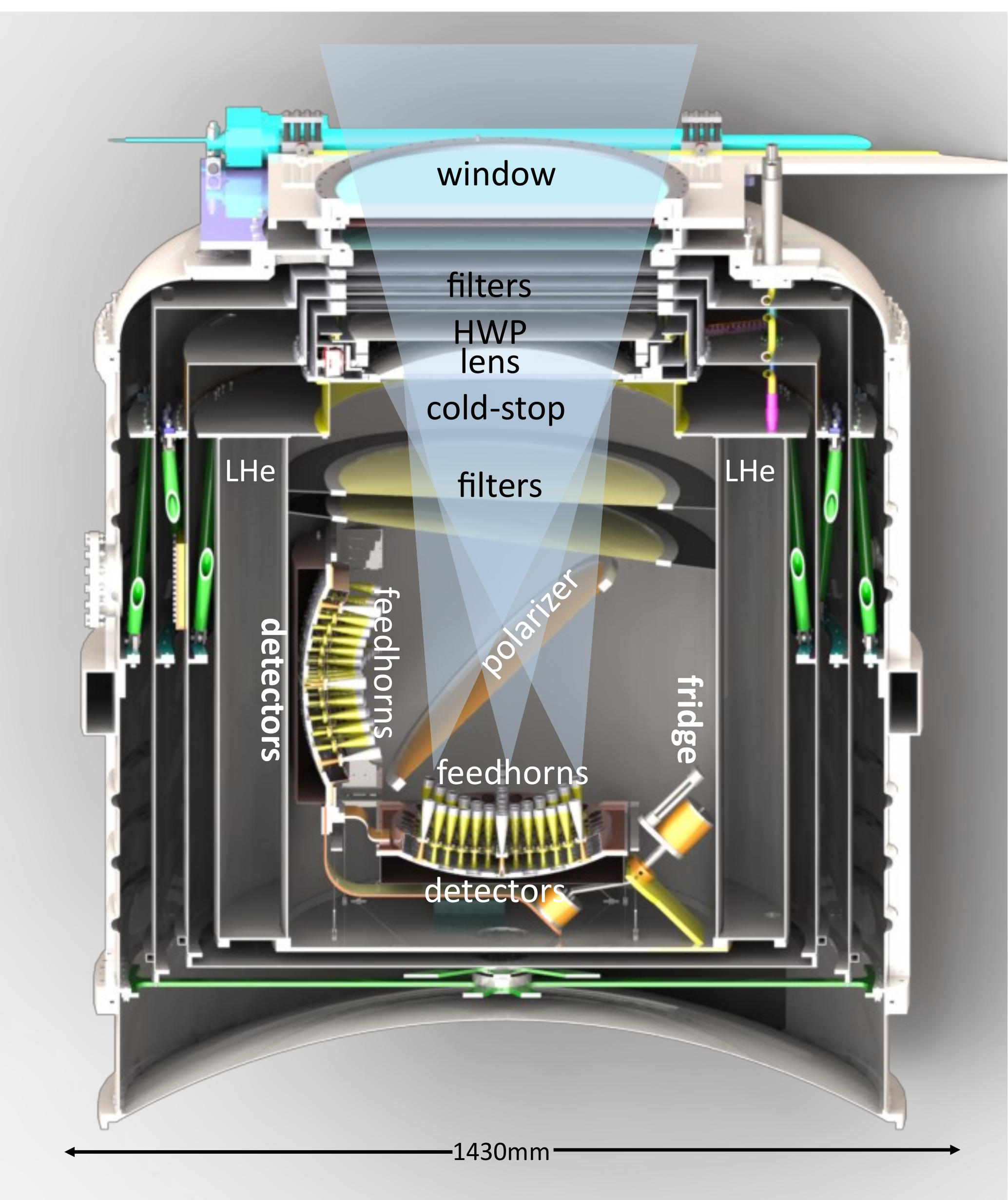}
   \caption{LSPE-SWIPE cryostat and optical system.
   Radiation enters from the top 
   window, passes across filters, 
   HWP, lens, \new{cold-stop}, and other filters.
   It then is split by the large wire grid and collected in the 
   two curved focal planes.
   } \label{fig:swipe_optics}%
    \end{figure}

The optical system of LSPE-SWIPE (figure~\ref{fig:swipe_optics}) consists in a single-lens, 
\SI{490}{mm} aperture refractor telescope, focusing incoming radiation 
on two large curved focal planes, split by a large wire grid (WG) polarizer. \new{Polarization modulation is obtained by a cryogenic \SI{500}{mm} wide rotating Half-Wave Plate which is placed skywards of the lens, below the window and the warm thermal filters.} 
\new{
The large plano-convex lens, attached to the 1.6\,K stage of the cryostat, 
is made of High Density Polyethylene (HDPE), ensuring a very good transmittance across the bands and limited dielectric losses at high frequencies. Typical dielectric properties of HDPE are a refractive index $n=1.52$ 
and a loss tangent of $\sim 5\times 10^{-4}$. The baseline optical design is based on these numbers, but further optimization will be performed after characterizing at low temperatures a
sample from the same batch that will be actually used to build the lens.
A layer of anti-reflection coating based on porous PTFE will be deployed on the lens surfaces in order to minimize reflection losses.
Full-aperture IR-blocking filters are arranged on each available thermal sink along the path from the window to the lens. Two more such filters are placed at 1.6\,K 
on the path to the focal plane directly below the lens. These are designed to cut most of the radiation emitted out of band by the HWP, its rotation mechanism and the lens itself. 
The final stage of spectral selection and band refinement is performed directly on the $300$\,mK stage, where small-aperture packs of bandpass and low-pass filters are mounted on the mouths of the pixel horns.}
Each focal plane is populated with 163 multi-moded horn antennas, 
each feeding a spider-web Transition Edge Sensor (TES) 
bolometer. 
    
The configuration fulfills our requirements with a low cross-polarization ($<1\%$) and a controlled instrumental polarization  
(including an absorption component $<0.04\%$ 
and an extremely low, and stable, emitted component by the use of a cold telescope). 
These \new{values} are reached at the edge of the corrected focal plane for all the 3 bands and they are negligible on axis. 
Besides the 490\,mm diameter lens \new{the system adopts a 460\,mm 
diameter cold aperture stop close to the lens (corresponding to an entrance pupil of 487\,mm diameter) and located on the opposite side of the lens with respect to the polarization modulator.}
The FOV, 20$^\circ$ wide, is split by a \SI{500}{mm} diameter 45$^\circ$ tilted wire grid in 2 curved focal planes (CFP\_T and \_R) \SI{300}{mm} in diameter, \new{with a resulting focal ratio of 1.75}.
The full optical system is kept at cryogenic temperature in the LSPE-SWIPE cryostat, in order to minimize its radiative loading on the detectors and to mitigate the signals due to the \new{thermal emittance of the rotating HWP}. 

\new{The design rationale of the SWIPE focal plane is based on a trade-off between the sensitivity and the angular resolution of the instrument, by trying to maximize the power collection efficiency of each detector at the selected resolution. This requirement clearly sets a constraint for the size of the focal plane region which must be covered by a single detector, and ultimately determines the collection area of an individual sensitive element of the receiver. 
In order to further improve the pixel efficiency, detectors are coupled to their corresponding focal plane pixel through multi-moded feedhorns \citep{legg2016}: large-aperture smooth-walled conical horn antennas feed the detectors by matching freespace radiation to multi-moded circular waveguides located at the horn throat. The waveguides select a frequency-dependent number of propagating modes (i.e. solutions to the propagation problem as constrained by the boundary conditions set by the waveguide geometry) so that for any given geometrical aperture of the horn, higher-order waves contribute in shaping the beam response of the horns. This results in a more flat-top (and broader) beam profile, with an overall higher illumination efficiency of the system pupil, and therefore a higher pixel throughput within the portion of its field of view which is used to collect radiation from the sky.}

\begin{figure}[!t]
   \centering
    \begin{tabular}{cc}
    \includegraphics[width=7.3cm]{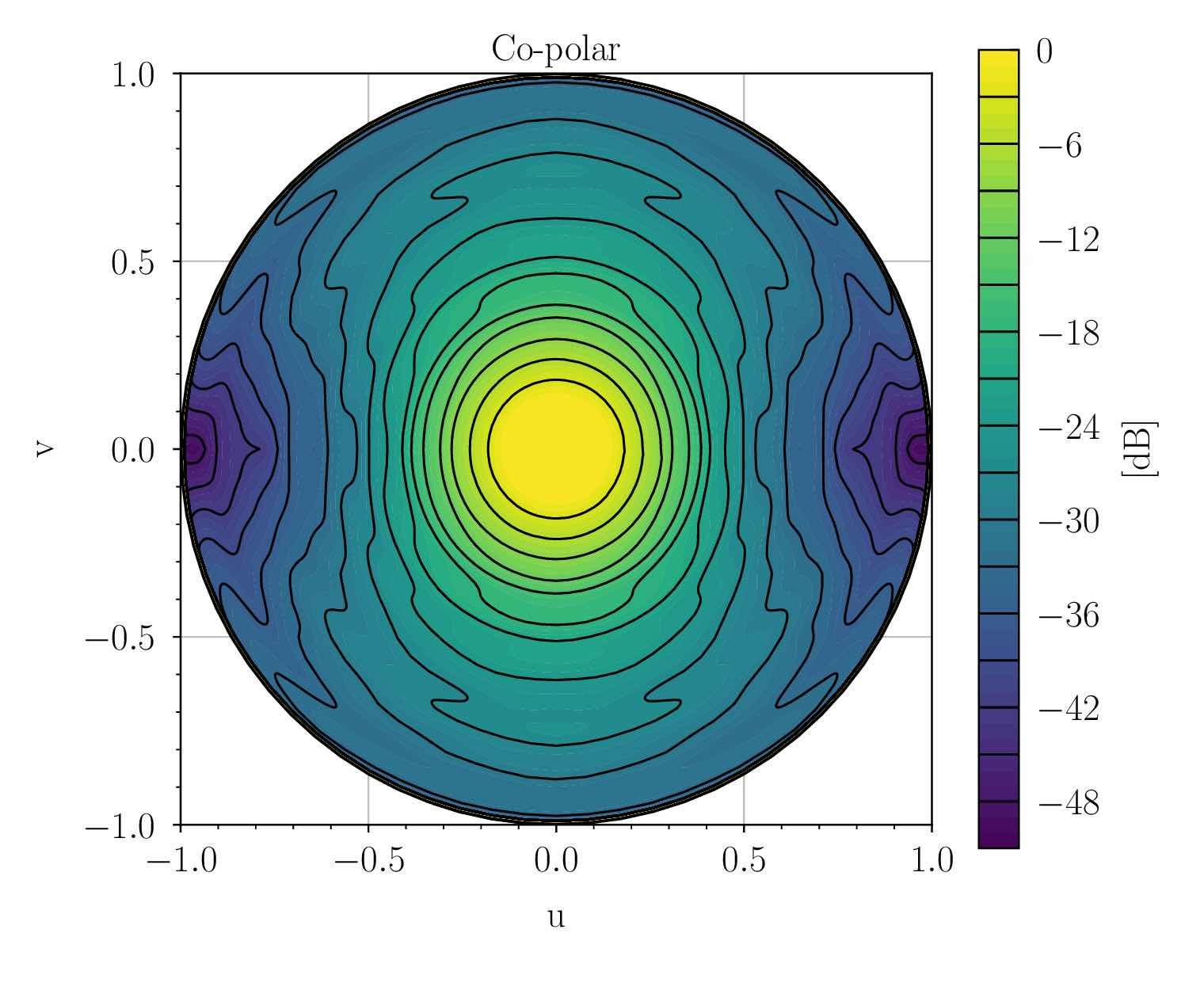}  
    &
    \includegraphics[width=7.3cm]{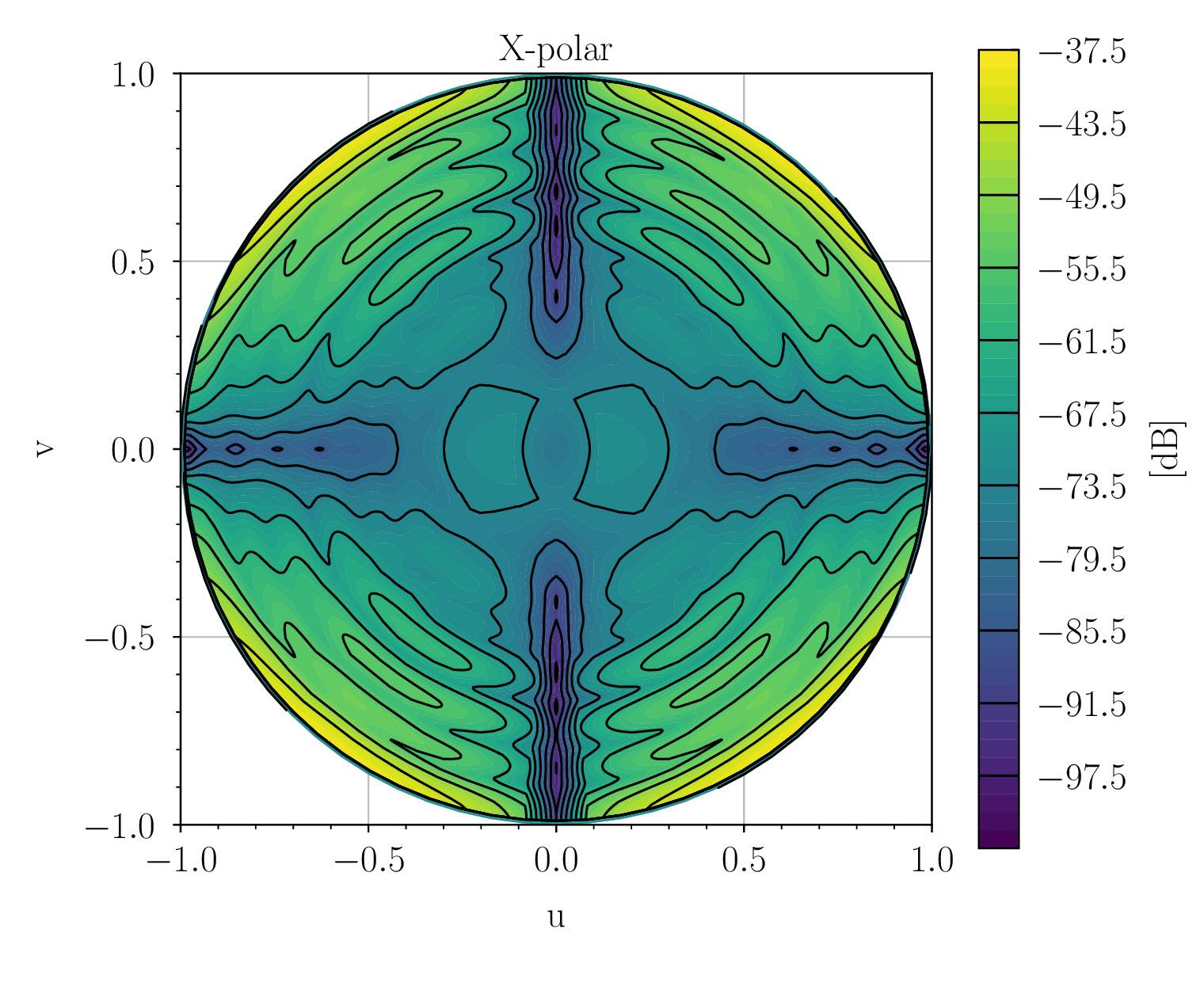}\\
    \includegraphics[width=7.3cm]{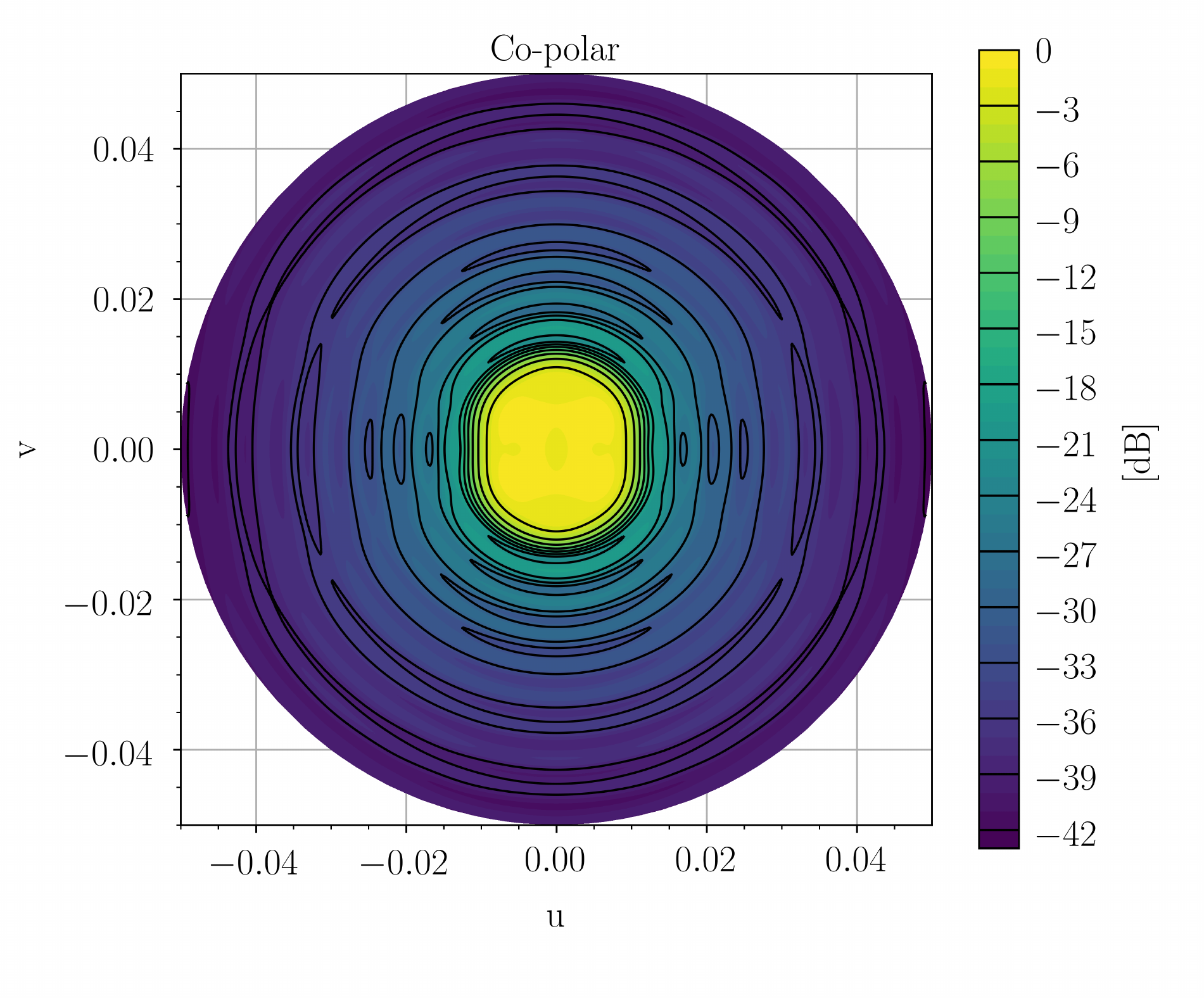}  
    &
    \includegraphics[width=7.3cm]{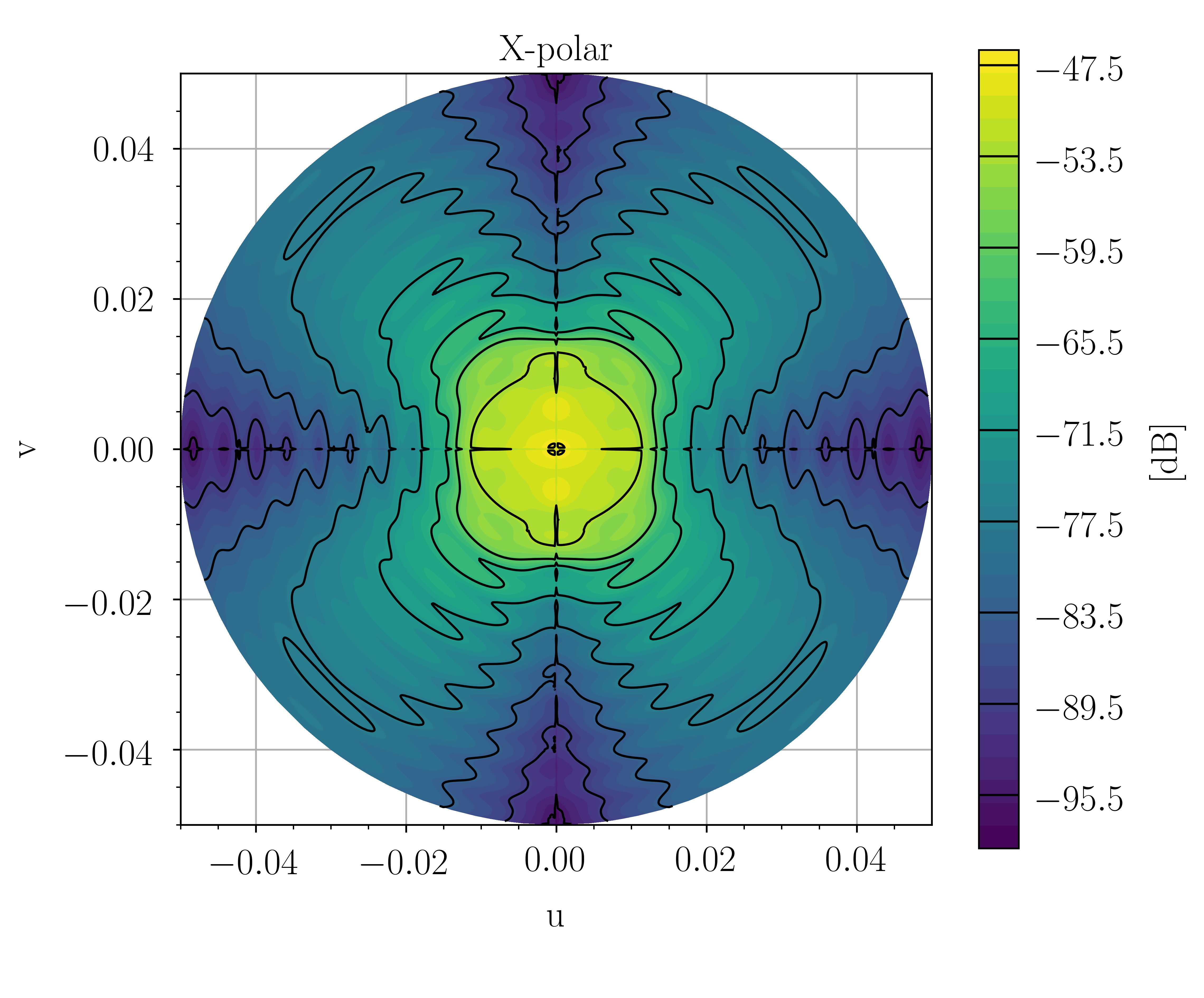}\\
    \end{tabular}
   \caption{\new{{\it Top panels: }} simulated beam response of the SWIPE multi-moded horns. Radiation is propagated 
   into the far field with linear polarization parallel to the $v$ axis in the plane of the horn aperture, corresponding to the $\phi=\SI{90}{\degree}$ axis in the far field. 
   \new{{\it Bottom panels: }simple physical optics simulation of the SWIPE telescope main beam, for a pixel at the center of the focal plane. The simulation includes the feed pattern shown above, the aperture stop and the HDPE lens. A perfectly absorbing tube and no further obstructions apart from the cold aperture stop have been assumed.}
   Co-polar and cross-polar patterns have been calculated according to Ludwig's third definition in \cite{ludwig} and are mapped 
   \new{as a function of} $u=\sin(\theta)\cos(\phi)$ and $v=\sin(\theta)\sin(\phi)$, \newb{where 
   $(\theta,\phi)$ are the standard spherical coordinates, and 
   the telescope bore-sight is pointing at $\theta=0$}. Contours are shown every \SI{3}{dB} for the co-polar patterns and every \SI{6}{dB} for the cross-polar patterns.} \label{fig:swipe_horn_beams}
\end{figure}

Under the assumption that radiation detection is based on purely incoherent processes on the detector absorber, the phase relation
among the coupled modes is not relevant to determine the coupling efficiency. 
Therefore, electromagnetic modeling of the horn-waveguide assembly can be easily performed by solving one reverse-propagation problem per each of the coupled modes selected by the waveguide. 
A far-field calculation of the field solution at the horn aperture then yields the individual contribution of each mode to the horn response, and the full multi-moded response is then computed 
as a power summation over the coupled modes.
This operation has been performed through the Ansys HFSS\footnote{\url{https://www.ansys.com}} software, and the calculated beam profile for the SWIPE horns is shown in figure~\ref{fig:swipe_horn_beams}. Here the contributions from the individual modes have been evenly weighted, as expected under energy equipartition conditions, and confirmed by numerical simulation of the absorber/cavity sub-system (see section~\ref{sec:detectors}).
A measurement of the feed angular response
is reported in \cite{Columbro_LSPE_feed_measure}.

Integration of the numerically evaluated profile times the horn effective area yields a value very close to $A_\text{eff}\Omega_\text{tot}=N_\text{modes}(r_\text{wg},\nu) \lambda^2$, where $N_\text{modes}(r_\text{wg},\nu)$ is the number of propagating wave solutions (i.e. modes with imaginary wavenumber) in a cylindrical waveguide of radius $r_\text{wg}$ at frequency $\nu$, and $\lambda$ is the free-space wavelength of monochromatic radiation. This result is expected \new{in the few-modes regime} and under equipartition conditions, where each coupled mode provides \new{the same fraction of the total working throughput}. 
In addition, since we use a full-field polarizer to split polarization in two independent focal planes, the polarization properties of the individual pixel assembly
are irrelevant for the end-to-end performance evaluation. 
Therefore, no concern arises due to the co-polar and cross-polar response behavior of the horns.

In order to simplify the design and production cycle of the horns, no additional optimization is performed at the pixel level. 
Instead, further suppression of power at large angles from the sky is obtained by heavily over-illuminating 
the cold aperture stop (with an edge taper of \SI{-10}{dB} at \SI{145}{GHz}). 
The multi-moded beam of each horn thus ensures a very uniform illumination pattern of the telescope lens, maximizing the aperture efficiency of the telescope, while unwanted power pickup in the horn sidelobes is mitigated through implementation of cold, stable, highly absorptive surfaces inside the telescope tube. Additional large-angle pickup due to strong beam truncation at the aperture will be mitigated through an absorptive external baffle.

This multi-moded approach ensures an optimal trade-off between the need for a conspicous number of independent focal plane elements and the net sensitivity of the individual pixels. This comes at the price of a lower angular resolution of the receiver, which is acceptable since the main observational target of SWIPE is polarization detection at large scales, from $\sim\ang{2}$ to one third of the full sky. 

\subsubsection{Polarization modulator}

In order to modulate the polarized component of the signal,
LSPE-SWIPE adopts a 
Stokes polarimeter based on a Half-Wave Plate built 
of metal mesh metamaterials. This technology has been 
developed by the Astronomy Instrumentation Group 
at the Department of Physics and Astronomy of the Cardiff University~\cite{Pisano1}.
 The mesh HWP consists of anisotropic metal grids, stacked together and embedded into polypropylene, which mimic the behaviour of a birefringent plate~\cite{Pisano2,pisano3}. 
The geometry and the spacing of the grids are chosen in such a way to provide high in-band transmission (above 95\%) 
and high polarization modulation efficiency (at 98\% level) across all the bands.

Due to the requirements of cryogenic temperature and continuous rotation of the HWP, we selected a superconducting magnetic bearing (SMB) \citep{Matsumura2016,Johnson2017,2020_Columbro_SPIE} as the technology to spin the HWP and to modulate the polarized signal at a sufficiently high rate 
(\new{the nominal values for SWIPE are \SI{0.5}{Hz} for 
HWP spin rate and 
\SI{2}{Hz} modulation rate, as 
derived in appendix \ref{app:swipe_scanning_params}} ). 
An innovative frictionless clamp/release device \citep{Columbro2018}, based on electromagnetic actuators, keeps the rotor in position at room temperature, and releases it below the superconductive transition temperature, when magnetic levitation works properly.
A simple method to measure the temperature and levitation height of the HWP rotating at cryogenic temperatures was 
developed specifically for LSPE-SWIPE~\cite{PdB_levitation_measurement2020}.

   \begin{figure}[t]
   \centering
   \includegraphics[width=5.5cm]{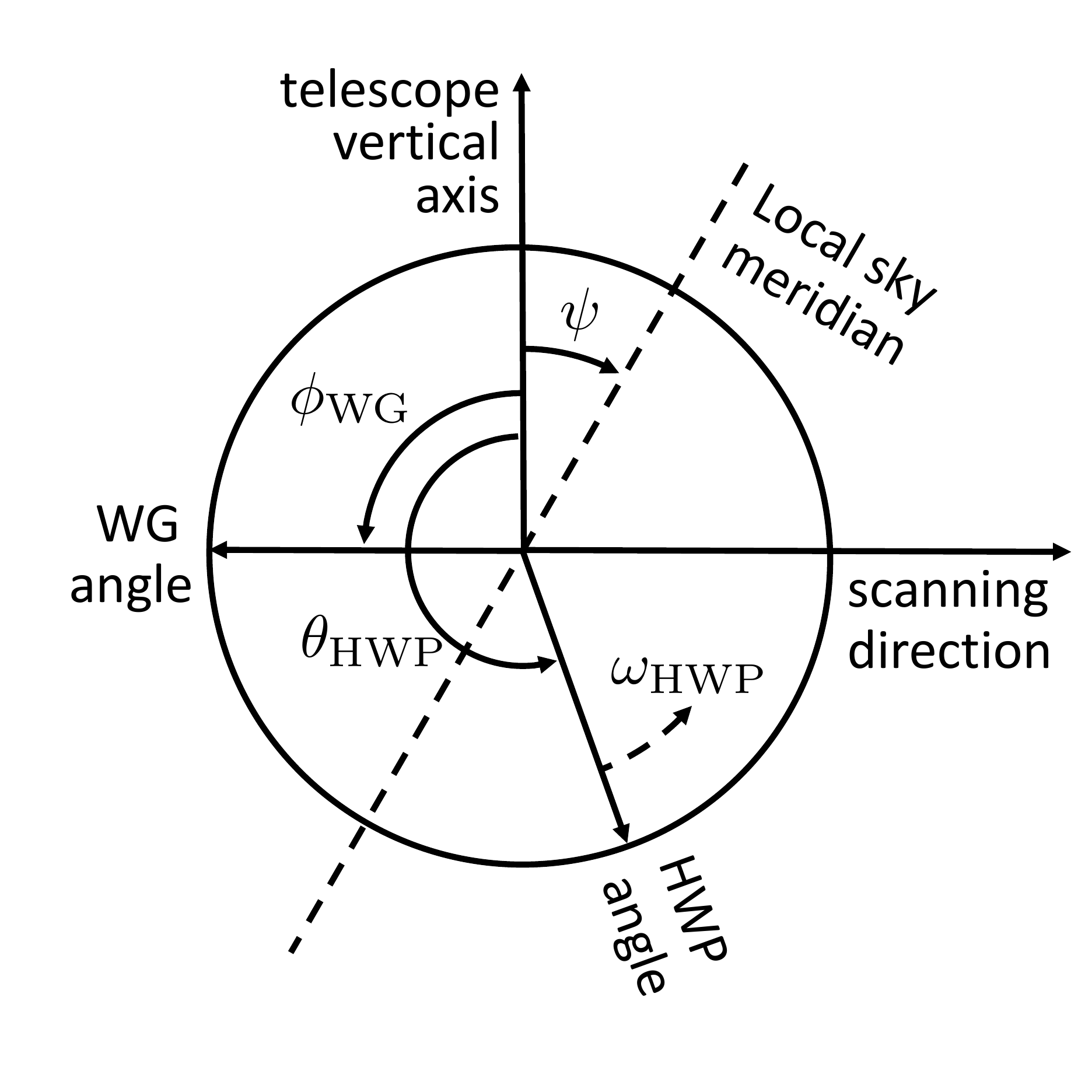}
   \caption{
   LSPE-SWIPE Stokes polarimeter angles as seen from the boresight.
   The dashed line is the instantaneous local sky meridian; the telescope vertical
   axis is tilted by an angle $\psi$ with 
   respect to the sky meridian, and is orthogonal 
   to the scanning direction. The wire grid inside the cryostat is oriented at 
   an angle $\phi_\text{WG}$ orthogonal to the vertical axis; 
   the HWP is spinning with angular velocity $\omega_\text{HWP}$, and 
   forms an angle $\theta_\text{HWP} = \omega_\text{HWP}t$ with the telescope vertical
   axis. 
   \label{fig:swipe_earth}
   }
              \label{fig:swipe_observing}%
    \end{figure}

\new{In an ideal Stokes polarimeter,}
the power hitting the detector 
can be computed as the first element of the Stokes vector obtained from the combination of Mueller matrices, taking into account both the rotating HWP and the WG polarizer:
$$
S_\text{out}=
M_\text{rot}^{-1}(\phi_\text{WG})
M_\text{WG}
M_\text{rot}(\phi_\text{WG})
M_\text{rot}^{-1}(\theta_\text{HWP})
M_\text{HWP}
M_\text{rot}(\theta_\text{HWP})
M_\text{rot}(\psi)
S_\text{sky}
$$
where
$S_\text{sky}$ is the Stokes vector $(I, Q, U, V)$ of the observed direction in the sky;
$M_\text{rot}$ is the rotation Mueller matrix;
$M_\text{HWP}$ is the HWP Mueller matrix;
$M_\text{WG}$ is the wire grid Mueller matrix;
$\psi$ is the angle between the local meridian and the telescope
vertical axis;
$\theta_\text{HWP}=\omega_\text{HWP}t$ is the rotation angle,
with respect to telescope vertical axis,
of the HWP which rotates with $\omega_\text{HWP}$
angular velocity;
$\phi_\text{WG}$ is the wire grid rotation angle with respect to 
telescope vertical axis (\ang{0} or \ang{90} in the case of SWIPE, 
for reflected and transmitted radiation); and 
$S_\text{out}$ is the resulting Stokes vector, of which the $I_\text{out}$ term
is the power hitting the detector.
Figure~\ref{fig:swipe_observing} illustrates the angles definition. 
Expanding the equation, we have
\begin{equation}\label{eq:swipe_modulation}
I_\text{out} = 
\frac 12
\left(
A I_\text{sky}(\hat n(t)) + 
B Q_\text{sky}(\hat n(t))+
C U_\text{sky}(\hat n(t))
\right)
\end{equation}
with
\begin{eqnarray*}
A &=&1\\
B &=&\cos(4 \omega_\text{HWP}t + 2 (\psi(t)-\phi_\text{WG}))\\
C &=&\sin(4 \omega_\text{HWP}t + 2 (\psi(t)-\phi_\text{WG}))
\end{eqnarray*}
where $\hat n(t)$ is the observed 
direction, and we have made explicit the time dependence. 

\new{
The HWP angular velocity is constrained by the detectors' time constant, while the payload scanning speed \newb{is constrained} by the telescope angular
response. 
The derivation of baseline parameter for LSPE-SWIPE is described in Appendix~\ref{app:swipe_scanning_params}, and the 
results are reported in table~\ref{tab:swipe-obspar}.
Notably, the scanning speed is 
$\omega_\text{payload} \simeq \SI{0.7}{\degree.s^{-1}}$ and the 
HWP angular velocity is
$\omega_\text{HWP}=\SI{3.14}{rad.s^{-1}}$.
}

   \begin{figure}[t!]
   \centering
   \includegraphics[width=.39\textwidth]{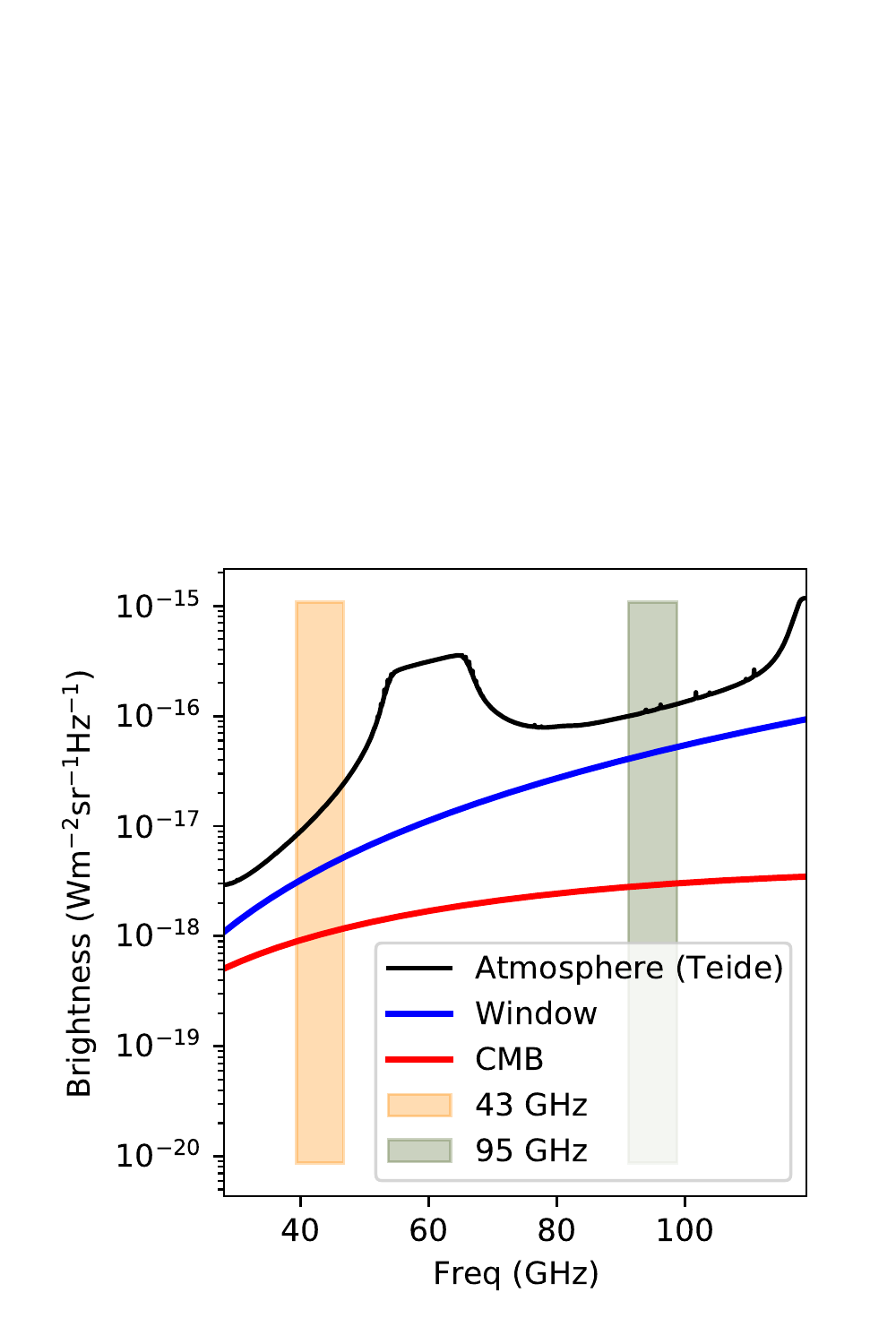}
   \includegraphics[width=.59\textwidth]{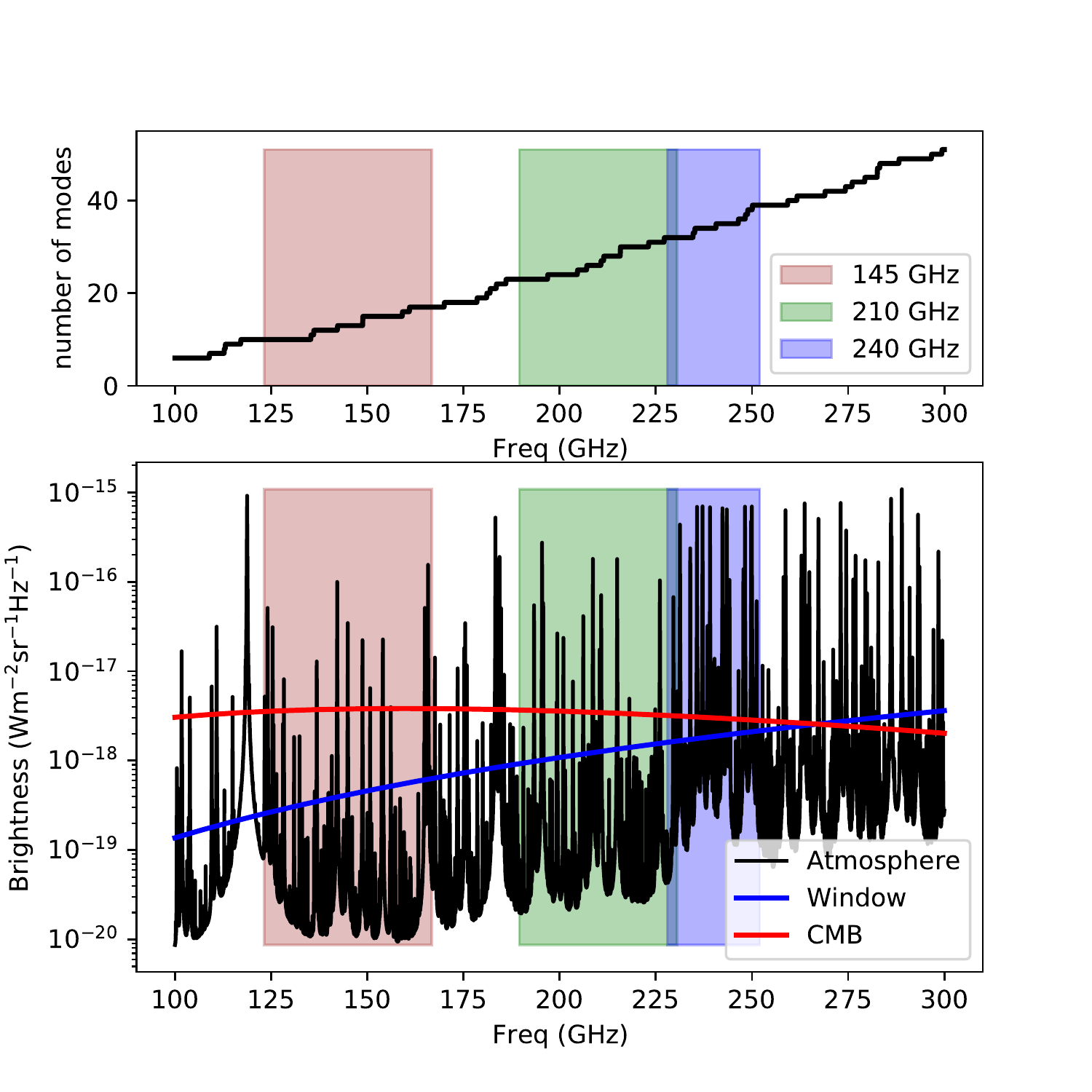}
   \caption{
   \emph{Left}: \new{LSPE-\Strip\ bands and expected brightness from 
   atmosphere, cryostat window and CMB}. 
   \emph{Top-right}: LSPE-SWIPE number of modes coupled to 
   the multi-moded optical assembly, as function of 
   frequency. The three bands centered at 145, 210, \SI{240}{GHz} are also shown. 
   \emph{Bottom-right}: LSPE-SWIPE bands and expected brightness from 
   atmosphere \new{at balloon altitude}, cryostat window and CMB. \new{For both \Strip\ and SWIPE the brightness 
   values don't include telescope and coupling efficiency.}
              \label{fig:swipe_bands}}%
    \end{figure}
   
\subsubsection{Detectors}\label{sec:detectors}

LSPE-SWIPE adopts TES detectors. 
In order to take advantage of the multi-moded 
coupling, radiative power propagated from the feedhorns into the mode-filtering waveguides must be absorbed by the detector with 
\new{as low an impedance mismatch as possible
} for all the propagated modes. One way to fulfill this requirement is to compress the effective wavelengths of the coupled modes into a narrower bandwidth by progressively re-enlarging the waveguide cross-section into a larger terminated cavity (flared waveguide), where a 15\,mm large spider-web absorber collects the power for detection by the TES. 
This solution has been validated through HFSS, providing a mode-dependent, frequency-dependent $S_{11}$ scattering parameter\footnote{Input port reflection coefficient.}
evaluation of the pixel assembly along the path from the waveguide to the absorber. The relative $S_{11}$-parameter dispersion for the \SI{150}{GHz} band is about 2\% over the coupled modes and frequencies, with an average return loss of $-\SI{22.6}{dB}$ when the cavity termination is set to a quarter of the average free-space wavelength of the band collected by the detector, and the absorber impedance is $\sim\SI{300}{\ohm}$. 
This result, to be validated also through experimental verification of the pixel performance, is used here to support the hypothesis that the main impact of the broadband performance evaluation for SWIPE is the variable number of modes $N_\text{modes}(r_\text{wg}, \nu)$ coupled by the waveguide 
when fed with broadband radiation. Figure~\ref{fig:swipe_bands} illustrates the 
coupled modes as a function of frequency, and the selected bands; in the bottom-right panel, it shows the power \new{entering} the system, 
with contribution from the CMB, the atmosphere and the cryostat window \new{(for \Strip\ in 
the bottom-left panel)}.
These are the input to the noise calculation analysis reported in section~\ref{sec:SWIPE_noise} and 
in table~\ref{tab:swipe-noise}.\\

The TES bolometer is a single Si chip $15\times\SI{15}{mm^2}$ with Au absorber deposited on a central free-standing Si$_{3}$N$_{4}$ membrane,
\SI{1}{\micro m} thick and \SI{10}{mm} diameter. 
After the TES and Au absorber film have \new{been} grown, the membrane is first etched in \new{the} shape of a 8\,mm diameter circular spider-web supported by 32 narrow legs and then suspended by means of Deep Reactive Ion Etching of the silicon beneath. 
The TES is located aside the circular spider-web and is in strong electronic contact with the external perimeter of the gold absorber. The TES consists of 120\,nm of a Au-Ti bilayer, which is manufactured taking care to maintain a process temperature profile below \SI{100}{\celsius}, to ensure a superconducting to normal transition at $T_c = 500-\SI{550}{mK}$. In fact, it has been observed that high process temperatures reduce $T_c$ towards its bulk value of $350-\SI{400}{mK}$, as demonstrated in \cite{tiTctuning}.
These operating temperatures represent an optimal compromise between the SWIPE's bath temperature of \SI{300}{mK} and the detector saturation limits due to the high optical power (order of \SI{10}{pW}) of the multi-mode configuration. 
The thermal conductance $G$, in the range of $65-\SI{100}{pW.K^{-1}}$, 
was measured in the first prototypes that have been operated at a base temperatures of about \SI{350}{mK}. 
The effective time constants in \new{the} Electro-Thermal Feedback (ETF) regime were evaluated from the frequency response 
\new{to a sinusoidal}
 sweep excitation to be  $20-\SI{33}{ms}$, about a factor $2-3$ larger than the ones expected by the model. In these working points the thermal fluctuation noise equivalent power, NEP, is about \SI{3e-17}{W.Hz^{-1/2}} (see section~\ref{sec:SWIPE_noise} for details).

   \begin{figure}[!t]
   \centering
      \begin{tabular}{p{0.6\textwidth} p{0.4\textwidth}}
  \vspace{-9cm} \hspace*{-0.5cm}
  \includegraphics[width=0.79\textwidth]{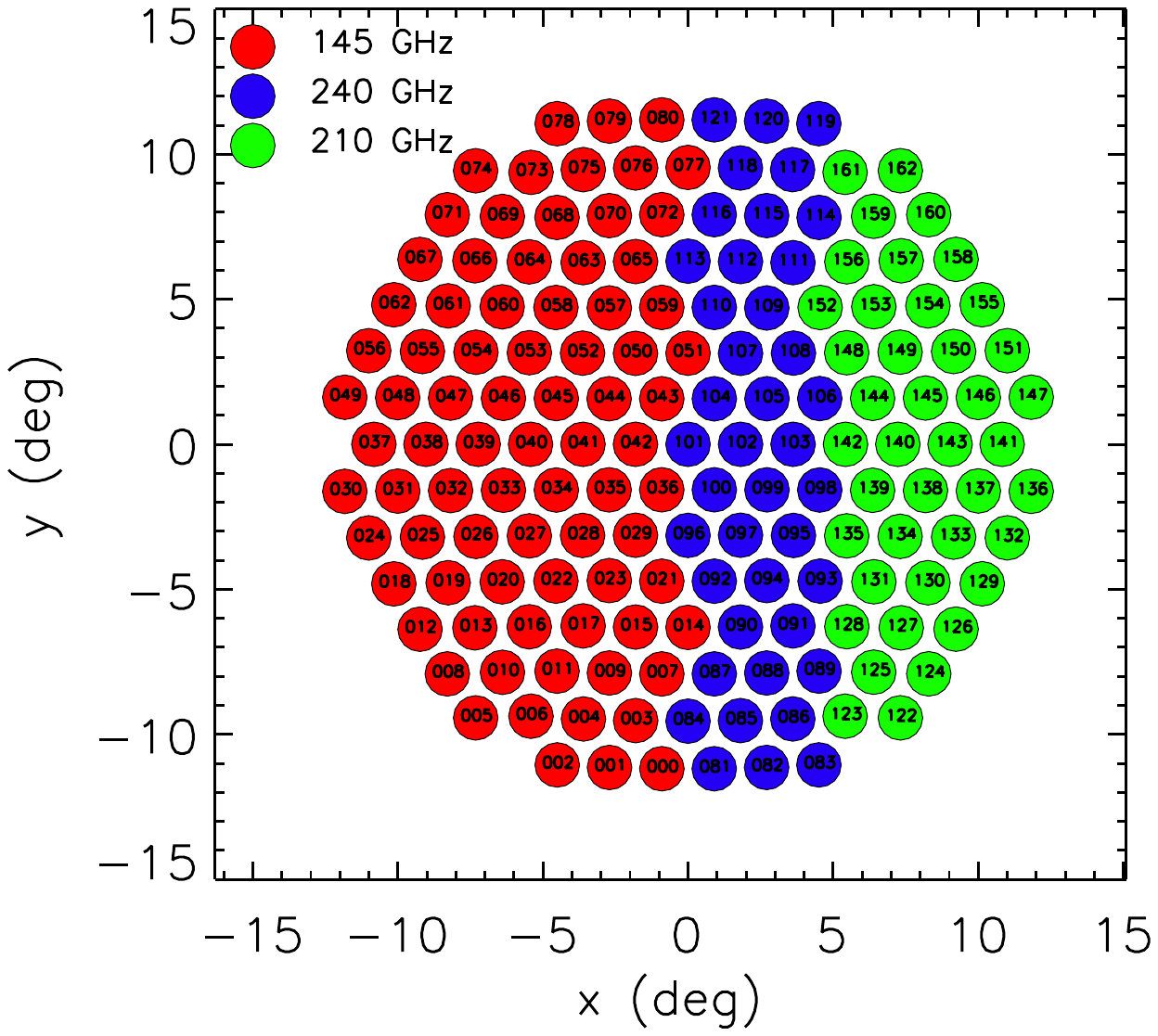} &
  \vspace{0pt} \hspace*{-1.5cm} 
  \includegraphics[width=0.39\textwidth]{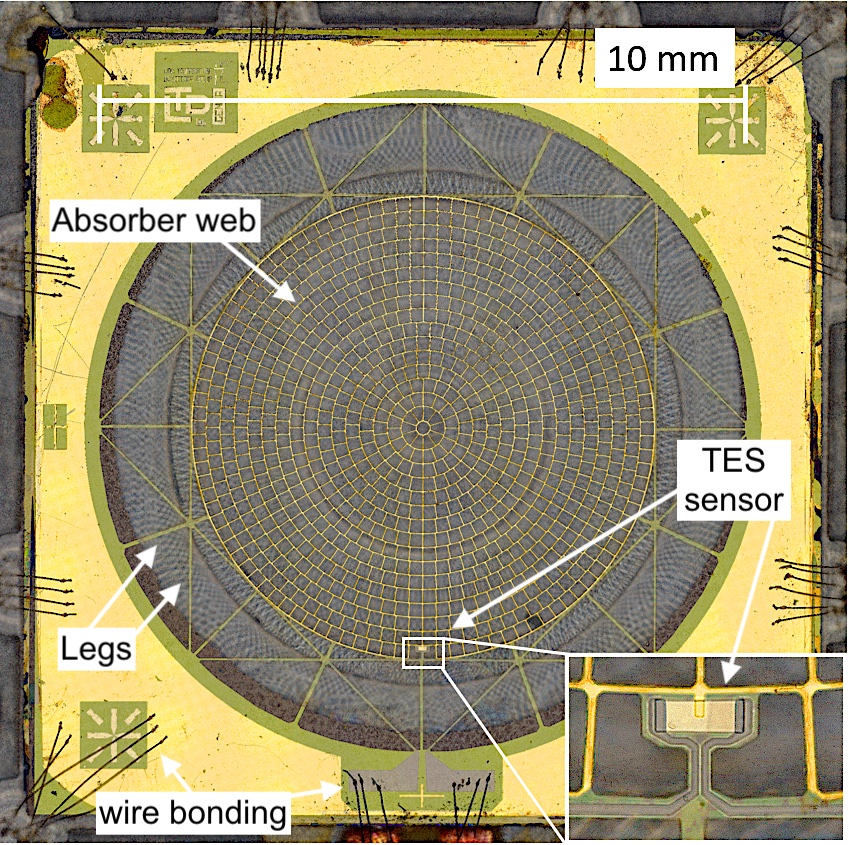}
\end{tabular}
   \caption{\emph{Left}: distribution of detectors in one of the two equivalent focal planes.  The payload rotates so that the scanning direction is along the x-axis. \emph{Right}: a LSPE-SWIPE large spider-web TES bolometer integrated in the backshort of the microwave cavity. 
   }
              \label{fig:swipe_focal_plane}%
    \end{figure}

The left panel of figure~\ref{fig:swipe_focal_plane} shows the distribution 
of detectors in one of the two equivalent focal planes.
The payload rotates so that the scanning direction 
is along the $x$ axis in the figure. The right panel of figure~\ref{fig:swipe_focal_plane} 
shows the LSPE-SWIPE large spider-web TES bolometer integrated in the backshort of the microwave cavity.

\subsubsection{Readout}
The 326 TES bolometers are read-out by Superconducting Quantum Interference Devices (SQUIDs) using a Frequency-Domain Multiplexing scheme (FDM), with each DC-SQUID sensing 16 TES~\citep{2018_Vaccaro}.
In the FDM scheme a group of detectors is readout with a single SQUID by connecting in parallel several RLC chains in which R is given by the TES variable resistance, and the LC filters define different frequencies. A  single signal containing all the different frequencies is therefore needed to bias all TESs simultaneously. The detectors modulate the  signal which is in turn  sent to the SQUID input, amplified and demodulated by digital electronics.
\new{The multiplexing tones are in the range from 100~kHz to 2~MHz, to be faster than the bolometers' response but below the cut-off frequency of the readout, the latter determined mainly by the length of the cables inside the cryostat. In this range we can safely accommodate 16 tones that readout TESs coupled to detectors at all the three bands. Some channels are coupled to blind detectors and/or calibration resistor to monitor gain fluctuations of the readout chain.}

\begin{figure}[!t]
\begin{center}
\includegraphics[width=0.65\columnwidth]{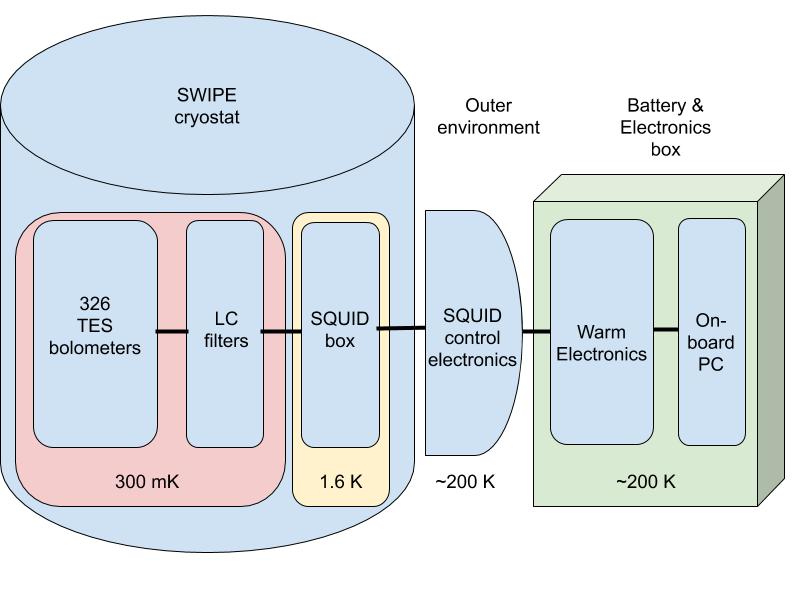}
\caption{\label{fig:readout}\new{Schematic} of the LSPE-SWIPE readout with the indication of the temperatures of the different stages.}    
\end{center}
\end{figure}

The readout electronic chain is composed of a \emph{cold} section inside the cryostat, at the same temperature of the detectors, and a \emph{warm} section, outside the cryostat. 
The entire chain is composed, going from the lowest to the highest temperature, of the LC filters and the bias resistors board, the SQUIDs boxes, the SQUID control unit and the warm electronics (see figure~\ref{fig:readout}).

\textbf{LC filters}. The LC filters necessary to the frequency domain multiplexing are assembled on dedicated boards placed at \SI{300}{mK} in close proximity to the TESs. The filters are composed of a niobium inductor fabricated by optical lithography coupled to a commercial 
Surface Mount Device (SMD) capacitor~\cite{Vaccaro:2019hmz}.
Given the inductance $L\approx\SI{15}{\micro H}$, the capacitors are chosen in order to give resonance frequencies in the \SI{200}{kHz} to \SI{2}{MHz} range.
Bolometers are connected to the LC board with shielded twisted pair wires. Furthermore the TES bias resistor is also placed on the board to minimize Johnson noise.
Each board hosts three LC chains to read a quarter of focal plane. Four such boards are used for each focal plane for a total of eight boards. 

\textbf{SQUID boxes}. Each LC board is connected with a custom low-inductance flat cable to a SQUID box placed at \SI{1.6}{K}. Each box is used to thermalize and shield three SQUIDs, for a total of 24 SQUIDs for the two focal planes. 
SQUIDs convert the modulated current signal into a modulated and amplified voltage signal which in turn is sent to a further amplification stage outside the cryostat.
The SQUIDs that are baselined for LSPE/SWIPE are 6-stage SQUID arrays from VTT \new{(model K3B)} with critical current 
$I_c \approx \SI{65}{\micro A}$, 
input inductance
$2$\,nH  and a transimpedance of 
\new{$30$ to \SI{50}{V.A^{-1}} typically}.
The flux coupling and noise are $\SI{36}{\micro \ampere.\Phi_0^{-1}}$ and $\lesssim \SI{0.1}{\micro \Phi_0.Hz^{-1/2}}$ respectively
($\Phi_0$ being the magnetic flux quantum). 
Two different coupling strengths can be selected for the feedback coil: 
$M_f^{-1}\simeq \SI{40}{\micro \ampere.\Phi_0^{-1}}$ 
and $M_f^{-1}\simeq \SI{90}{\micro \ampere.\Phi_0^{-1}}$.
%

\textbf{SQUID Control Units}. 
The SQUID control units (SCUs) are placed outside the cryostat and perform the main following tasks: 
%
(i) they provide the SQUID bias signal (which will be set at the SQUID operating temperature and will be tuned in flight);
%
(ii) they linearize SQUID response by means of a flux-locked loop (FLL hereafter);
and (iii)
they host the amplification stage needed to amplify the SQUID output voltage before the digitizing
stage. The desired amplification is achieved in two stages, in order to obtain the desired bandwidth and to minimize the noise referred to \newb{the} amplifier's input. The SQUID output is first amplified by a very low noise preamplifier based on a discrete JFET (IF3602) input differential cascode architecture, followed by a low noise CMOS operational
amplifier (OPA301). 
The equivalent input noise density is 
\SI{0.6}{nV.Hz^{-1/2}}
and the bandwidth extends up to at least 2\,MHz~\cite{fontanelli}.

\textbf{Warm readout}. The warm readout boards contain the ADCs (LTM9001IV) and the DACs (LTC1668IG) that are
used to generate the sum of sinusoids to bias the TES detectors and to digitize the modulated output. 
They in turn perform the digital demodulation and the data compression. 
They perform these operations by means of a system-on-module board hosting a FPGA and an ARM microprocessor
(MitySOM 5CSX System-On-Module\footnote{\href{https://www.criticallink.com/product/mitysom-5csx/}{https://www.criticallink.com/product/mitysom-5csx/}}). 
Each board, with a single \new{SOC,  runs two 
readout chains,} therefore the complete readout system is
composed of a total of 12 boards in a 6U standard, 
placed in a custom aluminum crate that provides the 
mechanical support and dissipates the generated heat.
Each warm readout board builds the packets that are 
sent to the on-board computer to be assembled in one 
single event.

\section{Sensitivity of instruments}\label{sec:sensitivity}
\new{
Realistic simulations of the observations are obtained 
for LSPE by means of noise estimation for \Strip\ and SWIPE,
and propagated
from \newc{time-ordered data} to maps using the instrument simulators
described in detail in Appendix~\ref{app:instrument_sim}.
}

\subsection{LSPE-\Strip{} noise estimation}

We model the noise of the \Strip{} polarimeters as the sum of a white noise plus a $1/f^\alpha$ component, so that the post-detection power spectrum can be written as:

\begin{equation}
    P(f) = \sigma^2\left[1+\left(\frac{f_\mathrm{knee}}{f}\right)^\alpha\right],
\end{equation}
where the knee frequency, $f_\mathrm{knee}$, is the frequency where the white noise and the $1/f^\alpha$ components contribute equally ($P(f_\mathrm{knee}) = 2\sigma^2$). Previous experience (QUIET, WMAP, \Planck-LFI) shows that this simple model provides a very good first-order description of the noise properties of  HEMT-based coherent devices.
The standard deviation of the white noise component of the $Q$ and $U$ Stokes parameters measured by each \Strip{} polarimeter in antenna temperature is given by:
\begin{equation}
    \Delta T_\mathrm{r.m.s.} = \frac{1}{\sqrt{2}}\frac{T_\mathrm{sys}}{\sqrt{\Delta\nu\,\tau}},
    \label{eq_strip_white_noise}
\end{equation}
where $T_\mathrm{sys}$ is the total intensity detected by the polarimeters (sky signals, emissions from the optical components and receiver noise temperature), $\Delta\nu$ is the receiver bandwidth and $\tau$ is the integration time. 
The factor $1/\sqrt{2}$ in equation~\ref{eq_strip_white_noise} results from the polarimeter correlation architecture and it is explained in \citep{QUIET2012a} and section~4 of \citep{Cleary2010}.
In table~\ref{tab_strip_white_noise_properties} we detail the budget leading to \new{the} current estimate of the average receiver white noise performance.

\begin{table}[!t]
 \begin{center}
    \begin{tabular}
    {l S[table-format=3.2] S[table-format=3.2]}
        \hline
        \hline
        \rule{0pt}{2ex}                                                  & \SI{43}{GHz} &  \SI{95}{GHz} \\
        \hline
        \rule{0pt}{3ex}\textbf{Sky signals in antenna temperature} & &\\		
        Atmospheric emission at Zenith $\left(\SI{}{K_{RJ}}\right)^1$\dotfill            & 16.3	    & 19.0\\
        CMB $\left(\SI{}{K_{RJ}}\right)$\dotfill                                                  & 1.8      & 1.1 \\
        \rule{0pt}{3ex}\textbf{Noise contributions} & &\\
        Mirror emission $\left(\SI{}{K_{RJ}}\right)^2$\dotfill                                 & 3.0	    & 3.0\\
        Window	$\left(\SI{}{K_{RJ}}\right)^3$\dotfill                                          & 3.0      & 8.0\\
        Filters $\left(\SI{}{K_{RJ}}\right)^3$\dotfill                                          & 2.0      & 3.0\\
        Feed system $\left(\SI{}{K_{RJ}}\right)^4$\dotfill                                      & 0.5	    & 0.5\\
        Polarimeter noise $\left(\SI{}{K_{RJ}}\right)^5$\dotfill                                & 34.0     & 104.2\\
        \rule{0pt}{3ex}System temperature$^6$, $T_\mathrm{sys}$ $\left(\SI{}{K_{RJ}}\right)$\dotfill & 61.7	    & 140.0\\
        \rule{0pt}{3ex}\textbf{1-second sensitivity per polarimeter}$^7$ & &\\
        Antenna temperature $\left(\SI{}{\micro K_{RJ}.s^{1/2}}\right)$
        \dotfill	              & 514.6  & 1139.5\\
        Thermod. temperature $\left(\SI{}{\micro K_{CMB}.s^{1/2}}\right)$
        \dotfill  & 539.7  & 1431.4\\
        \hline
    \end{tabular}
    \end{center}
    \footnotesize
    $^1$Simulated with \emph{am} Atmospheric Model code, based on partial water vapor measurements\\
    $^2$Assumes \SI{300}{K} physical temperature and 1\% mirror emissivity\\
    $^3$Estimated using electromagnetic simulations (\SI{60}{mm} window thickness)\\
    $^4$Assumes \SI{20}{K} physical temperature and $\sim\SI{0.1}{dB}$ insertion loss\\
    $^5$Measured during unit-level tests\\
    $^6$Calculated assuming a constant \newb{zenith angle} of \SI{20}{\degree} during the whole survey.\\
    $^7$Calculated assuming the receiver bandwidth reported in table~\ref{tab:instruments} and a constant \newb{zenith angle} of \SI{20}{\degree} during the whole survey.
        \caption{\label{tab_strip_white_noise_properties}White noise sensitivity budget of \Strip{} polarimeters}

\end{table}

We now discuss briefly the low-frequency properties of the noise spectrum and show how the expected impact from $1/f^\alpha$ noise components is small. In our measurements we expect two main sources of noise fluctuations on long time scales: (i) fluctuations in the receiver gain and (ii) variations in the atmospheric load. Both contribute to the $1/f^\alpha$ shape of the noise spectrum at low frequencies.
\Strip{} polarimeters have a very low susceptibility to gain fluctuations and $1/f^\alpha$ noise contributes in polarization only at frequencies less than few tens of mHz. 

This stability is the result of the differential nature of the receiver that allows one to recover the $Q$ and $U$ Stokes parameters by differentiating signals having essentially the same intensity, thus effectively canceling out common modes. The penalty is that these detectors are practically blind to the CMB total intensity, as these measurements retain all the common-mode fluctuations and are characterized by knee frequencies of the order of several Hz. 

If we assume zero or negligible polarization in the atmospheric signal, \newb{then} we can neglect, to first order, also the effect of $1/f^\alpha$ fluctuations in the atmospheric load. These will contaminate polarization measurements only through any leakage from total intensity to polarization that could be present in our receivers \newb{and that will be caused mainly by asymmetries in the polarizer-OMT system}. \newb{Considering} that our current estimates \newb{from OMT laboratory measurements \cite{Virone2014}} indicate a leakage of the order of $\sim 0.01\%$ \newb{this effect is likely to be negligible. We are currently working on a simulation framework that will model the expected fluctuations in the atmosphere brightness temperature and their effect of the sky polarization measurement and will allow us to quantify the impact of this systematic effect.}

\subsection{LSPE-SWIPE noise \new{estimation}}
\label{sec:SWIPE_noise}

\new{The LSPE-SWIPE detector's noise is given by the combination of photon noise,  detector thermal noise, 
readout electronics noise, and \newb{the} effect of cosmic rays.
}

The {\bf photon noise} is computed assuming that the
incoming radiation is the composition of: 
(i)~the CMB, a \SI{2.725}{K} black-body; 
(ii)~the residual atmosphere, as computed from 
the {\em am} Atmospheric 
Model\footnote{\url{https://doi.org/10.5281/zenodo.640645}} \cite{paine_scott_2019_3406496} 
assuming a pessimistic residual ambient pressure of 
\SI{10}{\milli\bar} and a zenith angle of \SI{45}{\degree}; 
(iii)~the cryostat window, as a \SI{240}{K} grey-body with 
emissivity computed assuming a layer of 
Mylar~\cite{1604073}, thickness $t = \SI{1}{mil}$ ($\sim\SI{25.4}{\micro\metre}$)
with $n_r = 1.57$ and loss tangent 
$\tan \delta = 2.25\times 10^{-3}$. 
The window emissivity is computed as 
$\varepsilon_{\text{window}, \nu} = 1-\exp(-2 \pi n_r (t/\lambda) \tan\delta) $, 
where $\lambda$ is the wavelength \cite{Lamb1996}. 
\new{The loading from the IR filters, lens, HWP and other
cryogenic elements is computed to be negligible with respect to CMB, window and atmosphere.}
Following \citep{1986_Lamarre}, 
for each component, 
the power on the detector is computed as 
\begin{equation}\label{eq:W_SWIPE}
P = \int  f_\nu \eta A \Omega  I_\nu d \nu
\end{equation}
where 
$f_\nu$ defines the filter pass-band; 
$\eta$ is the instrument efficiency, which 
\new{includes a factor 0.5 to take into account
the selection of one polarization by means of the wire grid};
$I_\nu$ is the spectral brightness (\SI{}{W.m^{-2}.sr^{-1}.Hz^{-1}}) of the component; 
$A\Omega$ is the throughput,  estimated
as $N_\text{modes} \lambda^2$, with $N_\text{modes}$
the number of electromagnetic modes coupled to 
each detector.
The photon noise equivalent 
power in \SI{}{W.Hz^{-1/2}} of a beam filling source is computed as:
\begin{equation}
\text{NEP}^2_\text{ph} =
2\int f_\nu \eta
A \Omega  I_\nu h \nu
\left(1+ \frac{f_\nu \eta c^2 I_\nu}{h\nu^3}
\right)d\nu .
\end{equation}
The total photon noise $\text{NEP}_\text{ph-total}$
is the quadrature sum of the photon noise from 
the CMB, $\text{NEP}_\text{ph-CMB}$, 
the atmosphere, $\text{NEP}_\text{ph-atm}$, and 
the window $\text{NEP}_\text{ph-window}$.

The {\bf detector's thermal noise} also depends on the 
power load. The higher the load, the 
higher must be the thermal conductivity $G$ which links
the detector to the thermal bath, in order to avoid 
the transitioning of the TES to normal state. 
Following \citep{2016gualtieri}, the \new{detector's} thermal noise
is computed as 
\begin{equation}\label{eq:NEP_det}
    \text{NEP}_\text{detector} = \sqrt{4 k_B T_c^2 G F}
\end{equation}
where $k_B$ is the Boltzmann constant,
$T_c =\SI{550}{mK}$ is the critical temperature,
and $F=\left[0.5;1\right]$ takes into account non-equilibrium 
effects in TES (we
assume a pessimistic $F=1$).
The optimal thermal conductivity is
\begin{equation}\label{eq:G}
G = \frac{n P_\text{sat} T_c^{n-1}}{T_c^n - T_\text{bath}^n}
\end{equation}
where $n = 3.2$ takes into account the thermal dependence of 
the conductivity, $T_\text{bath}=\SI{300}{mK}$ is the temperature
of the thermal bath, and $P_\text{sat} = 2.5\, P_\text{total}$ is the 
saturation power, with a 2.5 safety factor ($P_\text{total}$ being the total 
\newb{optical} power on the 
detector).  
Given that the detectors are all built with the 
same characteristics, we set the detector 
noise (equation~\ref{eq:NEP_det}) using the  highest
value among the thermal conductivity of the 
3 bands, $G_\text{max}$.
With this highest value, we compute the typical detector noise,
NET$_\text{detector,max}$.
Combining equation~\ref{eq:NEP_det} with \ref{eq:G} and \ref{eq:W_SWIPE}, 
it can be noted \new{that} $\text{NEP}_\text{detector}$ is proportional 
to $(A\Omega)^{1/2}$.

\begin{table}[!t]
    \centering
    \begin{tabular}{lccccc}
    \hline \hline
    Source & Value  & Value at SQUID & Factor to &  Note & Noise on detector\\
    & at source & input $\left(\SI{}{pA.Hz^{-1/2}}\right)$ & SQUID input &  &  $\left(\SI{}{aW.Hz^{-1/2}}\right)$\\
    \hline
    SQUID noise  & \SI{10}{pA.Hz^{-1/2}}  & $10$ & 1 & (a) & 10 \\
    DAC LTC1668  & \SI{50}{pA.Hz^{-1/2}}
    & $<5$ & $1/10$ & (b) & $<5$\\
    Preamplifier & \SI{0.6}{nV.Hz^{-1/2}} & $6$  & $\sim 100$~V/A & (c) & 6 \\
    Cabling      & \SI{0.3}{nV.Hz^{-1/2}}
    & $3$  & $\sim 100$~V/A & (c) & 3 \\
    Bias resistor   & \SI{2.6}{pA.Hz^{-1/2}}
     & 2.6  &1 & (d) & $\sim 2.5-4.0$ \\
    \hline
    Total & & & & (e) & $15-20$\\
    \hline
    \end{tabular}
    \caption{Contribution of selected readout electronics noise source\new{s} to the NEP budget. 
    (a) Typical noise of SQUID arrays currently being considered for the LSPE readout, given by the flux noise multiplied by the SQUID input coil coupling; 
    (b) the DAC noise is reduced at the SQUID input by a suitable resistive divider; 
    (c) the noise of the warm preamplifier is \new{referenced to} the SQUID input by using the SQUID transimpedence; the number reported here is the typical for the 6-series array SQUID being considered for the readout; 
    (d) current noise of the bias resistor at \SI{300}{mK}, with $R_b=\SI{0.1}{\ohm}$, and assuming the normal state resistance $R_N=\SI{1.0}{\ohm}$, and a TES resistance around $R_N/2$ in ETF; 
    (e) assumes a TES responsivity of $-\sqrt{2}/V_{\rm bias}$ with $V_{\rm bias} \approx 1  - 2$\,\SI{}{\micro V}: the noise at squid input has \new{to be divided by the responsivity} to get the NEP at detector.} 
    \label{tab:electronic-noise}
\end{table}

\begin{table}[tp]
\begin{center}
\begin{tabular}{l c c c}
\hline
\hline
\textbf{Band (GHz)}\dotfill& 145 & 210 & 240 \\
\hline
bandwidth\dotfill &30\%&20\%&10\%\\
$N_\text{modes}$$^1$\dotfill &[10;13.1;17] &  [23;27.0;32] &[32;34.5;39]\\
$A \Omega$ $\left(\SI{}{m^{2}.sr}\right)$\dotfill&\multicolumn{3}{c}{$N_\text{modes} \lambda^2$}\\
efficiency $\eta$\dotfill&0.3&0.25&0.25\\
\hline
\rule{0pt}{3ex}\textbf{Power on cryostat entrance} & & &  \\
$P_\text{CMB}$ $\left(\SI{}{pW}\right)$\dotfill&9.1 & 7.7 & 3.9\\
$P_\text{atm}$ $\left(\SI{}{pW}\right)$\dotfill&0.9 &  1.9 &  9.8\\
$P_\text{window}$ $\left(\SI{}{pW}\right)$\dotfill&1.0 & 2.8 & 2.4 \\
$P_\text{total}$ $\left(\SI{}{pW}\right)$\dotfill&11.0 & 12.4 & 16.1\\
\hline
\rule{0pt}{3ex}\textbf{Power on detector} & & & \\
$P_\text{total-detector}$ $\left(\SI{}{pW}\right)$\dotfill&3.3 & 3.1 & 4.0\\
\hline
\rule{0pt}{3ex}\textbf{Noise on detector} & & & \\
NEP$_\text{ph-CMB}$ $\left(\SI{}{aW.Hz^{-1/2}}\right)$\dotfill &23.5 & 23.3 & 17.6\\
NEP$_\text{ph-atm}$ $\left(\SI{}{aW.Hz^{-1/2}}\right)$\dotfill& 8.4 & 12.3 & 34.1\\
NEP$_\text{ph-window}$ $\left(\SI{}{aW.Hz^{-1/2}}\right)$\dotfill&7.8& 14.2 & 13.9\\
NEP$_\text{ph-total}$ $\left(\SI{}{aW.Hz^{-1/2}}\right)$\dotfill&26.1 &29.9 & 40.8\\
$G$ $\left(\SI{}{pW.K^{-1}}\right)$ \dotfill &56.1 & 52.7 &  68.4\\
$G_\text{max}$ $\left(\SI{}{pW.K^{-1}}\right)$ \dotfill &\multicolumn{3}{c} {68.4} \\ 
NEP$_\text{detector}$ $\left(\SI{}{aW.Hz^{-1/2}}\right)$\dotfill&30.6 & 29.7 & 33.8\\
NEP$_\text{detector,max}$ $\left(\SI{}{aW.Hz^{-1/2}}\right)$\dotfill&\multicolumn{3}{c} {33.8}\\
NEP$_\text{readout}$ $\left(\SI{}{aW.Hz^{-1/2}}\right)$\dotfill&\multicolumn{3}{c} {20}\\
NEP$_\text{total}$ $\left(\SI{}{aW.Hz^{-1/2}}\right)$\dotfill&  47.2 & 49.4  &56.6\\
\hline
\rule{0pt}{3ex}\textbf{Optical noise} & & & \\
NEP$_\text{optical-total}$ $\left(\SI{}{aW.Hz^{-1/2}}\right)$\dotfill& 157 & 197 & 226\\
NET $\left(\SI{}{\micro K_{CMB}.s^{1/2}}\right)$ \dotfill&11.4 & 12.3 & 26.2\\
margin $m$ $\left(\%\right)$ \dotfill &5 &20 &20 \\
NET$_\text{eff}$  $\left(\SI{}{\micro K_{CMB}.s^{1/2}}\right)$\dotfill&12.6 & 15.6 & 31.4\\
\hline 
 \multicolumn{4}{p{11.5 cm}}
            {\footnotesize \newb{$^1$The number of modes varies across the band 
            (see figure~\ref{fig:swipe_bands}), with less
            modes in the lower side of the band, and more modes in the higher side. 
            The three values in the square brackets indicate the number of modes at the minimum frequency of each band, the average across the band, and number at the maximum frequency of each band.}}
\end{tabular}
\end{center}
\caption{LSPE-SWIPE detectors radiative power and noise estimation. See text for details. 
\label{tab:swipe-noise}
}
\end{table}


\new{The {\bf readout electronics chain} is designed to}
keep its noise NEP$_{\rm readout}\lesssim \SI{20}{aW.Hz^{-1/2}}$, 
sub-dominant with respect to 
photon noise $\text{NEP}_\text{ph-total}$ and 
detector thermal noise $\text{NEP}_\text{detector}$.
\new{The known noise sources are first evaluated 
at their origin, then converted to values at 
the SQUID input}
by applying the appropriate conversion factors, 
\new{and finally converted} to equivalent noise 
on the detector by means of the TES current responsivity 
$R_I$.
To do so, we take into account the low frequency limit of $R_I$ in \new{the case of} strong electro-thermal feedback, i.e. $R_I\simeq -\sqrt{2}/V_{\rm bias}$, where $V_{\rm bias}$ is the TES bias voltage, and the $\sqrt 2$ factor originates from the AC bias in the multiplexing scheme (see the discussion in the appendix of~\cite{Dobbs:2011px} for further details).
The numbers quoted in table~\ref{tab:electronic-noise} are obtained assuming the expected voltage bias of $V_{\rm bias} \approx 1-\SI{2}{\micro V}$.
We take into consideration the following noise sources: the SQUID current noise, the DAC current noise, the SQUID preamplifier noise and the Johnson noise of \new{bias resistor and} cabling between the cold (inside the cryostat) and the warm (outside) section of the electronics. In table~\ref{tab:electronic-noise} we quote their typical values together with the factor needed for comparison at the SQUID input. See~\cite{tartariLTD18} for further details on the assumed noise model.
The total readout current noise NEP$_{\rm readout}$ is computed as the quadrature sum of these contributions.

The total noise equivalent power, NEP$_\text{total}$, is 
computed by quadrature sum of the 
photon noise from different sources, the detector
thermal noise and the readout noise. 
The optical noise, which converts the noise 
on the detector to noise at the instrument 
aperture, taking efficiency into account, 
is computed as 
\begin{equation*}
\text{NEP}_\text{optical-total} = \text{NEP}_\text{total}/\eta
\end{equation*}
Results of this calculation \new{are} converted to 
\SI{}{\micro K_{CMB}.s^{1/2}}
as:
\begin{equation}\label{eq:NET}
\text{NET} = \frac{T_\text{CMB}}
{\int f_\nu A\Omega B(T_\text{CMB}) \frac{x e^x}{e^x-1}dx}
\frac{\text{NEP}_\text{optical-total}}{\sqrt 2}
\end{equation}
where $B(T_\text{CMB})$ is the CMB black-body brightness, 
$x=h\nu/(k_B T_\text{CMB})$ is the reduced frequency, and the factor
$1/\sqrt 2$ takes into account the 
conversion from 
\SI{}{\micro K_{CMB}.Hz^{-1/2}}
to
\SI{}{\micro K_{CMB}.s^{1/2}}.
Notably, the photon noise NEP (as the thermal noise) 
is proportional to 
$(A\Omega)^{1/2}$, while the NET
is inverse proportional to $(A\Omega)^{1/2}$ and thus to 
$N_\text{modes}^{1/2}$. 
This is the advantage of multi-moded detectors:
higher photon noise, which relaxes \new{the} detector noise requirement, and lower NET.
Background and noise calculation results are
reported in table~\ref{tab:swipe-noise}.
In order to take into account possible effects such as 
contamination by cosmic rays (see next section), atmospheric
background variation, detectors yield, detector excess noise,
and other unexpected effects, we 
also report the margin $m$ value and the effective 
noise, 
$$
\text{NET}_\text{eff} = \text{NET}(1+m),
$$ 
which we use as input 
in the instrument simulator, for the results reported in section~\ref{sec:results}.
The margin $m$ is not the same
for all channels, due to the largest
uncertainty in the atmospheric modeling in the highest frequencies. 
\new{In SWIPE, the $1/f$ noise term has negligible impact, due to the
polarization modulation by means of the HWP. 
Measurements of the detector noise power spectra show the typical 
behaviour of $1/f$ evident at low frequency, on top of a  flat spectrum, 
with an high frequency roll-off due to
the detectors' time constants cut-off. 
}

\subsubsection{Cosmic rays rate}

TES detectors are sensitive to any form of energy 
deposited on the absorber, including the effect 
of cosmic rays. The flux of primary cosmic rays in the upper atmosphere is fairly well known, as well as its dependence on the latitude and on the solar cycles. We evaluated the expected rate of interactions by cosmic rays in the upper atmosphere (altitude \SI{40}{km}) along the orbit of SWIPE by using the measured fluxes and simulating the interactions of primary protons and alphas on the SWIPE cryostat, instrument and focal plane.  An energy-integrated flux of \SI{1.5}{particles.cm^{-2}.s^{-1}} is obtained at the minimum of the solar activity cycle, decreasing by a factor $\approx 2.5$ at the solar activity maximum.

We assume that cosmic rays release a signal in the bolometers whenever they interact with the gold-plated spider-web structure. By using the geometrical characteristics of our spider-web bolometers (diameter \SI{8}{mm}, fill factor $8\%$) we estimate an interaction rate of \SI{60}{mHz} per bolometer, giving roughly a \SI{1}{Hz} rate of interaction per readout chain, 
\new{given that} the bolometers \new{are} multiplexed in groups of 16.
The \SI{60}{mHz} rate 
is reasonable once compared with the (inverse of the) bolometer time constant, nevertheless suitable algorithms for cosmic ray hit identification and removal must be implemented. 
These algorithms also subtract the long tail after the glitch in the data. 
In a typical case, after a glitch, it is impossible to recover the 
first part of the tail, equal to $[5;10]\,\tau_\text{LP}$. With a rate of one event 
every \SI{16.7}{s} and a time constant 
$\tau_\text{LP} =\SI{30}{ms}$, this correspond 
to removing between 1 and 2\% of the data, well within our margins.

\section{Systematic effects and calibration}\label{sec:systematics}
In this section we present the most relevant systematic effects for the 
two instruments, with particular focus on the systematic effects
critical for the measurement of the CMB polarization. We also set
requirements on the knowledge of the most important instrumental 
parameters, and discuss the calibration plans.

\subsection{LSPE-\Strip{} systematic effects}
\label{sec_strip_systematic_effects}


    Here we provide a brief summary of the \Strip\ susceptibility to systematic effects, deferring to forthcoming papers a more detailed treatment. 

\paragraph{Systematic effects budget.}

    \new{We start by setting a top-level requirement on the maximum uncertainty from systematic effects on a single sky 
    pixel having the size of the \newb{Q-band} optical angular resolution. In general we want this uncertainty to be much less than that imposed
    by the white noise. Following the approach already adopted for Planck-LFI \cite{planck2013-p02a,Planck2015LFIsystematics} we set this limit to 5\% of the white noise level as a goal and 10\% of the white noise level as a requirement. In table~\ref{tab_strip_systematics_budget} we provide a list of systematic effects that could affect Strip polarimetric measurements and goal/requirement values for the upper limit in the systematic uncertainty.} \newb{The detailed breakdown has been defined according to our best current knowledge of the instrumental properties.}
    
    \begin{table}[t!]
        \begin{center}
            \begin{tabular}{l c c}
            \hline\noalign{\smallskip}
                \new{\Strip\ systematic effect} & \new{Goal} & \new{Requirement}\\
                                & \new{(\SI{}{\micro K})} & \new{(\SI{}{\micro K})} \\
                \hline
                \hline
                 \new{$I\rightarrow Q/U$ leakage\dotfill}                    & \new{0.030} & \new{0.050} \\
                 \new{$Q\rightarrow U$ and $U\rightarrow Q$ leakage\dotfill} & \new{0.020} & \new{0.030} \\
                 \new{Polarization angle uncertainty\dotfill}                & \new{0.010} & \new{0.030} \\
                 \new{$1/f$ noise\dotfill}                                   & \new{0.015} & \new{0.050} \\
                 \new{Far sidelobes\dotfill}                                 & \new{0.030} & \new{0.060} \\
                 \new{Pointing\dotfill}                                      & \new{0.010} & \new{0.030} \\
                 \new{Scan synchronous signals\dotfill}                      & \new{0.010} & \new{0.030} \\
                 \new{Other periodic signals\dotfill}                        & \new{0.001} & \new{0.003} \\
                 \new{Calibration-dependent effects\dotfill}                 & \new{0.010} & \new{0.030} \\
                 \noalign{\smallskip}\hline
                 \new{Total (quadrature sum$^1$)\dotfill}              & \new{0.053} & \new{0.114} \\
                 \noalign{\smallskip}\hline
                 \multicolumn{3}{p{9cm}}{\newb{\footnotesize{$^1$The quadrature sum results from the assumption that the various effects are uncorrelated. This assumption will be tested by detailed end-to-end simulations that are currently ongoing and that will be reported in a dedicated paper.}}}
            \end{tabular}
            
            \caption{\label{tab_strip_systematics_budget}\new{\Strip\ systematic effects budget. The numbers indicate the maximum systematic uncertainty on a pixel size equal to the angular resolution.}}
        \end{center}
    \end{table}


\paragraph{Polarimetric effects.}

    \Strip\ polarimeters are based on the QUIET design\new{, which provides} significant advantage: (i) the $Q$ and $U$ Stokes parameters are measured directly for each horn in the focal plane, instead of being recovered through the inversion of a condition matrix, (ii) the system is unaffected by gain and bandpass mismatches between the two acquisition lines of the same polarimeter, as well by as unbalances in phase switch states, and (iii) $1/f$ noise and other common-mode effects are efficiently removed from $Q$/$U$ timelines thanks to double demodulation.
    
    The most important polarization effect in the polarimetric chain is \newc{the leakage from total intensity to polarization that is caused by non ideal performance of the polarizer-OMT assembly. In particular the transmission imbalance, $\delta L_\mathrm{pol}$, of the two electrical ports of the polarizer cause a leakage $\mathcal{L}_{I\rightarrow Q} = \delta L_\mathrm{pol}$, while the OMT cross-polarization, $X_\mathrm{OMT}$, causes a leakage $\mathcal{L}_{I\rightarrow Q} = 2\sqrt{X_\mathrm{OMT}}$. Considering the combined effect of $\delta L_\mathrm{pol}$ and $X_\mathrm{OMT}$ we obtain $\mathcal{L}_{I\rightarrow Q} = \delta L_\mathrm{pol.} + 2 L_\mathrm{pol} \sqrt{X_\mathrm{OMT}}$, where $L_\mathrm{pol}$ is the polarizer average transmission.}
    
    \newc{If we consider the averaged measured OMT and polarizer cross-polarization and amplitude imbalance, $X_\mathrm{OMT}\sim -55$\,dB and $\delta L_\mathrm{pol}\sim 0.01$\,dB, we obtain a leakage term $\mathcal{L}_{I\rightarrow Q}\sim 0.5\%$. Notice that we have negligible leakage from $I$ to $U$, as also reported in \cite{quiet2011}. The reader will find further details about the polarimeter mathematical model and polarizer-OMT measurements is a series of technical papers about Strip that is currently in preparation for submission to JINST.}
    
    
    Another possible source of systematic effects is the difference in the bandpass among the various polarimeters. In fact, the polarimeters average the incoming signal over the bandpass, so that if the bandpasses are different and the source is not a black-body (as it contains, for example, the Galactic synchrotron emission) we have a residual systematic effect in the final, averaged map.
    We have performed simulations using bandpasses measured in the laboratory, a synthetic sky with CMB, Galactic synchrotron and dust emissions, and Monte Carlo realizations of the instrumental noise. Our results show that the angular power spectrum of the residual effect in polarization is about three orders of magnitude below the noise level, so that we can neglect it.
    
    Other imperfections are either compensated for by design (e.g. gain unbalance), or generate a leakage between $U$ and $Q$ that we estimate to be $\lesssim 1\%$ on the basis of the measured and simulated parameters of the various components in the polarimetric chain.

\paragraph{Thermal/electrical fluctuations.}

    Variations in temperature and bias voltages will generate common-mode fluctuations in the total intensity signal that will be canceled by the double demodulation. Only temperature variations in the feedhorn-OMT system can, in principle, leave a small residual in the $Q$ and $U$ parameters because of the leakage from intensity to polarization caused by the front-end cross-polarization. This residual effect is expected to be negligible and we will control its impact during data analysis by exploiting the instrument temperature housekeeping data.

\paragraph{Fluctuations in the atmosphere.}

    The atmosphere impacts CMB polarization measurements from the ground in two ways: (i) its average brightness temperature increases the \new{white} noise level of the measurements and (ii) it is a source of low-frequency noise \new{due to the correlation structures in the water vapor bubbles \cite{church1995predicting}.}
    
    Regarding the atmospheric load, we have estimated an average brightness temperature of \SI{16.3}{K} at \SI{43}{GHz} and \SI{19.0}{K} at \SI{95}{GHz} (see table~\ref{tab_strip_white_noise_properties}). This estimate is based on simulations carried out with the \emph{am} \new{Atmospheric} Model code
    using precipitable water vapor (PWV) measurements collected in 2018.
    
    Brightness temperature fluctuations in the atmosphere are 
    caused by PWV variations
    that follow the typical sub-tropical seasonal modulation.
    The effect of these fluctuations are canceled to first order in the polarization data by the pseudo correlation architecture of the \Strip\ polarimeters. A small fraction of these intensity fluctuations, however, leaks into $Q$ and $U$ because of the non-zero cross polarization of the polarizer-OMT assembly. Although this fraction is small \newc{($\sim 0.5\%$)} we are developing a Monte Carlo simulations to estimate their impact on polarization measurements
    (see appendix \ref{app:strip_atmosphere}).

\paragraph{Stray-light.}

    We define stray-light as the overall signal detected by the instrument from directions outside the main beam. The origin of these signals, detected by the optics sidelobes, can be astrophysical (e.g. the Galaxy), terrestrial (the emissions from the ground) and instrumental (e.g. the emissions from the telescope enclosure shields).
    The sidelobes can contribute to a spurious polarization in two ways: (i) by detecting directly a polarized signal far from the main beam (from the sky, from the Earth and from the Sun) and (ii) by converting a total intensity emission to polarization due to the cross-polar response of the \newc{telescope-feed} system. 
    
    Regarding the first point, our preliminary estimates based on the simulated beam far sidelobes show that spurious polarization detected directly from the sky is less than $\sim\SI{3}{nK}$ and, therefore, negligible. The assessment of the polarized input from the Earth is more difficult, because of the lack of data on the polarization properties of the microwave Earth emissions. Using Earth brightness temperature data measured at \SI{37}{GHz} by the Special Sensor Microwave/Imager instrument on board the Defense Meteorological Satellite Program\footnote{\url{http://www.remss.com/missions/ssmi}} we estimated an upper limit of \SI{0.1}{K} of polarized emission from the Earth potentially entering the telescope far sidelobes. We also estimated that with the current shielding this contribution should be maintained below $\sim\SI{0.05}{\micro K}$ in the scientific data.
    
    To avoid Sun contamination during daytime we will discard data where the Sun is at an angular distance less than \SI{10}{\degree} from the telescope line-of-sight. Our simulations show that this fraction corresponds to about 15\% of the data and is included in our duty cycle computation.  When the Sun is farther than \SI{10}{\degree} its emission will be detected by the beam far sidelobes that are at the level of about $-\SI{100}{dB}$, enough to dilute this signal to negligible levels.
    
    Regarding the intensity-to-polarization leakage \newc{caused by the Strip optics} we have considered the input from the sky and from temperature variations in the optical enclosure. Considering the \SI{44}{GHz} \Planck{} sky maps combined with the \newc{$-\SI{40}{dB}$} upper limit \newc{of the cross-polar beam (see inset table of figure~\ref{fig_strip_beams})} we find that the sky contributes with a spurious polarization of $\sim\SI{0.01}{\micro K}$. The polarization systematic effect induced by optical enclosure temperature fluctuations is not a concern, provided that we will be able to measure and decorrelate these fluctuations from the data.

\paragraph{Main beam asymmetry.} 

    Asymmetry in the main beams is a source of leakage from intensity to polarization that can be corrected in the power spectra, provided that one knows the main beams with percent precision down to about $-\SI{25}{dB}$. In \Strip\ we further control this effect ``in hardware'', thanks to the very symmetric optical response of the telescope crossed-Dragonian design that guarantees an average beam ellipticity less than 1\% with a corresponding cross-polar discrimination better than $-\SI{40}{dB}$ \newc{(see, again, the inset table of figure~\ref{fig_strip_beams})}.

\paragraph{Pointing effects.}

    \Strip\ will implement a night optical star tracker that will allow us to reconstruct the pointing with a precision of $\SI{15}{\arcsec}$ or better, 
    \newc{resulting in negligible} pointing systematic effects.
    As a reference, the precision reached by \Planck\ for the \SI{44}{GHz} LFI channel was \SI{27}{\arcsec} for the pointing reconstructed from the nominal Jupiter scans and \SI{19}{\arcsec} for the pointing reconstructed from the deep Jupiter scans \citep[section~5.3]{Planck2015LFIsystematics}. This precision was enough to guarantee the scientific performance and the impact of errors in the pointing reconstruction could be considered negligible \citep[figure~8]{Planck2015LFIbeams}.

\paragraph{Calibration effects.}

    An important source of systematic effects is the uncertainty in the instrument calibration parameters: (i) the photometric calibration (also named ``responsivity'') that converts the raw time-ordered-data into brightness temperature units, (ii) the beam pattern and (iii) the polarization angle, that defines the reference frame in which $Q$ and $U$ are measured by the polarimeters. We summarize our calibration strategies in section.~\ref{sec_strip_calibration}.
    
    To first order we will manage calibration effects ``in-hardware'', i.e., by achieving high precision in the measurement of the instrument calibration parameters. We will measure the photometric constant with a relative precision better than 1\% exploiting a combination of natural and artificial sources, and the main beams down to $-\SI{25}{dB}$ using a source placed on an Unmanned Aerial Vehicle (UAV).
    
    The polarization angle is known from the mechanical disposition of the feedhorns in the focal plane and its uncertainty is limited by mechanical tolerances and thermo-elastic variations during cooldown. Previous experience with \Planck-LFI \citep[section~2.1.3]{planck2014-a04} show that \new{this} uncertainty is less than \SI{0.5}{\degree}.
    
\begin{figure}[!t]
    \begin{center}
        \includegraphics[width=14cm]{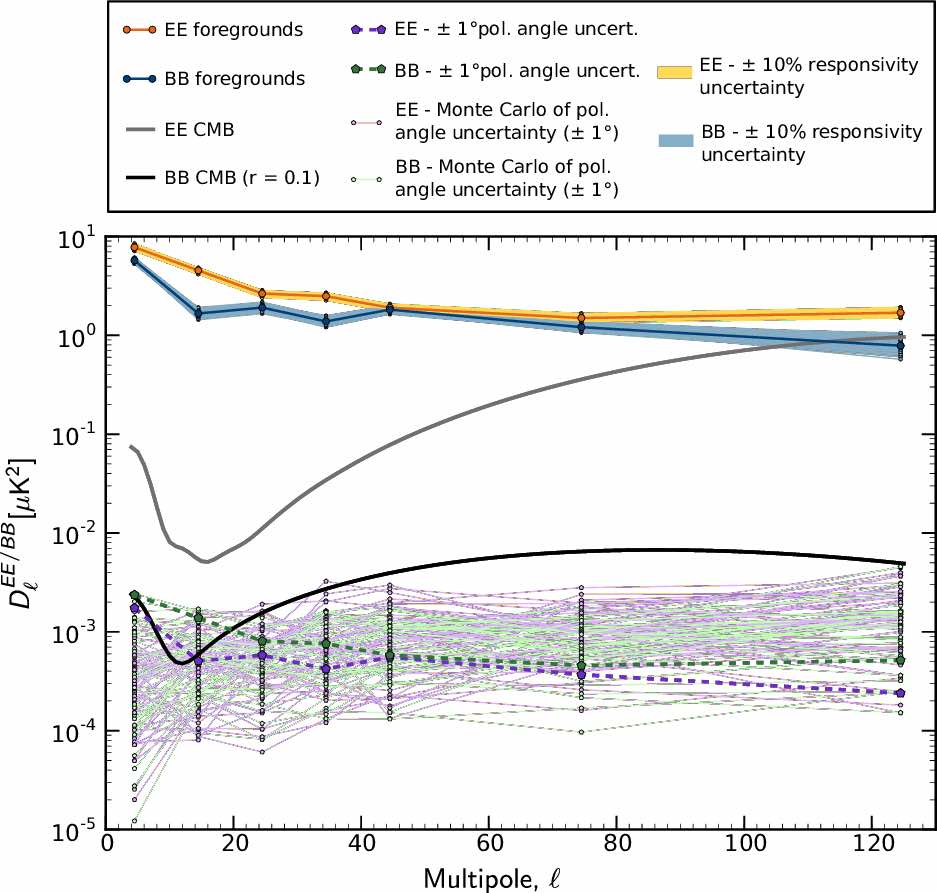}
        \end{center}
        \caption{\label{fig_calib_polangle_systematics}Power spectra evaluating the impact of polarization angle and photometric calibration uncertainty on \Strip\ measurements.}
    \end{figure}

    In figure~\ref{fig_calib_polangle_systematics} \citep[adapted from][]{Krachmalnicoff2015} we show our estimate of the impact of photometric calibration and polarization angle uncertainty on the power spectra measured by \Strip. The colored areas highlight the effect of $\pm 10\%$ uncertainty in the photometric calibration (a highly conservative estimate), while the bundle of purple and green lines is the result of a Monte Carlo simulation of a $\pm\SI{1}{\degree}$  uncertainty in the polarization angle. \new{From this figure we see that the expected level of contamination is much less than the synchrotron power spectra (yellow and blue curves) and they are therefore negligible considering the role of the 43\,GHz channel that is to remove synchrotron contamination from the data of the high frequency channels.}


\subsection{LSPE-\Strip{} calibration}
\label{sec_strip_calibration}

    We briefly describe here how we will measure the three main instrumental calibration parameters: (i) the photometric constant, (ii) the beam pattern and (iii) the polarization angle.

\paragraph{Photometric calibration.}

    We distinguish here two steps in the determination of the photometric constant: (i) ``absolute calibration'', i.e. the determination of the absolute value of the photometric constant and, (ii) ``relative calibration'', the measurement of time variations in the instrument responsivity caused by gain fluctuations. 
    
    Because the architecture of the \Strip\ polarimeters does not allow stable measurements in total intensity, for absolute calibration we must rely on bright polarized sources with known flux. The Crab Nebula is undoubtedly the best flux calibrator at these frequencies and we have shown that with one day of data it is possible to achieve a precision $\lesssim 10\%$, while \new{a} few weeks will be enough to approach a precision of $2-3\%$ \citep{Montresor2012}.
    
    Regarding relative calibration, we will measure instrumental gain changes with a stable signal generated by two thermally stabilized microwave generators (one in Q-band and the other in W-band) installed in the optical assembly. Fluctuations currently measured in our laboratory are less than 0.2\%, which will allow us to achieve a relative calibration with an overall precision better than 1\%.
    
\paragraph{Beam calibration.}

    The presence of 1/$f$ noise in the total intensity data measured by \Strip\ limits our ability to exploit natural point-sources, like Jupiter, to calibrate the main beams. 
    Therefore, we have developed an artificial calibrator system based on a microwave source placed on an \newb{unmanned aerial vehicle (UAV)}~\citep{Paonessa202012} that will fly over the \Strip\ telescope during the commissioning campaign, and will allow us to measure the main beams with the required precision (1\% at $-\SI{25}{dB}$).
    
\paragraph{Polarization angle.}

    The \Strip\ polarimeters measure directly the Stokes parameters $Q$ and $U$ in a coordinate system defined by the mechanical layout of the polarizer-OMT assembly \citep{Virone2014} and, ultimately, by the orientation of the OMT output waveguides.
    This means that the knowledge of the polarization angle is determined by the mechanical design and limited to about \SI{0.5}{\degree} by mechanical tolerances and by possible thermoelastic variations during cooldown.


\subsection{LSPE-SWIPE systematic effects and calibration}
\label{sec:swipe_sys_and_cal}

\newb{The minimization and control 
of systematic effects is a critical 
aspect for instrumentation designed for measurement of the  
CMB B-mode polarization.
In this section we discuss the main 
potential systematic effects in 
LSPE-SWIPE, we estimate their impact into the 
scientific target, and set the requirement on 
relevant instrumental parameters. 
Specifically, 
(i) 
in section~\ref{subsec:swipe_sys_requirements} we set some basic requirements from 
unavoidable instrumental properties, order of magnitude estimation,
and literature results; 
(ii)
in section~\ref{subsec:HWPsyst} we derive requirements for polarization angle and time 
response knowledge running the pixel-based tensor-to-scalar ratio estimation
pipeline (detailed in section~\ref{subsec:likelihood}) for a 
coherent rotation of the polarization in the sky; 
(iii)
in section~\ref{sec:swipe_sys} 
we derive requirements on the systematic effects produced
by the HWP at the frequency of the plate rotation, and its
harmonics, including a requirement on the knowledge 
of the HWP angular velocity and of the HWP angle,
in terms of random error and systematic offset; this is 
obtained by a full end-to-end simulation of the LSPE-SWIPE observation,
using the instrument simulator, 
prior of running the tensor-to-scalar 
ratio estimation pipeline; 
(iv)
finally in section~\ref{subsec:swipe-calibration} we summarize the 
requirements and briefly describe the calibration philosophy.}

\subsubsection{LSPE-SWIPE optical parameters requirements}
\label{subsec:swipe_sys_requirements}

LSPE-SWIPE is designed to minimize instrumental polarization. This is the spurious signal resulting
from the measurement of unpolarized radiation. 
In the case of CMB, the amount of unpolarized radiation coming from
the sky is overwhelming with respect to the polarized signal, minimizing instrumental
polarization is the most important driver of instrument design. 
\newb{In the case of SWIPE, some level of instrumental polarization is inevitably
generated by the cryostat
window, in case of incident radiation not orthogonal
to the window surface itself. 
This is due to 
multiple reflections in an isotropic dielectric 
slab~\cite{salatino2017}. 
For the off-axis detectors, considering the physical 
properties of the thin window reported in 
section~\ref{sec:SWIPE_noise}, 
this instrumental polarization will be up to 0.04\%.
Despite being a small value, it will produce a constant
polarized signal at the level of \SI{1}{mK} from the unpolarized CMB monopole, up to \SI{2}{\micro K} spurious polarization
from CMB dipole, and less than \SI{25}{nK} from CMB anisotropy.}
The \newb{high} constant signal is treated as an offset, 
\newb{removed} in the data analysis. Its stability depends
on the stability of the gain of the electronics and of the responsivity of the detectors and the \newb{effect of the instability} is not
synchronous with the observed sky. \newb{The detailed removal technique 
is described in section~\ref{sec:swipe_sys}}. 
\newb{Beyond this inevitable term,} instrumental polarization is reduced by system design, with an
optical system close to on-axis,  avoiding mirrors in favor of lenses, and using the polarization modulator as the first optical element 
(except window and thermal filters), thus relaxing
significantly the requirements on the following optical components.

The second parameter to be considered is cross-polarization,  defined as the
response of a polarimeter to an input signal polarized in direction orthogonal to the nominal
polarimeter direction. Cross-polarization results in leakage of E-mode into B-mode. 
Our requirement is that the maximum acceptable level of cross-polarization is below 2\%.
This is achieved again by means of an accurate optical design,
\newb{as described in section~\ref{sec:swipe_optics}}.

The third parameter to be considered is the ellipticity of the main beam (detector angular response 
in the sky). Spinning of
the Half-Wave Plate allows \newc{the system} to observe the same sky region with the same beam orientation, and
different polarimeter orientation. This strongly mitigates the ellipticity requirement, \newb{which was tested with simulations up to 25\% without a relevant impact}, 
and differential 
ellipticity among different detectors. 

Correct measurement of the angles of the polarimeters is crucial to avoid leakage from E-mode into
B-mode, and to avoid contamination in the measurement of fundamental physics effects such as
cosmic \newc{birefringence}. 
In this context, our system is characterized by the presence of a single, large
wire grid polarizer, defining the reference system for polarization measurements. With this design
the system is similar to an ideal polarimeter, and the angle of the single large wire 
grid polarimeter
can be accurately measured. 
The requirement on the polarimeter angle 
measurement is set in the next section.

Spectral matching among detectors 
has been historically a problem for instruments without a polarization modulator, just comparing
two independent measurements of the orthogonal polarization components. 
In LSPE-SWIPE we use a Stokes polarimeter, where the same detector measures both polarizations,
alternated by means of the \newb{rotating} HWP. 
In this configuration the most important requirement is
that the waveplate has high modulation efficiency over the detection bandwidth of the focal plane it
serves. In our system a single waveplate covers all the bands from 120 to \SI{260}{GHz}. This
requires over 70\% bandwidth for the waveplate, a goal certainly reachable with significant
accuracy, by means of metamaterials \citep[see][]{Pisano:06}.

\new{
Other relevant effects are the impact of polarized optical sidelobes, and   
the terms in the optical system Muller matrix converting radiation intensity 
into polarization. 
These effects are potentially critical for 
polarization estimation, even in presence of a HWP, and must 
be controlled by an effort in the calibration of the integrated system and in the 
data analysis. 
We note that simulations and data analysis of 
BICEP-2~\cite{2010ApJ...711.1141T} have shown that the polarized sidelobes 
induce a contamination at a level 400 times below
the $r=0.1$ target, thus also below the target of LSPE-SWIPE which has a very similar shielding scheme.
Regarding the spurious terms in the Mueller matrix,
the critical elements are $M^{4\theta}_{IQ}$ and $M^{4\theta}_{IU}$ 
that \newc{induce} a leakage modulated
at $4\omega_\text{HWP}t$ in equation~\ref{eq:swipe_modulation}. 
It has been shown~\cite{2018_Imada_IEEE} that a 10\degree\ tilted 
plate can induce these kind of terms
at the level of $10^{-4}$, which contaminates the B-mode  
below the level relevant for LSPE-SWIPE; 
the Atacama B-mode Search (ABS) experiment
\cite{2018JCAP...09..005K}
has measured intensity to polarization
leakage of order $7\times 10^{-4}$, 
corresponding to a systematic error contribution of $r < 0.01$, even without attempting to measure and remove the leakage 
signal~\cite{2016RScI...87i4503E}.
It is also possible to measure some of these parameters, and 
compensate their effect
adopting a mapmaking such as the one described in 
\cite{2010ApOpt..49.6313B}, which includes 
the spurious terms in the mapmaking equation. 
This approach requires a level 
of calibration of the intensity-to-polarization 
terms in the Mueller matrix terms $M_{IQ/U}^{4\theta}$
of the order of 
$10^{-4}$, which we set 
as \newc{a} calibration requirement}. 

\subsubsection{Polarization angle and detector time response requirements}\label{subsec:HWPsyst}

One of most problematic systematic effects for any B-mode probe
is the presence of a systematic error $\Delta \alpha$ in the polarization angle reconstruction.
Such an error induces
a $Q$ to $U$ rotation, and an E to B-mode leakage \cite{2009Pagano_malapola}:
\begin{equation}\label{eq:EB_roation}
\left\{
\begin{aligned}
C_\ell^{EE, obs} = C_\ell^{EE} \cos^2(2\Delta \alpha) + C_\ell^{BB} \sin^2(2\Delta \alpha)\\
C_\ell^{BB, obs} = C_\ell^{BB} \cos^2(2\Delta \alpha) + C_\ell^{EE} \sin^2(2\Delta \alpha).
\end{aligned}
\right.
\end{equation}
In order to set the requirement on the knowledge of SWIPE wire grid and 
polarization angle, we have run an estimation of the likelihood 
of $r$, as described in section~\ref{subsec:likelihood}, for a 
sky with a rotation of the E, B space as in equation~\ref{eq:EB_roation}. 
\begin{figure}[t!] 
   \centering
   \includegraphics[width=0.65\textwidth]{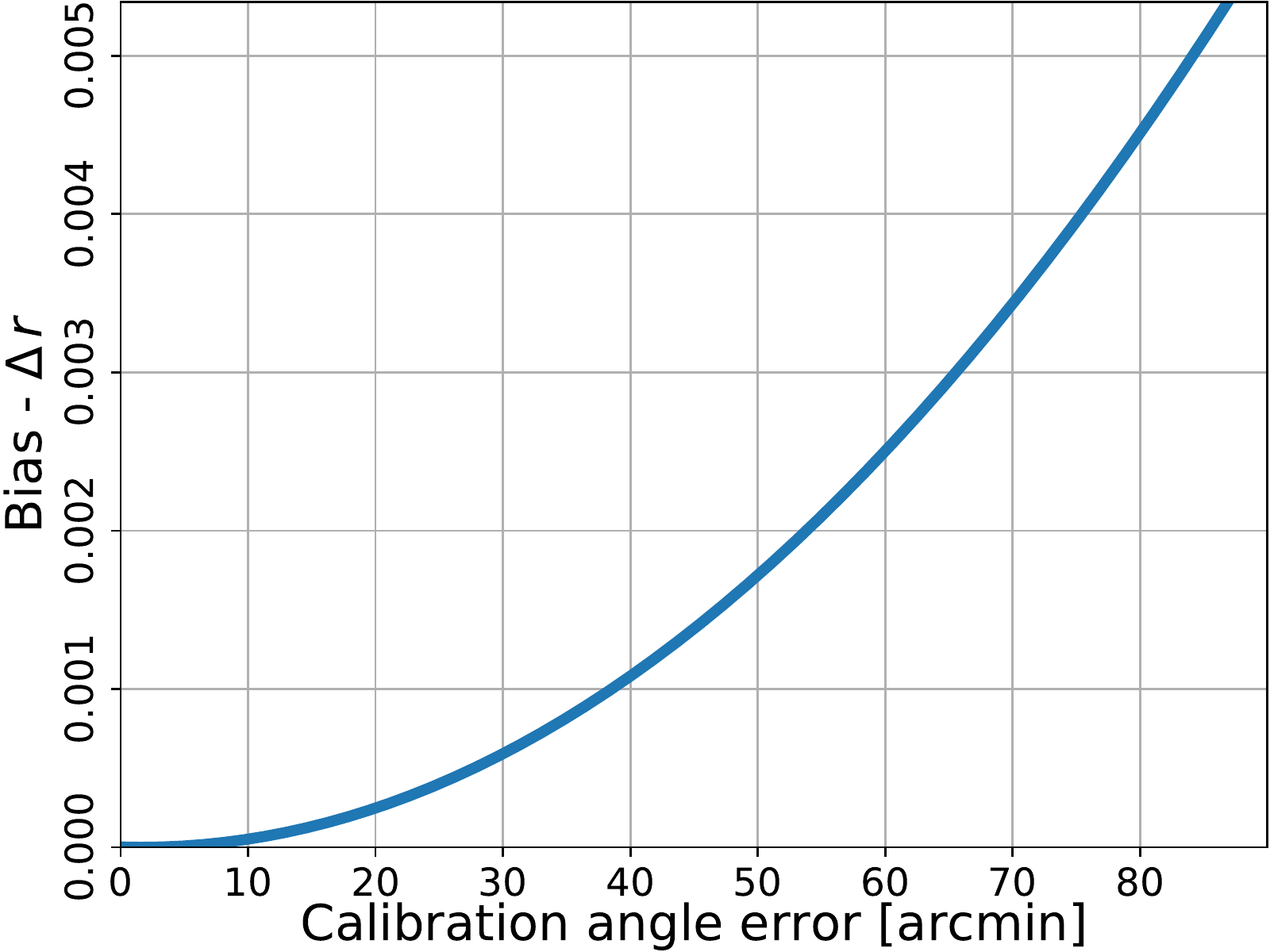}
   \caption{
   Bias in $r$ due to a rotation in the polarization angles. 
   \newb{The bias is computed by adding a rotation to the 
   polarization angles, and estimating the tensor-to-scalar 
   ratio $r$ from pixel based likelihood as described in section~\ref{subsec:likelihood}.}
   }
    \label{fig:swipe_angle_error}%
\end{figure}
The result is reported in figure~\ref{fig:swipe_angle_error}, in 
terms of a bias in $r$, if the recovered polarization is rotated by 
a given angle. 
In the case of LSPE-SWIPE the rotation may be due to a rotation 
of the Wire Grid, by an angle $\Delta \phi_\text{WG}$
or by a rotation of the HWP, by an angle
$2 \Delta \theta_\text{HWP}$. 
From figure~\ref{fig:swipe_angle_error}, we can set the requirements:
\begin{equation}\label{eq:swipe_angles_requirements}
\begin{aligned}
\Delta \phi_\text{WG} < \SI{40}{\arcmin}/\sqrt{2} = \SI{28}{\arcmin} \\
\Delta \theta_\text{HWP} < \SI{20}{\arcmin}/\sqrt{2} = \SI{14}{\arcmin}
\end{aligned}
\end{equation}
so that the uncorrelated combination of the two errors
is $\Delta \alpha = \SI{40}{\arcmin}$, which 
produces a bias $\Delta r \sim 0.001$, 
corresponding to 10\% of the uncertainty in $r$. 
These requirements are valid for the angles' knowledge, in the 
case \new{that} this is the only uncertainty in the system. 
In section~\ref{sec:swipe_sys} we consider the combination 
of a number of systematic effects, relevant for the 
measurement of the polarization. With that joint analysis,
we set more stringent requirements as reported in 
table~\ref{tab:swipe_sys_reqs}.

The TES detectors of SWIPE have an intrinsic time response. Their temporal transfer function
$H(\omega)$
can be approximated by a single pole low pass filter, as in equation~\ref{eq:swipe_tes_tf}.
\new{As described in appendix~\ref{app:swipe_tc}, an error in the knowledge 
of the temporal transfer function phase has the 
same effect of an error on the knowledge
of the HWP angle. 
The requirements on transfer function phase and the time constant knowledge can then 
be derived from the requirement on the angles reconstruction.
In appendix~\ref{app:swipe_tc}, we derive}:
\begin{eqnarray*}
&& \Delta \Phi\simeq \SI{42}{\arcmin} =  12 \times10^{-3}\,\text{rad}\\
&&    \Delta \tau_\text{LP} \simeq \SI{1.5}{ms}
\end{eqnarray*}
\new{where  
$\Delta \Phi$ is the error in the knowledge of the temporal transfer function phase
in the range $[3\div5]\omega_\text{HWP}$, and 
$\Delta \tau_\text{LP}$ is the corresponding error in the knowledge of the detector's time constant;} 
Also in this case, the requirement will be more stringent
if considered jointly with other effects, 
as presented in table~\ref{tab:swipe_sys_reqs}.
The HWP angle error and the error on the 
phase of the time transfer function can 
be disentangled and calibrated by 
spinning the HWP at different angular 
velocities, or different 
directions, both in flight and 
\new{during ground calibration. }

\subsubsection{HWP synchronous systematic effects: mitigation and requirements }
\label{sec:swipe_sys}

Besides errors in the polarization angle reconstruction, 
discussed in the previous section, another critical 
contamination in HWP based polarimeters is the 
generation of spurious signals at the frequency 
of the plate rotation, or its harmonics. 
Small differences ($\sim 10^{-3}$) in the absorption coefficients 
along the ordinary and extraordinary axes of the HWP produce a polarized emission. This radiation is modulated at 
twice the HWP spin frequency, $2f_\text{HWP}$, when \newc{it} is transmitted by the polarizer but could also be reflected by the polarizer, 
and back by the HWP, and induce a spurious signal at $4f_\text{HWP}$, the same frequency as the sky polarized signal \citep{Columbro2019}.
By simulations, the $2f_\text{HWP}$ contribution produces 
an equivalent temperature fluctuation of $\sim 1-\SI{10}{mK_{CMB}}$
while the $4f_\text{HWP}$ contribution is $<\SI{5}{\micro K_{CMB}}$ if the HWP temperature is kept below \SI{10}{K}. 
This last term is completely negligible in comparison to the 
instrumental polarization requirement set at 0.04\%, corresponding to 
a polarization signal of \SI{1}{mK}. 
These spurious signals must be removed by dedicated 
data analysis techniques. 
For the case of SWIPE, we developed a specific pipeline to deal 
with any spurious term synchronous with the HWP spin
frequency and harmonics. 
This is based on application of notch filters, centered
at the contaminated frequencies, and an iterative 
map-making to recover the removed signal, 
\new{as described in Appendix~\ref{app:iterative_map_making}}. 
\new{
In order to assess the efficacy of this 
technique, we performed a set of}
simulations, using  a multi-notch filter, i.e. a chain of notch filters, each with different central frequency.
The payload rotation period is set to \SI{8.6}{min} in accordance with equation~\ref{eq:swipe_period}  
and the notch filters width is set to $\Delta f=\SI{1}{mHz}$. 
HWP synchronous systematic effects are introduced at $\text{1}f_\text{HWP}$, $\text{2}f_\text{HWP}$,$\text{3}f_\text{HWP}$, $\text{4}f_\text{HWP}$ and $\text{5}f_\text{HWP}$ 
with amplitude \SI{10}{mK}, \SI{10}{mK}, \SI{1}{mK}, \SI{1}{mK}
and \SI{1}{mK} respectively. 
The same frequencies are used as centers of a 
stack of 5 notch filters. 
\newb{
Notably, the application of these 5 notch filters
and the iterative map-making impact the polarization 
map with a r.m.s of the residual map of 
order $\SI{5e-12}{K_{CMB}}$, showing that 
the procedure has unitary transfer function at all angular scales.}
\begin{figure}[!t]
\begin{centering}
\includegraphics[width=13.3cm]{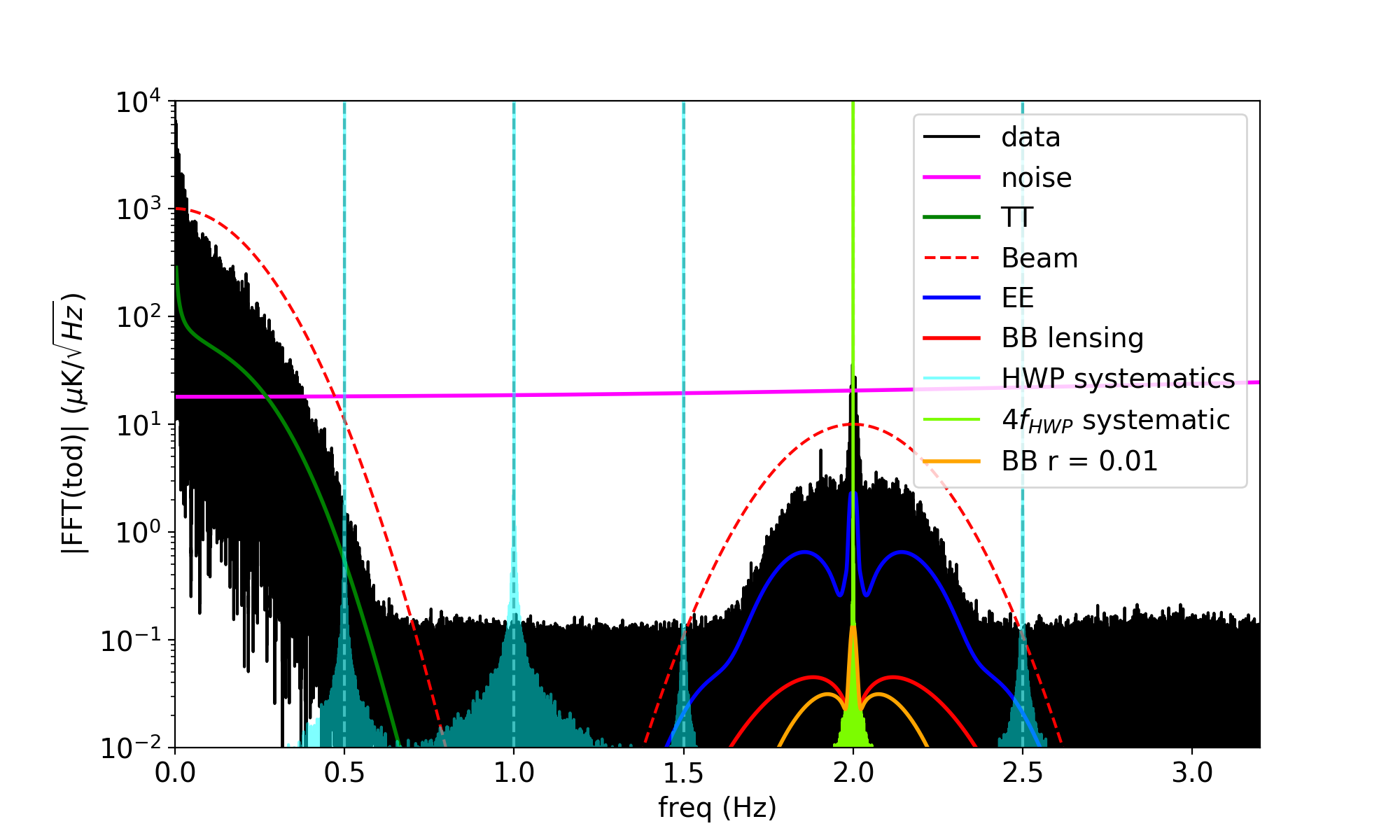}
\includegraphics[width=13.3cm]{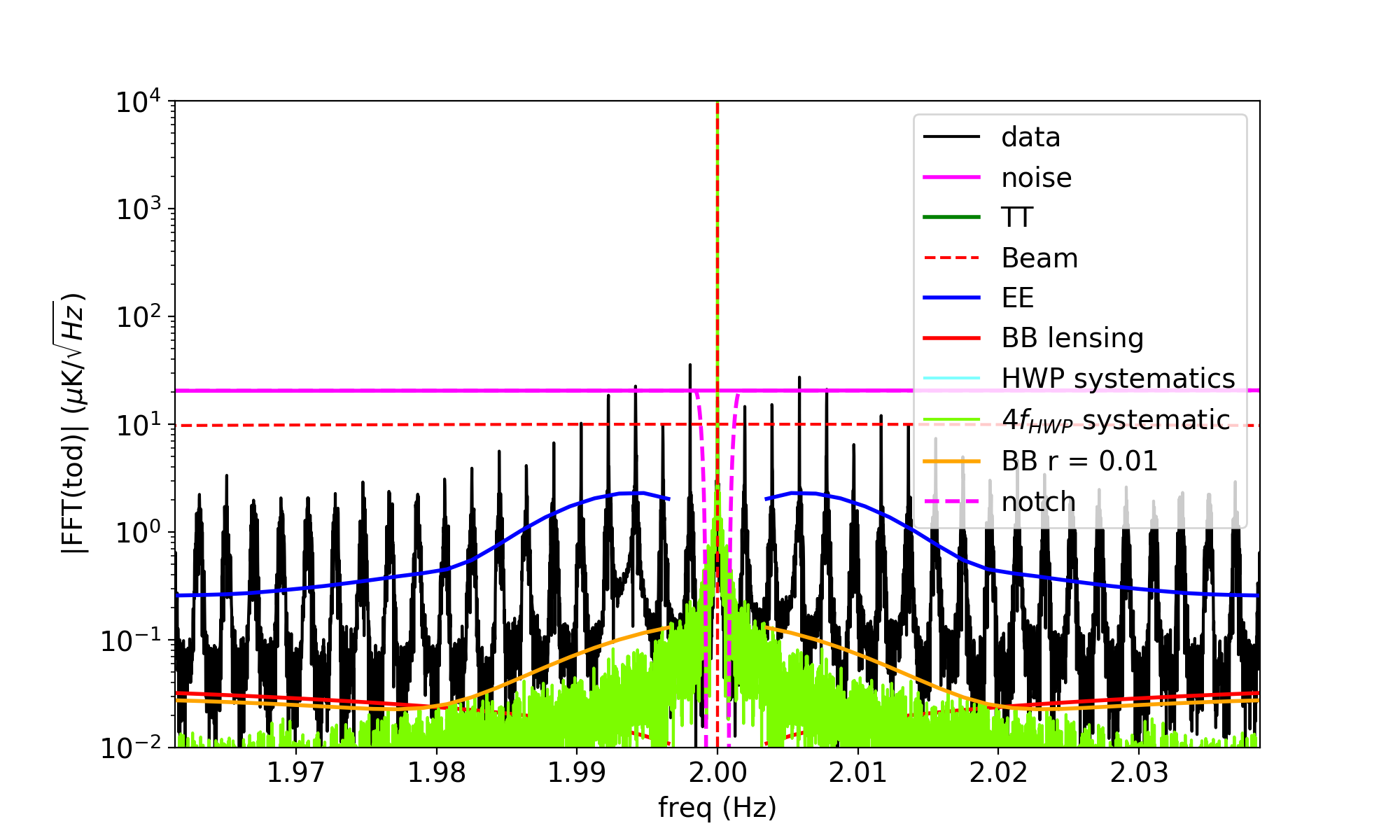}
\caption{\new{LSPE-SWIPE frequency spectrum}  (\emph{top}), and zoom-in 
near modulation frequency (\emph{bottom}),
for a 16 hours noise-free \new{CMB-only simulated timestream}. 
The black curve represents the data;
CMB temperature data are centered around 0 frequency, and polarization data
around $4f_\text{HWP}$. The magenta line is the noise for a single
detector at \SI{145}{GHz}. The magenta dashed line, is the noise multiplied by the notch filter.
Vertical dashed lines represent harmonics of the HWP spin
frequency.
The dark green curve is the expected signal for a temperature CMB
angular power spectrum, blue curve for E-mode power spectrum, 
red for B-mode (\new{lensing} only) and orange for inflationary B-mode. 
The light-green curve, visible in the bottom plot, is a systematic effect
at $4f_\text{HWP}$, with an amplitude of \SI{1}{mK}, spread in 
frequency due to the uncertainty in the HWP angular velocity 
$\sigma_{\omega_\text{HWP}}/\omega_\text{HWP}=0.6 \times 10^{-6}$.
The cyan clear curves are the systematic effects at 
1, 2, 3, 5$f_\text{HWP}$, as discussed in 
section~\ref{sec:swipe_sys}.
Since the signal is quasi-periodic, with period $T_\text{payload}$, 
its Fourier transform peaks at the modulation frequency $4f_\text{HWP}$
and then in frequency shifts equal to $\Delta f = 1/T_\text{payload}$, 
clearly visible in the bottom figure. 
\label{fig:swipe_fdata1}
}
\end{centering}
\end{figure}
\begin{figure}[!t]
\begin{centering}
    \includegraphics[width=11.4cm]{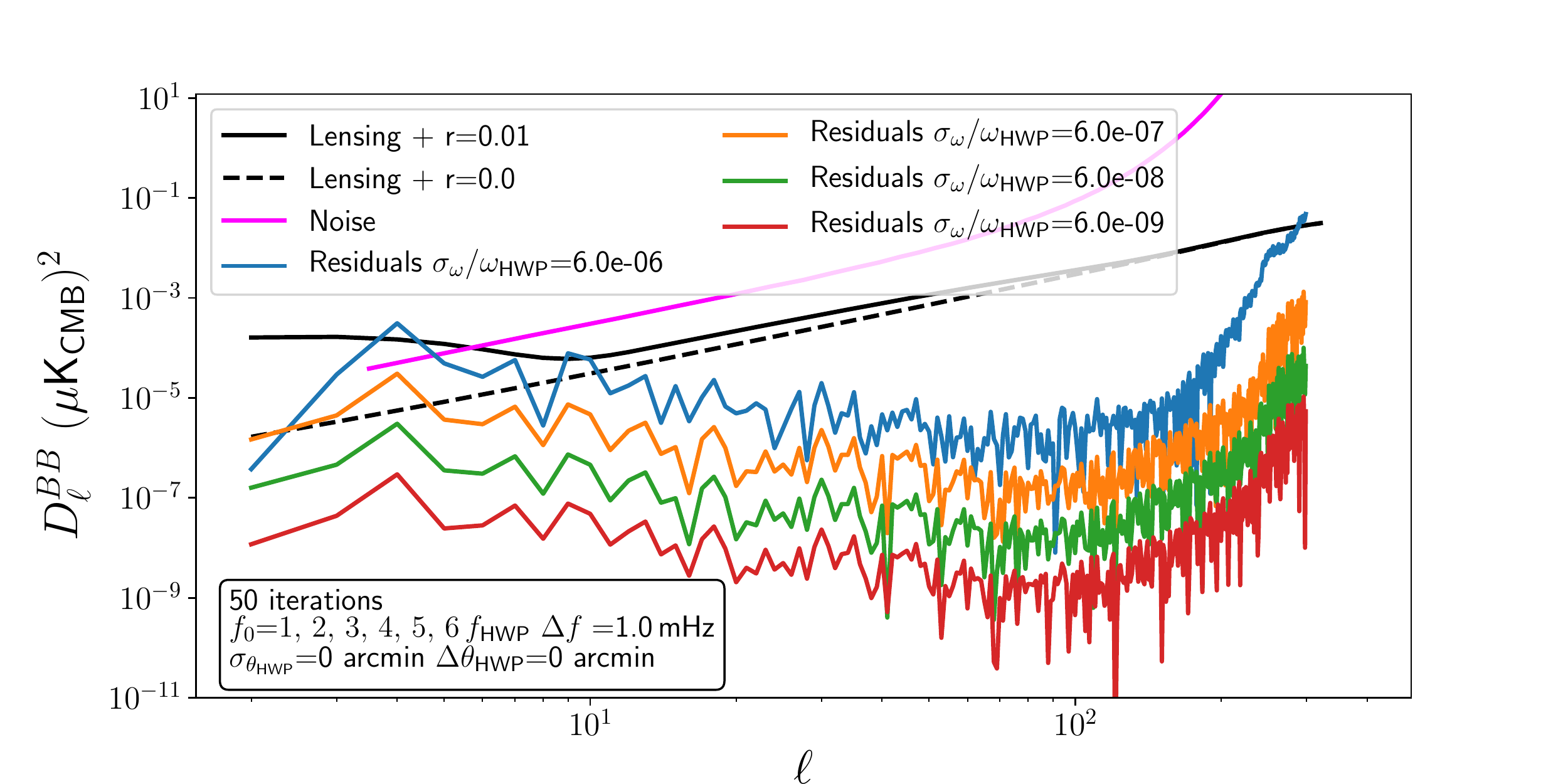}
\includegraphics[width=11.4cm]{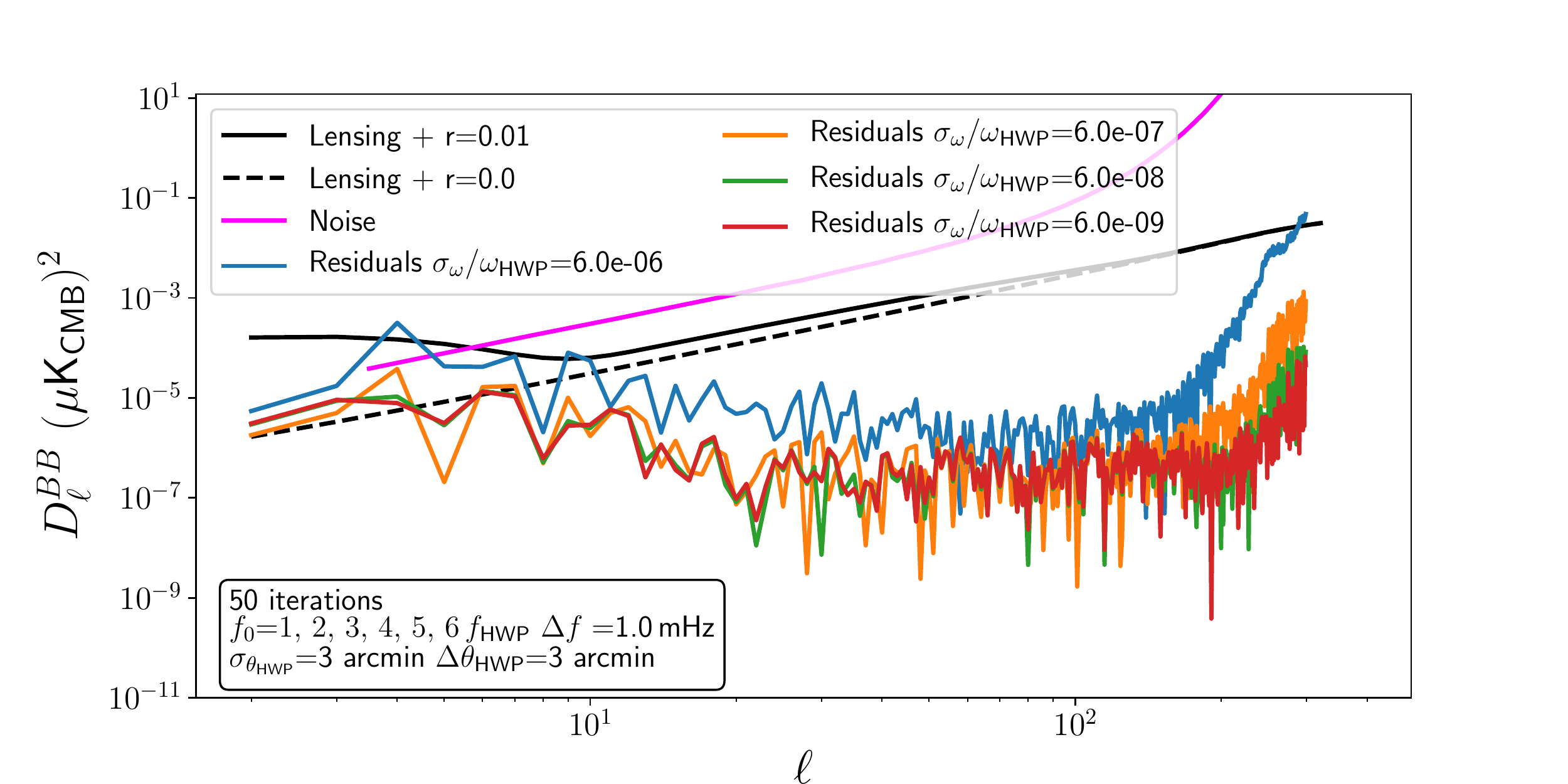}
\includegraphics[width=11.4cm]{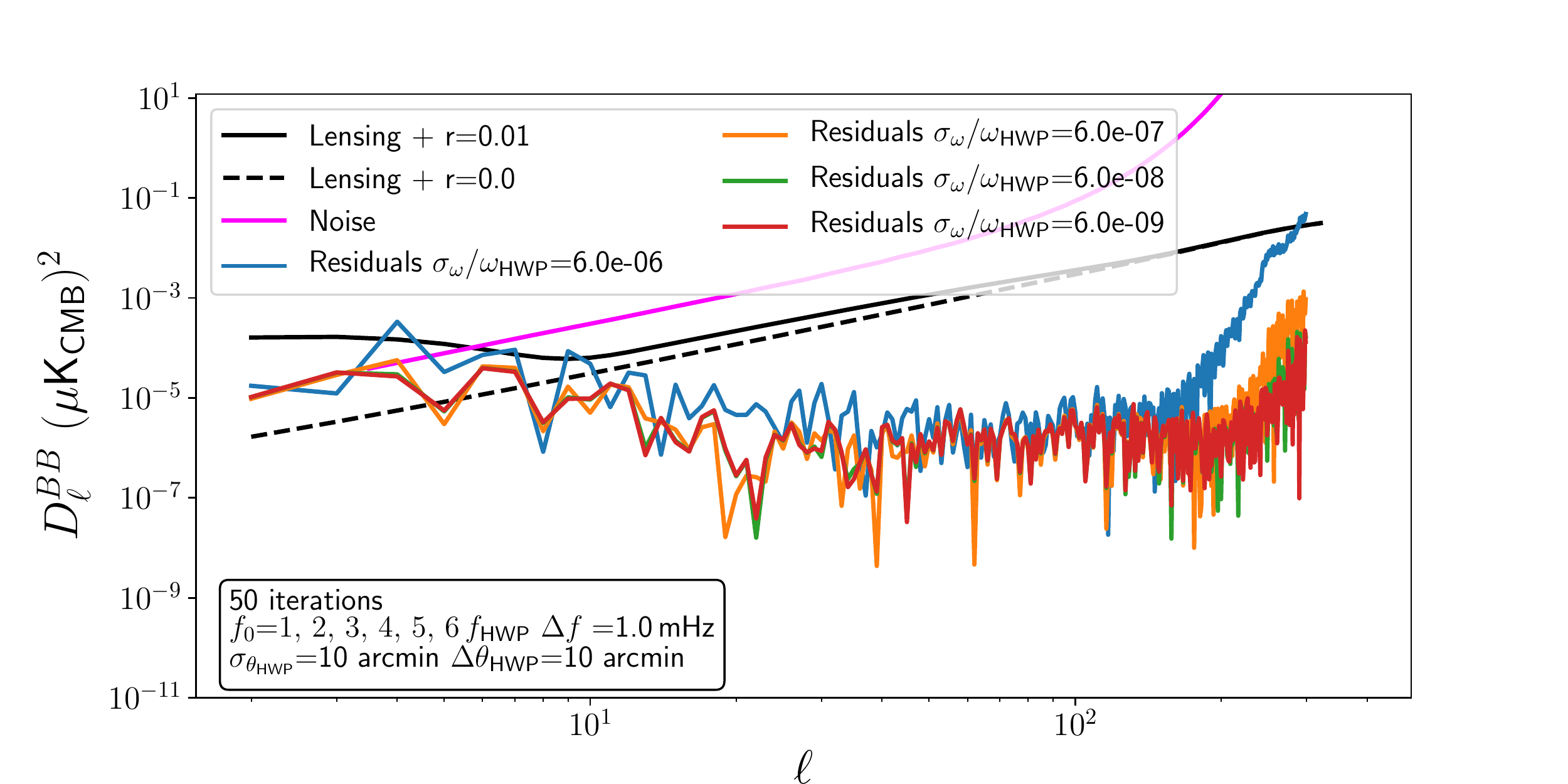}
    \caption{
    Impact of HWP synchronous 
    systematic effects at $\text{1}f_\text{HWP}$, $\text{2}f_\text{HWP}$, $\text{3}f_\text{HWP}$, $\text{4}f_\text{HWP}$ and $\text{5}f_\text{HWP}$, 
    with amplitude \SI{10}{mK}, \SI{10}{mK}, \SI{1}{mK}, \SI{1}{mK},
and \SI{1}{mK} respectively.
    Black lines: B-mode angular power spectra 
    for $r=0$ (dashed) and $r=0.01$ (continuous). 
    Coloured lines: 
    residual of B-mode power spectra of simulated map with 
    HWP synchronous systematic effects and notch filters, 
    with respect to simulated map without systematic effects nor filters. 
    The three panels are for the cases:
(\emph{top}) 
with HWP synchronous systematic effects, and HWP instability; 
(\emph{mid}) with HWP synchronous systematic effects, HWP instability, 
$\Delta \theta_\text{HWP}=\SI{3}{\arcmin}$\ and 
$\sigma_{\theta_\text{HWP}}=\SI{3}{\arcmin}$;
(\emph{bottom}) with HWP synchronous systematic effects, HWP instability, 
$\Delta \theta_\text{HWP}=\SI{10}{\arcmin}$ and 
$\sigma_{\theta_\text{HWP}}=\SI{10}{\arcmin}$;
in all panels, there are 4 continuous lines, 
\newb{corresponding to} different \newb{levels of 
HWP instability with} angular speed uncertainty: 
$\sigma_{\omega_\text{HWP}}/\omega_\text{HWP}=[0.6\times 10^{-8}, 0.6\times 10^{-7},
0.6\times 10^{-6},0.6\times 10^{-5}]$.
The magenta line is the noise power spectrum after component separation. 
}
    \label{fig:BBresiduals}
    \end{centering}
\end{figure}
In order to have a more realistic simulation, 
this contamination has been 
combined with:
\begin{itemize}
    \item an {\bf error in the measurement of the HWP angle}. 
    This is defined
    by the uncertainty with which we can readout the HWP angle.
    This is described by the parameter 
    $\sigma_{\theta_\text{HWP}}$. 
    We have explored 
    $\sigma_{\theta_\text{HWP}} = [0, 3, 10]~\SI{}{\arcmin}$.
    Our measurement precision, with Kalman filter approach
    \footnote{
    \new{The Kalman filter combines the dynamic model of the 
    polarization modulation mechanism, 
    the physical properties of the system, and multiple sequential
    measurements to make an estimate of the varying quantities that is better than the estimate obtained by using only 
    measurements.
    The filter works on discrete sampling, and combines a dynamical
    prediction of the next position with its noisy measurement.
    The combination is done using proper weights, that 
    take optimally into account the noise level and the quality 
    of the dynamical prediction. 
    }
    }, 
    is of order 
    $\sigma_{\theta_\text{HWP}} \leq \SI{0.1}{\arcmin}$;
    \item an {\bf offset in the knowledge of the HWP angle}. 
    This is described by the parameter 
    $\Delta \theta_\text{HWP}$.
    We explored the values $\Delta \theta_\text{HWP}=[0, 3, 10]~\SI{}{\arcmin}$;
    \item an {\bf instability in the HWP rotation rate}, modeled
    as a noise in the angular velocity of the HWP.
    This is described in terms of relative error by the parameter
    $\sigma_{\omega_\text{HWP}}/\omega_\text{HWP}$.
    The angular velocity samples are simulated as
    \begin{equation}\label{eq:hwp_istability}
    \omega_i = \omega_{HWP} \left(1+\frac{\sigma_{\omega_\text{HWP}}}{\omega_\text{HWP}}
    n_i  \sqrt{\frac{f_\text{sampling}}{f_\text{HWP}/\text{dpr}}}
    \right)
    \end{equation}
    where $n_i$ is a sample of a normal distributed random number, $f_\text{sampling}$ is the 
    simulation sampling rate \new{(100\,Hz in our case)}, $f_\text{HWP}$ is the 
    HWP rotation frequency, and ${\rm dpr}=64$ is the number 
    of angle measurements of the HWP in a revolution. 
    \new{The ratio}
    $f_\text{sampling}/(f_\text{HWP}/\text{dpr})$ 
    corresponds to the number of simulated samples between two subsequent HWP measured positions (data per revolution). 
    The simulation with equation~\ref{eq:hwp_istability} results
    \new{in} the same angular drift between two HWP \new{angle measurements},
    independently of the sampling rate of the simulation. 
    We explored the values 
    $\sigma_{\omega_\text{HWP}}/\omega_\text{HWP}=[0,0.6\times 10^{-8},0.6\times 10^{-7}, 0.6\times 10^{-6}, 0.6\times 10^{-5} ]$.
    Using
    the Kalman filter approach, we have measured
    a precision in the determination of the HWP 
    angular velocity
    of order $\sigma_{\omega_\text{HWP}}/\omega_\text{HWP} = 2\times10^{-6}$.
\end{itemize}
The major contribution from HWP spin rate instability 
is due to the fact that the various HWP synchronous
effects are not at a single frequency anymore, but are
spread in frequency due to the instability, thus
reducing the efficacy of the notch filter; 
this is partially compensated by measuring this
instability, and filtering in \newb{the} angle domain instead 
of \newb{the} time domain, but with some limitation coming
from the uncertainty in the angular velocity measure,
$\sigma_{\omega_\text{HWP}}/\omega_\text{HWP}$.

\begin{figure}[t!] 
   \centering
   \includegraphics[width=0.65\textwidth]{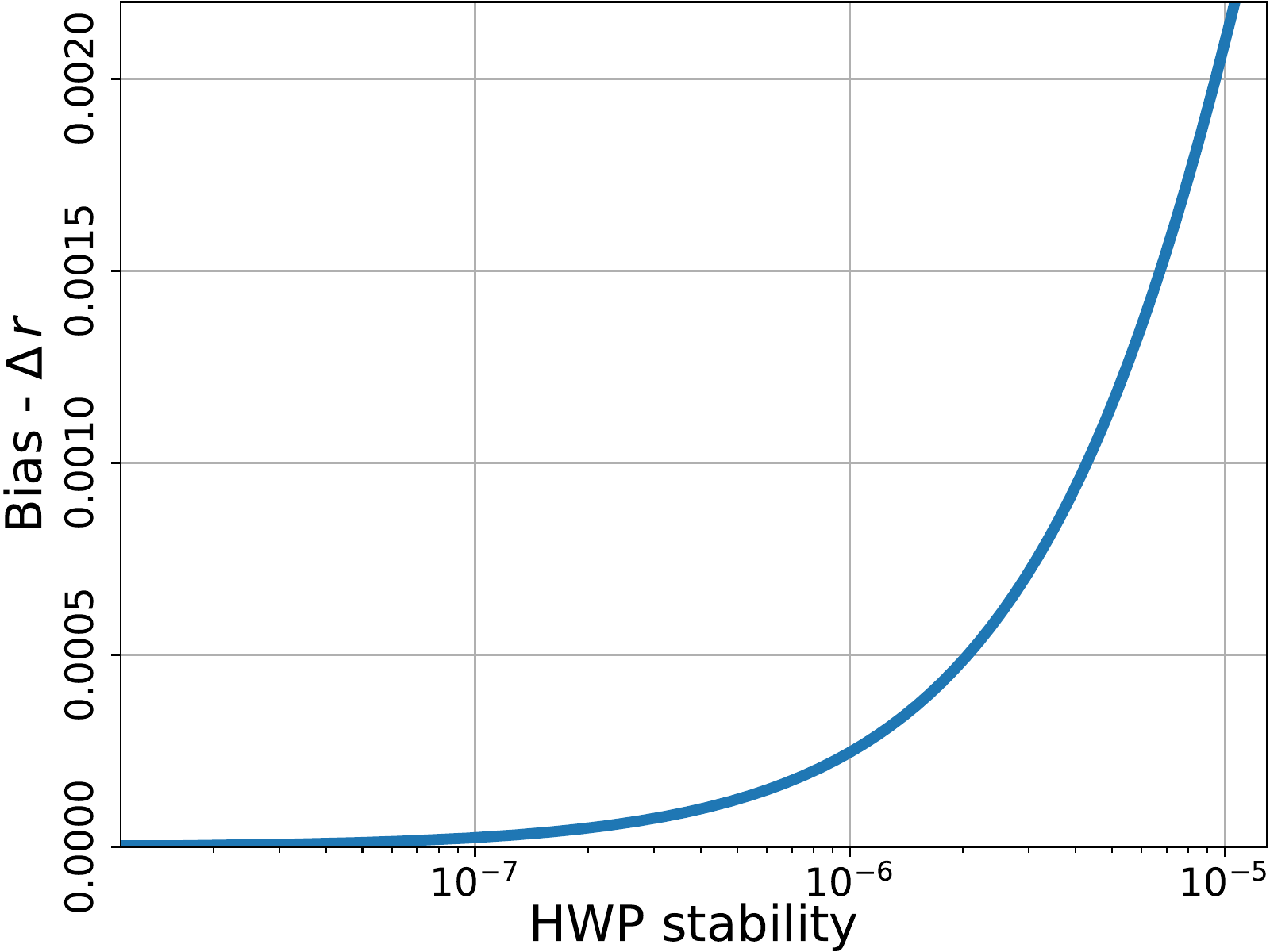}
   \caption{
   Bias in $r$ due to instability of the HWP, combined with 
   instrumental polarization, HWP angle offset, and HWP angle
   measurement error. 
   }
    \label{fig:swipe_r_HWP_systematics}%
\end{figure}

Figure~\ref{fig:swipe_fdata1} 
illustrates the 
\new{LSPE-SWIPE frequency spectrum} 
of
a \SI{16}{hrs} noise-free timeline, for simulations 
of CMB and systematic effects \new{(see caption for details)}. 
Figure~\ref{fig:BBresiduals} presents the results of this analysis in terms of B-mode angular power spectra. 
The black lines are the B-mode angular power spectra in case of 
$r = 0$ and $r=0.01$, which is the limit of our sensitivity. 
The coloured lines represent the residual power spectra
of the case with systematic effects and notch filter, versus the ideal case, without systematic effects nor filters.
Notably, in this figure we consider a combination 
of several systematic effects: instrumental polarization, 
HWP angle errors, HWP angle offset, uncertainty in the 
measurement of the HWP angular velocity. 
Applying \newb{the} $r$ estimation pipeline (see section~\ref{subsec:likelihood}) 
to the maps contaminated by the combinations of systematic effects just described, we obtain a bias in $r$ as reported in 
figure~\ref{fig:swipe_r_HWP_systematics}. The maps used to 
produce values in this figure are simulated with
$\sigma_{\theta_\text{HWP}} = \SI{10}{\arcmin}$;
$\Delta\theta_\text{HWP}=\SI{10}{\arcmin}$;
synchronous systematic effects at
[1,2,3,4,5]\;$f_\text{HWP}$ with 
amplitude [10,10,1,1,1]~\SI{}{mK} respectively;
and $\sigma_\omega / \omega_\text{HWP}$
as in the abscissas.

\subsubsection{LSPE-SWIPE calibration}
\label{subsec:swipe-calibration}

From the analysis reported in section~\ref{sec:swipe_sys_and_cal} and 
subsections, we set the requirements reported in
table~\ref{tab:swipe_sys_reqs}, for the most critical systematic 
effects for LSPE-SWIPE polarization measurement, considered jointly. 
\newb{In the table the B-mode r.m.s contamination is computed from 
the r.m.s. of the anisotropy (for intensity to polarization leakage) 
or from the r.m.s. of the E-mode (for the E- to B-mode leakage) multiplied
by the leakage factor. The r.m.s. is computed from the angular power spectrum as
\begin{equation*}
\sqrt{\langle \Delta T^2 \rangle} = \sqrt{\frac{1}{4\pi} \sum (2\ell+1) C_\ell B_\ell^2 }
\end{equation*}
The r.m.s values reported in the table are computed for 
a uniform systematic effect on the full focal plane. 
In practice each detector will have a different effect, and 
for some of the systematic effects
cancelations will occur due to redundancy. 
On the mission data,}
the impact of the residual systematic effects will be 
assessed by estimation of the bias in the angular power
spectra, by means of end-to-end Monte Carlo simulations. 

    \begin{table}[!t]
        \begin{center}
            \begin{tabular}{l l l l}
            \hline\noalign{\smallskip}
                LSPE-SWIPE Parameter &  Requirement & \newb{B-mode } & \newb{reference} \\
                &&\newb{(r.m.s.)}&\\
                \hline
                \hline
Instrumental polarization$^{1,2}$ \dotfill  &   $<4\times10^{-4}$ & 18\,nK$_\text{CMB}$ &\newb{\ref{subsec:swipe_sys_requirements}} \\
Cross polarization$^2$      \dotfill    &   $<0.02$ & 10\,nK$_\text{CMB}$ &\newb{ \ref{subsec:swipe_sys_requirements}}\\
Polarization angle recovery \dotfill&  $<\SI{40}{\arcmin}$ & 12\,nK$_\text{CMB}$ & \newb{\ref{subsec:HWPsyst}} \\
WG angle error   \dotfill    & $\Delta \phi_\text{WG} < \SI{20}{\arcmin}$ & 6\,nK$_\text{CMB}$& \newb{\ref{subsec:HWPsyst}, \ref{sec:swipe_sys}}\\
HWP angle offset    \dotfill        & $\Delta \theta_\text{HWP} < \SI{10}{\arcmin}$ & 6\,nK$_\text{CMB}$& \newb{\ref{subsec:HWPsyst}, \ref{sec:swipe_sys}}\\
HWP angle noise \dotfill            & $\sigma_{\theta_\text{HWP}} < \SI{10}{\arcmin}$ & 6\,nK$_\text{CMB}$& \newb{\ref{subsec:HWPsyst}, \ref{sec:swipe_sys}}\\
Time constant knowledge$^2$   \dotfill  & $\Delta {\tau_\text{LP}} < \SI{1.0}{ms}$ & 6\,nK$_\text{CMB}$& \newb{\ref{subsec:HWPsyst}, \ref{sec:swipe_sys}}\\
HWP angular velocity measurement \dotfill & $\sigma_{\omega_\text{HWP}}/\omega_\text{HWP} < 5 \times 10^{-6}$  & 6\,nK$_\text{CMB}$ & \newb{\ref{sec:swipe_sys}} \\
Mueller matrix $I\rightarrow Q/U$ terms knowledge & $\Delta {M^{4\theta}_{IQ}},\Delta {M^{4\theta}_{IU}}< 10^{-4} $ & 6\,nK$_\text{CMB}$& \newb{ \citep{2018_Imada_IEEE,2018JCAP...09..005K}} \\
\noalign{\smallskip}\hline
 \multicolumn{4}{p{15 cm}}
            {\footnotesize \newb{$^1$ This is an inevitable 
            term given by the effect of the radiation crossing 
            the cryostat window with a tilted angle, and is 
            so large only for the most off-axis detectors.
            The equivalent B-mode r.m.s. value is obtained multiplying the CMB
            anisotropy r.m.s by the coefficient, divided 
            by $\sqrt 2$, assuming equal E-, B-mode distribution. }}\\
  \multicolumn{4}{p{15 cm}}
            {\footnotesize \newb{           
            $^2$ The r.m.s. contribution is for a single detector
            and is not expected to be correlated among detectors. 
            The impact will be reduced due to redundancy 
            and cancelations. 
            }}
            \end{tabular}
            \caption{\label{tab:swipe_sys_reqs}LSPE-SWIPE main
            systematic effects \new{and calibration} 
            requirements.
            These requirements are derived considered the 
            various effects jointly.
            $\Delta$-s represent maximum offsets between true and measured values, $\sigma$-s
            represent fluctuations between the true values and the measured ones.
\newb{The B-mode r.m.s. contamination is computed from leakage
of intensity \newc{or} E-mode r.m.s into B-mode. This is only an order of 
magnitude estimation of the real impact into $r$ estimation, given that the real contamination depends on the distribution of the leakage in the B-mode angular power spectrum.} \newb{Some of these results are derived by a order of magnitude 
analysis in section~\ref{subsec:swipe_sys_requirements}, 
some from literature papers, while the 
most critically related to the use of the HWP are derived
from the simulations described in sections~\ref{subsec:HWPsyst}
and~\ref{sec:swipe_sys}
(see the reference column). 
            } }
        \end{center}
    \end{table}

As in the case of \Strip, the SWIPE calibration will be performed 
in multiple stages:
\begin{enumerate}
    \item at sub-system level: components will be tested individually 
    in order to define specific properties. These components include
    optical filters, HWP, horns, detectors, readout electronics;
    \item at system level: the integrated system, will undergo a long
    list of calibration tests
    \new{on the ground}. These include test of the polarization 
    properties of the integrated system as a function of frequency, 
    band-integrated polarization properties, angle dependent polarization
    properties, band-pass definition, angular response (by means of a 
    far field thermal source). It is worth noting that the properties
    of the instrument are not expected to change from ground to the
    stratosphere, given that the thermal configuration is the same;
    the major change will be in the different background which is expected
    to modify the detector responsivity, to be confirmed in flight; 
    \item during observation: the payload will
    undergo a limited number of tests for verification \new{of the ground-based calibration parameters.} 
    In particular,
    these tests will be updating detector responsivity, updating the 
    pointing direction of each 
    detector with respect to the 
    telescope reference frame, and 
    confirming polarization properties of the system, by 
    observation of the Crab nebula~\cite{2020aumont}, and by minimization
    of the E-B modes correlation. 
\end{enumerate}


\section{Results}\label{sec:results}

In this section we describe the component separation and the likelihood methods, and we \new{forecast} the main 
results of LSPE in terms of cosmological parameters. 
It is assumed that the systematic effects are within the
\newc{requirements} defined in section~\ref{sec:systematics}.
\new{
In each of the maps the noise is 
estimated using the NETs from tables~\ref{tab_strip_white_noise_properties} and~\ref{tab:swipe-noise},
 projected into maps by means of the 
 instrument simulators described 
 in appendix~\ref{app:instrument_sim}.
}


\subsection{Component separation}\label{sec:component_separation}

Component separation is a key element in CMB data analysis, and it turns out to be particularly challenging for the extraction of CMB polarization (see e.g. \citep{2005_Tucci,buzzelli_migliaccio2018,2020A&A...641A...4P}). In particular, diffuse Galactic dust and synchrotron emissions are the most relevant foregrounds in 
polarization. 
For the analysis presented in this paper, we consider the component separation apparatus represented by the \texttt{ForeGroundBuster}\footnote{\url{https://github.com/fgbuster/fgbuster} }, which is currently used to assess the foreground cleaning capabilities of a number of CMB B-mode probes \citep{Campeti:2019ylm,Stompor:2016hhw}. 
The method fits, in each pixel observed by both \Strip\ and SWIPE, for CMB signal, amplitude \newc{and} spectral \newc{index} of synchrotron, temperature, amplitude and \newc{spectral index} of dust. 
In our analysis, we include thermal dust and synchrotron as polarized foregrounds. \new{We exploit the publicly available package {\tt Python Sky Model} (PySM)\footnote{\url{https://github.com/bthorne93/PySM\_public}} which generates full sky simulated foregrounds in intensity and polarization, and consider the {\tt d0s0} configuration outlined below \citep{pysm}.} 
The synchrotron spectral brightness is modeled as a power law in frequency with a constant spectral index $\beta_s= -3$: $I_s(\nu,\hat{n}) = A_{s}(\hat{n}) (\nu/\nu_0)^{\beta_{s}}$, where $A_{s}(\hat{n})$ is the synchrotron amplitude. 
The dust component is modeled as a grey body, i.e. an almost thermal component at a temperature of $T_{d}=20$\,K, heated back by starlight \new{which is represented as an emissivity factor scaling as a power law in frequency, with} spectral index $\beta_{d}=1.54$: $I_d(\nu, \hat{n}) = \tau_{0}(\hat{n})(\nu/\nu_0)^{\beta_{d}}B(\nu,T_{d})$. The component separation procedure is performed only in polarization and recovers the value of the spectral indices $\beta_s, \beta_d$ \new{which are not varying in the sky}, as well as the amplitude of the synchrotron signal $A_s(\hat{n})$ and dust optical depth $\tau_{0}(\hat{n})$, \new{which vary with the sky direction}. 
\newb{The assumption of uniform spectral indices
 could lead to biases on the estimation of the tensor-to-scalar ratio. 
 This is true, in particular, for experiments targeting $r \simeq 10^{-3}$, for which  the estimation of the foreground parameters must be done in the various regions of the sky independently.
For experiments targeting higher values of $r$, such as LSPE, the impact of this spatial variation is expected to be less dramatic, leading to a bias
lower than or comparable to the statistical uncertainties. This result is confirmed by the SO paper \cite{2019JCAP_SO}, 
section 3.4.2, in particular in Table 4, where the first row of the table reports the case of fitting a sky model built with varying spectral indices, assuming uniform spectral indices in the reconstruction.  This results in a bias in $r$ of order $\Delta r \simeq 2\times10^{-3}$.  
}

In addition to the LSPE bands, we also consider the observations of the \Planck\ satellite between 30 and \SI{353}{GHz} \citep{planck2016-l01}, and the ones of QUIJOTE at \SI{11}{GHz} \citep{quijote,2019QUIJOTE}. \new{At each of the frequencies corresponding to these probes, and separately for the $Q$ and $U$ Stokes parameters, we generate and add foregrounds and CMB using the PySM, and also add noise realizations according to the SWIPE, \Strip, Planck and QUIJOTE sensitivities. We convolve all maps with a Gaussian beam in order to reach a common \SI{85}{arcmin} FWHM, \newb{which} corresponds to the largest beam associated to the LSPE channels. All maps are generated using \healpix \footnote{Hierarchical Equal Latitude Pixelization} at  $N_\text{side} = 128$. We did not consider any pixel-pixel correlation property of the noise.}
\new{The data model which is implemented in \texttt{ForeGroundBuster} and used in this paper is $d(\nu) \equiv \tens{A}s + n$, where $d(\nu)$ contains measured signal at each frequency $\nu$, $s$ are the maps of the different components, $\tens{A}$ is the mixing matrix which contains the parametric model to fit, and $n$ represents the noise in the maps. The parametric component separation process consists in obtaining an estimate $\tilde{s}=\tens{W}d$ of the components, by means of a kernel operator $\tens{W}$.}
Therefore, the key element of component separation is the $\tens{W}$ matrix, which is the linear operator that mixes
the frequency maps in the component maps, taking into account
the sensitivity and the contribution of each frequency 
to each astrophysical component. The elements $W_{i,j}$ of the 
$\tens{W}$ matrix (often referred as {\em weights}) admit negative values for frequencies that must 
be subtracted in order to solve for the astrophysical component. Frequency bands and weights for each component  are shown in table~\ref{tab:CS_weights_baseline}.
\begin{figure}[t]
   \centering
   \includegraphics[width=0.8\textwidth]{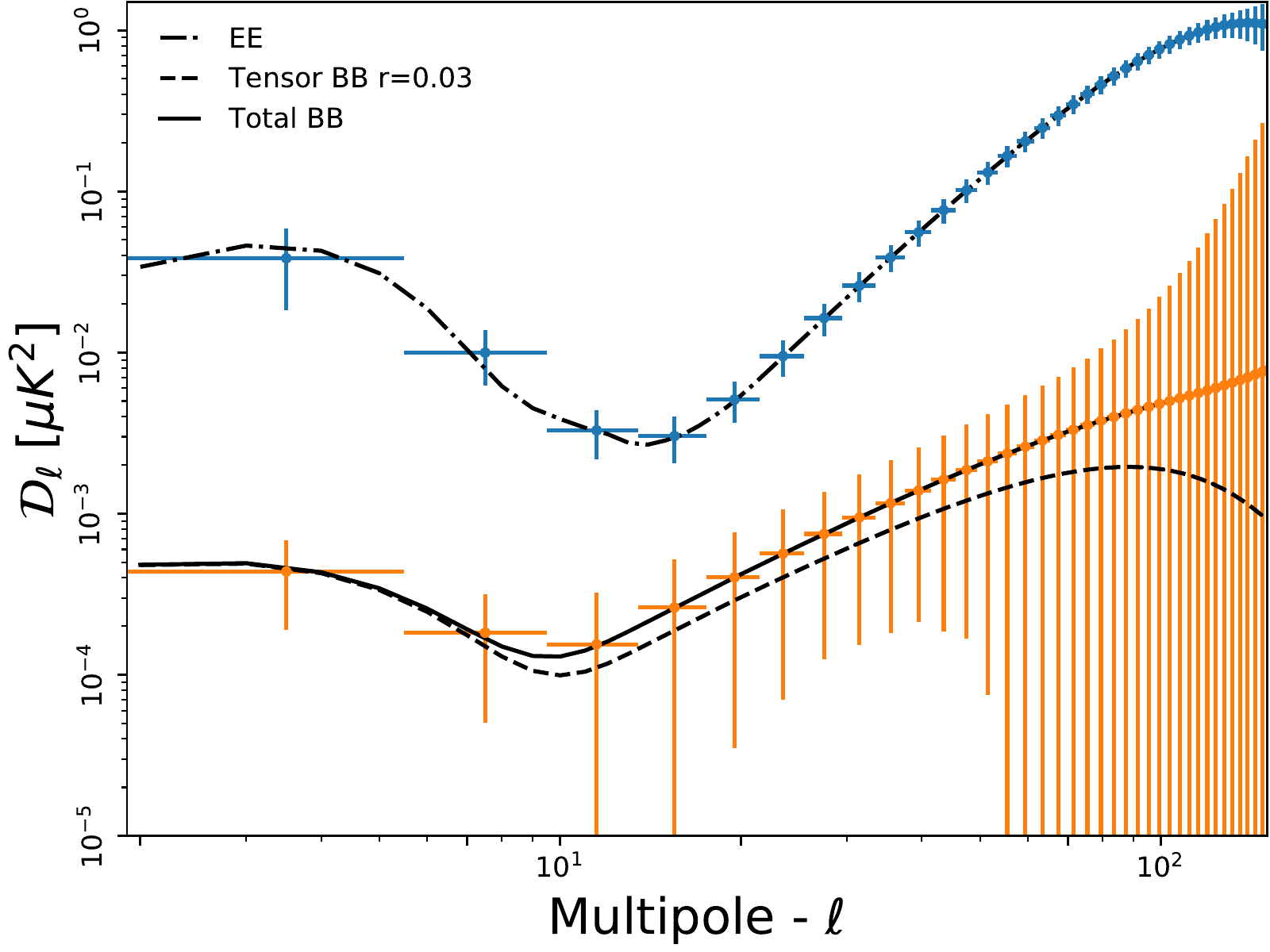}
   \caption{CMB E-mode (blue) and B-mode (orange) power spectra averaged over 1000 simulations after Component separation \new{in the baseline case of table~\ref{tab:CS_weights_baseline}}. }
    \label{fig:EE_and_BB_power}%
    \end{figure}
From this table it is clear that the 145\,GHz channel is the most important one for reconstructing the CMB,
clearly adding sensitivity to the currently available datasets. 
The table also quantitatively shows relevance 
\newc{of higher and lower frequency bands} 
for fitting and subtracting foregrounds. 
As an illustration, figure~\ref{fig:EE_and_BB_power} shows the polarization CMB power spectra \citep{Alonso:2018jzx} obtained averaging 1000 simulations after component separation.

In table~\ref{tab:CS_weights_minimal}, we show a minimal \new{case}, where we have used just the \SI{30}{GHz} channel of \Planck\ and LSPE frequencies. Moreover, table~\ref{tab:spectral_indexes} shows the accuracy of the component separation in terms of dust and synchrotron spectral indices. 

\begin{table}[t]
\centering
    \begin{tabular}{c|c|c|c|c}
        \hline
        \hline
        Band (GHz) & Probes & w$_{\rm CMB}$ $\scriptstyle\times 10^{3}$ & w$_{\rm Dust}$ $\scriptstyle\times 10^{3}$ & w$_{\rm Synch}$ $\scriptstyle\times 10^{3}$\\
         \hline
        11 & QUIJOTE & $-$1.1 & 0.24 & 56 \\
        30 & \Planck & 2.5 & $-$1.1 & 18 \\
        43 & Strip & 4.4 & $-$1.9 & 8.0 \\
        44 & \Planck & 1.9 & $-$0.82 & 3.2 \\
        70 & \Planck & 2.8 & $-$1.1 & 0.86 \\
        100 & \Planck & 14 & $-$5.3 & 0.41 \\
        143 & \Planck & 26 & $-$7.4 & $-$2.2 \\
        145 & SWIPE & 1200 & $-$330 & $-$110 \\
        210 & SWIPE & $-$130 & 200 & 5.6 \\
        217 & \Planck & $-$7.1 & 8.5 & 0.48 \\
        240 & SWIPE & $-$150 & 130 & 14\\
        353 & \Planck & $-$9.9 & 6.5 & 1.2 \\

        \hline
    \end{tabular}
        \caption{Component separation weights for each component in each channel.}
    \label{tab:CS_weights_baseline}
\end{table}

\begin{table}[t]
\centering
\begin{tabular}{c|c|c|c|c}
        \hline
        \hline
        Band (GHz) & Probes & w$_{\rm CMB}$ $\scriptstyle\times 10^{3}$ & w$_{\rm Dust}$ $\scriptstyle\times 10^{3}$ & w$_{\rm Synch}$ $\scriptstyle\times 10^{3}$ \\
         \hline
        30 & \Planck & $-$15 & 2.7 & 870 \\
        43 & Strip & $-$2.6 & $-$0.45 & 390 \\
        145 & SWIPE & 1400 & $-$410 & $-$1600 \\
        210 & SWIPE & $-$190 & 240 & 28 \\
        240 & SWIPE & $-$200 & 160 & 340 \\
        \hline
    \end{tabular}
        \caption{Component separation weights for each component in each channel.}
    \label{tab:CS_weights_minimal}
\end{table}

\begin{table}[t]
\centering
    \begin{tabular}{c|c|c}
        \hline
        \hline
        Parameter & Mean & $\sigma$ \\
         \hline
        $\beta_d$ & 1.539 & 0.001 \\
        $\beta_s$ & $-$2.999 & 0.002 \\
        \hline
    \end{tabular}
        \caption{Dust and synchrotron spectral indices obtained by parametric component separation.
        The component separation algorithm fits for a single 
        value in each map. The uncertainties are derived from the 
        standard deviation of 1000 realizations of the noise in the maps. }
    \label{tab:spectral_indexes}
\end{table}

\subsection{Likelihood}\label{subsec:likelihood}

The likelihood used in the parameter estimation is based on maps of Stokes parameters $T,Q,U$ in \healpix\ format. For the temperature map we assume perfect component separation outside a Galactic masks with \SI{2}{\micro K} per pixel of white noise\footnote{In temperature we assume signal dominated observations. The white noise added is only necessary for regularizing the inversion of the temperature block of the TQU covariance matrix, see e.g. \cite{planck2016-l05}}. The polarization maps after the component separation procedure, described in the previous section, are modeled as a sum of CMB signal, instrumental Gaussian noise and foreground Gaussian residuals. 
\new{This \newb{modeling} is consistent with analyses performed on current data at large 
angular scales~\cite{planck2016-l05} and with \newc{forecasts} performed on other forthcoming data~\cite{2019JCAP_SO}.} 
In this scenario the full likelihood expression reads 
\begin{equation}
\mathcal{P}(\vec{m}|C_{\ell})=\frac{1}{2\pi|\tens{C}(C_{\ell})|^{1/2}}
\exp\left(-\frac{1}{2}\vec{m}^{\tens{T}}\,\left[\tens{C}(C_{\ell})\right]^{-1}\vec{m}\right)\, ,
\label{eq:pbLike}
\end{equation}
where $\vec{m}\equiv{T,Q,U}$ is the data vector and $\tens{C}$ is total covariance matrix defined as the sum of signal and noise parts as 
\begin{equation}
  \tens{C}(C_{\ell})=\sum_{\ell=2}^{\ell_{\rm max}} \sum_{XY} \frac{2\ell+1}{4\pi} \, B_\ell^2 C_{\ell}^{XY} \tens{P}_{\ell}^{XY}+\tens{N},
\end{equation}
here $B_\ell$ is the beam window function, $\tens{P}_{\ell}^{XY}$ are the associated Legendre polynomials, as defined in \citep{Tegmark:2001zv}, and \tens{N} is the pixel-pixel noise covariance matrix.

In order to speed up the computation we perform the likelihood evaluation on lower resolution maps, still able to keep the full potentiality of LSPE maps. We consider two resolutions, \healpix\ $\Nside=16$ \newb{($\ell_\text{max}=32$)}, 
with a Gaussian beam of $\text{FWHM}=\SI{440}{\arcmin}$, which allows us to measure the E-mode reionization peak \newb{used for 
optical depth $\tau$ estimation in section~\ref{sec:optical_depth}},
and \healpix\ $\Nside=64$ \newb{($\ell_\text{max}=128$)}, with a Gaussian beam of $\text{FWHM}=\SI{110}{\arcmin}$, capable of measuring both the reionization and recombination peaks of the B-mode \newc{spectrum},
used for tensor-to-scalar ratio $r$ and cosmic birefringence estimation in sections~\ref{subsec:r_likelihood} and~\ref{subsec:biref}.

The likelihood analysis is performed simultaneously on a Monte Carlo of 1000 CMB, noise and residual foreground realizations. For each realization we estimate the reionization optical depth $\tau$ and the tensor-to-scalar ratio $r$. 
For each instrumental configuration the LSPE uncertainty on $\tau$ and $r$ is computed taking the average over the 1000 realizations of the $\log\left(\mathcal{P}(\vec{m}|C_{\ell})\right)$. 
In this way we efficiently take care of the scatter due to cosmic variance and instrumental noise. 
The other $\Lambda$CDM parameters are not sampled in this analysis, nonetheless we verified that opening the parameter exploration to full $\Lambda$CDM, and including a high-$\ell$ likelihood with noise performance compatible with \Planck, provides equivalent results.

\subsection{Reionization optical depth constraints}
\label{sec:optical_depth}

Measuring the polarization at very large scales, in particular the so-called reionization bump, allows 
\newc{constraining}
the Thompson scattering optical depth $\tau$. 
LSPE provides a cosmic variance limited measure of the polarization signal at very large scales ($\ell\lsim 20$) on $\sim 35\%$ of the sky. For the analysis presented here we conservatively consider a smaller portion, $f_{\rm sky,cmb}\simeq 25\%$, removing regions close to the Galactic plane potentially contaminated by residual foregrounds. The LSPE sensitivity \newc{over such a} sky fraction overcomes the current best estimates provided by \Planck\ HFI, i.e. $\tau=0.059\pm0.006$ \citep{Pagano:2019tci}, reaching $1-\sigma$ error on $\tau$ of $\sim 0.004$.

\begin{figure}[!tb]
   \centering
   \includegraphics[width=0.48\textwidth]{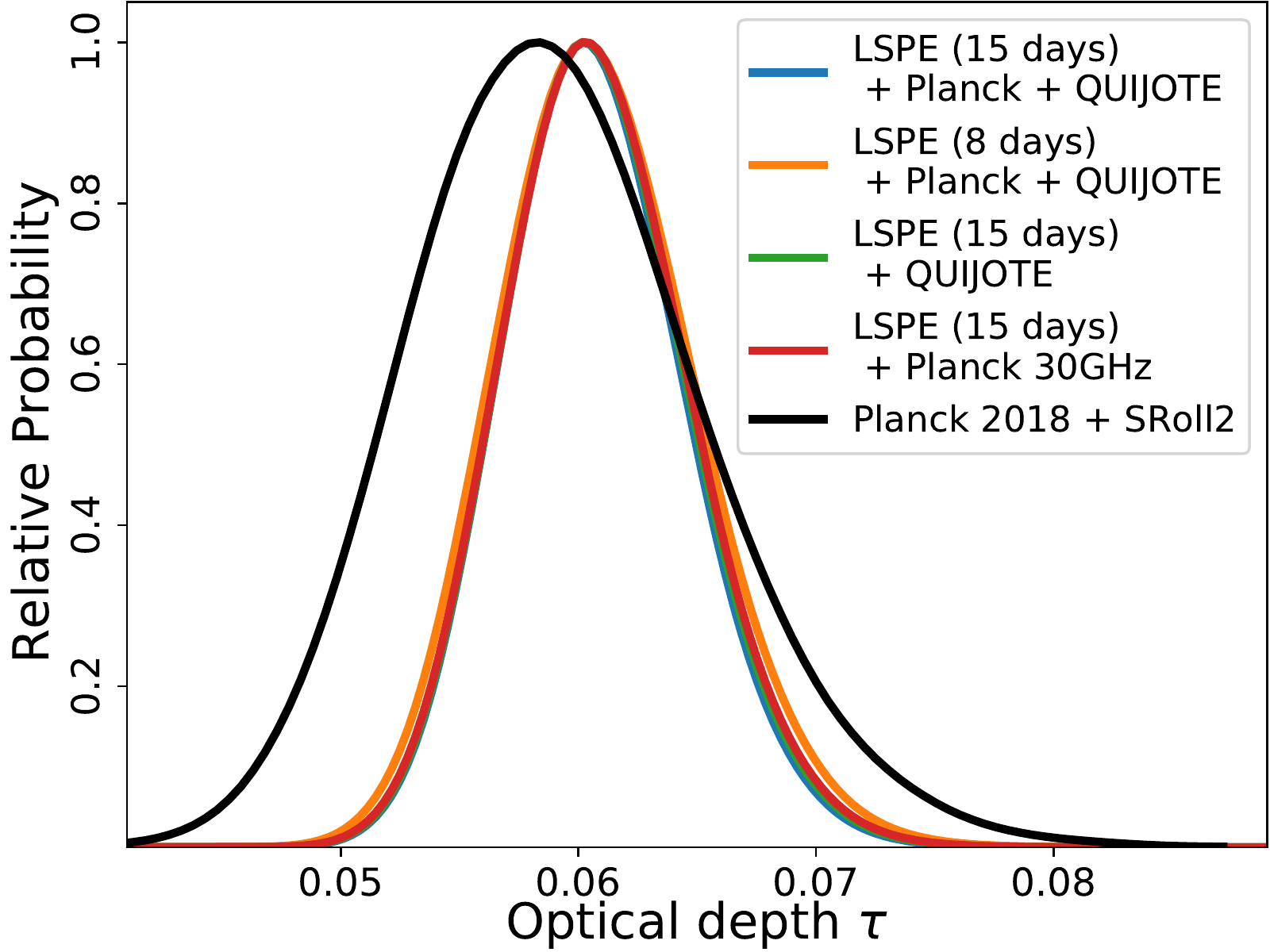}~~~
   \includegraphics[width=0.48\textwidth]{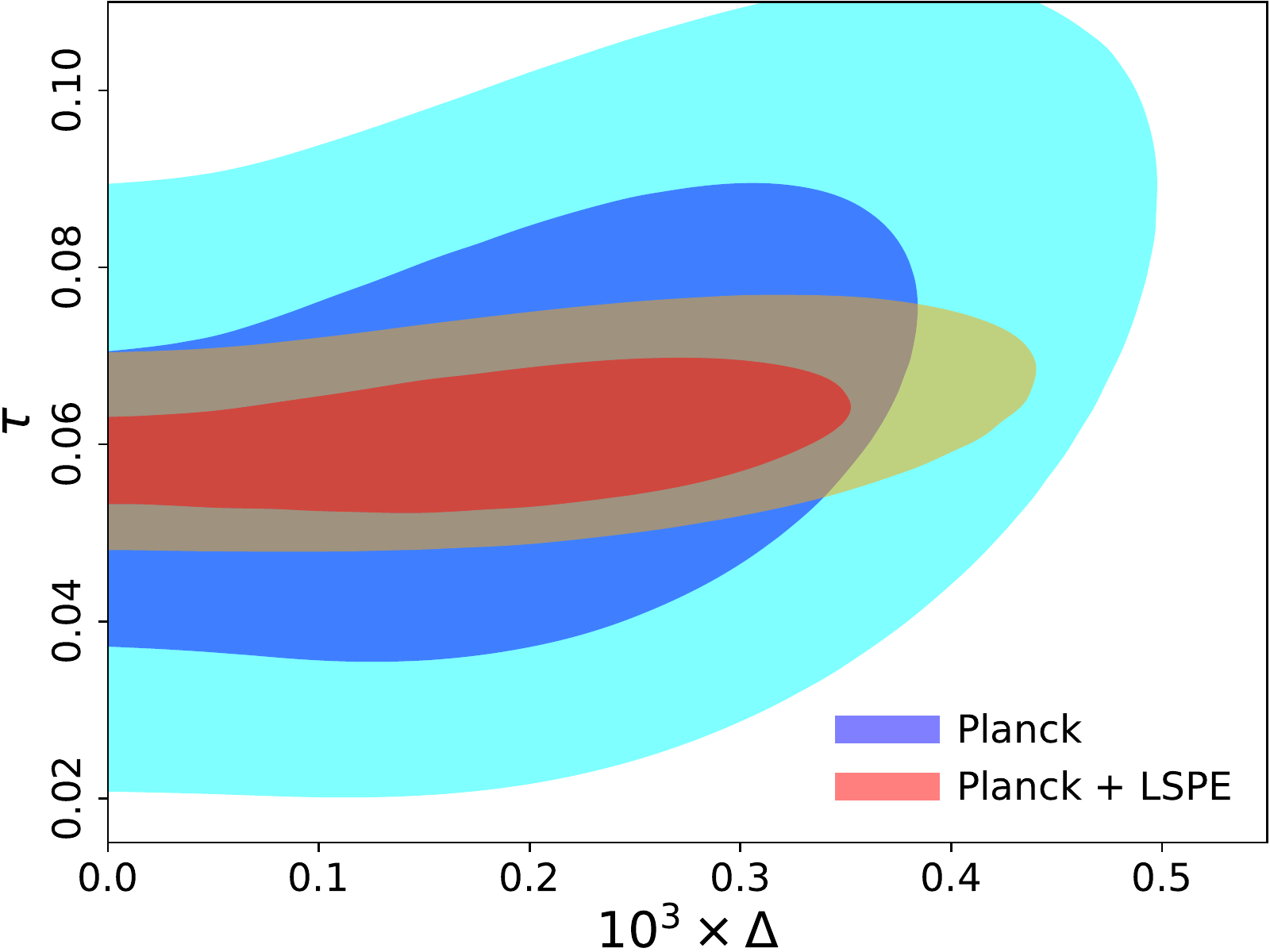}\\
   \caption{\emph{Left}: posterior probability for optical depth 
   $\tau$. The colored lines show different component separation configurations, see text for details. The black line shows the current best estimate on $\tau$.
    \emph{Right}: Joint posterior probability for scale parameter $\Delta$ in (\SI{}{Mpc^{-1}}) and optical depth $\tau$. Blue and cyan are 68\% and 95\% CL
    for \Planck; red and orange for \Planck\ and LSPE.
    \label{fig:posteriors_tau}
    }%
\end{figure}

The constraints on the reionization optical depth $\tau$ are reported in table~\ref{tab:tauerrors} and figure~\ref{fig:posteriors_tau} for different data combinations. For \Strip\ we consider \SI{2}{years} of observations, for SWIPE we explore two possibilities, \SI{15}{days} (SWIPE~15) or \SI{8}{days} of observation (SWIPE~8). \Planck\ and QUIJOTE are assumed with their nominal observational strategies and only considered in the portion of sky in common with LSPE. The $\tau$ constraints are rather stable showing that even after 8 days of SWIPE observations we reach the mission goal. Furthermore, even in a minimal configuration which considers only LSPE plus \Planck\ \SI{30}{GHz} as additional synchrotron tracer, the $\tau$ measure does not change substantially, showing that we are not heavily dependent on the usage of external datasets.

\begin{table}[tb]
\begin{center}
\begin{tabular}{l|c}
\hline \hline
Data Combination & $\sigma_{\tau}$ \\
\hline
\Strip\ + SWIPE 15 + QUIJOTE + \Planck\ & 0.0037 \\
\Strip\ + SWIPE ~~8 + QUIJOTE + \Planck\ & 0.0040 \\
\Strip\ + SWIPE 15 + QUIJOTE & 0.0038 \\
\Strip\ + SWIPE 15 + \Planck\ 30GHz & 0.0038 \\
\hline
\end{tabular}
\end{center}
\caption{Forecasted $1-\sigma$ errors on $\tau$ for different data \newc{combinations} obtained \newc{by} marginalizing over $\ln(10^{10}A_{\rm s})$. SWIPE 15 and SWIPE ~~8 stand respectively for 15 and \SI{8}{days} of mission time, in both cases the effective time used for the sky survey is reduced by \SI{1}{day} used for calibration and ancillary operations.}
\label{tab:tauerrors}
\end{table}%

LSPE can also provide valuable information on the study of one of the most discussed anomalies related to CMB, i.e. the lack of power at large angular scales in the anisotropy power spectrum \citep{Monteserin:2007fv,Cruz:2010ud,Gruppuso:2013xba,Schwarz:2015cma,Natale:2019dqm}. Entering in details about modeling and possible constraining techniques is beyond the scope of this paper. 
Here we only want to show 
how, in the context of a specific model such as the one predicting early departure from slow-roll inflation \citep[see e.g.][]{Dudas:2012vv,Kitazawa:2014dya}, LSPE provides valuable constraints, 
being able to break completely the remaining degeneracy that such models still have with the reionization optical depth $\tau$ in the current CMB data \citep{Gruppuso:2015xqa,Gruppuso:2017nap}.
As an example, the model described in \citep{Gruppuso:2015zia} modifies the primordial scalar power spectrum according to: 
\begin{equation}
P_{\Delta}(k)=A_s\frac{\left(k/k_*\right)^3}{\left[\left(k/k_*\right)^2+\left(\Delta/k_*\right)^2\right]^{2-\frac{n_s}{2}}},\nonumber
\end{equation}
where $k$ is the primordial perturbation wavenumber in \SI{}{Mpc^{-1}},
$A_s$ and $n_s$ are respectively the amplitude and the tilt of scalar perturbations, $k_*=\SI{0.05}{Mpc^{-1}}$ is the pivot scale, and $\Delta$ 
is a characteristic scale\footnote{For the forecast presented in this paper we choose $\Delta=\SI{0.0002}{Mpc^{-1}}$ as fiducial value.} 
which breaks the power-law at very low wavenumbers damping both temperature and polarization power spectra at low multipoles. In this particular case, as shown in right panel of figure~\ref{fig:posteriors_tau}, LSPE improves the \Planck\ constraint, substantially canceling the \newb{degeneracy} with $\tau$.



\subsection{Tensor-to-scalar ratio constraints}\label{subsec:r_likelihood}

The angular resolution of LSPE and the observational strategy allow measuring simultaneously both the reionization and the recombination \newc{peaks} of the primordial B-mode \newc{spectrum}. 
This makes LSPE an extremely complete and unique instrument observing a region of the sky not entirely visible from the southern hemisphere. Nevertheless the relatively small sky fraction usable, if compared with a satellite mission, limits our sensitivity at very large scales. 
In table~\ref{tab:rerrors} we report the constraints on $r$ for the different data combinations; two input $r$ values have been considered, i.e., $r=0.03$ and $r=0$.
The aggregate sensitivity allows \newc{detecting} $r=0.03$ with $3-\sigma$ significance for different data combinations. In this case limiting the mission time of SWIPE to \SI{8}{days} induces a non-negligible effect, reducing the $r$ significance down to $\sim2.3\,\sigma$, still within the mission requirements. 
In \newb{the} case of no primordial B-mode, 
the combination of \Strip\ and SWIPE with both \Planck\ and QUIJOTE as foreground tracers \newb{sets an upper limit
$r<0.015$ at 95\% confidence level
($r<0.024$ in the case of \SI{8}{days} 
of SWIPE observations).
}

\begin{table}[!t]
\begin{center}
\begin{tabular}{l|c|c}
\hline \hline
Data Combination & $\sigma_{\rm r}$ &95\%cl\\
\hline
\Strip\ + SWIPE 15 + QUIJOTE + \Planck\ & 0.0093 & 0.015\\
\Strip\ + SWIPE ~~8 + QUIJOTE + \Planck\ & 0.013 & 0.024\\
\Strip\ + SWIPE 15 + QUIJOTE & 0.0098 & 0.016\\
\Strip\ + SWIPE 15 + \Planck\ 30GHz & 0.010 & 0.018 \\
\hline
\end{tabular}
\end{center}
\caption{Expected sensitivity on the tensor-to-scalar ratio $r$. The second column shows $1-\sigma$ errors assuming an input $r=0.03$. The third column shows $95\%$ c.l. upper limits assuming no tensor B-modes (i.e. $r=0$).}
\label{tab:rerrors}
\end{table}%

\begin{figure}[!t]
   \centering
      \includegraphics[width=0.48\textwidth]{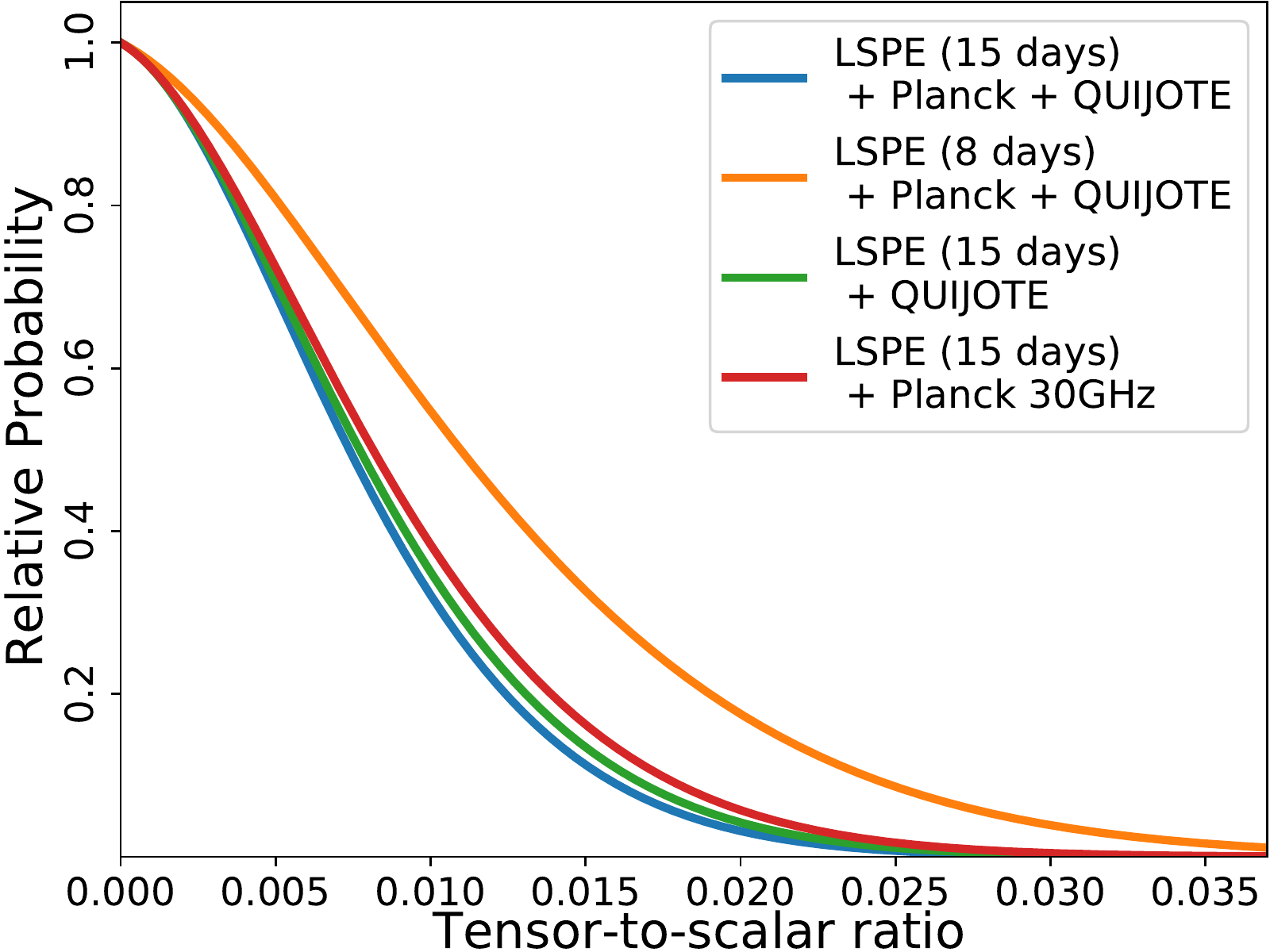}
      ~
   \includegraphics[width=0.48\textwidth]{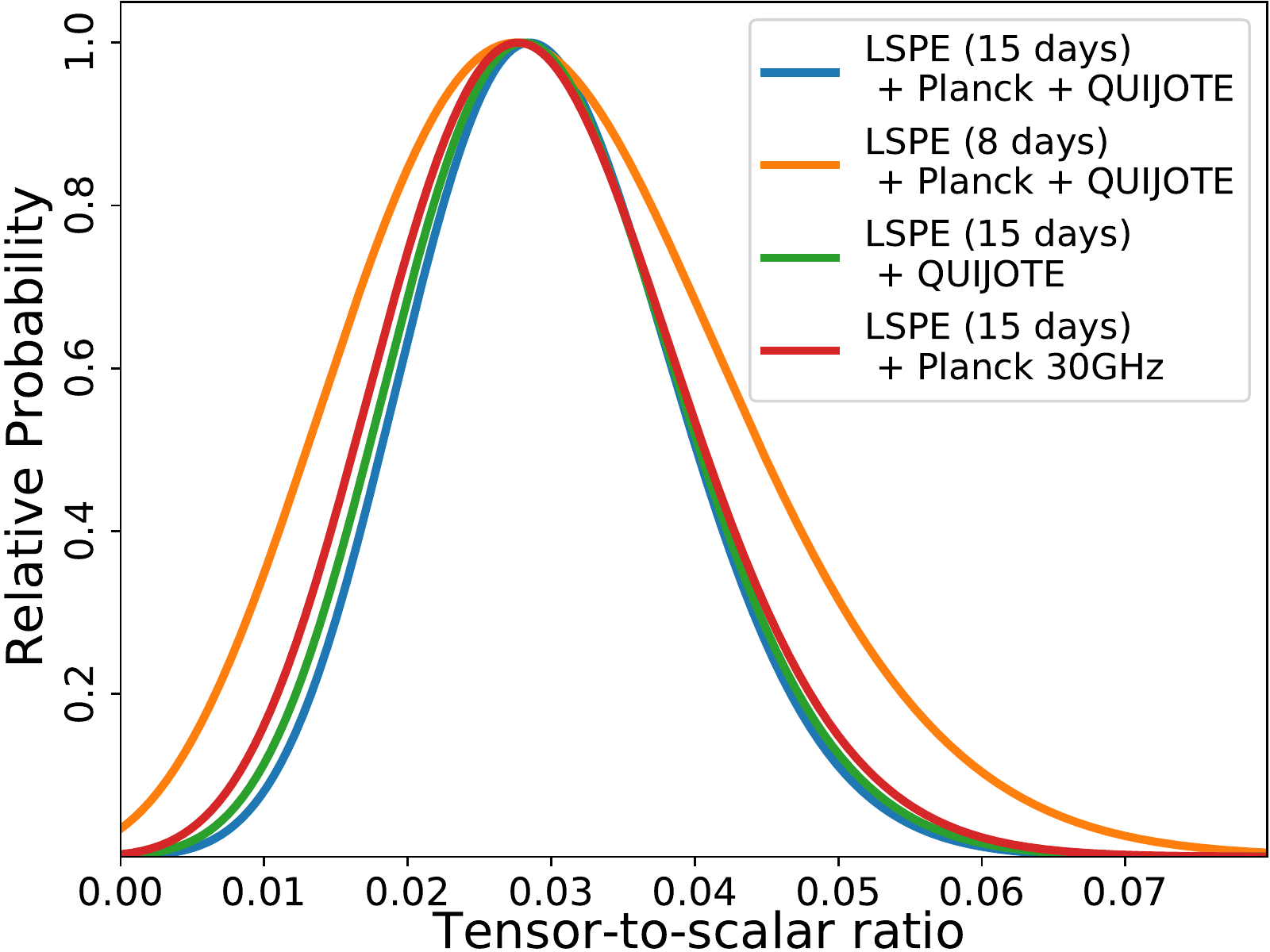}
   \caption{Posterior probability for tensor-to-scalar ratio 
   $r$ in case of $r=0$ (\emph{left}) and $r = 0.03$ (\emph{right}). The colored lines show different component separation configurations, see text for details.
    \label{fig:posteriors_r}
    }%
\end{figure}


\subsection{Constraints on cosmic birefringence}
\label{subsec:biref}

\begin{figure}[!t]
   \centering
   \includegraphics[width=0.45\textwidth]{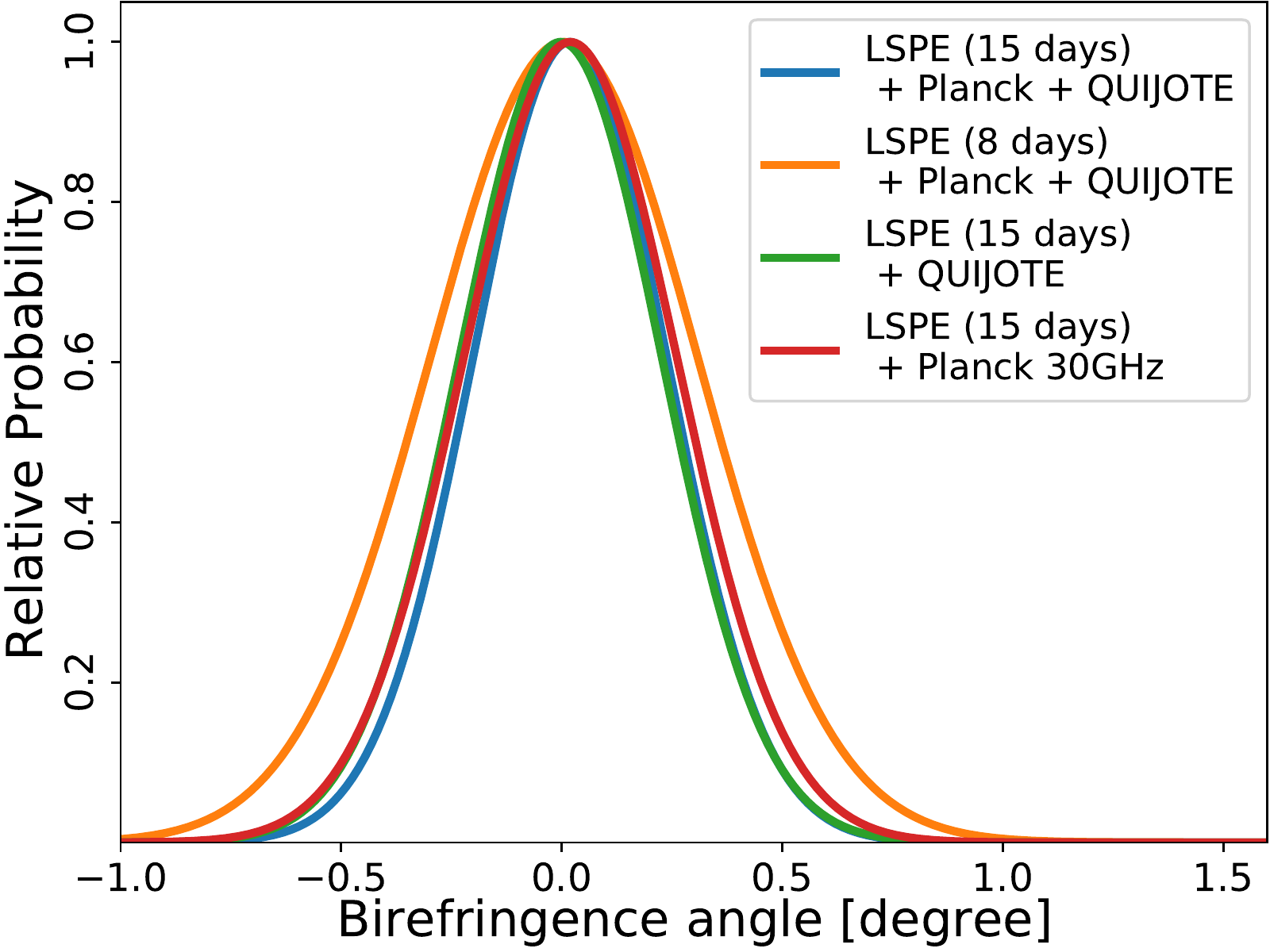}
   \caption{Posterior probability for cosmic birefringence angle obtained with the pixel-based likelihood.
    \label{fig:posteriors_alpha}
    }%
\end{figure}

CMB polarization data can also be used to probe cosmic birefringence (CB), i.e., the \newc{in-vacuum} rotation of the plane of linear polarization during propagation \citep{Carroll:1989vb}. In this section we focus on isotropic birefringence rotation, see e.g., \citep{Liu:2006uh,Feng:2006dp,Gubitosi:2009eu,Gruppuso:2015xza,Gruppuso:2016nhj,planck2014-a23}. For those measurements the calibration of the polarization angle of polarimeters is a key aspect, since miscalibration of such angle is completely degenerate with the rotation induced by CB, see section~\ref{subsec:HWPsyst} and references 
therein\footnote{A new method has been proposed recently which aims at breaking the degeneracy between birefringence angle and instrumental polarization angle \citep{Minami:2019ruj,Minami:2020xfg}.}. 
Assuming negligible calibration error on the polarization angle, we can constrain CB angle, $\alpha_\text{CB}$, with 
\new{the} same technique used to constrain $\tau$ and $r$, i.e. a pixel-based approach for the likelihood estimation.
Another possible approach is to use the so called D-Estimators, as defined in e.g. \citep{Gruppuso:2016nhj}, which employ TB and EB power spectra. 
In table~\ref{tab:alphaerrors} we report constraints of $\alpha_\text{CB}$ for different data \newc{combinations} for both the approaches mentioned above, in figure~\ref{fig:posteriors_alpha} we show the posteriors obtained with the pixel-based method. LSPE data will 
constrain uniform birefringence angle down to \SI{0.2}{\degree}, improving by a factor $3$ the current best estimate \citep{planck2014-a23}.

\begin{table}[t]
\begin{center}
\begin{tabular}{l|c|c}
\hline \hline
Data Combination & $\sigma_{\alpha_\text{CB}}^{\rm PB}$& $\sigma_{\alpha_\text{CB}}^{\rm DE}$ \\
\hline
\Strip\ + SWIPE 15 + QUIJOTE + \Planck\ & 0.22 & 0.19\\
\Strip\ + SWIPE ~~8 + QUIJOTE + \Planck\ & 0.30 & 0.29\\
\Strip\ + SWIPE 15 + QUIJOTE & 0.23 & 0.21\\
\Strip\ + SWIPE 15 + \Planck\ \SI{30}{GHz} & 0.24 & 0.22 \\
\hline
\end{tabular}
\end{center}
\caption{Forecasted sensitivity on the cosmic birefringence angle $\alpha_\text{CB}$ in \newc{degrees}. The second and third columns show $1-\sigma$ errors obtained with the pixel-based approach and D-Estimators approach, respectively.}
\label{tab:alphaerrors}
\end{table}%

\section{Conclusion}\label{sec:conclusion}
The Large Scale Polarization Explorer is a program 
dedicated to the \newb{measurement} of the CMB polarization and its
B-mode component in particular. 
We have presented the instruments' design and a detailed 
forecast of its performance. 
LSPE can put an upper limit to the tensor-to-scalar ratio
at the level of \new{$r<0.015$ at 95\% confidence level, and
\newc{can} detect a 
\newc{signal corresponding to} $r = 0.03$ with 
99.7\% confidence}. 
Moreover, LSPE can improve constraints on other parameters, 
like the optical depth of the Universe to the CMB, $\tau$, and
the rotation angle originated by a cosmic birefringence. 
This analysis is obtained by a full set of end-to-end simulations, 
including 
detailed noise estimation, 
instrument observations \new{(for each detector)}, 
\new{contamination by foregrounds,}
map-making,
component separation \new{with realistic foreground residuals}, and 
cosmological parameter extraction \new{with pixel-based likelihood}.
\new{
The assumption about systematic effects 
is that their contribution can be reduced to below the noise 
level by means of stringent requirements at design level and by system level calibration.}
We also present techniques for the control and removal of HWP
synchronous systematic effects, \new{including the case of 
HWP rotational instability. }

With its rotating HWP, LSPE-SWIPE represents 
an important pathfinder of the forthcoming LiteBIRD mission from the point of view of the instrument requirements,
instrument design, calibration, 
control of systematic effects, and data analysis.

\newpage
\appendix
\section{\new{LSPE-SWIPE scanning and modulation parameters}}
\label{app:swipe_scanning_params}

\begin{table}[tp]
\begin{center}
\begin{tabular}{l r l }
\hline
\hline
Parameter\dotfill& variable & value \\
\hline
Precession period \dotfill  & $T_\text{Earth}$       & $\sim\SI{24}{hrs}$\\
Precession velocity \dotfill&$\omega_\text{Earth}$  & $\sim\SI{73}{\micro rad.s^{-1}}$  \\
Precession angle$^1$ \dotfill   &$\alpha_\text{p}$    & $\sim12-\SI{23}{\degree}$  \\
Payload spin period \dotfill        & $T_\text{payload}$    &\SI{8.6}{min}\\
Payload spin rate \dotfill          &$f_\text{payload}$     &\SI{1.93}{mHz} \\
Payload spin velocity \dotfill      & $\omega_\text{payload}$&\SI{12.1}{mrad.s^{-1}}
$\simeq$\SI{0.7}{\degree.s^{-1}}\\
Zenith angle \dotfill        &$\beta$                &\SI{45}{\degree} \\
Altitude range (combined with FOV)$^2$\dotfill      &$\Delta \beta$         & $\pm\SI{10}{\degree}\pm\SI{10}{\degree}$\\
Max in sky speed$^3$\dotfill    &$\omega_\text{payload} \sin \beta_\text{max}$ & $\sim \SI{0.63}{\degree.s^{-1}}$ \\
HWP period\dotfill          & $T_\text{HWP}$        & \SI{2.0}{s}\\
HWP rate \dotfill           &$f_\text{HWP}$         & \SI{0.5}{Hz}\\
HWP velocity\dotfill        &$\omega_\text{HWP}$    & \SI{3.14}{rad.s^{-1}}\\
Modulation period\dotfill   &$T_\text{mod} = T_\text{HWP}/4$   & \SI{0.5}{s}\\
Modulation rate\dotfill     &$f_\text{mod} = 4f_\text{HWP}$   & \SI{2.0}{Hz}\\
Modulations per FWHM$^4$ \dotfill& $N_\text{mod}$     &   4.5\\
\hline
\end{tabular}
\end{center}
    \footnotesize
    $^1$Depending on launch latitude \\
    $^2$The first interval is due to telescope altitude range, the second to the FOV aperture\\
    $^3$Maximum scanning speed of a detector in the sky\\
    $^4$Number of modulation periods in a beam FWHM
\caption{LSPE-SWIPE baseline parameters of observation and modulation strategy.
See figure~\ref{fig:strategies}, right panel, 
for reference. 
\label{tab:swipe-obspar}}
\end{table}

The driving parameter to define the modulator spin rate 
is the lowest value between detector 
time constant, cut-off frequency and maximum modulation frequency.
The TES developed for LSPE-SWIPE have a typical time constant 
$\tau_\text{LP}=\SI{30}{ms}$. 
To first approximation, this can be modeled as a single pole
low-pass filter, with transfer function 
\begin{equation}\label{eq:swipe_tes_tf}
    H(\omega) = \frac{1}{1+j\omega \tau_\text{LP}} = \frac{1}{1+j 2 \pi f \tau_\text{LP}},
\end{equation}
($\omega$ being the angular frequency here) 
with cut-off frequency $f_{\tau_\text{LP}} = 1/(2 \pi \tau_\text{LP}) =\SI{5.3}{Hz}$. 
The HWP can spin up to $f_\text{HWP}=\SI{1.5}{Hz}$, which corresponds to a modulation frequency
$f_\text{mod} = 4 f_\text{HWP} = \SI{6}{Hz}$. 
The limiting term is then the \new{detector's time constant}. 
In order to limit the sensitivity degradation due to the 
low-pass filtering, considering that most of the polarization 
signal lies in the $[3, 5]f_\text{HWP}$ range, we set the 
HWP spin rate to $f_\text{HWP}=\SI{0.5}{Hz}$. Most of the polarization 
signal lies then in the  $[1.5, 2.5]$\,Hz
range. 
The transfer function attenuation at $f_\text{max}=\SI{2.5}{Hz}$ is
\begin{equation*}
|H(f_\text{max})| = \frac 1{\sqrt{1+(2 \pi f_\text{max} \tau_\text{LP})^2}}
= 0.9
\end{equation*}
In order to set the payload angular velocity,
we approximate the angular response as a Gaussian profile with standard
deviation $\sigma_b=\theta_\text{FWHM}/(2{\sqrt {2\ln 2}})$; given a scanning speed $\omega_\text{payload}$,
we convert the angular width into a temporal width
$\sigma_t = \sigma_b  /(\omega_\text{payload}\sin \beta_\text{max})$ 
where
$\sin \beta_\text{max} = 0.9$ accounts for the altitude projection effect;
this can be converted into a frequency width
\begin{equation*}
    \sigma_f = \frac{1}{2 \pi \sigma_t} = \frac{\omega_\text{payload} \sin \beta_\text{max}}{2\pi \sigma_b};
\end{equation*}
we require that $3\sigma_f \leq f_\text{HWP}$,
so that 99.7\% of the signal lies in the 
$[3,5]\,f_\text{HWP}$ range.
The condition is then:
 \begin{eqnarray}
     3\frac{\omega_\text{payload}\sin\beta_\text{max}}{2 \pi \sigma_b }&&\leq f_\text{HWP}\nonumber\\
     \omega_\text{payload} && \leq\frac{2\pi\sigma_b f_\text{HWP}}{3\sin\beta_\text{max}}
     =\SI{0.695}{\degree.s^{-1}}\nonumber \\
     T_\text{payload} && \geq 8.6\,\text{min}\label{eq:swipe_period}.
 \end{eqnarray}
As a baseline, we adopt $T_\text{payload} =8.6$\,min. 
A beam FWHM is covered $N_\text{mod}$ times the HWP modulation period, with 
\begin{equation*}
    N_\text{mod} = \frac{\theta_\text{FWHM}}{\omega_\text{payload}\sin \beta_\text{max}}4f_\text{HWP} = 4.5
\end{equation*}
\new{The complete list of observing parameters for LSPE-SWIPE is reported in table~\ref{tab:swipe-obspar}.}

\section{\new{Instrument simulators}}
\label{app:instrument_sim}

Simulators are key elements for instrument design and for data analysis. 
In the design phase, they allow to predict the scientific performance of the 
instruments and the impact of systematic effects. In the data analysis phase, 
they allow to run Monte Carlo realizations of the observations, which are 
necessary to estimate instrumental biases, to measure transfer functions, 
and to propagate uncertainties. 

The instrument simulator of LSPE-\Strip{} is written in the Julia\footnote{\url{https://julialang.org/}} language~\cite{bezanson2017julia} and takes advantage of Message Passing Interface (MPI) libraries to parallelize the computation. It is a modular package  containing the following components:
\begin{itemize}
    \item Instrument database containing the configuration of the focal plane and the characteristics of each polarimetric chain to be integrated in the instrument;
    \item Pointing generation: starting from the configuration and behavior in time of the telescope motors, it produces a timestream of pointing information;
    \item White noise and $1/f$ noise generation;
    \item Destriping;
    \item Map-making.
\end{itemize}
Being based on a dynamic language like Julia, every module can either be called interactively in a Jupyter\footnote{\url{https://jupyter.org/}} notebook or run on a High Performance Computing  (HPC) cluster.

The instrument simulator of LSPE-SWIPE consists in a parallel Fortran-90 code 
which takes as input:
\begin{itemize} 
\item the number of detectors;
\item the detector's positions in the focal plane; 
\item the detector's noise, in terms of NET and $1/f$ knee frequency;
\item the mission starting date and duration;
\item the angular response in the sky of each detector, as a 2d matrix; this
is convolved in pixel space, in a radius specified also as input parameter;
\item the HWP operation strategy (stepping, spinning, spinning rate, stepping period);
\item the level of HWP synchronous systematic effects, as a signal in \SI{}{\micro K_{CMB}} at the HWP spinning frequency and its harmonics;
\item HWP angle offset, angular velocity instability, and error in angle measurement;
\item timeline filter (high-pass, low-pass, band-pass, notch-filter);
\item map-making algorithm details, as simple re-binning, or iterative
destriping.
\end{itemize}
It generates in output:
\begin{itemize} 
\item timeline of each detector;
\item maps of each detector;
\item map of combined detectors;
\item coverage map;
\item noise covariance $3 \times 3$ matrix for each observed pixel.
\end{itemize}

\new{
Both simulators can take as input sky maps of any kind. In the analysis presented in this paper, the sky maps have been generated with \healpix\ code, 
for CMB maps, and PySM for foregrounds maps, as detailed in 
section~\ref{sec:component_separation}.
The atmospheric noise is not included in the simulations, but only the 
atmospheric background power and the corresponding white noise.
Noise correlation among different detectors can be included in the 
SWIPE simulator, but are not included in this analysis.
Similarly, realistic beams (angular response) of the telescope 
can be used, including sidelobes, but only symmetrical main-beams have
been considered in this work. In particular for SWIPE, the presence of the HWP
strongly attenuates the impact of beam asymmetries. On the other side, 
beam sidelobes are potentially strong sources of contamination 
if not properly measured or modeled, and removed from timelines and maps. 
}
\section{Atmospheric fluctuations estimation for \Strip}
\label{app:strip_atmosphere}
\begin{figure}[t!]
        \begin{center}
            \includegraphics[width=\textwidth]{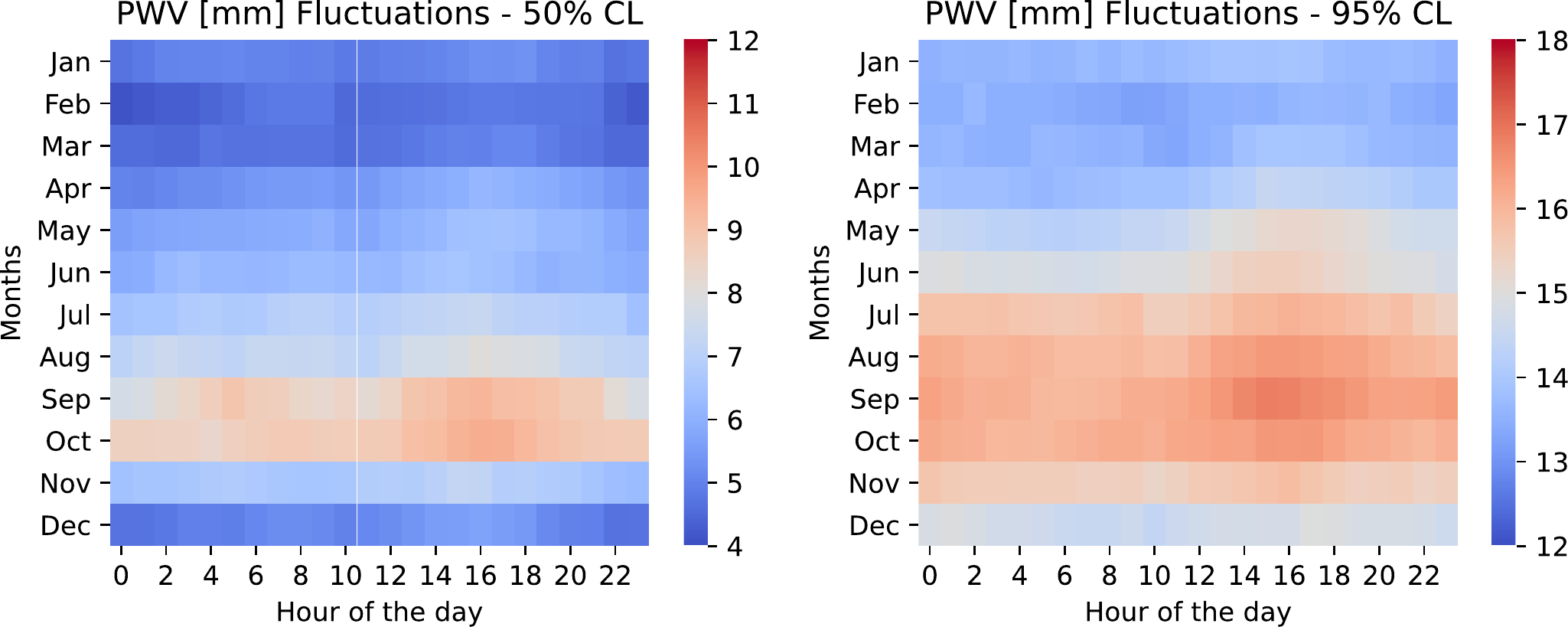}
            \caption{\label{fig_atmosphere_example_cumulative_distribution}Daily and seasonal fluctuations of the PWV in mm above Pico del Teide - Tenerife. We can appreciate the low daily PWV fluctuations compared to the seasonal ones.}
        \end{center}
    \end{figure}

    \new{The atmospheric fluctuations are very difficult to model and forecast. For this reason, a statistical approach is mandatory. Since we are interested in assessing the atmospheric transparency and its medium-time fluctuations (seasonal and day/night cycles), the atmospheric model has to rely on the statistical fluctuations of the PWV,
    TS (Surface Temperature), and $P_0$ (Surface Pressure).}   
    \new{For each hour of each month, we have created the cumulative distribution functions (CDFs\footnote{The CDF is the integral of the probability density function, PDF. We used the CDF instead of the PDF derived from the data because is smoother and is not biased by the binning.}) of weather parameters used by the \emph{am} software to evaluate the atmospheric brightness temperature. To build the CDFs we have used the ERA-5 reanalysis dataset\footnote{\url{https://www.ecmwf.int}}, which estimates the atmosphere's history using a numerical model to assimilate historical measurements.}
    \new{We then use the simulation-framework Cmb Atmospheric Library (\texttt{CAL} \cite{stefano_mandelli_2021_4439199}), to estimate the contribution of the seasonal atmospheric fluctuations. In figure \ref{fig_atmosphere_example_cumulative_distribution} we show the statistical water vapor seasonal and daily fluctuations that result from 40 years, hourly time sampled, of ERA-5 reanalysis data. The PWV fluctuations are presented at two different CL: 50\% and 95\%. The first one is the most representative because it does not contain the samples with bad weather conditions, when the telescope does not observe, while the second one represents the average Tenerife atmosphere behavior.}
    \new{We are working on completing the final results about the 
    fluctuation of the atmospheric temperature. This argument will be the 
    main topic of a dedicated paper that will be released in late 2021.}

\section{\new{LSPE-SWIPE time response knowledge requirements}}
\label{app:swipe_tc}

The TES detectors of LSPE-SWIPE have an intrinsic time response. Their temporal transfer function
$H(\omega)$
can be approximated by a single pole low pass filter, as in equation~\ref{eq:swipe_tes_tf}.
More precisely, the time transfer function 
is characterized by an amplitude,
$|H|$ (gain effect) 
and a phase $\Phi(H)$ (time delay effect). 
The time delay is 
\begin{equation}\nonumber
    \Delta t = 
    \frac{\Phi(H)}{\omega},
\end{equation}
where $\omega=2\pi f$ is the signal angular
frequency
of interest, 3 to $5 \omega_\text{HWP}$ in our case.
During this time, the HWP moves of an angle 
\begin{equation*}
    \Delta \theta_\text{HWP} = 
    \omega_\text{HWP}\Delta t =\Phi(H) \frac{\omega_\text{HWP}}{\omega}.
\end{equation*}
This time delay is deconvolved in the analysis 
pipeline. An error in the knowledge 
of the transfer function phase has the 
same effect of an error on the knowledge
of the HWP angle. Given the requirement on the 
HWP angle knowledge, $\Delta \theta_\text{HWP}$, we can set the requirement 
on the knowledge of the transfer function phase as:
\begin{equation}\label{eq:swipe_phase_error}
    \Delta \Phi = \frac{\omega}{\omega_\text{HWP}}
    \Delta \theta_\text{HWP} = 3
    \Delta \theta_\text{HWP} = \SI{42}{\arcmin} =  12\times10^{-3}\,\text{rad}
\end{equation}
where we have
considered that the frequencies of interest 
range from $3 f_\text{HWP}$. 
We can express this requirement 
in terms of the level of knowledge of the time
constant $\tau_\text{LP}$ of a single pole low-pass filter, considering that
$    \tan \Phi = -\omega \tau_\text{LP}$. 
An error on the time constant $\Delta {\tau_\text{LP}}$
generates an error on the phase
\begin{equation}\nonumber
    \Delta \Phi = \frac{\omega}{(\omega \tau_\text{LP})^2+1}
    \Delta {\tau_\text{LP}}.
\end{equation}
Inverting
\begin{equation}\nonumber
    \Delta{\tau_\text{LP}} = \frac{(\omega \tau_\text{LP})^2+1}{\omega}
    \Delta \Phi = \frac{(\omega \tau_\text{LP})^2+1}{\omega_\text{HWP}}\Delta \theta_\text{HWP},
\end{equation}
where we have used equation~\ref{eq:swipe_phase_error}.
With a time constant $\tau_\text{LP}=\SI{30}{ms}$ and 
a frequency in the 1.5 to \SI{2.5}{Hz} range, 
this means a requirement:
\begin{equation}
    \Delta{\tau_\text{LP}} \simeq \SI{1.5}{ms}
\end{equation}
(relative error $\Delta{\tau_\text{LP}}/\tau_\text{LP} =5\% $).
Also in this case, the requirement will be more stringent
if considered jointly with other effects. 

\section{\new{LSPE-SWIPE iterative map-making}}
\label{app:iterative_map_making}

\new{
We briefly describe the notch filters and 
iterative map-making used 
by the SWIPE instrument simulator pipeline to
remove systematic effects appearing at the 
HWP spin frequency and harmonicsm, and 
recover the filtered signal.
}

The notch filters are defined as
\begin{equation}
    F(f) =
    \begin{cases}
    1&\text{if $f \leq f_1$}\\
    0.5 \left( 1+\cos\left(\pi\frac{f-f_1}{f_2-f_1}\right)
    \right)&
    \text{if $f_1 <f \leq f_2$}\\
    0&\text{if $f_2<f \leq f_3$}\\
    0.5 \left( 1-\cos\left(\pi\frac{f-f_3}{f_4-f_3}\right)
    \right)&
    \text{if $f_3 <f \leq f_4$}\\
       1&\text{if $f>f_4$}\\
    \end{cases}
\end{equation}
where $f_1=f_0-2\Delta f$, $f_2=f_0-\Delta f$, $f_3 = f_0+\Delta f$, and $f_4=f_0+2\Delta f$ are the
frequencies where the filter starts to drop, reaches 0, 
starts to rise, reaches 1 respectively; 
$f_0$ is the notch filter frequency, and $\Delta f$ is
the notch filter width. 
In simulations, we set $f_0$ to the HWP spin frequency
and harmonics, and $\Delta f=\SI{1}{mHz}$, lower than the
payload default spin rate, set at \SI{1.93}{mHz} (table~\ref{tab:swipe-obspar}).
The notch filter applied at 4 times the HWP spin rate, $4f_\text{HWP}$, can remove part of the signal at the largest scales 
in the sky. This happens despite \new{the fact that} the payload spin rate is above
the notch filter width. In fact, 
the sky signal is not exactly
periodic, and part of it is spread  
\new{in the interval $[4f_\text{HWP}\pm f_\text{payload}]$ 
, i.e. within the spin rate frequency from modulation frequency}. 
This signal is recovered by an iterative mapmaking process. 

In the following, $P$ is the pointing matrix, 
$M_P$ is the rebinning matrix; $\otimes$ means a filtering. 
The iterative procedure is 
applied to 
the timeline $d_j$ of the $j$-th detector:
\begin{itemize}
    \item each timeline is notch-filtered to remove
    the contamination $\tilde d_j = F \otimes d_j$;
    \item all filtered timelines are combined in a first 
    map, by simple rebin $m_{0} = \sum_j M_{P} \tilde d_j$;
    \item here starts the iteration, from $i=0$;
    \item a synthetic timeline is produced for each detector: 
    $h_{j} = P m_{i}$;
    \item the synthetic timeline is filtered by $1-F$, to recover the missing signal: $\tilde h_{j,i} = (1-F)\otimes h_{j,i}$;
    \item the filtered synthetic timeline is added to the 
    filtered original timeline:
    $d_{j,i} = \tilde d_j + \tilde h_{j,i}$;
    \item a new map is produced:
    $m_{i+1} = \sum_j M_{P} \tilde d_{j,i}$, 
    and the procedure is iterated.
\end{itemize}
Testing different numbers of iterations, we have found
that 50 iterations represents a good trade-off between 
computational time and residual signal in 
the map with filtered timelines.

\acknowledgments
The development of LSPE is supported by ASI (grant LSPE I/022/11/0) and INFN.
We acknowledge the usage of the \healpix \citep{2005_healpix} and am software packages, 
computational resources from NERSC, Cineca and
CNAF.
We acknowledge support from the COSMOS network (www.cosmosnet.it) through the ASI (Italian Space Agency) Grants 2016-24-H.0 and 2016-24-H.1-2018, and Sapienza University.
JARM, RGS and MPdT acknowledge financial support from the Spanish
Ministry of Science and Innovation (MICINN) under the project AYA2017-84185-P, 
and the European Union's Horizon 2020 research and innovation
programme under grant agreement number 687312 (RADIOFOREGROUNDS).
\newc{The authors gratefully acknowledge the time and expertise devoted to reviewing the manuscript by the referees. }


\bibliographystyle{JHEP} 
\bibliography{biblio} 

\providecommand{\href}[2]{#2}\begingroup\raggedright\begin{thebibliography}{100}

\bibitem{PhysRevLett.78.2054}
U.~Seljak and M.~Zaldarriaga, \emph{Signature of gravity waves in the
  polarization of the microwave background},
  \href{https://doi.org/10.1103/PhysRevLett.78.2054}{\emph{Phys. Rev. Lett.}
  {\bfseries 78} (1997) 2054}.

\bibitem{doi:10.1146/annurev-astro-081915-023433}
M.~Kamionkowski and E.D.~Kovetz, \emph{{The Quest for B Modes from Inflationary
  Gravitational Waves}},
  \href{https://doi.org/10.1146/annurev-astro-081915-023433}{\emph{Annu. Rev.
  Astron. Astr.} {\bfseries 54} (2016) 227}.

\bibitem{planck2016-l01}
{Planck Collaboration}, N.~{Aghanim}, Y.~{Akrami}, F.~{Arroja}, M.~{Ashdown},
  J.~{Aumont} et~al., \emph{{Planck 2018 results. I. Overview and the
  cosmological legacy of Planck}},
  \href{https://doi.org/10.1051/0004-6361/201833880}{\emph{\aap} {\bfseries
  641} (2020) A1} [\href{https://arxiv.org/abs/1807.06205}{{\ttfamily
  1807.06205}}].

\bibitem{2021tristram}
M.~{Tristram}, A.J.~{Banday}, K.M.~{G{\'o}rski}, R.~{Keskitalo},
  C.R.~{Lawrence}, K.J.~{Andersen} et~al., \emph{{Planck constraints on the
  tensor-to-scalar ratio}},
  \href{https://doi.org/10.1051/0004-6361/202039585}{\emph{\aap} {\bfseries
  647} (2021) A128} [\href{https://arxiv.org/abs/2010.01139}{{\ttfamily
  2010.01139}}].

\bibitem{bicep2018}
P.A.R.~{Ade}, Z.~{Ahmed}, R.W.~{Aikin}, K.D.~{Alexander}, D.~{Barkats},
  S.J.~{Benton} et~al., \emph{{Constraints on Primordial Gravitational Waves
  Using Planck, WMAP, and New BICEP2/Keck Observations through the 2015
  Season}}, \href{https://doi.org/10.1103/PhysRevLett.121.221301}{\emph{{Phys.
  Rev. Lett.}} {\bfseries 121} (2018) 221301}
  [\href{https://arxiv.org/abs/1810.05216}{{\ttfamily 1810.05216}}].

\bibitem{2010_arcidiacono}
M.~Archidiacono, A.~Cooray, A.~Melchiorri and S.~Pandolfi, \emph{Cmb neutrino
  mass bounds and reionization},
  \href{https://doi.org/10.1103/PhysRevD.82.087302}{\emph{Phys. Rev. D}
  {\bfseries 82} (2010) 087302}.

\bibitem{2015_allison}
R.~Allison, P.~Caucal, E.~Calabrese, J.~Dunkley and T.~Louis, \emph{Towards a
  cosmological neutrino mass detection},
  \href{https://doi.org/10.1103/PhysRevD.92.123535}{\emph{Phys. Rev. D}
  {\bfseries 92} (2015) 123535}.

\bibitem{2016SPIE.9914E..0SG}
J.A.~{Grayson}, P.A.R.~{Ade}, Z.~{Ahmed}, K.D.~{Alexander}, M.~{Amiri},
  D.~{Barkats} et~al., \emph{{BICEP3 performance overview and planned Keck
  Array upgrade}},  in \emph{Millimeter, Submillimeter, and Far-Infrared
  Detectors and Instrumentation for Astronomy VIII}, W.S.~{Holland} and
  J.~{Zmuidzinas}, eds., vol.~9914 of \emph{Society of Photo-Optical
  Instrumentation Engineers (SPIE) Conference Series}, p.~99140S, July, 2016,
  \href{https://doi.org/10.1117/12.2233894}{DOI}
  [\href{https://arxiv.org/abs/1607.04668}{{\ttfamily 1607.04668}}].

\bibitem{2020_moncelsi}
L.~Moncelsi, P.A.R.~Ade, A.~Z, M.~Amiri, D.~Barkats, R.B.~Thakur et~al.,
  \emph{{Receiver development for BICEP Array, a next-generation CMB
  polarimeter at the South Pole}},  in \emph{Millimeter, Submillimeter, and
  Far-Infrared Detectors and Instrumentation for Astronomy X}, J.~Zmuidzinas
  and J.-R.~Gao, eds., vol.~11453, pp.~189 -- 206, International Society for
  Optics and Photonics, SPIE, 2020,
  \href{https://doi.org/10.1117/12.2561995}{DOI}.

\bibitem{2016SPIE.9914E..1KH}
K.~{Harrington}, T.~{Marriage}, A.~{Ali}, J.W.~{Appel}, C.L.~{Bennett},
  F.~{Boone} et~al., \emph{{The Cosmology Large Angular Scale Surveyor}},  in
  \emph{Millimeter, Submillimeter, and Far-Infrared Detectors and
  Instrumentation for Astronomy VIII}, W.S.~{Holland} and J.~{Zmuidzinas},
  eds., vol.~9914 of \emph{Society of Photo-Optical Instrumentation Engineers
  (SPIE) Conference Series}, p.~99141K, July, 2016,
  \href{https://doi.org/10.1117/12.2233125}{DOI}
  [\href{https://arxiv.org/abs/1608.08234}{{\ttfamily 1608.08234}}].

\bibitem{2016JLTP..184..805S}
A.~{Suzuki}, P.~{Ade}, Y.~{Akiba}, C.~{Aleman}, K.~{Arnold}, C.~{Baccigalupi}
  et~al., \emph{{The Polarbear-2 and the Simons Array Experiments}},
  \href{https://doi.org/10.1007/s10909-015-1425-4}{\emph{J. Low Temp. Phys.}
  {\bfseries 184} (2016) 805}
  [\href{https://arxiv.org/abs/1512.07299}{{\ttfamily 1512.07299}}].

\bibitem{2020PhRvD.101l2003S}
J.T.~{Sayre}, C.L.~{Reichardt}, J.W.~{Henning}, P.A.R.~{Ade}, A.J.~{Anderson},
  J.E.~{Austermann} et~al., \emph{{Measurements of B -mode polarization of the
  cosmic microwave background from 500 square degrees of SPTpol data}},
  \href{https://doi.org/10.1103/PhysRevD.101.122003}{\emph{\prd} {\bfseries
  101} (2020) 122003} [\href{https://arxiv.org/abs/1910.05748}{{\ttfamily
  1910.05748}}].

\bibitem{2014_spt-3g}
B.A.~Benson, P.A.R.~Ade, Z.~Ahmed, S.W.~Allen, K.~Arnold, J.E.~Austermann
  et~al., \emph{{SPT-3G: a next-generation cosmic microwave background
  polarization experiment on the South Pole telescope}},  in \emph{Millimeter,
  Submillimeter, and Far-Infrared Detectors and Instrumentation for Astronomy
  VII}, W.S.~Holland and J.~Zmuidzinas, eds., vol.~9153, pp.~552 -- 572,
  International Society for Optics and Photonics, SPIE, 2014,
  \href{https://doi.org/10.1117/12.2057305}{DOI}.

\bibitem{2020_aiola_act}
S.~{Aiola}, E.~{Calabrese}, L.~{Maurin}, S.~{Naess}, B.L.~{Schmitt},
  M.H.~{Abitbol} et~al., \emph{{The Atacama Cosmology Telescope: DR4 maps and
  cosmological parameters}},
  \href{https://doi.org/10.1088/1475-7516/2020/12/047}{\emph{\jcap} {\bfseries
  2020} (2020) 047} [\href{https://arxiv.org/abs/2007.07288}{{\ttfamily
  2007.07288}}].

\bibitem{2020_choi_actC}
S.K.~{Choi}, M.~{Hasselfield}, S.-P.P.~{Ho}, B.~{Koopman}, M.~{Lungu},
  M.H.~{Abitbol} et~al., \emph{{The Atacama Cosmology Telescope: a measurement
  of the Cosmic Microwave Background power spectra at 98 and 150 GHz}},
  \href{https://doi.org/10.1088/1475-7516/2020/12/045}{\emph{\jcap} {\bfseries
  2020} (2020) 045} [\href{https://arxiv.org/abs/2007.07289}{{\ttfamily
  2007.07289}}].

\bibitem{2019JCAP_SO}
P.~{Ade}, J.~{Aguirre}, Z.~{Ahmed}, S.~{Aiola}, A.~{Ali}, D.~{Alonso} et~al.,
  \emph{{The Simons Observatory: science goals and forecasts}},
  \href{https://doi.org/10.1088/1475-7516/2019/02/056}{\emph{J. Cosmol.
  Astropart. P.} {\bfseries 2019} (2019) 056}
  [\href{https://arxiv.org/abs/1808.07445}{{\ttfamily 1808.07445}}].

\bibitem{groundbird}
S.~{Oguri}, J.~{Choi}, T.~{Damayanthi}, M.~{Hattori}, M.~{Hazumi},
  H.~{Ishitsuka} et~al., \emph{{GroundBIRD: Observing Cosmic Microwave
  Polarization at Large Angular Scale with Kinetic Inductance Detectors and
  High-Speed Rotating Telescope}},
  \href{https://doi.org/10.1007/s10909-015-1420-9}{\emph{J. Low Temp. Phys.}
  {\bfseries 184} (2016) 786}.

\bibitem{2020arXiv201102213H}
J.C.~{Hamilton}, L.~{Mousset}, E.S.~{Battistelli}, M.A.~{Bigot-Sazy},
  P.~{Chanial}, R.~{Charlassier} et~al., \emph{{QUBIC I: Overview and
  ScienceProgram}}, {\emph{arXiv e-prints} (2020) arXiv:2011.02213}
  [\href{https://arxiv.org/abs/2011.02213}{{\ttfamily 2011.02213}}].

\bibitem{2020arXiv200812619T}
{The CMB-S4 Collaboration}, {:}, K.~{Abazajian}, G.E.~{Addison}, P.~{Adshead},
  Z.~{Ahmed} et~al., \emph{{CMB-S4: Forecasting Constraints on Primordial
  Gravitational Waves}}, {\emph{arXiv e-prints} (2020) arXiv:2008.12619}
  [\href{https://arxiv.org/abs/2008.12619}{{\ttfamily 2008.12619}}].

\bibitem{2018JLTP..193.1112G}
R.~{Gualtieri}, J.P.~{Filippini}, P.A.R.~{Ade}, M.~{Amiri}, S.J.~{Benton},
  A.S.~{Bergman} et~al., \emph{{SPIDER: CMB Polarimetry from the Edge of
  Space}}, \href{https://doi.org/10.1007/s10909-018-2078-x}{\emph{J. Low Temp.
  Phys.} {\bfseries 193} (2018) 1112}
  [\href{https://arxiv.org/abs/1711.10596}{{\ttfamily 1711.10596}}].

\bibitem{2016SPIE.9914E..1JG}
N.N.~{Gandilo}, P.A.R.~{Ade}, D.~{Benford}, C.L.~{Bennett}, D.T.~{Chuss},
  J.L.~{Dotson} et~al., \emph{{The Primordial Inflation Polarization Explorer
  (PIPER)}},  in \emph{Millimeter, Submillimeter, and Far-Infrared Detectors
  and Instrumentation for Astronomy VIII}, W.S.~{Holland} and J.~{Zmuidzinas},
  eds., vol.~9914 of \emph{Society of Photo-Optical Instrumentation Engineers
  (SPIE) Conference Series}, p.~99141J, July, 2016,
  \href{https://doi.org/10.1117/12.2231109}{DOI}
  [\href{https://arxiv.org/abs/1607.06172}{{\ttfamily 1607.06172}}].

\bibitem{2019PICO}
S.~{Hanany}, M.~{Alvarez}, E.~{Artis}, P.~{Ashton}, J.~{Aumont}, R.~{Aurlien}
  et~al., \emph{{PICO: Probe of Inflation and Cosmic Origins}},  in
  \emph{Bulletin of the American Astronomical Society}, vol.~51, p.~194, Sept.,
  2019 [\href{https://arxiv.org/abs/1908.07495}{{\ttfamily 1908.07495}}].

\bibitem{2019JLTP..194..443H}
M.~{Hazumi}, P.A.R.~{Ade}, Y.~{Akiba}, D.~{Alonso}, K.~{Arnold}, J.~{Aumont}
  et~al., \emph{{LiteBIRD: A Satellite for the Studies of B-Mode Polarization
  and Inflation from Cosmic Background Radiation Detection}},
  \href{https://doi.org/10.1007/s10909-019-02150-5}{\emph{J. Low Temp. Phys.}
  {\bfseries 194} (2019) 443}.

\bibitem{2020LiteBIRD-JLTP}
H.~{Sugai}, P.A.R.~{Ade}, Y.~{Akiba}, D.~{Alonso}, K.~{Arnold}, J.~{Aumont}
  et~al., \emph{{Updated Design of the CMB Polarization Experiment Satellite
  LiteBIRD}}, \href{https://doi.org/10.1007/s10909-019-02329-w}{\emph{J. Low
  Temp. Phys.} (2020) } [\href{https://arxiv.org/abs/2001.01724}{{\ttfamily
  2001.01724}}].

\bibitem{2012_pdb_lspe}
P.~{de Bernardis}, S.~{Aiola}, G.~{Amico}, E.~{Battistelli}, A.~{Coppolecchia},
  A.~{Cruciani} et~al., \emph{{SWIPE: a bolometric polarimeter for the
  Large-Scale Polarization Explorer}},  in \emph{Millimeter, Submillimeter, and
  Far-Infrared Detectors and Instrumentation for Astronomy VI}, vol.~8452 of
  \emph{\procspie}, p.~84523F, Sept., 2012,
  \href{https://doi.org/10.1117/12.926569}{DOI}
  [\href{https://arxiv.org/abs/1208.0282}{{\ttfamily 1208.0282}}].

\bibitem{2012_aiola}
S.~{Aiola}, G.~{Amico}, P.~{Battaglia}, E.~{Battistelli}, A.~{Ba{\'o}}, P.~{de
  Bernardis} et~al., \emph{{The Large-Scale Polarization Explorer (LSPE)}},  in
  \emph{Ground-based and Airborne Instrumentation for Astronomy IV}, vol.~8446
  of \emph{\procspie}, p.~84467A, Sept., 2012,
  \href{https://doi.org/10.1117/12.926095}{DOI}.

\bibitem{2012SPIE_bersanelli}
M.~{Bersanelli}, A.~{Mennella}, G.~{Morgante}, M.~{Zannoni}, G.~{Addamo},
  A.~{Baschirotto} et~al., \emph{{A coherent polarimeter array for the Large
  Scale Polarization Explorer (LSPE) balloon experiment}},  in
  \emph{Ground-based and Airborne Instrumentation for Astronomy IV}, vol.~8446
  of \emph{\procspie}, p.~84467C, Sept., 2012,
  \href{https://doi.org/10.1117/12.925688}{DOI}
  [\href{https://arxiv.org/abs/1208.0164}{{\ttfamily 1208.0164}}].

\bibitem{ltd18lamagna}
L.~{Lamagna}, G.~{Addamo}, P.A.R.~{Ade}, C.~{Baccigalupi}, A.M.~{Baldini},
  P.M.~{Battaglia} et~al., \emph{{Progress report on the Large Scale
  Polarization Explorer}},
  \href{https://doi.org/10.1007/s10909-020-02454-x}{\emph{J. Low Temp. Phys.}
  (2020) arXiv:2005.01187} [\href{https://arxiv.org/abs/2005.01187}{{\ttfamily
  2005.01187}}].

\bibitem{Incardona2020}
F.~Incardona, \emph{{Observing the polarized Cosmic Microwave Background from
  the Earth : scanning strategy and polarimeters test for the LSPE / STRIP
  instrument}}, Ph.D. thesis, University of Milan, 2020.

\bibitem{incardona_spie2018}
F.~{Incardona}, M.~{Benetti}, M.~{Bersanelli}, C.~{Franceschet}, D.~{Maino},
  A.~{Mennella} et~al., \emph{{Preliminary scanning strategy analysis for the
  LSPE-STRIP instrument}},  in \emph{Society of Photo-Optical Instrumentation
  Engineers (SPIE) Conference Series}, vol.~10708, p.~107082F, Jul, 2018,
  \href{https://doi.org/10.1117/12.2315005}{DOI}.

\bibitem{Krachmalnicoff2018}
N.~{Krachmalnicoff}, E.~{Carretti}, C.~{Baccigalupi}, G.~{Bernardi},
  S.~{Brown}, B.M.~{Gaensler} et~al., \emph{{S-PASS view of polarized Galactic
  synchrotron at 2.3 GHz as a contaminant to CMB observations}},
  \href{https://doi.org/10.1051/0004-6361/201832768}{\emph{Astron. Astrophys.}
  {\bfseries 618} (2018) A166}
  [\href{https://arxiv.org/abs/1802.01145}{{\ttfamily 1802.01145}}].

\bibitem{SPIE-OT}
J.A.~{Castro-Almaz{\'a}n}, C.~{Mu{\~n}oz-Tu{\~n}{\'o}n},
  B.~{Garc{\'\i}a-Lorenzo}, G.~{P{\'e}rez-Jord{\'a}n}, A.M.~{Varela} and
  I.~{Romero}, \emph{{Precipitable Water Vapour at the Canarian Observatories
  (Teide and Roque de los Muchachos) from routine GPS}},  in \emph{Observatory
  Operations: Strategies, Processes, and Systems VI}, vol.~9910 of
  \emph{Society of Photo-Optical Instrumentation Engineers (SPIE) Conference
  Series}, p.~99100P, Jul, 2016,
  \href{https://doi.org/10.1117/12.2232646}{DOI}.

\bibitem{tenerife}
C.M.~{Guti{\'e}rrez}, R.~{Rebolo}, R.A.~{Watson}, R.D.~{Davies}, A.W.~{Jones}
  and A.N.~{Lasenby}, \emph{{The Tenerife Cosmic Microwave Background Maps:
  Observations and First Analysis}},
  \href{https://doi.org/10.1086/308246}{\emph{Astrophys. J.} {\bfseries 529}
  (2000) 47} [\href{https://arxiv.org/abs/astro-ph/9903196}{{\ttfamily
  astro-ph/9903196}}].

\bibitem{bartol}
B.~{Femen{\'\i}a}, R.~{Rebolo}, C.M.~{Guti{\'e}rrez}, M.~{Limon} and
  L.~{Piccirillo}, \emph{{The Instituto de Astrof{\'\i}sica de Canarias-Bartol
  Cosmic Microwave Background Anisotropy Experiment: Results of the 1994
  Campaign}}, \href{https://doi.org/10.1086/305549}{\emph{Astrophys. J.}
  {\bfseries 498} (1998) 117}
  [\href{https://arxiv.org/abs/astro-ph/9711225}{{\ttfamily
  astro-ph/9711225}}].

\bibitem{jboiac}
D.L.~{Harrison}, J.A.~{Rubi{\~n}o-Martin}, S.J.~{Melhuish}, R.A.~{Watson},
  R.D.~{Davies}, R.~{Rebolo} et~al., \emph{{A measurement at the first acoustic
  peak of the cosmic microwave background with the 33-GHz interferometer}},
  \href{https://doi.org/10.1046/j.1365-8711.2000.03762.x}{\emph{Mon. Not. Roy.
  Astron. Soc.} {\bfseries 316} (2000) L24}
  [\href{https://arxiv.org/abs/astro-ph/0004357}{{\ttfamily
  astro-ph/0004357}}].

\bibitem{cosmosomas}
S.~{Fern{\'a}ndez-Cerezo}, C.M.~{Guti{\'e}rrez}, R.~{Rebolo}, R.A.~{Watson},
  R.J.~{Hoyland}, S.R.~{Hildebrandt} et~al., \emph{{Observations of the cosmic
  microwave background and galactic foregrounds at 12-17GHz with the COSMOSOMAS
  experiment}},
  \href{https://doi.org/10.1111/j.1365-2966.2006.10505.x}{\emph{Mon. Not. Roy.
  Astron. Soc.} {\bfseries 370} (2006) 15}
  [\href{https://arxiv.org/abs/astro-ph/0601203}{{\ttfamily
  astro-ph/0601203}}].

\bibitem{vsa}
R.A.~{Watson}, P.~{Carreira}, K.~{Cleary}, R.D.~{Davies}, R.J.~{Davis},
  C.~{Dickinson} et~al., \emph{{First results from the Very Small Array - I.
  Observational methods}},
  \href{https://doi.org/10.1046/j.1365-8711.2003.06338.x}{\emph{Mon. Not. Roy.
  Astron. Soc.} {\bfseries 341} (2003) 1057}
  [\href{https://arxiv.org/abs/astro-ph/0205378}{{\ttfamily
  astro-ph/0205378}}].

\bibitem{quijote}
J.A.~{Rubi{\~n}o-Mart{\'\i}n}, R.~{Rebolo}, M.~{Aguiar},
  R.~{G{\'e}nova-Santos}, F.~{G{\'o}mez-Re{\~n}asco}, J.M.~{Herreros} et~al.,
  \emph{{The QUIJOTE-CMB experiment: studying the polarisation of the galactic
  and cosmological microwave emissions}},  in \emph{Ground-based and Airborne
  Telescopes IV}, vol.~8444 of \emph{Society of Photo-Optical Instrumentation
  Engineers (SPIE) Conference Series}, p.~84442Y, Sep, 2012,
  \href{https://doi.org/10.1117/12.926581}{DOI}.

\bibitem{2020groundbirdJLTD}
K.~{Lee}, J.~{Choi}, R.T.~{G{\'e}nova-Santos}, M.~{Hattori}, M.~{Hazumi},
  S.~{Honda} et~al., \emph{{GroundBIRD: A CMB Polarization Experiment with MKID
  Arrays}}, \href{https://doi.org/10.1007/s10909-020-02511-5}{\emph{Journal of
  Low Temperature Physics} {\bfseries 200} (2020) 384}
  [\href{https://arxiv.org/abs/2011.07705}{{\ttfamily 2011.07705}}].

\bibitem{TAYLOR2006993}
A.C.~Taylor, \emph{{Clover - A B-mode polarization experiment}},
  \href{https://doi.org/https://doi.org/10.1016/j.newar.2006.09.026}{\emph{New
  Astron. Rev.} {\bfseries 50} (2006) 993 }.

\bibitem{2018_franceschet}
C.~{Franceschet}, S.~{Realini}, A.~{Mennella}, G.~{Addamo}, A.~{Ba{\'u}},
  P.M.~{Battaglia} et~al., \emph{{The STRIP instrument of the Large Scale
  Polarization Explorer: microwave eyes to map the Galactic polarized
  foregrounds}},  in \emph{Millimeter, Submillimeter, and Far-Infrared
  Detectors and Instrumentation for Astronomy IX}, vol.~10708 of \emph{Society
  of Photo-Optical Instrumentation Engineers (SPIE) Conference Series},
  p.~107081G, Jul, 2018, \href{https://doi.org/10.1117/12.2313558}{DOI}.

\bibitem{2019_realini}
S.~{Realini}, C.~{Franceschet} and A.~{Mennella}, \emph{{Modelling the
  radiation pattern of a dual circular polarization system}},
  \href{https://doi.org/10.1088/1748-0221/14/03/P03005}{\emph{J. Instrum.}
  {\bfseries 14} (2019) P03005}.

\bibitem{peverini2015}
O.A.~Peverini, G.~Virone, F.~{Del Torto}, C.~Franceschet, F.~Villa, M.~Lumia
  et~al., \emph{{Q-band antenna-feed system for the Large Scale Polarization
  Explorer balloon experiment}},  in \emph{2015 International Conference on
  Electromagnetics in Advanced Applications (ICEAA)}, pp.~883--886, IEEE, sep,
  2015, \href{https://doi.org/10.1109/ICEAA.2015.7297240}{DOI}.

\bibitem{deltorto_2011}
F.~{Del Torto}, M.~Bersanelli, F.~Cavaliere, A.~{De Rosa}, O.~D'Arcangelo,
  C.~Franceschet et~al., \emph{{W-band prototype of platelet feed-horn array
  for CMB polarisation measurements}},
  \href{https://doi.org/10.1088/1748-0221/6/06/P06009}{\emph{J. Instrum.}
  {\bfseries 6} (2011) 6009} [\href{https://arxiv.org/abs/1107.1157}{{\ttfamily
  1107.1157}}].

\bibitem{eom2006}
S.Y.~Eom and Y.B.~Korchemkin, \emph{{A New Comb Circular Polarizer Suitable for
  Millimeter-Band Application}},
  \href{https://doi.org/10.4218/etrij.06.0206.0110}{\emph{ETRI Journal}
  {\bfseries 28} (2006) 656}.

\bibitem{Virone2014}
G.~Virone, O.A.~Peverini, M.~Lumia, G.~Addamo and R.~Tascone, \emph{{Platelet
  Orthomode Transducer for Q-Band Correlation Polarimeter Clusters}},
  \href{https://doi.org/10.1109/TMTT.2014.2325793}{\emph{IEEE T. Microw.
  Theory} {\bfseries 62} (2014) 1487}.

\bibitem{quiet2012b}
{QUIET Collaboration}, D.~Araujo, C.~Bischoff, A.~Brizius, I.~Buder, Y.~Chinone
  et~al., \emph{{Second season QUIET observations: Measurements of the cosmic
  microwave background polarization power spectrum at 95 GHz}},
  \href{https://doi.org/10.1088/0004-637X/760/2/145}{\emph{Astrophys. J.}
  {\bfseries 760} (2012) } [\href{https://arxiv.org/abs/1207.5034}{{\ttfamily
  1207.5034}}].

\bibitem{chen2014}
Y.-L.~Chen, T.~Chiueh and H.-F.~Teng, \emph{A 77-118 {GHz} {RESONANCE}-{FREE}
  {SEPTUM} {POLARIZER}},
  \href{https://doi.org/10.1088/0067-0049/211/1/11}{\emph{Astrophys. J., Suppl.
  Ser.} {\bfseries 211} (2014) 11}.

\bibitem{quiet2011}
{QUIET Collaboration}, C.~Bischoff, A.~Brizius, I.~Buder, Y.~Chinone, K.~Cleary
  et~al., \emph{{{First Season QUIET Observations: Measurements of Cosmic
  Microwave Background Polarization Power Spectra at 43 GHz in the Multipole
  Range $25\le\ell\le 475$}}},
  \href{https://doi.org/10.1088/0004-637X/741/2/111}{\emph{Astrophys. J.}
  {\bfseries 741} (2011) 111}
  [\href{https://arxiv.org/abs/1012.3191}{{\ttfamily 1012.3191}}].

\bibitem{iarocci2008}
A.~{Iarocci}, P.~{Benedetti}, F.~{Caprara}, A.~{Cardillo}, F.~{di Felice},
  G.~{di Stefano} et~al., \emph{{PEGASO: An ultra light long duration
  stratospheric payload for polar regions flights}},
  \href{https://doi.org/10.1016/j.asr.2007.05.079}{\emph{Adv. Space Res.}
  {\bfseries 42} (2008) 1633}.

\bibitem{peterzen_memsait2008}
S.~{Peterzen}, S.~{Masi}, P.~{Dragoy}, R.~{Ibba} and D.~{Spoto}, \emph{{Long
  Duration Balloon flights development. (Italian Space Agency)}},
  {\emph{\memsai} {\bfseries 79} (2008) 792}.

\bibitem{2019EPJWC_MASI}
{Masi, Silvia}, {Coppolecchia, A.}, {Battistelli, E.}, {de Bernardis, P.},
  {Columbro, F.}, {D\'{}Alessandro, G.} et~al., \emph{Balloon-borne cosmic
  microwave background experiments},
  \href{https://doi.org/10.1051/epjconf/201920901046}{\emph{EPJ Web Conf.}
  {\bfseries 209} (2019) 01046}.

\bibitem{debernardisIAUS2013}
P.~{de Bernardis}, S.~{Masi} and {OLIMPO and LSPE Teams}, \emph{{Precision CMB
  measurements with long-duration stratospheric balloons: activities in the
  Arctic}},  in \emph{Astrophysics from Antarctica}, M.G.~{Burton}, X.~{Cui}
  and N.F.H.~{Tothill}, eds., vol.~288 of \emph{IAU Symposium}, pp.~208--213,
  Jan., 2013, \href{https://doi.org/10.1017/S1743921312016894}{DOI}.

\bibitem{wipica2018}
F.~{Piacentini}, A.~{Coppolecchia}, P.~{de Bernardis}, G.~{Di Stefano},
  A.~{Iarocci}, L.~{Lamagna} et~al., \emph{{Winter long duration stratospheric
  balloons from Polar regions}}, {\emph{arXiv e-prints} (2018)
  arXiv:1810.05565} [\href{https://arxiv.org/abs/1810.05565}{{\ttfamily
  1810.05565}}].

\bibitem{1993AA...271..683D}
P.~{de Bernardis}, E.~{Aquilini}, A.~{Boscaleri}, M.~{de Petris}, M.~{Gervasi},
  L.~{Martinis} et~al., \emph{{ARGO: a balloon-borne telescope for measurements
  of the millimeter diffuse sky emission}}, {\emph{Astron. Astrophys.}
  {\bfseries 271} (1993) 683}.

\bibitem{2003ApJS..148..527C}
B.P.~{Crill}, P.A.R.~{Ade}, D.R.~{Artusa}, R.S.~{Bhatia}, J.J.~{Bock},
  A.~{Boscaleri} et~al., \emph{{BOOMERANG: A Balloon-borne Millimeter-Wave
  Telescope and Total Power Receiver for Mapping Anisotropy in the Cosmic
  Microwave Background}},
  \href{https://doi.org/10.1086/376894}{\emph{Astrophys. J., Suppl. Ser.}
  {\bfseries 148} (2003) 527}
  [\href{https://arxiv.org/abs/astro-ph/0206254}{{\ttfamily
  astro-ph/0206254}}].

\bibitem{2006AA...458..687M}
S.~{Masi}, P.A.R.~{Ade}, J.J.~{Bock}, J.R.~{Bond}, J.~{Borrill}, A.~{Boscaleri}
  et~al., \emph{{Instrument, method, brightness, and polarization maps from the
  2003 flight of BOOMERanG}},
  \href{https://doi.org/10.1051/0004-6361:20053891}{\emph{Astron. Astrophys.}
  {\bfseries 458} (2006) 687}
  [\href{https://arxiv.org/abs/astro-ph/0507509}{{\ttfamily
  astro-ph/0507509}}].

\bibitem{2002APh....17..101B}
A.~{Beno{\^\i}t}, P.~{Ade}, A.~{Amblard}, R.~{Ansari}, E.~{Aubourg},
  J.~{Bartlett} et~al., \emph{{Archeops: a high resolution, large sky coverage
  balloon experiment for mapping cosmic microwave background anisotropies}},
  \href{https://doi.org/10.1016/S0927-6505(01)00141-4}{\emph{Astropart. Phys.}
  {\bfseries 17} (2002) 101}
  [\href{https://arxiv.org/abs/astro-ph/0106152}{{\ttfamily
  astro-ph/0106152}}].

\bibitem{2005ESASP.590..581M}
S.~{Masi}, M.~{Calvo}, L.~{Conversi}, P.~{de Bernardis}, M.~{de Petris}, G.~{de
  Troia} et~al., \emph{{A balloon-borne survey of the mm/sub-mm sky: OLIMPO}},
  in \emph{17th ESA Symposium on European Rocket and Balloon Programmes and
  Related Research}, B.~{Warmbein}, ed., vol.~590 of \emph{ESA Special
  Publication}, pp.~581--586, Aug., 2005.

\bibitem{Paiella_2019}
A.~Paiella, A.~Coppolecchia, L.~Lamagna, P.~Ade, E.~Battistelli,
  M.G.~Castellano et~al., \emph{{Kinetic inductance detectors for the {OLIMPO}
  experiment: design and pre-flight characterization}},
  \href{https://doi.org/10.1088/1475-7516/2019/01/039}{\emph{J. Cosmol.
  Astropart. P.} {\bfseries 2019} (2019) 039}.

\bibitem{2019JCAP...07..003M}
S.~{Masi}, P.~{de Bernardis}, A.~{Paiella}, F.~{Piacentini}, L.~{Lamagna},
  A.~{Coppolecchia} et~al., \emph{{Kinetic Inductance Detectors for the OLIMPO
  experiment: in-flight operation and performance}},
  \href{https://doi.org/10.1088/1475-7516/2019/07/003}{\emph{J. Cosmol.
  Astropart. P.} {\bfseries 2019} (2019) 003}
  [\href{https://arxiv.org/abs/1902.08993}{{\ttfamily 1902.08993}}].

\bibitem{1990SPIE.1341...58B}
A.~Boscaleri, V.~Venturi and R.~Colzi, \emph{{Time-domain computer simulation
  program as first step of a full digital high-precision pointing system for
  platform in balloon-borne remote sensing}},  in \emph{Infrared Technology
  XVI}, I.J.~Spiro, ed., vol.~1341, pp.~58 -- 65, International Society for
  Optics and Photonics, SPIE, 1990,
  \href{https://doi.org/10.1117/12.23076}{DOI}.

\bibitem{1990SPIE.1304..127B}
A.~Boscaleri, \emph{{A time domain design technique for high precision full
  digital pointing system in balloon-borne remote infrared sensing}},  in
  \emph{Acquisition, Tracking, and Pointing IV}, S.~Gowrinathan, ed.,
  vol.~1304, International Society for Optics and Photonics, SPIE, 1990,
  \href{https://doi.org/10.1117/12.2322204}{DOI}.

\bibitem{1994MeScT...5..190B}
A.~{Boscaleri}, V.~{Venturi} and D.~{Tirelli}, \emph{{The ARGO experiment
  pointing system as an example for other single-axis platform pointing
  systems}}, \href{https://doi.org/10.1088/0957-0233/5/2/016}{\emph{Meas. Sci.
  Technol.} {\bfseries 5} (1994) 190}.

\bibitem{2003RScI...74.4169N}
F.~{Nati}, P.~{de Bernardis}, A.~{Iacoangeli}, S.~{Masi}, A.~{Benoit} and
  D.~{Yvon}, \emph{{A fast star sensor for balloon payloads}},
  \href{https://doi.org/10.1063/1.1602961}{\emph{Rev. Sci. Instrum.} {\bfseries
  74} (2003) 4169}.

\bibitem{2007AA...467.1313M}
J.F.~{Mac{\'\i}as-P{\'e}rez}, G.~{Lagache}, B.~{Maffei}, K.~{Ganga},
  A.~{Bourrachot}, P.~{Ade} et~al., \emph{{Archeops in-flight performance, data
  processing, and map making}},
  \href{https://doi.org/10.1051/0004-6361:20065258}{\emph{Astron. Astrophys.}
  {\bfseries 467} (2007) 1313}
  [\href{https://arxiv.org/abs/astro-ph/0603665}{{\ttfamily
  astro-ph/0603665}}].

\bibitem{1994Cryo...34.1001P}
P.~{Palumbo}, E.~{Aquilini}, P.~{Cardoni}, P.~{de Bernardis}, A.~{De Ninno},
  L.~{Martinis} et~al., \emph{{Balloon-borne $^{3}$He cryostat for millimetre
  bolometric photometry}},
  \href{https://doi.org/10.1016/0011-2275(94)90093-0}{\emph{Cryogenics}
  {\bfseries 34} (1994) 1001}.

\bibitem{1999Cryo...39..217M}
S.~{Masi}, P.~{Cardoni}, P.~{de Bernardis}, F.~{Piacentini}, A.~{Raccanelli}
  and F.~{Scaramuzzi}, \emph{{A long duration cryostat suitable for balloon
  borne photometry}},
  \href{https://doi.org/10.1016/S0011-2275(99)00018-1}{\emph{Cryogenics}
  {\bfseries 39} (1999) 217}.

\bibitem{2016ExA....42..199B}
J.P.~{Bernard}, P.~{Ade}, Y.~{Andr{\'e}}, J.~{Aumont}, L.~{Bautista}, N.~{Bray}
  et~al., \emph{{PILOT: a balloon-borne experiment to measure the polarized FIR
  emission of dust grains in the interstellar medium}},
  \href{https://doi.org/10.1007/s10686-016-9506-1}{\emph{Exp. Astron.}
  {\bfseries 42} (2016) 199}.

\bibitem{Coppo2020}
A.~Coppolecchia, L.~Lamagna, S.~Masi, P.~Ade, G.~Amico, E.~Battistelli et~al.,
  \emph{The long duration cryogenic system of the olimpo balloon-borne
  experiment: design and in-flight performance},
  \href{https://doi.org/https://doi.org/10.1016/j.cryogenics.2020.103129}{\emph{Cryogenics}
  (2020) 103129}.

\bibitem{2016SPIE.9912E..65C}
G.~{Coppi}, P.~{de Bernardis}, A.J.~{May}, S.~{Masi}, M.~{McCulloch},
  S.J.~{Melhuish} et~al., \emph{{Developing a long duration $^{3}$He fridge for
  the LSPE-SWIPE instrument}},  in \emph{\procspie}, vol.~9912 of \emph{Society
  of Photo-Optical Instrumentation Engineers (SPIE) Conference Series},
  p.~991265, SPIE (2016), \href{https://doi.org/10.1117/12.2232448}{DOI}.

\bibitem{2017zilic}
K.~Zilic, A.~Aboobaker, F.~Aubin, C.~Geach, S.~Hanany, N.~Jarosik et~al.,
  \emph{A double vacuum window mechanism for space-borne applications},
  \href{https://doi.org/10.1063/1.4981814}{\emph{Rev. Sci. Instrum.} {\bfseries
  88} (2017) }.

\bibitem{1983IAUS..104..135D}
G.~{dall'Oglio}, P.~{de Bernardis}, S.~{Masi} and F.~{Melchiorri},
  \emph{{Measurement of the 3K Cosmic Background Noise in the Far Infrared}},
  in \emph{Early Evolution of the Universe and its Present Structure},
  G.O.~{Abell} and G.~{Chincarini}, eds., vol.~104 of \emph{IAU Symposium},
  p.~135, Jan., 1983.

\bibitem{macias2007}
J.F.~{Mac{\'{\i}}as-P{\'e}rez}, G.~{Lagache}, B.~{Maffei}, K.~{Ganga},
  A.~{Bourrachot}, P.~{Ade} et~al., \emph{{Archeops in-flight performance, data
  processing, and map making}},
  \href{https://doi.org/10.1051/0004-6361:20065258}{\emph{Astron. Astrophys.}
  {\bfseries 467} (2007) 1313}
  [\href{https://arxiv.org/abs/astro-ph/0603665}{{\ttfamily
  astro-ph/0603665}}].

\bibitem{DAlessandro:UHMW}
G.~{D'Alessandro}, A.~{Paiella}, A.~{Coppolecchia}, M.G.~{Castellano},
  I.~{Colantoni}, P.~{de Bernardis} et~al., \emph{{Ultra high molecular weight
  polyethylene: Optical features at millimeter wavelengths}},
  \href{https://doi.org/10.1016/j.infrared.2018.02.008}{\emph{Infrared Phys.
  Technol.} {\bfseries 90} (2018) 59}
  [\href{https://arxiv.org/abs/1803.05228}{{\ttfamily 1803.05228}}].

\bibitem{legg2016}
S.~{Legg}, L.~{Lamagna}, G.~{Coppi}, P.~{de Bernardis}, G.~{Giuliani},
  R.~{Gualtieri} et~al., \emph{{Development of the multi-mode horn-lens
  configuration for the LSPE-SWIPE B-mode experiment}},  in \emph{Millimeter,
  Submillimeter, and Far-Infrared Detectors and Instrumentation for Astronomy
  VIII}, vol.~9914 of \emph{Proceeding SPIE}, p.~991414, July, 2016,
  \href{https://doi.org/10.1117/12.2232400}{DOI}.

\bibitem{Columbro_LSPE_feed_measure}
F.~Columbro, P.G.~Madonia, L.~Lamagna, E.S.~Battistelli, A.~Coppolecchia,
  P.~de~Bernardis et~al., \emph{Swipe multi-mode pixel assembly design and beam
  pattern measurements at cryogenic temperature},
  \href{https://doi.org/10.1007/s10909-020-02396-4}{\emph{J. Low Temp. Phys.}
  {\bfseries 199} (2020) 312}.

\bibitem{ludwig}
A.C.~{Ludwig}, \emph{{The definition of cross polarization.}},
  \href{https://doi.org/10.1109/TAP.1973.1140406}{\emph{IEEE T. Antenn.
  Propag.} {\bfseries 21} (1973) 116}.

\bibitem{Pisano1}
G.~{Pisano}, C.~{Tucker}, P.A.R.~{Ade}, P.~{Moseley} and M.W.~{Ng}, \emph{Metal
  mesh based metamaterials for millimetre wave and thz astronomy applications},
   in \emph{2015 8th UK, Europe, China Millimeter Waves and THz Technology
  Workshop (UCMMT)}, pp.~1--4, 2015,
  \href{https://doi.org/10.1109/UCMMT.2015.7460631}{DOI}.

\bibitem{Pisano2}
G.~{Pisano}, M.~{Ng}, V.~{Haynes} and B.~{Maffei}, \emph{A broadband metal-mesh
  half-wave plate for millimetre wave linear polarisation rotation},  in
  \emph{Progress In Electromagnetics Research M}, vol.~25, pp.~101--114, 2012,
  \href{https://doi.org/10.2528/PIERM12051410}{DOI}.

\bibitem{pisano3}
G.~{Pisano}, P.~{Ade}, C.~{Tucker} and M.W.~{Ng}, \emph{Large bandwidth mesh
  half-wave plates for millimetre and thz wave astronomy},  in \emph{2015 40th
  International Conference on Infrared, Millimeter, and Terahertz waves
  (IRMMW-THz)}, pp.~1--1, 2015.

\bibitem{Matsumura2016}
T.~{Matsumura}, H.~{Kataza}, S.~{Utsunomiya}, R.~{Yamamoto}, M.~{Hazumi} and
  N.~{Katayama}, \emph{Design and performance of a prototype polarization
  modulator rotational system for use in space using a superconducting magnetic
  bearing}, \href{https://doi.org/10.1109/TASC.2016.2533584}{\emph{IEEE Trans.
  Appl. Supercond.} {\bfseries 26} (2016) 1}.

\bibitem{Johnson2017}
B.~Johnson, F.~Columbro, D.~Araujo, M.~Limon, B.~Smiley, G.~Jones et~al.,
  \emph{A large-diameter hollow-shaft cryogenic motor based on a
  superconducting magnetic bearing for millimeter-wave polarimetry},
  \href{https://doi.org/10.1063/1.4990884}{\emph{Rev. Sci. Instrum.} {\bfseries
  88} (2017) }.

\bibitem{2020_Columbro_SPIE}
F.~{Columbro}, P.~{de Bernardis}, L.~{Lamagna}, S.~{Masi}, A.~{Paiella},
  F.~{Piacentini} et~al., \emph{{A polarization modulator unit for the mid- and
  high-frequency telescopes of the LiteBIRD mission}},  vol.~11443 of
  \emph{Society of Photo-Optical Instrumentation Engineers (SPIE) Conference
  Series}, p.~114436Z, Dec., 2020,
  \href{https://doi.org/10.1117/12.2577818}{DOI}.

\bibitem{Columbro2018}
F.~Columbro, P.~de~Bernardis and S.~Masi, \emph{A clamp and release system for
  superconductive magnetic bearings},
  \href{https://doi.org/10.1063/1.5035332}{\emph{Rev. Sci. Instrum.} {\bfseries
  89} (2018) }.

\bibitem{PdB_levitation_measurement2020}
P.~de~Bernardis, F.~Columbro, S.~Masi, A.~Paiella and G.~Romeo, \emph{A simple
  method to measure the temperature and levitation height of devices rotating
  at cryogenic temperatures},
  \href{https://doi.org/10.1063/5.0005498}{\emph{Rev. Sci. Instrum.} {\bfseries
  91} (2020) 045118}
  [\href{https://arxiv.org/abs/https://doi.org/10.1063/5.0005498}{{\ttfamily
  https://doi.org/10.1063/5.0005498}}].

\bibitem{tiTctuning}
D.~Vaccaro, B.~Siri, A.M.~Baldini, M.~Biasotti, F.~Cei, V.~Ceriale et~al.,
  \emph{{Tuning the $T_C$ of Titanium Thin Films for Transition-Edge Sensors by
  Annealing in Argon}},
  \href{https://doi.org/10.1007/s10909-018-2090-1}{\emph{J. Low Temp. Phys.}
  {\bfseries 193} (2018) 1122}.

\bibitem{2018_Vaccaro}
D.~Vaccaro, A.M.~Baldini, F.~Cei, L.~Galli, M.~Grassi, D.~Nicol\`o et~al.,
  \emph{"{The FDM readout for the LSPE/SWIPE TES bolometers}"},  in
  \emph{Millimeter, Submillimeter, and Far-Infrared Detectors and
  Instrumentation for Astronomy IX}, J.~Zmuidzinas and J.-R.~Gao, eds.,
  vol.~10708, pp.~732 -- 742, International Society for Optics and Photonics,
  \procspie, 2018, \href{https://doi.org/10.1117/12.2310148}{DOI}.

\bibitem{Vaccaro:2019hmz}
D.~Vaccaro, A.~Baldini, F.~Cei, L.~Galli, M.~Grassi, D.~Nicol\`o et~al.,
  \emph{{A frequency domain multiplexing system to readout the TES bolometers
  on the LSPE/SWIPE experiment}},
  \href{https://doi.org/10.1016/j.nima.2018.10.116}{\emph{Nucl. Instrum. Meth.
  A} {\bfseries 936} (2019) 169}.

\bibitem{fontanelli}
F.~Fontanelli, M.~Biasotti, A.~Bevilacqua and F.~Siccardi, \emph{{The front-end
  electronics of the LSPE-SWIPE experiment}},  in \emph{Space Telescopes and
  Instrumentation 2016: Optical, Infrared, and Millimeter Wave}, H.A.~MacEwen,
  G.G.~Fazio, M.~Lystrup, N.~Batalha, N.~Siegler and E.C.~Tong, eds.,
  vol.~9904, pp.~1633 -- 1638, International Society for Optics and Photonics,
  \procspie, 2016, \href{https://doi.org/10.1117/12.2232859}{DOI}.

\bibitem{QUIET2012a}
C.~Bischoff, A.~Brizius, I.~Buder, Y.~Chinone, K.~Cleary, R.N.~Dumoulin et~al.,
  \emph{{{THE} Q/U {IMAGING} {EXPERIMENT} {INSTRUMENT}}},
  \href{https://doi.org/10.1088/0004-637x/768/1/9}{\emph{Astrophys. J.}
  {\bfseries 768} (2013) 9}.

\bibitem{Cleary2010}
K.~Cleary, \emph{{Coherent polarimeter modules for the QUIET experiment}},  in
  \emph{Society of Photo-Optical Instrumentation Engineers (SPIE) Conference
  Series}, vol.~7741 of \emph{Society of Photo-Optical Instrumentation
  Engineers (SPIE) Conference Series}, jul, 2010,
  \href{https://doi.org/10.1117/12.857673}{DOI}.

\bibitem{paine_scott_2019_3406496}
S.~Paine, \emph{The am atmospheric model},  Sept., 2019.
\newblock 10.5281/zenodo.3406496.

\bibitem{1604073}
M.N.~{Afsar}, K.A.~{Korolev}, L.~{Subramanian} and I.I.~{Tkachov},
  \emph{{Complex Dielectric Measurements of Materials at Q- Band, V- Band and
  W- Band Frequencies with High Power Sources}},  in \emph{2005 IEEE
  Instrumentationand Measurement Technology Conference Proceedings}, vol.~1,
  pp.~82--87, 2005.

\bibitem{Lamb1996}
J.W.~Lamb, \emph{Miscellaneous data on materials for millimetre and
  submillimetre optics}, \href{https://doi.org/10.1007/BF02069487}{\emph{J.
  Infrared. Millim. W.} {\bfseries 17} (1996) 1997}.

\bibitem{1986_Lamarre}
J.M.~{Lamarre}, \emph{{Photon noise in photometric instruments at far-infrared
  and submillimeter wavelengths}},
  \href{https://doi.org/10.1364/AO.25.000870}{\emph{\ao} {\bfseries 25} (1986)
  870}.

\bibitem{2016gualtieri}
R.~{Gualtieri}, E.S.~{Battistelli}, A.~{Cruciani}, P.~{de Bernardis},
  M.~{Biasotti}, D.~{Corsini} et~al., \emph{{Multi-mode TES Bolometer
  Optimization for the LSPE-SWIPE Instrument}},
  \href{https://doi.org/10.1007/s10909-015-1436-1}{\emph{J. Low Temp. Phys.}
  {\bfseries 184} (2016) 527}
  [\href{https://arxiv.org/abs/1602.07744}{{\ttfamily 1602.07744}}].

\bibitem{Dobbs:2011px}
M.~Dobbs et~al., \emph{{Frequency Multiplexed SQUID Readout of Large Bolometer
  Arrays for Cosmic Microwave Background Measurements}},
  \href{https://doi.org/10.1063/1.4737629}{\emph{Rev. Sci. Instrum.} {\bfseries
  83} (2012) 073113} [\href{https://arxiv.org/abs/1112.4215}{{\ttfamily
  1112.4215}}].

\bibitem{tartariLTD18}
A.~Tartari, A.M.~Baldini, F.~Cei, L.~Galli, M.~Grassi, D.~Nicol{\`o} et~al.,
  \emph{Development and testing of the fdm read-out of the tes arrays aboard
  the lspe/swipe balloon-borne experiment},
  \href{https://doi.org/10.1007/s10909-020-02431-4}{\emph{J. Low Temp. Phys.}
  {\bfseries 199} (2020) 212}.

\bibitem{planck2013-p02a}
{\sorthelp{Planck Collaboration 2014C}}{Planck Collaboration III},
  \emph{{\textit{Planck} 2013 results. III. LFI systematic uncertainties}},
  \href{https://doi.org/10.1051/0004-6361/201321574}{\emph{Astron. Astrophys.}
  {\bfseries 571} (2014) A3} [\href{https://arxiv.org/abs/1303.5064}{{\ttfamily
  1303.5064}}].

\bibitem{Planck2015LFIsystematics}
{Planck Collaboration}, P.A.R.~Ade, J.~Aumont, C.~Baccigalupi, A.J.~Banday,
  R.B.~Barreiro et~al., \emph{{Planck 2015 results. III. LFI systematic
  uncertainties}},
  \href{https://doi.org/10.1051/0004-6361/201526998}{\emph{Astron. Astrophys.}
  {\bfseries 594} (2016) A3}.

\bibitem{church1995predicting}
S.~Church, \emph{Predicting residual levels of atmospheric sky noise in
  ground-based observations of the cosmic background radiation}, {\emph{Mon.
  Not. Roy. Astron. Soc.} {\bfseries 272} (1995) 551}.

\bibitem{Planck2015LFIbeams}
{Planck Collaboration}, P.A.R.~Ade, N.~Aghanim, M.~Ashdown, J.~Aumont,
  C.~Baccigalupi et~al., \emph{{Planck 2015 results. IV. Low Frequency
  Instrument beams and window functions}},
  \href{https://doi.org/10.1051/0004-6361/201525809}{\emph{Astron. Astrophys.}
  {\bfseries 594} (2016) A4}.

\bibitem{planck2014-a04}
{\sorthelp{Planck Collaboration 2015C}}{Planck Collaboration III},
  \emph{{\textit{Planck} 2015 results. III. LFI systematic uncertainties}},
  \href{https://doi.org/10.1051/0004-6361/201526998}{\emph{Astron. Astrophys.}
  {\bfseries 594} (2016) A3}
  [\href{https://arxiv.org/abs/1507.08853}{{\ttfamily 1507.08853}}].

\bibitem{Krachmalnicoff2015}
N.~Krachmalnicoff, \emph{{Challenges for Present and Future Cosmic Microwave
  Background Observations: Systematic Effects and Foreground Emission in
  Polarization}}, Ph.D. thesis, University of Milan, 2015.

\bibitem{Montresor2012}
G.~Montresor, \emph{{A responsivity calibration strategy for the LSPE-STRIP
  balloon experiment}},  Master's thesis, University of Milan, 2012.

\bibitem{Paonessa202012}
F.~Paonessa, G.~Virone, L.~Ciorba, G.~Addamo, M.~Lumia, G.~Dassano et~al.,
  \emph{{Design and verification of a Q-Band test source for UAV-based
  radiation pattern measurements}},
  \href{https://doi.org/10.1109/TIM.2020.3031127}{\emph{IEEE Trans. Instrum.
  Meas.} {\bfseries 69} (2020) 9366}.

\bibitem{salatino2017}
M.~Salatino, P.~de~Bernardis and S.~Masi, \emph{Modeling transmission and
  reflection mueller matrices of dielectric half-wave plates},
  \href{https://doi.org/10.1007/s10762-016-0320-7}{\emph{Journal of Infrared,
  Millimeter, and Terahertz Waves} {\bfseries 38} (2017) 215}.

\bibitem{Pisano:06}
G.~Pisano, G.~Savini, P.A.R.~Ade, V.~Haynes and W.K.~Gear, \emph{Achromatic
  half-wave plate for submillimeter instruments in cosmic microwave background
  astronomy: experimental characterization},
  \href{https://doi.org/10.1364/AO.45.006982}{\emph{Appl. Opt.} {\bfseries 45}
  (2006) 6982}.

\bibitem{2010ApJ...711.1141T}
Y.D.~{Takahashi}, P.A.R.~{Ade}, D.~{Barkats}, J.O.~{Battle}, E.M.~{Bierman},
  J.J.~{Bock} et~al., \emph{{Characterization of the BICEP Telescope for
  High-precision Cosmic Microwave Background Polarimetry}},
  \href{https://doi.org/10.1088/0004-637X/711/2/1141}{\emph{Astrophys. J}
  {\bfseries 711} (2010) 1141}
  [\href{https://arxiv.org/abs/0906.4069}{{\ttfamily 0906.4069}}].

\bibitem{2018_Imada_IEEE}
H.~Imada, T.~Matsumura, R.~Takaku, G.~Patanchon, H.~Ishino, Y.~Sakurai et~al.,
  \emph{Instrumentally induced spurious polarization of a multi-layer half wave
  plate for a cmb polarization observation},  pp.~61--67, Jan., 2018.

\bibitem{2018JCAP...09..005K}
A.~{Kusaka}, J.~{Appel}, T.~{Essinger-Hileman}, J.A.~{Beall}, L.E.~{Campusano},
  H.-M.~{Cho} et~al., \emph{{Results from the Atacama B-mode Search (ABS)
  experiment}},
  \href{https://doi.org/10.1088/1475-7516/2018/09/005}{\emph{\jcap} {\bfseries
  2018} (2018) 005} [\href{https://arxiv.org/abs/1801.01218}{{\ttfamily
  1801.01218}}].

\bibitem{2016RScI...87i4503E}
T.~{Essinger-Hileman}, A.~{Kusaka}, J.W.~{Appel}, S.K.~{Choi}, K.~{Crowley},
  S.P.~{Ho} et~al., \emph{{Systematic effects from an ambient-temperature,
  continuously rotating half-wave plate}},
  \href{https://doi.org/10.1063/1.4962023}{\emph{Rev. Sci. Instrum.} {\bfseries
  87} (2016) 094503} [\href{https://arxiv.org/abs/1601.05901}{{\ttfamily
  1601.05901}}].

\bibitem{2010ApOpt..49.6313B}
S.A.~{Bryan}, T.E.~{Montroy} and J.E.~{Ruhl}, \emph{{Modeling dielectric
  half-wave plates for cosmic microwave background polarimetry using a Mueller
  matrix formalism}}, \href{https://doi.org/10.1364/AO.49.006313}{\emph{\ao}
  {\bfseries 49} (2010) 6313}
  [\href{https://arxiv.org/abs/1006.3359}{{\ttfamily 1006.3359}}].

\bibitem{2009Pagano_malapola}
L.~{Pagano}, P.~{de Bernardis}, G.~{de Troia}, G.~{Gubitosi}, S.~{Masi},
  A.~{Melchiorri} et~al., \emph{{CMB polarization systematics, cosmological
  birefringence, and the gravitational waves background}},
  \href{https://doi.org/10.1103/PhysRevD.80.043522}{\emph{\prd} {\bfseries 80}
  (2009) 043522} [\href{https://arxiv.org/abs/0905.1651}{{\ttfamily
  0905.1651}}].

\bibitem{Columbro2019}
F.~{Columbro}, E.S.~{Battistelli}, A.~{Coppolecchia}, G.~{D'Alessandro}, P.~{de
  Bernardis}, L.~{Lamagna} et~al., \emph{{The short wavelength instrument for
  the polarization explorer balloon-borne experiment: Polarization modulation
  issues}}, \href{https://doi.org/10.1002/asna.201913566}{\emph{Astron. Nachr.}
  {\bfseries 340} (2019) 83}
  [\href{https://arxiv.org/abs/1904.01891}{{\ttfamily 1904.01891}}].

\bibitem{2020aumont}
J.~{Aumont}, J.F.~{Mac{\'\i}as-P{\'e}rez}, A.~{Ritacco}, N.~{Ponthieu} and
  A.~{Mangilli}, \emph{{Absolute calibration of the polarisation angle for
  future CMB B-mode experiments from current and future measurements of the
  Crab nebula}},
  \href{https://doi.org/10.1051/0004-6361/201833504}{\emph{Astron. Astrophys.}
  {\bfseries 634} (2020) A100}
  [\href{https://arxiv.org/abs/1911.03164}{{\ttfamily 1911.03164}}].

\bibitem{2005_Tucci}
M.~Tucci, E.~Martínez-González, P.~Vielva and J.~Delabrouille, \emph{{Limits
  on the detectability of the CMB B-mode polarization imposed by foregrounds}},
  \href{https://doi.org/10.1111/j.1365-2966.2005.09123.x}{\emph{Mon. Not. Roy.
  Astron. Soc.} {\bfseries 360} (2005) 935}
  [\href{https://arxiv.org/abs/https://academic.oup.com/mnras/article-pdf/360/3/935/3206389/360-3-935.pdf}{{\ttfamily
  https://academic.oup.com/mnras/article-pdf/360/3/935/3206389/360-3-935.pdf}}].

\bibitem{buzzelli_migliaccio2018}
A.~{Buzzelli}, M.~{Migliaccio}, G.~{de Gasperis}, P.~{de Bernardis}, S.~{Masi}
  and N.~{Vittorio}, \emph{{Impact of polarized foregrounds on LSPE-SWIPE
  observations}},  in \emph{J. Phys. Conf. Ser.}, vol.~956, p.~012002, Jan.,
  2018, \href{https://doi.org/10.1088/1742-6596/956/1/012002}{DOI}.

\bibitem{2020A&A...641A...4P}
{Planck Collaboration}, Y.~{Akrami}, M.~{Ashdown}, J.~{Aumont},
  C.~{Baccigalupi}, M.~{Ballardini} et~al., \emph{{Planck 2018 results. IV.
  Diffuse component separation}},
  \href{https://doi.org/10.1051/0004-6361/201833881}{\emph{Astron. Astrophys.}
  {\bfseries 641} (2020) A4}
  [\href{https://arxiv.org/abs/1807.06208}{{\ttfamily 1807.06208}}].

\bibitem{Campeti:2019ylm}
P.~Campeti, D.~Poletti and C.~Baccigalupi, \emph{{Principal component analysis
  of the primordial tensor power spectrum}},
  \href{https://doi.org/10.1088/1475-7516/2019/09/055}{\emph{J. Cosmol.
  Astropart. P.} {\bfseries 1909} (2019) 055}
  [\href{https://arxiv.org/abs/1905.08200}{{\ttfamily 1905.08200}}].

\bibitem{Stompor:2016hhw}
R.~Stompor, J.~Errard and D.~Poletti, \emph{{Forecasting performance of CMB
  experiments in the presence of complex foreground contaminations}},
  \href{https://doi.org/10.1103/PhysRevD.94.083526}{\emph{Phys. Rev.}
  {\bfseries D94} (2016) 083526}
  [\href{https://arxiv.org/abs/1609.03807}{{\ttfamily 1609.03807}}].

\bibitem{pysm}
B.~Thorne, J.~Dunkley, D.~Alonso and S.~Naess, \emph{{The Python Sky Model:
  software for simulating the Galactic microwave sky}},
  \href{https://doi.org/10.1093/mnras/stx949}{\emph{Mon. Not. Roy. Astron.
  Soc.} {\bfseries 469} (2017) 2821}
  [\href{https://arxiv.org/abs/1608.02841}{{\ttfamily 1608.02841}}].

\bibitem{2019QUIJOTE}
J.A.~{Rubi{\~n}o-Mart{\'\i}n}, {Planck Collaboration} and {QUIJOTE
  Collaboration}, \emph{{Cosmology with the Cosmic Microwave Background: Latest
  Results from the PLANCK satellite and the QUIJOTE experiment}},  in
  \emph{Highlights on Spanish Astrophysics X}, B.~{Montesinos}, A.~{Asensio
  Ramos}, F.~{Buitrago}, R.~{Sch{\"o}del}, E.~{Villaver}, S.~{P{\'e}rez-Hoyos}
  et~al., eds., pp.~32--43, Mar., 2019.

\bibitem{Alonso:2018jzx}
{\scshape LSST Dark Energy Science} collaboration, \emph{{A unified
  pseudo-$C_\ell$ framework}},
  \href{https://doi.org/10.1093/mnras/stz093}{\emph{Mon. Not. Roy. Astron.
  Soc.} {\bfseries 484} (2019) 4127}
  [\href{https://arxiv.org/abs/1809.09603}{{\ttfamily 1809.09603}}].

\bibitem{planck2016-l05}
{\sorthelp{Planck Collaboration 2018E}}{Planck Collaboration V},
  \emph{{\textit{Planck} 2018 results. V. Power spectra and likelihoods}},
  {\emph{Astron. Astrophys., submitted} (2019) }
  [\href{https://arxiv.org/abs/1907.12875}{{\ttfamily 1907.12875}}].

\bibitem{Tegmark:2001zv}
M.~Tegmark and A.~de~Oliveira-Costa, \emph{{How to measure CMB polarization
  power spectra without losing information}},
  \href{https://doi.org/10.1103/PhysRevD.64.063001}{\emph{Phys. Rev.}
  {\bfseries D64} (2001) 063001}
  [\href{https://arxiv.org/abs/astro-ph/0012120}{{\ttfamily
  astro-ph/0012120}}].

\bibitem{Pagano:2019tci}
L.~Pagano, J.-M.~Delouis, S.~Mottet, J.-L.~Puget and L.~Vibert,
  \emph{{Reionization optical depth determination from Planck HFI data with ten
  percent accuracy}},
  \href{https://doi.org/10.1051/0004-6361/201936630}{\emph{Astron. Astrophys.}
  {\bfseries 635} (2020) A99}
  [\href{https://arxiv.org/abs/1908.09856}{{\ttfamily 1908.09856}}].

\bibitem{Monteserin:2007fv}
C.~Monteserin, R.B.B.~Barreiro, P.~Vielva, E.~Martinez-Gonzalez, M.P.~Hobson
  and A.N.~Lasenby, \emph{{A low CMB variance in the WMAP data}},
  \href{https://doi.org/10.1111/j.1365-2966.2008.13149.x}{\emph{Mon. Not. Roy.
  Astron. Soc.} {\bfseries 387} (2008) 209}
  [\href{https://arxiv.org/abs/0706.4289}{{\ttfamily 0706.4289}}].

\bibitem{Cruz:2010ud}
M.~Cruz, P.~Vielva, E.~Martinez-Gonzalez and R.B.~Barreiro, \emph{{Anomalous
  variance in the WMAP data and Galactic Foreground residuals}},
  \href{https://doi.org/10.1111/j.1365-2966.2010.18067.x}{\emph{Mon. Not. Roy.
  Astron. Soc.} {\bfseries 412} (2011) 2383}
  [\href{https://arxiv.org/abs/1005.1264}{{\ttfamily 1005.1264}}].

\bibitem{Gruppuso:2013xba}
A.~Gruppuso, P.~Natoli, F.~Paci, F.~Finelli, D.~Molinari, A.~De~Rosa et~al.,
  \emph{{Low Variance at large scales of WMAP 9 year data}},
  \href{https://doi.org/10.1088/1475-7516/2013/07/047}{\emph{J. Cosmol.
  Astropart. P.} {\bfseries 1307} (2013) 047}
  [\href{https://arxiv.org/abs/1304.5493}{{\ttfamily 1304.5493}}].

\bibitem{Schwarz:2015cma}
D.J.~Schwarz, C.J.~Copi, D.~Huterer and G.D.~Starkman, \emph{{CMB Anomalies
  after Planck}},
  \href{https://doi.org/10.1088/0264-9381/33/18/184001}{\emph{Class. Quant.
  Grav.} {\bfseries 33} (2016) 184001}
  [\href{https://arxiv.org/abs/1510.07929}{{\ttfamily 1510.07929}}].

\bibitem{Natale:2019dqm}
U.~Natale, A.~Gruppuso, D.~Molinari and P.~Natoli, \emph{{Is the lack of power
  anomaly in the CMB correlated with the orientation of the Galactic plane?}},
  \href{https://doi.org/10.1088/1475-7516/2019/12/052}{\emph{J. Cosmol.
  Astropart. P.} {\bfseries 1912} (2019) 052}
  [\href{https://arxiv.org/abs/1908.10637}{{\ttfamily 1908.10637}}].

\bibitem{Dudas:2012vv}
E.~Dudas, N.~Kitazawa, S.P.~Patil and A.~Sagnotti, \emph{{CMB Imprints of a
  Pre-Inflationary Climbing Phase}},
  \href{https://doi.org/10.1088/1475-7516/2012/05/012}{\emph{J. Cosmol.
  Astropart. P.} {\bfseries 1205} (2012) 012}
  [\href{https://arxiv.org/abs/1202.6630}{{\ttfamily 1202.6630}}].

\bibitem{Kitazawa:2014dya}
N.~Kitazawa and A.~Sagnotti, \emph{{Pre-inflationary clues from String
  Theory?}}, \href{https://doi.org/10.1088/1475-7516/2014/04/017}{\emph{J.
  Cosmol. Astropart. P.} {\bfseries 1404} (2014) 017}
  [\href{https://arxiv.org/abs/1402.1418}{{\ttfamily 1402.1418}}].

\bibitem{Gruppuso:2015xqa}
A.~Gruppuso, N.~Kitazawa, N.~Mandolesi, P.~Natoli and A.~Sagnotti,
  \emph{{Pre-Inflationary Relics in the CMB?}},
  \href{https://doi.org/10.1016/j.dark.2015.12.001}{\emph{Phys. Dark Univ.}
  {\bfseries 11} (2016) 68} [\href{https://arxiv.org/abs/1508.00411}{{\ttfamily
  1508.00411}}].

\bibitem{Gruppuso:2017nap}
A.~Gruppuso, N.~Kitazawa, M.~Lattanzi, N.~Mandolesi, P.~Natoli and A.~Sagnotti,
  \emph{{The Evens and Odds of CMB Anomalies}},
  \href{https://doi.org/10.1016/j.dark.2018.03.002}{\emph{Phys. Dark Univ.}
  {\bfseries 20} (2018) 49} [\href{https://arxiv.org/abs/1712.03288}{{\ttfamily
  1712.03288}}].

\bibitem{Gruppuso:2015zia}
A.~Gruppuso and A.~Sagnotti, \emph{{Observational Hints of a Pre--Inflationary
  Scale?}}, \href{https://doi.org/10.1142/S0218271815440083}{\emph{Int. J. Mod.
  Phys.} {\bfseries D24} (2015) 1544008}
  [\href{https://arxiv.org/abs/1506.08093}{{\ttfamily 1506.08093}}].

\bibitem{Carroll:1989vb}
S.M.~Carroll, G.B.~Field and R.~Jackiw, \emph{{Limits on a Lorentz and Parity
  Violating Modification of Electrodynamics}},
  \href{https://doi.org/10.1103/PhysRevD.41.1231}{\emph{Phys. Rev. D}
  {\bfseries 41} (1990) 1231}.

\bibitem{Liu:2006uh}
G.-C.~Liu, S.~Lee and K.-W.~Ng, \emph{{Effect on cosmic microwave background
  polarization of coupling of quintessence to pseudoscalar formed from the
  electromagnetic field and its dual}},
  \href{https://doi.org/10.1103/PhysRevLett.97.161303}{\emph{Phys. Rev. Lett.}
  {\bfseries 97} (2006) 161303}
  [\href{https://arxiv.org/abs/astro-ph/0606248}{{\ttfamily
  astro-ph/0606248}}].

\bibitem{Feng:2006dp}
B.~Feng, M.~Li, J.-Q.~Xia, X.~Chen and X.~Zhang, \emph{{Searching for CPT
  Violation with Cosmic Microwave Background Data from WMAP and BOOMERANG}},
  \href{https://doi.org/10.1103/PhysRevLett.96.221302}{\emph{Phys. Rev. Lett.}
  {\bfseries 96} (2006) 221302}
  [\href{https://arxiv.org/abs/astro-ph/0601095}{{\ttfamily
  astro-ph/0601095}}].

\bibitem{Gubitosi:2009eu}
G.~Gubitosi, L.~Pagano, G.~Amelino-Camelia, A.~Melchiorri and A.~Cooray,
  \emph{{A Constraint on Planck-scale Modifications to Electrodynamics with CMB
  polarization data}},
  \href{https://doi.org/10.1088/1475-7516/2009/08/021}{\emph{J. Cosmol.
  Astropart. P.} {\bfseries 0908} (2009) 021}
  [\href{https://arxiv.org/abs/0904.3201}{{\ttfamily 0904.3201}}].

\bibitem{Gruppuso:2015xza}
A.~Gruppuso, M.~Gerbino, P.~Natoli, L.~Pagano, N.~Mandolesi, A.~Melchiorri
  et~al., \emph{{Constraints on cosmological birefringence from Planck and
  Bicep2/Keck data}},
  \href{https://doi.org/10.1088/1475-7516/2016/06/001}{\emph{J. Cosmol.
  Astropart. P.} {\bfseries 1606} (2016) 001}
  [\href{https://arxiv.org/abs/1509.04157}{{\ttfamily 1509.04157}}].

\bibitem{Gruppuso:2016nhj}
A.~Gruppuso, G.~Maggio, D.~Molinari and P.~Natoli, \emph{{A note on the
  birefringence angle estimation in CMB data analysis}},
  \href{https://doi.org/10.1088/1475-7516/2016/05/020}{\emph{J. Cosmol.
  Astropart. P.} {\bfseries 1605} (2016) 020}
  [\href{https://arxiv.org/abs/1604.05202}{{\ttfamily 1604.05202}}].

\bibitem{planck2014-a23}
{\sorthelp{Planck Collaboration IntZX}}{Planck Collaboration Int. XLIX},
  \emph{{\textit{Planck} intermediate results. XLIX. Parity-violation
  constraints from polarization data}},
  \href{https://doi.org/10.1051/0004-6361/201629018}{\emph{Astron. Astrophys.}
  {\bfseries 596} (2016) A110}
  [\href{https://arxiv.org/abs/1605.08633}{{\ttfamily 1605.08633}}].

\bibitem{Minami:2019ruj}
Y.~Minami, H.~Ochi, K.~Ichiki, N.~Katayama, E.~Komatsu and T.~Matsumura,
  \emph{{Simultaneous determination of the cosmic birefringence and
  miscalibrated polarization angles from CMB experiments}},
  \href{https://doi.org/10.1093/ptep/ptz079}{\emph{Prog. Theor. Exp. Phys.}
  {\bfseries 2019} (2019) 083E02}
  [\href{https://arxiv.org/abs/1904.12440}{{\ttfamily 1904.12440}}].

\bibitem{Minami:2020xfg}
Y.~Minami, \emph{{Determination of miscalibrated polarization angles from
  observed CMB and foreground $EB$ power spectra: Application to partial-sky
  observation}}, \href{https://doi.org/10.1093/ptep/ptaa057}{\emph{Prog. Theor.
  Exp. Phys.} {\bfseries 2020} (2020) 063E01}
  [\href{https://arxiv.org/abs/2002.03572}{{\ttfamily 2002.03572}}].

\bibitem{bezanson2017julia}
J.~Bezanson, A.~Edelman, S.~Karpinski and V.B.~Shah, \emph{Julia: A fresh
  approach to numerical computing}, {\emph{SIAM review} {\bfseries 59} (2017)
  65}.

\bibitem{stefano_mandelli_2021_4439199}
S.~Mandelli, T.~Kisner, R.~Keskitalo, A.~Zonca and G.~Puglisi,
  \emph{cmbgroundbased/cal: First official release - v1.0},  Jan., 2021.
\newblock 10.5281/zenodo.4439199.

\bibitem{2005_healpix}
K.M.~{G{\'o}rski}, E.~{Hivon}, A.J.~{Banday}, B.D.~{Wandelt}, F.K.~{Hansen},
  M.~{Reinecke} et~al., \emph{{HEALPix: A Framework for High-Resolution
  Discretization and Fast Analysis of Data Distributed on the Sphere}},
  \href{https://doi.org/10.1086/427976}{\emph{Astrophys. J.} {\bfseries 622}
  (2005) 759} [\href{https://arxiv.org/abs/astro-ph/0409513}{{\ttfamily
  astro-ph/0409513}}].

\end{thebibliography}\endgroup

\end{document}